\documentclass[aps,prb,twocolumn,superscriptaddress,longbibliography,floatfix]{revtex4-2}
\usepackage{amsmath,amssymb,graphicx,bm,xcolor,dsfont,booktabs,multirow,array,longtable}
\usepackage[colorlinks=true,linkcolor=blue,citecolor=blue,urlcolor=blue]{hyperref}
\newcommand{\Tr}{\mathrm{Tr}}
\newcommand{\ii}{\mathrm{i}}
\newcommand{\one}{\mathds{1}}
\newcommand{\Z}{\mathbb{Z}}
\newcommand{\T}{\Theta}

\makeatletter\newcommand{\SMcite}[1]{\begingroup\@fileswfalse\cite{#1}\endgroup}\makeatother
\begin{document}

\title{Kitaev, semi-Dirac, and spinon quantum spin Hall liquids on the square-hexagon-dodecagon lattice: A parent projective symmetry group classification}

\author{Harsh Raj}
\affiliation{Department of Physics, Indian Institute of Technology Madras, Chennai 600036, India}
\author{Yasir Iqbal}
\email{yiqbal@physics.iitm.ac.in}
\affiliation{Department of Physics, Indian Institute of Technology Madras, Chennai 600036, India}

\begin{abstract}
The square-hexagon-dodecagon (SHD) lattice hosts an exactly solvable Kitaev spin liquid, while further-neighbor couplings frustrate its Heisenberg magnet into classically degenerate lines where quantum fluctuations can melt the N\'eel order. To place this state among the lattice's spin liquids, we classify fermionic parton mean-field states with singlet and spin--orbit-coupled triplet amplitudes by their projective symmetry groups (PSGs). As SHD sites lack site symmetry, its $\Z_2$ PSGs are the projective classes of $p6mm$ with time reversal, from which those of every single-orbit $p6mm$ lattice follow. In the idealized Kitaev model, where lattice operations do not rotate spins, these liquids carry a single $\Z_2$ PSG descending from the $SU(2)$ Lieb-flux state, and their projections are exact eigenstates. The PSG's sixfold-rotation sign, fixed by the dodecagon flux, doubles all Majorana bands, and the mirrors exclude a chiral phase in any uniform field, so only a sublattice-staggered three-spin term drives the non-Abelian Ising phase. Beyond the Kitaev state, we find a critical semi-Dirac spin liquid, with mean-field correlations $\propto r^{-3}$ along three directions and $r^{-6}$ otherwise, and extended Dirac phases with up to $48$ fermion flavors, among them an $N_f=12$ state whose PSG forbids every singlet mass. Out-of-plane Dzyaloshinskii--Moriya couplings gap the Dirac states of its $U(1)$ class into a mean-field spinon counterpart of the Kane--Mele quantum spin Hall insulator, with spin Chern number $1$ or $3$ and spin-carrying monopoles. Chiral states imply Kalmeyer--Laughlin, semion$\times$semion, and three-fermion orders, rotation-breaking ones show $\bm{q}\to-\bm{q}$ asymmetric neutron spectra, and sign-alternating fluxes break time reversal alone. Masses and monopoles in these many-flavor phases and the energetics of frustrated SHD Heisenberg models are natural next steps.
\end{abstract}

\maketitle

\tableofcontents

\section{Introduction}
\label{sec:intro}

Quantum spin liquids are phases of magnets in which quantum fluctuations prevent long-range order down to zero temperature and the elementary excitations carry fractional quantum numbers~\cite{SavaryBalents2017,ZhouKanodaNg2017,Broholm2020}. Kitaev's honeycomb model~\cite{Kitaev2006} is an exactly solvable example. Its Majorana excitations and emergent $\Z_2$ gauge field are known without approximation, and exactly solvable variants exist on other tricoordinated lattices, among them the star lattice, where time reversal is broken spontaneously~\cite{YaoKivelson2007,DusuelSchmidtVidalZaffino2008,dOrnellasKnolle2024}, the square-octagon and related lattices~\cite{YangZhouSun2007,Kells2011,BaskaranSanthoshShankar2009}, the Kekul\'e-modulated honeycomb lattice~\cite{QuinnBhattacharjeeMoessner2015}, and hyperbolic tilings~\cite{Mosseri2025,Dusel2025}. For frustrated Heisenberg magnets no such solution is available. The most widely used approximate description represents the spin by fermionic spinons coupled to an emergent gauge field~\cite{BaskaranZouAnderson1987,BaskaranAnderson1988,AffleckZouHsuAnderson1988,Wen2002,Wen2004}, and symmetric spin liquids are then distinguished by their projective symmetry group (PSG), the projective representation under which the spinons transform~\cite{Wen2002}. PSG classifications exist for the square lattice~\cite{Wen2002}, also with spin--orbit coupling~\cite{ReutherLeeAlicea2014}; for the triangular~\cite{WangVishwanath2006,ZhengMeiQi2015,Lu2016,BieriLhuillierMessio2016}, honeycomb~\cite{LuRan2011}, kagome~\cite{LuRanLee2011,BieriMessioBernuLhuillier2015}, star~\cite{ChoyKim2009,YangParamekantiKim2010}, and ruby~\cite{MaitySamajdarIqbal2024,MaityIqbalSamajdar2026} lattices, all with space group $p6mm$; for the stuffed honeycomb~\cite{SahooFlint2020}, square-octagon, maple-leaf, trellis, and Shastry--Sutherland~\cite{Maity2023,Sonnenschein2024,Chatterjee2026,Maity2025SS} lattices; and in three dimensions for the cubic~\cite{Sonnenschein2020}, diamond~\cite{Chauhan2023}, pyrochlore~\cite{LiuHalaszBalents2019,LiuHalaszBalents2021,SchneiderHalimehPunk2022,ChernKimCastelnovo2022,LiuBalentsIqbal2026}, hyperkagome~\cite{HuangKimLu2017,JinZhou2020}, hyperhoneycomb~\cite{HuangChoiKimLu2018}, hyper-hyperkagome~\cite{ChernKim2021}, and trillium~\cite{LiBiswasParameswaran2025} lattices.

The square-hexagon-dodecagon (SHD) lattice, the Archimedean tiling $4.6.12$~\cite{TomczakRichter1999,RichterSchulenburgHonecker2004}, carries both kinds of magnet. It is tricoordinated, and its Kitaev model, studied together with the Kitaev--Heisenberg model in Ref.~\cite{Barari2026}, is exactly solvable: all faces are even, and the model realizes a gapped toric-code spin liquid throughout its anisotropic phase diagram except at a single gapless point~\cite{Barari2026}. The nearest-neighbor Heisenberg model on the same lattice, whose bonds form a bipartite graph, is N\'eel ordered with a strongly reduced moment~\cite{TomczakRichter1999,Tomczak2001,Farnell2014,Farnell2018}. Further-neighbor couplings frustrate this order. In the classical limit they produce lines of couplings on which the lowest Luttinger--Tisza band is flat and the ground state is extensively degenerate, and a wedge of couplings in which the Luttinger--Tisza bound is generically not reached and, over most of the cut we studied, the lowest states found are nonregular noncoplanar or incommensurate (Sec.~\ref{sec:classical}). Whether quantum fluctuations turn these regimes into spin liquids is open, and the competition is not restricted to spin liquids. With twelve spins per unit cell, the lattice is not subject to a Lieb--Schultz--Mattis constraint~\cite{LiebSchultzMattis1961,Oshikawa2000,Hastings2004,PoWatanabeJianZaletel2017}. Its dodecagons, its hexagons, and its squares each partition the sites, so products of singlets on any one of these sets of clusters can preserve all lattice symmetries; such cluster-singlet states are natural competitors of a spin liquid. The same tiling is the carbon network of graphenylene, predicted theoretically~\cite{Song2013Graphenylene} and reported in synthesis~\cite{Du2017}, and of graphyne sheets derived from it~\cite{CostaMiranda2025}. It is also the net of a metal--organic framework~\cite{Eddaoudi2002} and has been realized in artificial spin systems~\cite{Pac2026}, and its electronic bands host nearly flat bands and a quantum spin Hall phase~\cite{CrastoDeLima2019,Pham2026}. None of these systems is a spin-$1/2$ Mott insulator, and we are not aware of a quantum magnet with this lattice. We ask where the Kitaev spin liquid sits among the fermionic spin liquids of the lattice, which other gapless and topological mean-field states the lattice supports with frustration or spin--orbit coupling, and which of them can be stable phases beyond mean-field theory.

We work in Abrikosov-fermion mean-field theory with spin-singlet and spin-triplet amplitudes on the six inequivalent bond orbits up to fourth neighbors; the triplet amplitudes describe the effect of spin--orbit coupling~\cite{DoddsBhattacharjeeKim2013,LiuHalaszBalents2021}. The mean-field amplitudes are free parameters constrained only by the PSG. The single-occupancy constraint is imposed on average through the Lagrange multipliers, and no ansatz is optimized for a specific spin Hamiltonian. A given PSG can therefore be gapped in one region of its amplitude space and gapless in another, and the spectral types reported below refer to such regions. The PSG also fixes the allowed directions of the Dzyaloshinskii--Moriya-like spinon hopping on each bond. These directions follow Moriya's rules~\cite{Moriya1960} or their inversion, depending on the projective sign of the bond-reversing symmetry.

The classification on this lattice is controlled by the absence of site symmetry. The twelve sites of the SHD unit cell occupy the general Wyckoff position of $p6mm$, so the space group acts freely on them; time reversal fixes every site, and its on-site condition decides which projective classes occur. For the invariant gauge group (IGG) $\Z_2$ the condition can be met for both signs of $\T^2$, and the $128$ $\Z_2$ PSGs are exactly the cohomology classes of $p6mm\times\Z_2^\T$ with coefficients in $\Z_2$. For the IGG $U(1)$ the on-site condition removes the classes in which time reversal squares to $+1$ while reversing the $U(1)$ rotation, $64$ of the $208$ discrete cohomology classes, and leaves $144$ discrete PSGs and $16$ one-parameter families with a continuous flux; there are $64$ $SU(2)$ PSGs. On lattices with nontrivial site symmetry the site stabilizers must in addition act through on-site projective representations, which removes some classes and splits others. The $\Z_2$ PSGs of every lattice whose sites form a single orbit of $p6mm$ therefore follow from the SHD classes by restricting each class to the stabilizer of a site and counting its on-site realizations. For the triangular, honeycomb, kagome, star, and ruby lattices this agrees with the earlier classifications where they overlap and organizes their PSGs by SHD class. The honeycomb lattice has $112$ algebraic PSGs, which include, in addition to the PSGs of Ref.~\cite{LuRan2011}, eight PSGs with $g_\T=\ii\tau^3$, all of which admit nonzero mean-field amplitudes. On the triangular lattice the reduction identifies pairs of PSGs that share all relation values and differ in the trace of an on-site matrix.

Within this classification the Kitaev spin liquids carry a single $\Z_2$ PSG throughout their phase diagram, in all three regions that differ in which lattice excitations are labeled as the $e$ and $m$ anyons. The Kitaev model on this lattice is an idealized model: every lattice operation preserves the bond colors, so no spin rotation accompanies it, unlike in spin--orbit-coupled magnets, where Kitaev couplings arise with Ising axes tied to the bond directions~\cite{ChaloupkaJackeliKhaliullin2010}. Its PSG descends from the $SU(2)$ state with Lieb's flux pattern through bond-dependent triplet amplitudes, and its Gutzwiller-projected ansatz is an exact eigenstate up to a global spin rotation, as for the parton descriptions of the honeycomb model~\cite{BurnellNayak2011,MandalShankarBaskaran2012}. The PSG accounts for two features of the exact solution. The twofold degeneracy of all Majorana bands, which Ref.~\cite{Barari2026} traced to $(C_2\T)^2=-1$, follows from the projective sign $\eta_{C_6}=-1$ of the sixfold rotation, which is set by the $\pi$ flux through the dodecagon at the rotation center. Because the mirrors preserve the bond colors, the model in a uniform magnetic field of any direction keeps a unitary mirror symmetry, which forces the chiral central charge and the thermal Hall conductivity to vanish; a uniform field cannot produce a chiral phase unless the mirrors are broken spontaneously. A sublattice-staggered three-spin term, odd under the mirrors and under time reversal but even under their product, drives the non-Abelian Ising phase. A $[111]$ field of opposite sign on the two sublattices generates this term only for field components small compared with the two-flux gaps, and the reversal of the Chern number found at larger three-spin couplings depends on the relative weights assigned to the three types of three-spin paths.

Away from the Kitaev point, the symmetric PSGs support gapless spinon spectra in regions of their amplitude space. Nodal-line $SU(2)$ spin liquids occur, with phase boundaries known in closed form. Dirac spin liquids occur with $N_f=8$, $12$, $24$, $36$, and $48$ flavors of two-component Dirac fermions, and with $N_f=72$ and $96$ in thin regions of one $SU(2)$ class. Extended phases with $N_f=48$ occur in two $U(1)$ classes. In U011-100101 projective translations, mirrors that act as particle--hole transformations, and the antiunitary twofold rotation pin the nodes to mirror lines and bound the flavor number by $48$ for amplitudes up to fourth neighbors, with one step of the bound established numerically; in U000-011001 time reversal induces a sublattice symmetry, which protects nodes at generic momenta and imposes no such bound. These are mean-field candidates for quantum electrodynamics in $2+1$ dimensions (QED$_3$) with many flavors~\cite{Hermele2004}. Their stability is decided by the fermion bilinears: the large-$N_f$ scaling dimension of the elementary monopole~\cite{Borokhov2002,Pufu2014} exceeds $6$ for $N_f\ge24$, so monopoles are irrelevant there, whereas the symmetry-allowed bilinears of these states have not been classified. For the $N_f=12$ Dirac state of the $U(1)$ class U001-000000 the PSG forbids every spin-singlet fermion mass, and its fate depends on its monopoles, whose large-$N_f$ dimension, $3.1$, lies close to the threshold $3$ of relevance, as for the $N_f=12$ Dirac spin liquid of the maple-leaf lattice~\cite{Zhang2026}. A semi-Dirac spin liquid, in which two Dirac nodes merge as at the boundary of the gapless phase of Kitaev's honeycomb model~\cite{Kitaev2006} and in the trellis-lattice spin liquid~\cite{Chatterjee2026}, lies on a critical hypersurface of codimension one between an $N_f=12$ Dirac phase and a gapped state; no symmetry of $p6mm$ or of its projective realizations pins such a twofold touching. Its mean-field spin correlations decay as $r^{-3}$ along three lattice directions and as $r^{-6}$ along all others. Interactions can modify such anisotropic critical points~\cite{Isobe2016}, and these exponents are mean-field values.

The topological states arise with spin--orbit coupling or broken time reversal. Spin-dependent imaginary spinon hopping, the counterpart of the Kane--Mele coupling~\cite{KaneMele2005a,KaneMele2005b}, turns $U(1)$ states with $S^z$ conservation into mean-field spinon quantum spin Hall insulators along two routes. In U001-000000, the $U(1)$ class without projective signs, out-of-plane Dzyaloshinskii--Moriya couplings gap Dirac parents and give spin Chern numbers $1$ and $3$, the two-dimensional counterpart of the spinon Fu--Kane--Mele model on the diamond lattice~\cite{FuKaneMele2007,Chauhan2023}. In U001-000001 a projective sign allows an out-of-plane Dzyaloshinskii--Moriya vector on the square diagonals, which inverts bands of a gapped parent. With spin--orbit coupling that breaks $S^z$ conservation, a random survey finds $\Z_2$-odd mean-field states in at least ten $U(1)$ classes. These mean-field states are not stable spin liquids in two dimensions. With conserved $S^z$ the monopoles of the gauge field carry $S^z$ equal to the spin Chern number, and the physical state orders, with dipolar, spin-nematic, or $(S^+)^3$ order for spin Chern numbers $1$, $2$, and $3$; with generic spin--orbit coupling the compact $U(1)$ gauge field confines~\cite{Polyakov1977}, unlike its three-dimensional counterpart on the diamond lattice. Chiral spin liquids on the SHD lattice require same-sublattice bonds. No chiral PSG of $p6mm$ combines a translation with time reversal, and of the four chiral classes only the one in which the mirrors are combined with time reversal admits nonzero Chern numbers. From the mean-field Chern numbers we infer Kalmeyer--Laughlin, semion$\times$semion, and three-fermion topological orders. Free-spinon thermal Hall plateaus equal the physical ones only for $\Z_2$ states; for a Kalmeyer--Laughlin state the gauge field halves the plateau to chiral central charge $|c|=1$, and for an $SU(2)$ state with Chern number $-3$ the expected value is $c=-21/5$ rather than the free-spinon value $-6$. The chiral states that break the sixfold rotation have no Chern number and show instead an asymmetry of the dynamical structure factor under $\bm{q}\to-\bm{q}$. Because no mirror of the lattice maps an odd loop onto itself, time reversal can also be broken by triangle fluxes of alternating sign with every lattice symmetry preserved.

Gapped $U(1)$ and $SU(2)$ mean-field states without a Chern--Simons term confine in two dimensions~\cite{Polyakov1977}. Among the states discussed here, the candidates for stable spin liquids are therefore the gapped $\Z_2$ states, which include the Kitaev spin liquids, the chiral states with nonzero Chern number, and the Dirac spin liquids, subject to the conditions on bilinears and monopoles stated above; the semi-Dirac state is a critical point. Sections~\ref{sec:lattice}--\ref{sec:parent} set up the lattice, the parton theory, the classification, and its reduction to other $p6mm$ lattices. Sections~\ref{sec:ansatz}--\ref{sec:kitaev} construct the symmetric and chiral states and analyze their spectra, the effect of spin--orbit coupling, and the Kitaev spin liquids, and Sec.~\ref{sec:conclusion} discusses the results and their limitations. Derivations are in the appendices and in the Supplemental Material~\cite{SM}.
\section{Lattice, Symmetries, and Bonds}
\label{sec:lattice}

\subsection{Geometry}

The SHD lattice is the Archimedean tiling $4.6.12$, also known as the truncated trihexagonal tiling, in which every vertex is shared by one square, one hexagon, and one dodecagon~\cite{TomczakRichter1999,RichterSchulenburgHonecker2004}. Its space group is the wallpaper group $p6mm$. The dodecagon centers occupy the Wyckoff position $1a$ (site symmetry $6mm$), the hexagon centers $2b$ ($3m$), and the square centers $3c$ ($2mm$). The twelve spins of a unit cell sit at the general position $12f$, whose site-symmetry group is trivial. The tight-binding density of states of this and the other Archimedean lattices is computed in Ref.~\cite{JosephBoettcher2025}.

\begin{figure*}
\centering
\includegraphics[width=\textwidth]{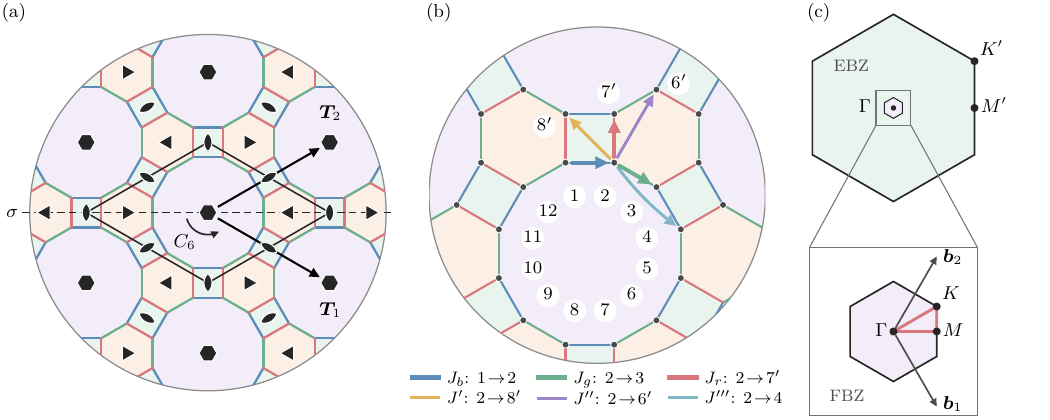}
\caption{(a) Square-hexagon-dodecagon lattice with the elements of $p6mm$: the primitive translations $\bm{T}_1$ and $\bm{T}_2$ and the unit cell (parallelogram), the sixfold rotation $C_6$ about the dodecagon center, and the mirror $\sigma:y\to-y$ (dashed line). Triangles and lenses mark the threefold and twofold rotation centers at the hexagon and square centers. Nearest-neighbor bonds are colored by orbit: square-dodecagon ($J_b$), hexagon-dodecagon ($J_g$), and square-hexagon ($J_r$) edges. (b) Reference unit cell with the sublattice labels $\mu=1,\dots,12$ and the reference bonds $i\to j$ of the six bond orbits up to fourth neighbors (Table~\ref{tab:bonds}); primed sites belong to the cell $(x,y)=(-1,1)$. (c) Extended Brillouin zone (EBZ) with the first Brillouin zone (FBZ) drawn to scale at its center, and a magnified FBZ with the reciprocal vectors $\bm{b}_{1,2}$ and the path $\Gamma\to K\to M\to\Gamma$ used for band structures. In units of $2\pi/A$, $\bm{b}_{1,2}=(1/\sqrt3,\mp1)$, $M=(1/\sqrt3,0)$, $K=(1/\sqrt3,1/3)$, $M'=9M$, and $K'=9K$; the EBZ refers to the commensurate realization of Appendix~\ref{app:bz}.}
\label{fig:lattice}
\end{figure*}

We use the ideal Archimedean geometry, with all nearest-neighbor bonds of unit length. The Bravais lattice is spanned by
\begin{equation}
\bm{T}_{1,2}=A\left(\tfrac{\sqrt3}{2},\mp\tfrac12\right),\qquad A=3+\sqrt3,
\label{eq:T}
\end{equation}
and a site is labeled by $i=(x,y,\mu)$, with $\bm{r}_i=x\bm{T}_1+y\bm{T}_2+\bm{\epsilon}_\mu$ and $\mu=1,\dots,12$. The twelve basis vectors lie on a regular dodecagon of circumradius $(\sqrt6+\sqrt2)/2$ centered at the origin, $\bm{\epsilon}_\mu$ at polar angle $105^\circ-30^\circ(\mu-1)$. With $c=(1+\sqrt3)/2$ and $d=1+\sqrt3/2$ they read
\begin{align}
\bm{\epsilon}_1&=\left(-\tfrac12,d\right), & \bm{\epsilon}_2&=\left(\tfrac12,d\right), & \bm{\epsilon}_3&=(c,c),\nonumber\\
\bm{\epsilon}_4&=\left(d,\tfrac12\right), & \bm{\epsilon}_5&=\left(d,-\tfrac12\right), & \bm{\epsilon}_6&=(c,-c),\nonumber\\
\bm{\epsilon}_7&=\left(\tfrac12,-d\right), & \bm{\epsilon}_8&=\left(-\tfrac12,-d\right), & \bm{\epsilon}_9&=(-c,-c),\nonumber\\
\bm{\epsilon}_{10}&=\left(-d,-\tfrac12\right), & \bm{\epsilon}_{11}&=\left(-d,\tfrac12\right), & \bm{\epsilon}_{12}&=(-c,c).
\label{eq:basis}
\end{align}
The geometry, the site labels, and the symmetry elements are shown in Fig.~\ref{fig:lattice}. The nearest-neighbor bonds form a bipartite graph, with sublattice $A$ formed by the odd and sublattice $B$ by the even values of $\mu$.

Gauge-invariant spectra depend on the geometry only through the connectivity of the lattice and the Bravais vectors, not through the basis vectors $\bm{\epsilon}_\mu$. Spin structure factors depend on the actual site positions, and we compute them with the ideal positions. These positions generate a dense subgroup of the plane, so that $S(\bm{q})$ is defined at every momentum but is not periodic on any finite extended zone. We display it in the extended Brillouin zone of a commensurate realization of the same tiling, constructed in Appendix~\ref{app:bz}, in which all sites lie on a triangular lattice of spacing $A/9$ and which has the same space group and the same bond graph; for the ideal positions this zone is a display window rather than a period cell. Two results are computed with the commensurate positions instead, the structure-factor asymmetry of Fig.~\ref{fig:chiral-patterns}(e) and the structure factors of the Kitaev model in Fig.~\ref{fig:kitaev-sq}.

\subsection{Symmetry Group}

The space group is generated by the two translations $T_1$, $T_2$, the rotation $C_6$ by $+60^\circ$ about the dodecagon center at the origin, and the mirror $\sigma$ that maps $y\to-y$. Their action on site labels is
\begin{align}
T_1&:(x,y,\mu)\mapsto(x+1,y,\mu),\nonumber\\
T_2&:(x,y,\mu)\mapsto(x,y+1,\mu),\nonumber\\
C_6&:(x,y,\mu)\mapsto(-y,x+y,[\mu-2]),\nonumber\\
\sigma&:(x,y,\mu)\mapsto(y,x,\mu_\sigma),
\label{eq:action}
\end{align}
where $[k]=((k-1)\bmod 12)+1$ and the mirror permutes the sublattices as $\mu_\sigma=((8-\mu)\bmod 12)+1$, that is, $1\leftrightarrow8$, $2\leftrightarrow7$, $3\leftrightarrow6$, $4\leftrightarrow5$, $9\leftrightarrow12$, and $10\leftrightarrow11$. Time reversal $\T$ acts trivially on the sites. The group $p6mm\times\Z_2^{\T}$ is defined by the thirteen relations
\begin{align}
&T_2T_1T_2^{-1}T_1^{-1}=T_1C_6T_1^{-1}T_2C_6^{-1}=T_2C_6T_1^{-1}C_6^{-1}=e,\nonumber\\
&T_2\sigma T_1^{-1}\sigma^{-1}=T_1\sigma T_2^{-1}\sigma^{-1}=e,\nonumber\\
&C_6^6=\sigma^2=(\sigma C_6)^2=\T^2=e,\nonumber\\
&O\,\T\,O^{-1}\T^{-1}=e\quad (O=T_1,T_2,C_6,\sigma).
\label{eq:relations}
\end{align}
Besides the commutation of the translations, the first two lines fix the point-group action on the translations: $C_6$ maps $\bm{T}_1\to\bm{T}_2$ and $\bm{T}_2\to\bm{T}_2-\bm{T}_1$, and $\sigma$ exchanges $\bm{T}_1$ and $\bm{T}_2$. The third line contains the point-group relations of $C_{6v}$ and of time reversal. The last line states that time reversal commutes with the four generators of $p6mm$. Since the site-symmetry group is trivial, every site is the image of the reference site $(0,0,1)$ under exactly one element of $p6mm$.

\subsection{Bonds}

\begin{table}
\caption{The six symmetry-inequivalent bond orbits up to fourth neighbors: square-dodecagon ($J_b$), hexagon-dodecagon ($J_g$), and square-hexagon ($J_r$) edges, square diagonals ($J'$), and second neighbors on a hexagon ($J''$) and on a dodecagon ($J'''$). The reference bond $i\to j$ starts in the unit cell $(0,0)$. Lengths are in units of the nearest-neighbor bond length; primed site labels belong to the unit cell at $(x,y)=(-1,1)$. The last column gives the element of $p6mm$ that maps the reference bond onto itself with its endpoints exchanged; $J''$ and $J'''$ bonds have no such element.}
\label{tab:bonds}
\begin{ruledtabular}
\begin{tabular}{lcccc}
bond & $i\to j$ & length & per cell & reversing element\\
\hline
$J_b$ & $1\to2$ & $1$ & 6 & $C_6^3\sigma$\\
$J_g$ & $2\to3$ & $1$ & 6 & $C_6^2\sigma$\\
$J_r$ & $2\to7'$ & $1$ & 6 & $T_1^{-1}T_2\sigma$\\
$J'$ & $2\to8'$ & $\sqrt2$ & 6 & $T_1^{-1}T_2C_6^3$\\
$J''$ & $2\to6'$ & $\sqrt3$ & 12 & --\\
$J'''$ & $2\to4$ & $2\cos15^\circ$ & 12 & --
\end{tabular}
\end{ruledtabular}
\end{table}

Up to fourth neighbors there are six symmetry-inequivalent bond orbits, listed in Table~\ref{tab:bonds} and drawn in Fig.~\ref{fig:lattice}(b). The three nearest-neighbor orbits $J_b$, $J_g$, and $J_r$ are the edges shared by a square and a dodecagon, a hexagon and a dodecagon, and a square and a hexagon, respectively; each is reversed by a mirror through its midpoint. The second neighbors are the diagonals $J'$ of the squares, reversed by the twofold rotation about the square center. The third and fourth neighbors, $J''$ and $J'''$, connect second neighbors on a hexagon and on a dodecagon, respectively. Each of these two orbits has twelve bonds per unit cell, as many as the order of the point group, so their bonds have trivial stabilizers. A symmetric mean-field amplitude on a $J''$ or $J'''$ bond is then constrained only by time reversal and by the projective phases discussed below. The nearest-neighbor bonds connect the two sublattices, whereas $J'$, $J''$, and $J'''$ connect sites of the same sublattice. Each further-neighbor bond closes a triangle with two nearest-neighbor bonds, $J_b$--$J_r$--$J'$, $J_g$--$J_r$--$J''$, or $J_b$--$J_g$--$J'''$, and these bonds are the sources of geometric frustration in a Heisenberg model on this lattice. The bonds $J'$, $J''$, and $J'''$ are exactly the pairs of sites at graph distance two. The next pairs in Euclidean distance are the diameters of the hexagons, six per unit cell at graph distance three, whose length $2$ exceeds the length $2\cos15^\circ\approx1.932$ of $J'''$ by only $3.5\%$; they are not included, so that the truncation at $J'''$ is a truncation in graph distance rather than in Euclidean distance. We write $J_1$ collectively for the three nearest-neighbor orbits, and $J_b=J_g=J_r=J_1$ when their amplitudes or couplings are equal.
\section{Parton Mean-Field Theory and Projective Symmetry Groups}
\label{sec:parton}

\subsection{Abrikosov Fermions and Mean-Field Ans\"atze}

We write each spin-$1/2$ operator in terms of Abrikosov fermions $f_{i\alpha}$~\cite{Abrikosov1965,BaskaranZouAnderson1987,AffleckZouHsuAnderson1988}, $\bm{S}_i=\frac12 f^\dagger_{i\alpha}\bm{\sigma}_{\alpha\beta}f_{i\beta}$, subject to the single-occupancy constraint $f^\dagger_{i\alpha}f_{i\alpha}=1$. It is convenient to collect the fermions of site $i$ into the matrix
\begin{equation}
\psi_i=\begin{pmatrix} f_{i\uparrow} & f^\dagger_{i\downarrow}\\ f_{i\downarrow} & -f^\dagger_{i\uparrow}\end{pmatrix},
\label{eq:psi}
\end{equation}
on which spin rotations act from the left and the $SU(2)$ gauge redundancy of the representation acts from the right, $\psi_i\to\psi_iW_i^\dagger$ with $W_i\in SU(2)$~\cite{AffleckZouHsuAnderson1988,DagottoFradkinMoreo1988,Wen2002}. We denote by $\sigma^a$ Pauli matrices in spin space and by $\tau^a$ Pauli matrices in gauge space, with $\sigma^0$ and $\tau^0$ the identity. The most general quadratic parton Hamiltonian that respects the gauge structure is~\cite{LiuHalaszBalents2021}
\begin{equation}
H_{\rm MF}=\sum_{i,j}\sum_{\alpha=0,x,y,z}\Tr\left[\sigma^\alpha\psi_i u^{(\alpha)}_{ij}\psi_j^\dagger\right]+\sum_i\Tr\left[\psi_i a_i\psi_i^\dagger\right],
\label{eq:HMF}
\end{equation}
where the sum runs over ordered pairs of sites. The singlet link matrix $u^{(0)}_{ij}$ and the three triplet link matrices $u^{(a)}_{ij}$ ($a=x,y,z$) take the forms
\begin{align}
u^{(0)}_{ij}&=\ii\chi^0_{ij}\tau^0+\chi^3_{ij}\tau^3+\Delta^1_{ij}\tau^1+\Delta^2_{ij}\tau^2,\nonumber\\
u^{(a)}_{ij}&=t^{a0}_{ij}\tau^0+\ii\left(t^{a1}_{ij}\tau^1+t^{a2}_{ij}\tau^2+t^{a3}_{ij}\tau^3\right),
\label{eq:links}
\end{align}
with real coefficients and $u^{(\alpha)}_{ji}=u^{(\alpha)\dagger}_{ij}$. The singlet coefficients $\chi^0$ and $\chi^3$ are the imaginary and real parts of the spin-independent hopping, $\Delta^{1,2}$ the singlet pairing amplitudes, and the triplet coefficients describe spin-dependent hopping and pairing. Appendix~\ref{app:bilinears} gives the explicit fermion bilinears. The Lagrange multipliers $a_i=a^k_i\tau^k$ ($k=1,2,3$) enforce the constraint on average. A set $\{u^{(\alpha)}_{ij},a_i\}$ is a mean-field ansatz. Under a gauge transformation it changes as
\begin{equation}
u^{(\alpha)}_{ij}\to W_i\, u^{(\alpha)}_{ij}W_j^\dagger,\qquad a_i\to W_i\, a_iW_i^\dagger,
\label{eq:gauge}
\end{equation}
and ans\"atze related by a gauge transformation describe the same physical state after projection onto the physical Hilbert space. Singlet ans\"atze describe spin-rotation-invariant states. Triplet amplitudes become allowed when spin--orbit coupling locks spin rotations to lattice operations; they describe, at mean-field level, the effect of Dzyaloshinskii--Moriya and other anisotropic exchange interactions~\cite{Dzyaloshinsky1958,Moriya1960,DoddsBhattacharjeeKim2013,LiuHalaszBalents2021}.

The invariant gauge group (IGG) of an ansatz is the group of gauge transformations that leave it invariant. It always contains $\{\pm\tau^0\}\cong\Z_2$. If the only other invariance is a $U(1)$ subgroup, which can be chosen as $\{e^{\ii\theta\tau^3}\}$ on every site, the singlet links lie in the span of $\ii\tau^0$ and $\tau^3$, the triplet links in the span of $\tau^0$ and $\ii\tau^3$, and the multipliers along $\tau^3$. If the IGG is the full $SU(2)$, the links are proportional to $\ii\tau^0$ (singlet) and $\tau^0$ (triplet) and all multipliers vanish. The IGG is the gauge group that remains unbroken at mean-field level and determines the low-energy gauge theory: $\Z_2$, $U(1)$, or $SU(2)$~\cite{Wen2002}.

\subsection{Projective Symmetry}

A physical symmetry need not leave an ansatz invariant; it suffices that it does so up to a gauge transformation. For a lattice operation $O$ with spin rotation $R_O\in SO(3)$, the ansatz is symmetric if there exists a gauge transformation $G_O(i)$ such that
\begin{align}
u^{(0)}_{O(i)O(j)}&=G_O(O(i))\,u^{(0)}_{ij}\,G_O^\dagger(O(j)),\nonumber\\
u^{(b)}_{O(i)O(j)}&=\sum_a\left[R_O\right]_{ba}G_O(O(i))\,u^{(a)}_{ij}\,G_O^\dagger(O(j)),
\label{eq:psg-spatial}
\end{align}
and similarly $a_{O(i)}=G_O(O(i))a_iG_O^\dagger(O(i))$. Time reversal, implemented in the gauge-transformed form of Ref.~\cite{Wen2002}, reverses the sign of every link and multiplier,
\begin{equation}
u^{(\alpha)}_{ij}=-G_\T(i)\,u^{(\alpha)}_{ij}\,G_\T^\dagger(j),\qquad a_i=-G_\T(i)\,a_i\,G_\T^\dagger(i),
\label{eq:psg-T}
\end{equation}
for the singlet and the triplet channels alike~\cite{LiuHalaszBalents2021}; the relative sign between hopping and spin is already contained in the forms~\eqref{eq:links}. In a spin--orbit-coupled system the spin rotation accompanying a lattice operation is the rotation part of that operation, acting on the spin as an axial vector. For the generators of $p6mm$ this gives $R_{T_1}=R_{T_2}=\one$, $R_{C_6}$ equal to the rotation by $+60^\circ$ about $\hat{z}$, and $R_\sigma=\mathrm{diag}(-1,1,-1)$ for the mirror $y\to-y$. The spin of an electron, and of a spinon, transforms under the double cover of these rotations, in which $C_6^6=\sigma^2=(\sigma C_6)^2=-1$. Because the triplet amplitudes transform with $R_O\in SO(3)$, these signs do not enter Eq.~\eqref{eq:psg-spatial}, and the classification of Sec.~\ref{sec:algebraic} is the same with and without spin--orbit coupling. With spin--orbit coupling the projective quantum numbers of a spinon under $C_6^6$, $\sigma^2$, and $(\sigma C_6)^2$ are the corresponding relation values of the PSG multiplied by $-1$. For time reversal, the choice $G_\T=\ii\tau^2$ in Eq.~\eqref{eq:psg-T} is the time reversal of the Abrikosov fermions, $f_\uparrow\to f_\downarrow$, $f_\downarrow\to-f_\uparrow$, without an additional gauge transformation, so that for a $\Z_2$ PSG the time reversal of a spinon squares to the relation value $\eta_\T$ of $\T^2$.

The combined operations $G_OO$ form the PSG of the ansatz~\cite{Wen2002}. Its elements compose as
\begin{equation}
G_{O_1O_2}(i)=G_{O_1}(i)\,G_{O_2}\big(O_1^{-1}(i)\big),
\label{eq:compose}
\end{equation}
so that every relation~\eqref{eq:relations} of the symmetry group becomes, in the PSG, an element of the IGG. Gauge-inequivalent solutions of these conditions are the algebraic PSGs. Each realizable algebraic PSG labels a distinct spin liquid at mean-field level. Two solutions are equivalent if they are related by a gauge transformation $G_O(i)\to W_iG_O(i)W^\dagger_{O^{-1}(i)}$ that preserves the IGG, or by multiplying a generator by an element of the IGG (a re-lifting).

\subsection{Gauge-Invariant Fluxes}
\label{sec:flux}

The gauge-invariant content of an ansatz is carried by the flux operators $P_C(i)=u_{ij}u_{jk}\cdots u_{li}$ of closed loops $C$ based at a site $i$~\cite{Wen2002}, here built from the singlet links. Two ans\"atze with the same bond moduli and on-site terms are gauge equivalent if and only if their flux operators for a generating set of loops, all based at a common site, are simultaneously conjugate. For a $\Z_2$ ansatz this requires the traces of single loops and of products of two and three loops; for $U(1)$ and $SU(2)$ ans\"atze, whose flux operators commute, the single-loop fluxes suffice. A loop of length $L$ gives $P_C=\ii^LQ_C$ with $Q_C\propto\cos\phi_C\,\tau^0+\ii\sin\phi_C\,\hat{\bm{n}}_C\cdot\bm{\tau}$. The IGG is the subgroup of $SU(2)$ that commutes with all based flux operators and with the on-site terms transported to the base site: $SU(2)$ if all $P_C\propto\tau^0$, $U(1)$ if the directions $\hat{\bm{n}}_C$ are collinear, and $\Z_2$ otherwise.

\begin{figure}
\centering
\includegraphics[width=\columnwidth]{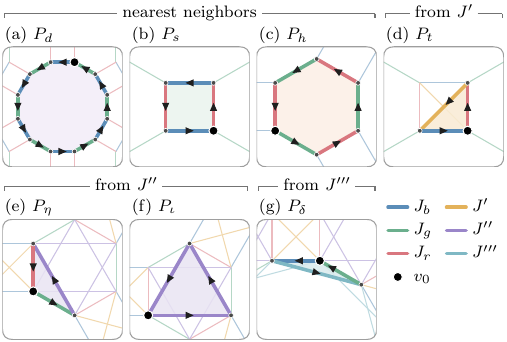}
\caption{Minimal set of gauge-invariant flux loops on the SHD lattice, one representative per symmetry orbit, all based at the reference site $v_0=(0,0,2)$ (black dot). The arrows give the order of the links in the flux operator $P_C(v_0)=u_{v_0j}u_{jk}\cdots u_{lv_0}$. Loop bonds are drawn bold and colored by type; the faint background shows the bond types present up to the shortest range at which the loop is needed, indicated above each panel. (a) The dodecagon $P_d$ ($J_b$, $J_g$), (b) square $P_s$ ($J_b$, $J_r$), and (c) hexagon $P_h$ ($J_g$, $J_r$) generate all loops for nearest-neighbor bonds. Every further-neighbor bond joins two sites of the same sublattice and closes a triangle: (d) the square triangle $P_t$ ($J_b$, $J_r$, $J'$), (e) the hexagon triangle $P_\eta$ ($J_g$, $J_r$, $J''$) and (f) the inner triangle $P_\iota$ of three $J''$ bonds, and (g) the dodecagon triangle $P_\delta$ ($J_b$, $J_g$, $J'''$). Three members of each triangle orbit in (d), (e), and (g) pass through $v_0$, one with $v_0$ at each corner; the two not drawn are the images of the drawn triangle under the $p6mm$ operations that map its other corners to $v_0$. $P_\iota$ is invariant under the threefold rotation about the hexagon center and is the only member of its orbit through $v_0$. Once $J'\neq0$, the square is the product of two square triangles through $v_0$, and once $J''\neq0$, the hexagon is generated by $P_\iota$ and the hexagon triangles. Panel (a) is drawn at half the scale of panels (b)--(g).}
\label{fig:loops}
\end{figure}

The number of independent loops per unit cell is $6$ for nearest-neighbor bonds, $12$ up to $J'$, $24$ up to $J''$, and $36$ up to $J'''$. A generating set of loop orbits, shown in Fig.~\ref{fig:loops}, consists of the square ($J_b,J_r$), hexagon ($J_g,J_r$), and dodecagon ($J_b,J_g$) for nearest-neighbor bonds. The further-neighbor orbits add triangles (Sec.~\ref{sec:lattice}): the square triangle ($J_b,J_r,J'$), the hexagon triangle ($J_g,J_r,J''$) together with the inner triangle of three $J''$ bonds, and the dodecagon triangle ($J_b,J_g,J'''$). Since all sites are related by symmetry, it suffices to consider loops through the reference site $(0,0,2)$. For $SU(2)$ and $U(1)$ ans\"atze the faces of a unit cell satisfy
\begin{equation}
\Phi_d+2\Phi_h+3\Phi_s=\Phi\pmod{2\pi},
\label{eq:fluxsum}
\end{equation}
where $\Phi_d$, $\Phi_h$, and $\Phi_s$ are the fluxes through a dodecagon, a hexagon, and a square, and $\Phi$ is the $U(1)$ flux per unit cell fixed by the PSG, with $\Phi=0$ or $\pi$ for $\eta_{12}=\pm1$ outside the sector $(0,1,0)$ and the continuous flux of Eq.~\eqref{eq:u1flux} in that sector.

Time reversal maps $P_C(i)\to(-1)^LG_\T(i)P_C(i)G^\dagger_\T(i)$. In a time-reversal-symmetric ansatz the flux operators of odd loops are therefore traceless, $\cos\phi_C=0$, i.e., $\phi_C=\pm\pi/2$ (a physical flux $0$ or $\pi$), while even loops are constrained only through $G_\T$. For an $SU(2)$ ansatz every flux operator is proportional to $\tau^0$. An odd loop then gives $P_C\propto\ii^L\tau^0$ (a physical flux $\pm\pi/2$). Such a flux changes sign under time reversal, so an $SU(2)$ ansatz with a nonzero odd loop breaks time reversal. Real hopping with flux $0$ or $\pi$ through a triangle is time-reversal symmetric, but its flux operators are proportional to $\tau^3$, which reduces the gauge structure to $U(1)$. Since every further-neighbor bond closes a triangle (Sec.~\ref{sec:lattice}), a time-reversal-symmetric $SU(2)$ state on the SHD lattice can contain further-neighbor bonds only in combinations that close no odd loop. Section~\ref{sec:ansatz} identifies the PSGs that allow this.

\subsection{Chiral Projective Symmetry Groups}

A chiral spin liquid breaks time reversal while every element of the space group is a symmetry either by itself or in combination with $\T$~\cite{KalmeyerLaughlin1987,WenWilczekZee1989,BieriLhuillierMessio2016}. The elements realized without time reversal form a subgroup $H\subseteq p6mm$ of index one or two, and the symmetry group of the state is the magnetic group $H\cup(p6mm\setminus H)\,\T$. Such subgroups are kernels of homomorphisms $\chi:p6mm\to\Z_2$. Applying $\chi$ to the relations $T_2C_6T_1^{-1}C_6^{-1}=e$ and $T_1C_6T_1^{-1}T_2C_6^{-1}=e$ gives $\chi(T_1)=\chi(T_2)$ and $\chi(T_2)=1$: both translations are commutators in $p6mm$. Neither translation can therefore be combined with time reversal, so no chiral spin liquid on a $p6mm$ lattice, the SHD lattice included, has a translation that is a symmetry only in combination with $\T$ ($\tau_t=0$ in the notation of Ref.~\cite{BieriLhuillierMessio2016}). This statement concerns translations only; it does not exclude scalar-chirality patterns that alternate in sign within the unit cell, which occur in classes A, C, and D (Sec.~\ref{sec:chiral}). The characters are fixed by their values on $C_6$ and $\sigma$, which leaves four chiral classes:
\begin{center}
\begin{tabular}{cccc}
class & $(\chi(C_6),\chi(\sigma))$ & broken & preserved with $\T$\\\hline
A & $(+,+)$ & $\T$ & --\\
B & $(+,-)$ & $\sigma$, $\T$ & $\sigma\T$\\
C & $(-,+)$ & $C_6$, $\T$ & $C_6\T$\\
D & $(-,-)$ & $C_6$, $\sigma$, $\T$ & $C_6\T$, $\sigma\T$
\end{tabular}
\end{center}
Class B is the analog of the Kalmeyer--Laughlin state, in which mirrors are combined with time reversal. Each magnetic group above is isomorphic to $p6mm$ as an abstract group, so the algebraic PSGs of each chiral class are in one-to-one correspondence with the PSGs of the space group alone. The character $\chi$ enters only through the sign of Eq.~\eqref{eq:psg-T}, which an antiunitary element $O\T$ imposes on the ansatz, $u^{(\alpha)}_{O(i)O(j)}=-G_{O\T}(O(i))\,[R_Ou^{(\cdot)}_{ij}]^{(\alpha)}\,G^\dagger_{O\T}(O(j))$~\cite{BieriLhuillierMessio2016} [Eq.~(29) there, extended here to triplet links].
\section{Algebraic Projective Symmetry Groups}
\label{sec:algebraic}

The derivation of the $\Z_2$ PSGs is summarized in Appendix~\ref{app:psg}; the complete derivations, including those of the $SU(2)$, $U(1)$, and space-group PSGs, are given in Secs.~SIII and~SIV of the Supplemental Material~\cite{SM}. Because $p6mm$ acts freely on the sites, a gauge choice brings the sublattice parts of all gauge transformations to a common form. The classification then reduces to the possible values of the thirteen relations~\eqref{eq:relations} in the IGG, modulo re-liftings. Time reversal, however, fixes every site: the stabilizer of a site in $p6mm\times\Z_2^\T$ is $\Z_2^\T$, and the time-reversal matrix $G_\T(i)$ must satisfy the on-site condition imposed by $\T^2$. For the IGG $\Z_2$ this condition, $G_\T(i)^2=\eta_\T$, can be met for both signs, and the algebraic PSGs are the classes of $H^2(p6mm\times\Z_2^\T,\Z_2)$~\cite{Wen2002,EssinHermele2013}. For a $U(1)$ IGG the on-site condition removes some classes of the corresponding cohomology group (Sec.~\ref{sec:alg-u1}).

\subsection{$\Z_2$ Projective Symmetry Groups}
\label{sec:alg-z2}

\begin{table}
\caption{Sublattice signs of the $\Z_2$ projective symmetry groups, Eq.~\eqref{eq:z2psg}. The same signs define the $SU(2)$ and $U(1)$ classes.}
\label{tab:z2}
\begin{ruledtabular}
\begin{tabular}{cccc}
$\mu$ & $g_{C_6}(\mu)$ & $g_\sigma(\mu)$ & $h(\mu)$\\
\hline
1 & $+1$ & $+1$ & $+1$\\
2 & $+1$ & $\eta_{C_6}\eta_{\sigma C_6}$ & $\eta_{\T\sigma}\eta_{\T C_6}$\\
3 & $+1$ & $\eta_{C_6}\eta_{\sigma C_6}\eta_\sigma$ & $\eta_{\T C_6}$\\
4 & $+1$ & $\eta_{C_6}\eta_\sigma$ & $\eta_{\T\sigma}$\\
5 & $+1$ & $\eta_{C_6}$ & $+1$\\
6 & $\eta_{C_6}$ & $\eta_{C_6}\eta_{\sigma C_6}$ & $\eta_{\T\sigma}\eta_{\T C_6}$\\
7 & $+1$ & $\eta_{C_6}\eta_{\sigma C_6}\eta_\sigma$ & $\eta_{\T C_6}$\\
8 & $+1$ & $\eta_\sigma$ & $\eta_{\T\sigma}$\\
9 & $+1$ & $\eta_{C_6}$ & $+1$\\
10 & $+1$ & $\eta_{C_6}\eta_{\sigma C_6}$ & $\eta_{\T\sigma}\eta_{\T C_6}$\\
11 & $\eta_{C_6}$ & $\eta_{C_6}\eta_{\sigma C_6}\eta_\sigma$ & $\eta_{\T C_6}$\\
12 & $+1$ & $\eta_{C_6}\eta_\sigma$ & $\eta_{\T\sigma}$
\end{tabular}
\end{ruledtabular}
\end{table}

In a suitable gauge every $\Z_2$ PSG on the SHD lattice takes the form
\begin{align}
G_{T_1}(x,y,\mu)&=\tau^0,\nonumber\\
G_{T_2}(x,y,\mu)&=\eta_{12}^{x}\,\tau^0,\nonumber\\
G_{C_6}(x,y,\mu)&=\eta_{12}^{\,x(x+1)/2+xy+y}\,g_{C_6}(\mu)\,\tau^0,\nonumber\\
G_\sigma(x,y,\mu)&=\eta_{12}^{\,xy}\,g_\sigma(\mu)\,\tau^0,\nonumber\\
G_\T(x,y,\mu)&=h(\mu)\,g_\T,
\label{eq:z2psg}
\end{align}
with $g_\T=\tau^0$ or $\ii\tau^2$ and the sublattice signs $g_{C_6}(\mu)$, $g_\sigma(\mu)$, and $h(\mu)$ given in Table~\ref{tab:z2}. The forms depend on seven signs,
\begin{equation}
\eta_{12},\quad \eta_{C_6},\quad \eta_\sigma,\quad \eta_{\sigma C_6},\quad \eta_{\T C_6},\quad \eta_{\T\sigma},\quad \eta_{\T},
\label{eq:z2labels}
\end{equation}
which are the values in the IGG of the relations $T_2T_1T_2^{-1}T_1^{-1}$, $C_6^6$, $\sigma^2$, $(\sigma C_6)^2$, $C_6\T C_6^{-1}\T^{-1}$, $\sigma\T\sigma^{-1}\T^{-1}$, and $\T^2$, with $\eta_\T=+1$ for $g_\T=\tau^0$ and $\eta_\T=-1$ for $g_\T=\ii\tau^2$. All seven are invariant under gauge transformations, which act on relation values by conjugation and so leave $\pm\tau^0$ unchanged, and under re-liftings, since each of these relations contains every generator an even number of times (Appendix~\ref{app:psg-conv}). They are independent, since Eqs.~\eqref{eq:z2psg} satisfy all thirteen relations for every choice of them. There are therefore
\begin{equation}
2^7=128
\label{eq:n128}
\end{equation}
gauge-inequivalent $\Z_2$ PSGs. The remaining relations take forced values, $\eta_{T_1\T}=\eta_{T_2\T}=+1$, $\eta_{T_1\sigma}=\eta_{T_2\sigma}$, and $\eta_{T_2C_6}\eta_{T_1\sigma}=\eta_{12}$, and two of them can be set to $+1$ by re-lifting the translations. The count agrees with the group-cohomology result $H^2(p6mm\times\Z_2^\T,\Z_2)=\Z_2^7$ (Appendix~\ref{app:psg}), as it must for the IGG $\Z_2$. Its seven generators are three classes from the point group $C_{6v}$ ($\eta_{C_6}$, $\eta_\sigma$, $\eta_{\sigma C_6}$), one from the translations ($\eta_{12}$, a $\pi$ flux per unit cell), two from characters of $p6mm$ that time reversal may carry ($\eta_{\T C_6}$, $\eta_{\T\sigma}$), and one from $\T^2$.

We label a $\Z_2$ PSG as $\mathrm{Z}n_{T_2}n_{\T\sigma}n_{\T C_6}n_\sigma n_{\sigma C_6}n_{C_6}(t)$, where each digit is $0$ for $+1$ and $1$ for $-1$, $n_{T_2}=\eta_{12}$, $n_{\T\sigma}=\eta_{\T\sigma}$, $n_{\T C_6}=\eta_{\T C_6}$, $n_\sigma=\eta_\sigma$, $n_{\sigma C_6}=\eta_{\sigma C_6}$, $n_{C_6}=\eta_{C_6}$, and $t=0$ ($1$) for $g_\T=\tau^0$ ($\ii\tau^2$). An additional sign $n_{C_6T_1}\equiv\eta_{T_1C_6}$, which multiplies $G_{C_6}$ by $n_{C_6T_1}^x$, is removed by the gauge transformation $W(x,y,\mu)=(-1)^{x+y}\tau^0$ together with re-liftings of the translations and carries no physical information (Step~8 of Appendix~\ref{app:psg}).

\subsection{$SU(2)$ Projective Symmetry Groups}
\label{sec:alg-su2}

For an $SU(2)$ ansatz in the canonical form $u^{(0)}_{ij}=\ii\chi_{ij}\tau^0$ the IGG consists of uniform $SU(2)$ rotations. Every PSG element can then be written as $G_O(i)=\epsilon_O(i)\,g_O$ with a sign $\epsilon_O(i)$ and a uniform $g_O\in SU(2)$ that is absorbed into the IGG. The sign structures are those of Eq.~\eqref{eq:z2psg} with Table~\ref{tab:z2}, but $\T^2$ is no longer an invariant because the IGG contains every candidate $g_\T$. This leaves $2^6=64$ $SU(2)$ PSGs, labeled $\mathrm{SU}n_{T_2}n_{\T\sigma}n_{\T C_6}n_\sigma n_{\sigma C_6}n_{C_6}$. Through Eq.~\eqref{eq:psg-T}, each of the four choices of $(\eta_{\T C_6},\eta_{\T\sigma})$ selects which bond orbits can carry a time-reversal-invariant $SU(2)$ ansatz (Sec.~\ref{sec:ansatz}).

\subsection{$U(1)$ Projective Symmetry Groups}
\label{sec:alg-u1}

With the IGG fixed to $\{e^{\ii\theta\tau^3}\}$ on all sites, every PSG element normalizes it, $G_O(i)=e^{\ii\phi_O(i)\tau^3}(\ii\tau^1)^{w_O}$, with $w_O=1$ when the symmetry reverses the sense of the $U(1)$ rotation. The integers $w_O$ are gauge invariant: conjugation by a PSG element is an automorphism of the IGG, and no gauge transformation that preserves the IGG changes this automorphism. The relations force $w_{T_1}=w_{T_2}=0$, which leaves eight sectors $(w_{C_6},w_\sigma,w_\T)$. Within each sector every PSG can be brought to the form
\begin{equation}
G_O(x,y,\mu)=G^{\Z_2}_O(x,y,\mu)\,(\ii\tau^1)^{w_O},
\label{eq:u1psg}
\end{equation}
with $G^{\Z_2}_O$ the sign forms of Eq.~\eqref{eq:z2psg} at $g_\T=\tau^0$, except in the sector $(0,1,0)$, which admits a continuous flux $\Phi$ through the unit cell:
\begin{align}
G_{T_2}(x,y,\mu)&=e^{\ii\Phi x\tau^3},\nonumber\\
G_{C_6}(x,y,\mu)&=e^{\ii\Phi[x(x+1)/2+xy+y]\tau^3}\,g_{C_6}(\mu),\nonumber\\
G_\sigma(x,y,\mu)&=e^{\ii\Phi xy\tau^3}\,g_\sigma(\mu)\,\ii\tau^1.
\label{eq:u1flux}
\end{align}
At $\Phi=0$ and $\pi$ these reduce to the forms~\eqref{eq:u1psg} with $\eta_{12}=\pm1$. Re-lifting by arbitrary $U(1)$ elements identifies some of the $\Z_2$ labels; in the sector $(0,0,0)$, for instance, only the product $\eta_\sigma\eta_{\sigma C_6}\eta_{C_6}$ survives (Sec.~SIV\,E of the Supplemental Material~\cite{SM}). The resulting independent labels and the number of classes in each sector are listed in Table~\ref{tab:u1}. In total there are $144$ $U(1)$ PSGs with discrete labels and $16$ one-parameter families. The rotation $C_6$ acts on a unit-cell flux as $\Phi\to(-1)^{w_{C_6}}\Phi$, the mirror reverses orientation and acts as $\Phi\to-(-1)^{w_\sigma}\Phi$, and time reversal acts as $\Phi\to(-1)^{w_\T}\Phi$. Of the eight sectors, only $(w_{C_6},w_\sigma,w_\T)=(0,1,0)$ has all three operations leaving a generic $\Phi$ invariant. Conjugation by the uniform gauge transformation $\ii\tau^1$ maps $\Phi\to-\Phi$, so the families are parametrized by $\Phi\in[0,\pi]$.

The relation values of a $U(1)$ PSG define a class of $H^2(p6mm\times\Z_2^\T,U(1))$, with the symmetry group acting on $U(1)$ through the integers $w_O$, but the $U(1)$ PSGs are not all of these classes. Because $\T$ fixes every site, $G_\T(i)$ is an on-site matrix, and for $w_\T=1$ it squares to $(e^{\ii\phi\tau^3}\ii\tau^1)^2=-\tau^0$ for every $\phi$. The classes with $\T^2=+\tau^0$ in the four sectors with $w_\T=1$ therefore have no representative on the SHD lattice [identity (k) in Sec.~SIV\,E of the Supplemental Material~\cite{SM}]. The $U(1)$ PSGs are the classes of $H^2$ that satisfy this on-site condition, with $\Phi$ and $-\Phi$ identified by the global gauge transformation $\ii\tau^1$. A direct computation of the cohomology group, carried out on a set of sites on which all of $p6mm\times\Z_2^\T$, time reversal included, acts freely, gives $16$, $32$, $16\times U(1)$, $32$, $32$, $32$, $32$, and $32$ classes in the sectors of Table~\ref{tab:u1}, that is, $208$ discrete classes and $16$ one-parameter families. The on-site condition removes half of the classes in each sector with $w_\T=1$ and leaves the $144$ discrete PSGs and $16$ families found above.

\begin{table}
\caption{$U(1)$ projective symmetry groups. Each sector is specified by $(w_{C_6},w_\sigma,w_\T)$; the labels listed are independent, and the others are fixed to $+1$ by re-liftings. The PSGs take the form~\eqref{eq:u1psg}, and in the sector $(0,1,0)$ the form~\eqref{eq:u1flux} with a continuous flux $\Phi$ replacing $\eta_{12}$.}
\label{tab:u1}
\begin{ruledtabular}
\begin{tabular}{ccc}
$(w_{C_6},w_\sigma,w_\T)$ & independent labels & number of PSGs\\
\hline
$(0,0,0)$ & $\eta_{12},\eta_{\T\sigma},\eta_{\T C_6},\eta_{C_6}$ & 16\\
$(0,0,1)$ & $\eta_{12},\eta_\sigma,\eta_{\sigma C_6},\eta_{C_6}$ & 16\\
$(0,1,0)$ & $\Phi,\eta_{\T\sigma},\eta_{\T C_6},\eta_\sigma,\eta_{\sigma C_6}$ & $16\times U(1)$\\
$(0,1,1)$ & $\eta_{12},\eta_\sigma,\eta_{\sigma C_6},\eta_{C_6}$ & 16\\
$(1,0,0)$ & $\eta_{12},\eta_{\T\sigma},\eta_{\T C_6},\eta_{\sigma C_6},\eta_{C_6}$ & 32\\
$(1,0,1)$ & $\eta_{12},\eta_\sigma,\eta_{\sigma C_6},\eta_{C_6}$ & 16\\
$(1,1,0)$ & $\eta_{12},\eta_{\T\sigma},\eta_{\T C_6},\eta_\sigma,\eta_{C_6}$ & 32\\
$(1,1,1)$ & $\eta_{12},\eta_\sigma,\eta_{\sigma C_6},\eta_{C_6}$ & 16
\end{tabular}
\end{ruledtabular}
\end{table}

\subsection{Projective Symmetry Groups of the Space Group}
\label{sec:alg-sg}

Removing time reversal from the analysis above gives $2^4=16$ $\Z_2$ PSGs labeled by $(\eta_{12},\eta_\sigma,\eta_{\sigma C_6},\eta_{C_6})$ and $16$ $SU(2)$ PSGs. The $U(1)$ PSGs fall into four sectors $(w_{C_6},w_\sigma)$: four in $(0,0)$, with labels $\eta_{12},\eta_{C_6}$; four one-parameter families in $(0,1)$, with labels $\eta_\sigma,\eta_{\sigma C_6}$ and flux $\Phi$; eight in $(1,0)$, with labels $\eta_{12},\eta_{\sigma C_6},\eta_{C_6}$; and eight in $(1,1)$, with labels $\eta_{12},\eta_\sigma,\eta_{C_6}$. Each of these, combined with one of the four characters of Sec.~\ref{sec:parton}, defines a chiral PSG.

Table~\ref{tab:counts} summarizes the classification. The counts agree with a solution of the PSG equations on finite lattice patches without gauge fixing.

\begin{table}
\caption{Number of algebraic projective symmetry groups on the SHD lattice. $U(1)$ entries give discrete classes plus one-parameter families with a continuous flux.}
\label{tab:counts}
\begin{ruledtabular}
\begin{tabular}{lccc}
symmetry group & $\Z_2$ & $U(1)$ & $SU(2)$\\
\hline
$p6mm\times\Z_2^\T$ & 128 & $144+16\times U(1)$ & 64\\
$p6mm$ & 16 & $20+4\times U(1)$ & 16
\end{tabular}
\end{ruledtabular}
\end{table}
\section{The SHD Lattice as the Parent of the $p6mm$ Classifications}
\label{sec:parent}

\begin{figure}
\centering
\includegraphics[width=\columnwidth]{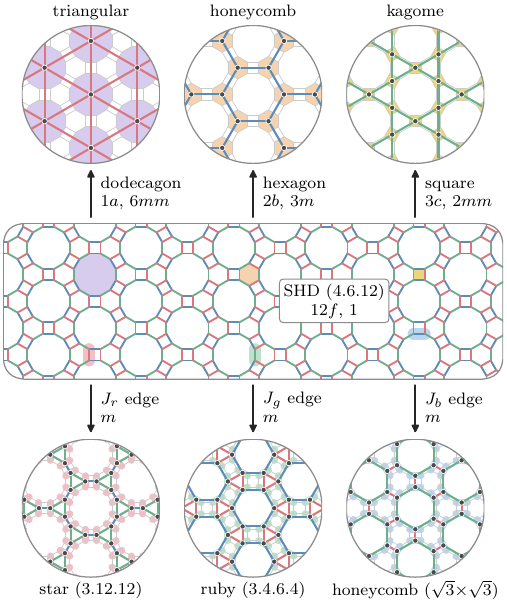}
\caption{The SHD lattice as the parent of the $p6mm$ spin lattices. Center: a patch of the SHD ($4.6.12$) lattice, whose spins occupy the general Wyckoff position $12f$ of $p6mm$ (trivial site symmetry $1$); bonds are colored by type, $J_b$ (square-dodecagon edge, blue), $J_g$ (hexagon-dodecagon edge, green), and $J_r$ (square-hexagon edge, red). Each outer panel shows the lattice obtained by shrinking every cluster of one kind to its centroid while keeping the space group: dodecagons give the triangular lattice ($1a$, site symmetry $6mm$), hexagons the honeycomb lattice ($2b$, $3m$), squares the kagome lattice ($3c$, $2mm$), and $J_r$, $J_g$, and $J_b$ edges give the star ($3.12.12$), ruby ($3.4.6.4$), and a honeycomb lattice with a $\sqrt{3}\times\sqrt{3}$ enlarged cell, respectively, all with six sites per unit cell on mirror lines (site symmetry $m$). The arrows are labeled by the collapsed cluster and by the Wyckoff position (for faces) and site-symmetry group of the resulting sites. One cluster of each kind is shaded in the central patch with the color used for all collapsed clusters (shaded faces or edges) in the corresponding outer panel. In the outer panels the parent SHD lattice is drawn in light gray, dots mark the new sites, and two new sites are joined whenever an SHD bond joined their clusters; each new bond carries the color of that SHD bond. All panels are drawn to the same scale on the same region of the plane, centered on a dodecagon, so that the derived lattices keep the bond lengths inherited from the SHD lattice with unit edges.}
\label{fig:parent}
\end{figure}

The triangular, honeycomb, kagome, star, and ruby lattices share the space group $p6mm$ with the SHD lattice and differ from it only in the Wyckoff position of the spins: $1a$ for the triangular lattice, $2b$ for the honeycomb lattice, $3c$ for the kagome lattice, and the position $6e$ on mirror lines, with site symmetry $m$, for the star (3.12.12) and ruby (3.4.6.4) lattices. Each of these lattices is obtained from the SHD lattice by shrinking a cluster of sites to a point while keeping the space group (Fig.~\ref{fig:parent}). The dodecagons shrink to the sites of the triangular lattice, the hexagons to those of the honeycomb lattice, and the squares to those of the kagome lattice, while shrinking the $J_r$, $J_g$, or $J_b$ edges gives the star lattice, the ruby lattice, or a honeycomb lattice with a $\sqrt3\times\sqrt3$ enlarged cell, respectively.

The contraction also relates the projective symmetry groups. Consider a lattice whose sites form a single orbit $p6mm/S$ with site-symmetry group $S$. An algebraic PSG with IGG $\Z_2$ on this lattice carries two pieces of data. Its relation values define a class $\eta\in H^2(p6mm\times\Z_2^\T,\Z_2)$. Its restriction to the elements that fix a site defines an on-site representation of $\tilde{S}$, the extension of $S\times\Z_2^\T$ by the IGG. This is a linear representation of $\tilde{S}$ by $SU(2)$ matrices that maps the nontrivial element of the IGG to $-\tau^0$. It is defined up to conjugation in $SU(2)$ and up to twisting by characters of $p6mm\times\Z_2^\T$, which are re-liftings. Conversely, a class $\eta$ together with such an on-site representation determines the PSG up to gauge equivalence (Appendix~\ref{app:parent}). The number of algebraic PSGs on the lattice is therefore
\begin{equation}
N_S=\sum_{\eta}\, n_S(\eta),
\label{eq:wyckoff}
\end{equation}
where $n_S(\eta)$ counts the inequivalent on-site realizations of the restriction of $\eta$ to $\tilde{S}$, which may be zero.

Equation~\eqref{eq:wyckoff} is the standard induction argument for representations of a group induced from a subgroup~\cite{Mackey1952,Brown1982}, applied to the extension of $p6mm\times\Z_2^\T$ defined by $\eta$ and to the stabilizer of a site. It holds for every lattice whose sites form a single orbit of $p6mm$, and for the IGG $\Z_2$. Lattices with several site orbits require one on-site representation per orbit, and for a $U(1)$ IGG the on-site condition removes classes already on the SHD lattice (Sec.~\ref{sec:alg-u1}). The classes $\eta$ are the mean-field counterparts of the symmetry-fractionalization classes of gapped $\Z_2$ spin liquids~\cite{EssinHermele2013,QiCheng2018}, and the on-site representations express how the Wyckoff position of the spins constrains them, a dependence that also underlies lattice-homotopy constraints on quantum magnets~\cite{PoWatanabeJianZaletel2017}.

For the SHD lattice $S$ is trivial: the stabilizer of a site in $p6mm\times\Z_2^\T$ is $\Z_2^\T$, and $\tilde{S}$ is generated by $\T$ and the IGG. Each class then has exactly one realization, $n_S(\eta)=1$, since the on-site condition $G_\T(i)^2=\eta_\T$ can be met for both signs, and the $\Z_2$ classification of the SHD lattice is $H^2(p6mm\times\Z_2^\T,\Z_2)$ itself. The same holds for any lattice whose sites occupy the general position $12f$ of $p6mm$; the SHD lattice is such a lattice, and it is in this sense that it serves as the parent of the $\Z_2$ classifications of the other lattices of Fig.~\ref{fig:parent}. For each of them, Eq.~\eqref{eq:wyckoff} organizes its PSGs by SHD class. Each SHD class $\eta$ is restricted to the stabilizer of the collapsed cluster, whose extension $\tilde{S}$ is constructed from the SHD PSG in Appendix~\ref{app:parent}, and contributes $n_S(\eta)$ PSGs on the derived lattice. The contraction therefore does not map PSGs to PSGs: an SHD class has no counterpart on the derived lattice if $n_S(\eta)=0$, one if $n_S(\eta)=1$, and several PSGs with the same relation values if $n_S(\eta)>1$. On the lattices considered here $n_S(\eta)\le2$.

The images of the representations of each $\tilde{S}$ in $SU(2)$ are cyclic or binary dihedral groups (Appendix~\ref{app:parent}). Evaluating Eq.~\eqref{eq:wyckoff} with these representations and the 128 SHD classes of Sec.~\ref{sec:alg-z2} gives Table~\ref{tab:parent}. A direct solution of the PSG equations on the triangular, honeycomb, and kagome lattices, with on-site matrices in the binary dihedral group of order 24 and classes separated by relation values and traces of on-site stabilizer elements, gives the same classes. The 38 triangular PSGs realize 30 distinct classes $\eta$, eight of them with $n_S(\eta)=2$, and the 112 honeycomb PSGs realize 80 classes, 32 of them with $n_S(\eta)=2$. On the kagome, star, and ruby lattices $n_S(\eta)\le1$, so that distinct PSGs have distinct relation values.

\begin{table}
\caption{Algebraic $\Z_2$ projective symmetry groups of $p6mm$ lattices obtained from the SHD classification by Eq.~\eqref{eq:wyckoff}. ``Central'' counts PSGs in which the on-site time-reversal matrix is $\pm\tau^0$. ``Realizable'' counts PSGs that admit a nonzero singlet mean-field amplitude on some bond or site; the other PSGs force every amplitude to vanish at all ranges. On the triangular lattice one realizable PSG first admits an amplitude on the thirteenth shell of neighbors (Sec.~\ref{sec:parent}). The column ``sites ($S$)'' gives the Wyckoff position of the spins and its site-symmetry group, and the last column the number of PSGs of the space group alone.}
\label{tab:parent}
\begin{ruledtabular}
\begin{tabular}{lccccc}
lattice & sites ($S$) & total & central & realizable & $p6mm$\\
\hline
SHD & $12f$ ($1$) & 128 & 64 & 112 & 16\\
triangular & $1a$ ($6mm$) & 38 & 14 & 20 & 14\\
honeycomb & $2b$ ($3m$) & 112 & 48 & 88 & 24\\
kagome & $3c$ ($2mm$) & 30 & 10 & 20 & 10\\
star & $6e$ ($m$) & 80 & 32 & 64 & 16\\
ruby & $6e$ ($m$) & 80 & 32 & 64 & 16
\end{tabular}
\end{ruledtabular}
\end{table}

The sites of the star and ruby lattices lie on the same family of mirror lines and are related by a one-to-one map that commutes with $p6mm$, so the two lattices have identical classifications: 80 algebraic PSGs, of which 64 are realizable. The $\sqrt3\times\sqrt3$ honeycomb lattice obtained by collapsing the $J_b$ edges has its sites on the other family of mirror lines, at $6d$ (site symmetry $m$). It has 80 algebraic PSGs, 32 of them central, and 16 PSGs of the space group alone.

For the kagome, triangular, honeycomb, star, and ruby lattices the counts can be set against earlier classifications. The kagome entries reproduce the 30 algebraic PSGs derived in Appendix~B of Ref.~\cite{LuRanLee2011}; the 10 with a central time-reversal matrix force every mean-field amplitude to vanish, and the other 20 are the spin liquids of Table~II there. The 10 space-group classes agree with the five point-group representations of Table~V of Ref.~\cite{BieriLhuillierMessio2016}, each with both values of the translation sign. For the triangular lattice the 14 space-group classes are the seven point-group representations of Table~II of Ref.~\cite{BieriLhuillierMessio2016}, each with both values of the translation sign. With time reversal there are 38 PSGs: 14 with a central and 24 with a noncentral time-reversal matrix (Appendix~\ref{app:parent}). The 14 central PSGs and four of the noncentral ones, Z000011(1), Z000101(1), and the two PSGs of the class Z010111(1), force every amplitude to vanish, so that 20 PSGs are realizable. One of them, Z100110(1), admits no amplitude up to the twelfth shell of neighbors and a nonzero amplitude on the thirteenth shell, at a distance $2\sqrt7$ in units of the lattice constant. Of the eight classes with $n_S(\eta)=2$, four are realized by pairs of noncentral and four by pairs of central PSGs; the two PSGs of a pair share all relation values and differ in the traces of their on-site rotation matrices. The 24 noncentral PSGs therefore carry 20 distinct sets of symmetry-fractionalization labels, the number of phases with distinct spinon quantum numbers listed in Table~I of Ref.~\cite{Lu2016} and of states listed in Table~II of Ref.~\cite{ZhengMeiQi2015}. The 20 realizable PSGs carry 17 distinct label sets, since the four noncentral PSGs without amplitudes account for three label sets.

For the honeycomb lattice we find 112 gauge-inequivalent algebraic PSGs, of which 88 are realizable. Reference~\cite{LuRan2011} lists 160 algebraic PSGs, of which at most 128 can be realized (Appendix~B there). The two lists correspond as follows (Appendix~\ref{app:parent}). The 32 entries with $\psi_3=-2\pi/3$ outside the family (II)(ii)(B)($\beta$) of Ref.~\cite{LuRan2011} are mapped onto the entries with $\psi_3=+2\pi/3$ by the global gauge transformation $\ii\tau^1$, and the eight entries of that family with $\psi_3=0$ are gauge equivalent to entries of the family (II)(ii)(A). For its 16 entries with $\psi_3=\pm2\pi/3$ the relation $(\sigma C_6)^2$ does not evaluate to $\pm\tau^0$, so these entries have no counterpart among the $\Z_2$ PSGs. The remaining 104 entries are pairwise inequivalent and give 104 of the 112 PSGs. The other eight have $g_\T=\ii\tau^3$ and realize the eight classes Z$n_{T_2}$011$n_{\sigma C_6}n_{C_6}$(1), with $\eta_\sigma=\eta_{\T C_6}=\eta_\T=-1$ and $\eta_{\T\sigma}=+1$, each with $n_S(\eta)=1$; all eight admit nonzero amplitudes. The same accounting relates the realizable PSGs: of the 128 entries that can be realized, 16 are the entries without a $\Z_2$ counterpart, 24 are related to other entries by $\ii\tau^1$, and 8 are equivalent to (II)(ii)(A) entries, which leaves 80, and the eight additional PSGs give the 88 realizable PSGs. For the star lattice, the Schwinger-boson analysis of Ref.~\cite{ChoyKim2009} finds 16 algebraic $\Z_2$ PSGs of the space group [Eq.~(13) there]; the fermionic classification without time reversal also has 16 classes, one for each class of $H^2(p6mm,\Z_2)$. The 64 $\Z_2$ PSGs of the ruby lattice listed in Table~II of Ref.~\cite{MaitySamajdarIqbal2024} are the 64 realizable classes of Table~\ref{tab:parent}. The other 16 classes have the same central matrix $G_\T=\tau^0$ (up to a global sign) on all sites, for which Eq.~\eqref{eq:psg-T} forces every amplitude to vanish.

Through Eq.~\eqref{eq:wyckoff}, a spin liquid on any lattice of Table~\ref{tab:parent} can be assigned its SHD class. States with the same class, on the same or on different lattices, share all symmetry-fractionalization labels but can differ in their on-site representations, as the triangular and honeycomb pairs show. This difference is gauge invariant, since the trace of the on-site matrix of a site-stabilizer element is unchanged by gauge transformations, and it changes the symmetric on-site terms and bond amplitudes, and hence the mean-field states, that the two PSGs allow.
\section{Symmetric Spin Liquids}
\label{sec:ansatz}

\subsection{Construction}

For a given PSG the mean-field amplitudes on a reference bond $i_0\to j_0$ are constrained only by time reversal and by the elements that map the bond onto itself or onto its reverse. With the stabilizers of Table~\ref{tab:bonds}, Eqs.~\eqref{eq:psg-spatial} and~\eqref{eq:psg-T} become linear conditions on the real coefficients of Eq.~\eqref{eq:links}, whose solutions are the symmetric amplitudes of the orbit; the other bonds of the orbit follow by applying the element of $p6mm$ that maps the reference bond onto them. For the $\Z_2$ and $SU(2)$ PSGs of Eq.~\eqref{eq:z2psg} this produces the sign rules collected in Table~S8 of the Supplemental Material~\cite{SM}. The allowed singlet and triplet amplitudes on the six reference bonds are fixed by a few signs of the PSG, through closed-form rules for all symmetric and chiral PSGs derived in Sec.~SV\,B of the Supplemental Material~\cite{SM} and summarized in Table~S9 there. Time reversal alone fixes the on-site fields $a_\mu$ in the same way, since no spatial element fixes a site.

An ansatz realizes a given IGG only if no larger group leaves it invariant. Self-consistent on-site terms do not reduce the IGG of the bond amplitudes. They are Lagrange multipliers, fixed by $\langle\psi_i^\dagger\tau^k\psi_i\rangle=0$, which are the stationarity conditions of the mean-field free energy with respect to $a_i$. At $T>0$ this free energy is strictly concave in the multipliers, so the self-consistent multipliers, whose existence we assume, are unique. It is also invariant under the IGG of the bond amplitudes, which acts on the multipliers by conjugation, $a_i\to W_ia_iW_i^\dagger$, and uniqueness then requires the multipliers to commute with that IGG. We take $T=0$ as the limit $T\to0^+$; at $T=0$ itself additional solutions with components along the generators of the IGG can exist, but they describe the same ground state. The IGG of a generic choice of the allowed bond amplitudes, propagated over a finite patch of the lattice, follows from the linear equations $W_iu_{ij}=u_{ij}W_j$ for quaternions $W_i$ on the largest connected component of the bond graph. A one-, two-, or four-dimensional solution space corresponds to $\Z_2$, $U(1)$, or $SU(2)$. In practice the coefficients of the symmetric amplitudes are drawn from a standard normal distribution, and the ansatz is built on the 19 unit cells (228 sites) with $|x|,|y|,|x+y|\le2$. The $W_i$ are propagated from a root along a spanning tree of the largest connected component of the bond graph, which leaves linear equations for the quaternion at the root; their solution dimension is the number of singular values below $10^{-9}\max(1,s_{\max})$, and the smaller dimension from two independent draws is kept. A PSG is counted as realized at a given bond range if generic symmetry-allowed bond amplitudes up to that range have exactly the IGG of the PSG on a connected bond graph. Ans\"atze whose nonzero bonds form disconnected clusters, such as isolated hexagons or dimers, describe decoupled clusters rather than spin liquids and are not counted. A PSG is called realized when generic symmetric bond amplitudes have exactly its IGG; this does not imply that these amplitudes solve the self-consistency equations of a particular spin Hamiltonian.

\subsection{Realized Spin Liquids}

\begin{table}
\caption{Number of PSGs realized as spin liquids (generic bond amplitudes with exactly the IGG of the PSG on a connected bond graph) as a function of the longest bond included, for singlet amplitudes only and, in parentheses, for singlet and triplet amplitudes. $U(1)$ entries count classes with discrete labels plus families with a continuous flux; the sixteen labels of the sector $(0,1,0)$ are counted only as families.}
\label{tab:realized}
\begin{ruledtabular}
\begin{tabular}{lccc}
range & $\Z_2$ & $U(1)$ & $SU(2)$\\
\hline
$J_b,J_g,J_r$ & 4 (16) & 4+1 (20+4) & 4 (16)\\
$+J'$ & 18 (96) & 14+3 (104+8) & 8 (32)\\
$+J''$ & 40 (96) & 38+3 (104+8) & 8 (32)\\
$+J'''$ & 88 (112) & 74+5 (124+12) & 16 (48)
\end{tabular}
\end{ruledtabular}
\end{table}

\begin{table*}
\caption{$\Z_2$ spin liquids realized with singlet amplitudes up to third neighbors ($J''$). For each PSG the table lists the gauge components allowed on the reference bond of every orbit, in the gauge of Eq.~\eqref{eq:z2psg} ($\tau^{1,3}$ and $\tau^{1,2,3}$ denote the spans of $\tau^1,\tau^3$ and of $\tau^1,\tau^2,\tau^3$), the allowed on-site field $a_\mu$, and the shortest bond range at which generic bond amplitudes have IGG $\Z_2$ on a connected bond graph (range $J_1$: nearest-neighbor bonds only). Amplitudes on all other bonds follow from the rules of Table~S8 of the Supplemental Material~\cite{SM}. Components related by a residual global gauge rotation that commutes with the PSG are not independent.}
\label{tab:z2ans}
\begin{ruledtabular}
\begin{tabular}{rlccccccc c}
No. & PSG & $J_b$ & $J_g$ & $J_r$ & $J'$ & $J''$ & $J'''$ & $a_\mu$ & range\\
\hline
1 & Z010000(0) & $\tau^{1,2,3}$ & $\tau^{1,2,3}$ & $\tau^{1,2,3}$ & $0$ & $0$ & $0$ & $0$ & $J_1$\\
2 & Z010001(0) & $\tau^{1,2,3}$ & $\tau^{1,2,3}$ & $\tau^{1,2,3}$ & $0$ & $0$ & $0$ & $0$ & $J_1$\\
3 & Z110000(0) & $\tau^{1,2,3}$ & $\tau^{1,2,3}$ & $\tau^{1,2,3}$ & $0$ & $0$ & $0$ & $0$ & $J_1$\\
4 & Z110001(0) & $\tau^{1,2,3}$ & $\tau^{1,2,3}$ & $\tau^{1,2,3}$ & $0$ & $0$ & $0$ & $0$ & $J_1$\\
5 & Z000000(1) & $\tau^{1,3}$ & $\tau^{1,3}$ & $\tau^{1,3}$ & $\tau^{1,3}$ & $\tau^{1,3}$ & $\tau^{1,3}$ & $\tau^{1,3}$ & $J'$\\
6 & Z001000(1) & $\tau^{2}$ & $\tau^{1,3}$ & $\tau^{1,3}$ & $\tau^{2}$ & $\tau^{1,3}$ & $\ii\tau^0,\tau^{2}$ & $\tau^{1,3}$ & $J'$\\
7 & Z001001(1) & $\tau^{2}$ & $\tau^{1,3}$ & $\tau^{1,3}$ & $\ii\tau^0$ & $\tau^{1,3}$ & $\ii\tau^0,\tau^{2}$ & $\tau^{1,3}$ & $J'$\\
8 & Z001010(1) & $\ii\tau^0$ & $\tau^{1,3}$ & $\tau^{1,3}$ & $\tau^{2}$ & $\tau^{1,3}$ & $\ii\tau^0,\tau^{2}$ & $\tau^{1,3}$ & $J'$\\
9 & Z001011(1) & $\ii\tau^0$ & $\tau^{1,3}$ & $\tau^{1,3}$ & $\ii\tau^0$ & $\tau^{1,3}$ & $\ii\tau^0,\tau^{2}$ & $\tau^{1,3}$ & $J'$\\
10 & Z011000(0) & $0$ & $\tau^{1,2,3}$ & $\tau^{1,2,3}$ & $\tau^{1,2,3}$ & $0$ & $\ii\tau^0,\tau^{1,2,3}$ & $0$ & $J'$\\
11 & Z011010(0) & $0$ & $\tau^{1,2,3}$ & $\tau^{1,2,3}$ & $\tau^{1,2,3}$ & $0$ & $\ii\tau^0,\tau^{1,2,3}$ & $0$ & $J'$\\
12 & Z100001(1) & $\tau^{1,3}$ & $\tau^{1,3}$ & $\tau^{1,3}$ & $\tau^{1,3}$ & $\tau^{1,3}$ & $\tau^{1,3}$ & $\tau^{1,3}$ & $J'$\\
13 & Z101000(1) & $\tau^{2}$ & $\tau^{1,3}$ & $\tau^{1,3}$ & $\ii\tau^0$ & $\tau^{1,3}$ & $\ii\tau^0,\tau^{2}$ & $\tau^{1,3}$ & $J'$\\
14 & Z101001(1) & $\tau^{2}$ & $\tau^{1,3}$ & $\tau^{1,3}$ & $\tau^{2}$ & $\tau^{1,3}$ & $\ii\tau^0,\tau^{2}$ & $\tau^{1,3}$ & $J'$\\
15 & Z101010(1) & $\ii\tau^0$ & $\tau^{1,3}$ & $\tau^{1,3}$ & $\ii\tau^0$ & $\tau^{1,3}$ & $\ii\tau^0,\tau^{2}$ & $\tau^{1,3}$ & $J'$\\
16 & Z101011(1) & $\ii\tau^0$ & $\tau^{1,3}$ & $\tau^{1,3}$ & $\tau^{2}$ & $\tau^{1,3}$ & $\ii\tau^0,\tau^{2}$ & $\tau^{1,3}$ & $J'$\\
17 & Z111001(0) & $0$ & $\tau^{1,2,3}$ & $\tau^{1,2,3}$ & $\tau^{1,2,3}$ & $0$ & $\ii\tau^0,\tau^{1,2,3}$ & $0$ & $J'$\\
18 & Z111011(0) & $0$ & $\tau^{1,2,3}$ & $\tau^{1,2,3}$ & $\tau^{1,2,3}$ & $0$ & $\ii\tau^0,\tau^{1,2,3}$ & $0$ & $J'$\\
19 & Z000001(1) & $\tau^{1,3}$ & $\tau^{1,3}$ & $\tau^{1,3}$ & $0$ & $\tau^{1,3}$ & $\tau^{1,3}$ & $\tau^{1,3}$ & $J''$\\
20 & Z000010(1) & $0$ & $\tau^{1,3}$ & $\tau^{1,3}$ & $\tau^{1,3}$ & $\tau^{1,3}$ & $\tau^{1,3}$ & $\tau^{1,3}$ & $J''$\\
21 & Z000100(1) & $\tau^{1,3}$ & $0$ & $0$ & $\tau^{1,3}$ & $\tau^{1,3}$ & $\tau^{1,3}$ & $\tau^{1,3}$ & $J''$\\
22 & Z010000(1) & $\tau^{2}$ & $\tau^{2}$ & $\tau^{2}$ & $\tau^{1,3}$ & $\tau^{1,3}$ & $\tau^{1,3}$ & $\tau^{1,3}$ & $J''$\\
23 & Z010010(1) & $\ii\tau^0$ & $\tau^{2}$ & $\tau^{2}$ & $\tau^{1,3}$ & $\tau^{1,3}$ & $\tau^{1,3}$ & $\tau^{1,3}$ & $J''$\\
24 & Z010100(1) & $\tau^{2}$ & $\ii\tau^0$ & $\ii\tau^0$ & $\tau^{1,3}$ & $\tau^{1,3}$ & $\tau^{1,3}$ & $\tau^{1,3}$ & $J''$\\
25 & Z010110(1) & $\ii\tau^0$ & $\ii\tau^0$ & $\ii\tau^0$ & $\tau^{1,3}$ & $\tau^{1,3}$ & $\tau^{1,3}$ & $\tau^{1,3}$ & $J''$\\
26 & Z011000(1) & $\tau^{1,3}$ & $\tau^{2}$ & $\tau^{2}$ & $\tau^{2}$ & $\tau^{1,3}$ & $\ii\tau^0,\tau^{2}$ & $\tau^{1,3}$ & $J''$\\
27 & Z011001(1) & $\tau^{1,3}$ & $\tau^{2}$ & $\tau^{2}$ & $\ii\tau^0$ & $\tau^{1,3}$ & $\ii\tau^0,\tau^{2}$ & $\tau^{1,3}$ & $J''$\\
28 & Z011100(1) & $\tau^{1,3}$ & $\ii\tau^0$ & $\ii\tau^0$ & $\tau^{2}$ & $\tau^{1,3}$ & $\ii\tau^0,\tau^{2}$ & $\tau^{1,3}$ & $J''$\\
29 & Z011101(1) & $\tau^{1,3}$ & $\ii\tau^0$ & $\ii\tau^0$ & $\ii\tau^0$ & $\tau^{1,3}$ & $\ii\tau^0,\tau^{2}$ & $\tau^{1,3}$ & $J''$\\
30 & Z100000(1) & $\tau^{1,3}$ & $\tau^{1,3}$ & $\tau^{1,3}$ & $0$ & $\tau^{1,3}$ & $\tau^{1,3}$ & $\tau^{1,3}$ & $J''$\\
31 & Z100011(1) & $0$ & $\tau^{1,3}$ & $\tau^{1,3}$ & $\tau^{1,3}$ & $\tau^{1,3}$ & $\tau^{1,3}$ & $\tau^{1,3}$ & $J''$\\
32 & Z100101(1) & $\tau^{1,3}$ & $0$ & $0$ & $\tau^{1,3}$ & $\tau^{1,3}$ & $\tau^{1,3}$ & $\tau^{1,3}$ & $J''$\\
33 & Z110001(1) & $\tau^{2}$ & $\tau^{2}$ & $\tau^{2}$ & $\tau^{1,3}$ & $\tau^{1,3}$ & $\tau^{1,3}$ & $\tau^{1,3}$ & $J''$\\
34 & Z110011(1) & $\ii\tau^0$ & $\tau^{2}$ & $\tau^{2}$ & $\tau^{1,3}$ & $\tau^{1,3}$ & $\tau^{1,3}$ & $\tau^{1,3}$ & $J''$\\
35 & Z110101(1) & $\tau^{2}$ & $\ii\tau^0$ & $\ii\tau^0$ & $\tau^{1,3}$ & $\tau^{1,3}$ & $\tau^{1,3}$ & $\tau^{1,3}$ & $J''$\\
36 & Z110111(1) & $\ii\tau^0$ & $\ii\tau^0$ & $\ii\tau^0$ & $\tau^{1,3}$ & $\tau^{1,3}$ & $\tau^{1,3}$ & $\tau^{1,3}$ & $J''$\\
37 & Z111000(1) & $\tau^{1,3}$ & $\tau^{2}$ & $\tau^{2}$ & $\ii\tau^0$ & $\tau^{1,3}$ & $\ii\tau^0,\tau^{2}$ & $\tau^{1,3}$ & $J''$\\
38 & Z111001(1) & $\tau^{1,3}$ & $\tau^{2}$ & $\tau^{2}$ & $\tau^{2}$ & $\tau^{1,3}$ & $\ii\tau^0,\tau^{2}$ & $\tau^{1,3}$ & $J''$\\
39 & Z111100(1) & $\tau^{1,3}$ & $\ii\tau^0$ & $\ii\tau^0$ & $\ii\tau^0$ & $\tau^{1,3}$ & $\ii\tau^0,\tau^{2}$ & $\tau^{1,3}$ & $J''$\\
40 & Z111101(1) & $\tau^{1,3}$ & $\ii\tau^0$ & $\ii\tau^0$ & $\tau^{2}$ & $\tau^{1,3}$ & $\ii\tau^0,\tau^{2}$ & $\tau^{1,3}$ & $J''$\\
\end{tabular}
\end{ruledtabular}
\end{table*}
\begin{table*}
\caption{$\Z_2$ spin liquids that first appear when the fourth-neighbor bonds $J'''$ are included; columns as in Table~\ref{tab:z2ans}.}
\label{tab:z2ansb}
\begin{ruledtabular}
\begin{tabular}{rlccccccc c}
No. & PSG & $J_b$ & $J_g$ & $J_r$ & $J'$ & $J''$ & $J'''$ & $a_\mu$ & range\\
\hline
1 & Z000011(1) & $0$ & $\tau^{1,3}$ & $\tau^{1,3}$ & $0$ & $\tau^{1,3}$ & $\tau^{1,3}$ & $\tau^{1,3}$ & $J'''$\\
2 & Z000101(1) & $\tau^{1,3}$ & $0$ & $0$ & $0$ & $\tau^{1,3}$ & $\tau^{1,3}$ & $\tau^{1,3}$ & $J'''$\\
3 & Z001000(0) & $\tau^{1,2,3}$ & $0$ & $0$ & $\tau^{1,2,3}$ & $0$ & $\ii\tau^0,\tau^{1,2,3}$ & $0$ & $J'''$\\
4 & Z001001(0) & $\tau^{1,2,3}$ & $0$ & $0$ & $\ii\tau^0$ & $0$ & $\ii\tau^0,\tau^{1,2,3}$ & $0$ & $J'''$\\
5 & Z001010(0) & $\ii\tau^0$ & $0$ & $0$ & $\tau^{1,2,3}$ & $0$ & $\ii\tau^0,\tau^{1,2,3}$ & $0$ & $J'''$\\
6 & Z001100(0) & $\tau^{1,2,3}$ & $0$ & $0$ & $\tau^{1,2,3}$ & $0$ & $\ii\tau^0,\tau^{1,2,3}$ & $0$ & $J'''$\\
7 & Z001100(1) & $\tau^{2}$ & $0$ & $0$ & $\tau^{2}$ & $\tau^{1,3}$ & $\ii\tau^0,\tau^{2}$ & $\tau^{1,3}$ & $J'''$\\
8 & Z001101(0) & $\tau^{1,2,3}$ & $0$ & $0$ & $\ii\tau^0$ & $0$ & $\ii\tau^0,\tau^{1,2,3}$ & $0$ & $J'''$\\
9 & Z001101(1) & $\tau^{2}$ & $0$ & $0$ & $\ii\tau^0$ & $\tau^{1,3}$ & $\ii\tau^0,\tau^{2}$ & $\tau^{1,3}$ & $J'''$\\
10 & Z001110(0) & $\ii\tau^0$ & $0$ & $0$ & $\tau^{1,2,3}$ & $0$ & $\ii\tau^0,\tau^{1,2,3}$ & $0$ & $J'''$\\
11 & Z001110(1) & $\ii\tau^0$ & $0$ & $0$ & $\tau^{2}$ & $\tau^{1,3}$ & $\ii\tau^0,\tau^{2}$ & $\tau^{1,3}$ & $J'''$\\
12 & Z001111(1) & $\ii\tau^0$ & $0$ & $0$ & $\ii\tau^0$ & $\tau^{1,3}$ & $\ii\tau^0,\tau^{2}$ & $\tau^{1,3}$ & $J'''$\\
13 & Z010001(1) & $\tau^{2}$ & $\tau^{2}$ & $\tau^{2}$ & $0$ & $\tau^{1,3}$ & $\tau^{1,3}$ & $\tau^{1,3}$ & $J'''$\\
14 & Z010011(1) & $\ii\tau^0$ & $\tau^{2}$ & $\tau^{2}$ & $0$ & $\tau^{1,3}$ & $\tau^{1,3}$ & $\tau^{1,3}$ & $J'''$\\
15 & Z010101(1) & $\tau^{2}$ & $\ii\tau^0$ & $\ii\tau^0$ & $0$ & $\tau^{1,3}$ & $\tau^{1,3}$ & $\tau^{1,3}$ & $J'''$\\
16 & Z010111(1) & $\ii\tau^0$ & $\ii\tau^0$ & $\ii\tau^0$ & $0$ & $\tau^{1,3}$ & $\tau^{1,3}$ & $\tau^{1,3}$ & $J'''$\\
17 & Z011001(0) & $0$ & $\tau^{1,2,3}$ & $\tau^{1,2,3}$ & $\ii\tau^0$ & $0$ & $\ii\tau^0,\tau^{1,2,3}$ & $0$ & $J'''$\\
18 & Z011010(1) & $0$ & $\tau^{2}$ & $\tau^{2}$ & $\tau^{2}$ & $\tau^{1,3}$ & $\ii\tau^0,\tau^{2}$ & $\tau^{1,3}$ & $J'''$\\
19 & Z011011(0) & $0$ & $\tau^{1,2,3}$ & $\tau^{1,2,3}$ & $\ii\tau^0$ & $0$ & $\ii\tau^0,\tau^{1,2,3}$ & $0$ & $J'''$\\
20 & Z011011(1) & $0$ & $\tau^{2}$ & $\tau^{2}$ & $\ii\tau^0$ & $\tau^{1,3}$ & $\ii\tau^0,\tau^{2}$ & $\tau^{1,3}$ & $J'''$\\
21 & Z011100(0) & $0$ & $\ii\tau^0$ & $\ii\tau^0$ & $\tau^{1,2,3}$ & $0$ & $\ii\tau^0,\tau^{1,2,3}$ & $0$ & $J'''$\\
22 & Z011110(0) & $0$ & $\ii\tau^0$ & $\ii\tau^0$ & $\tau^{1,2,3}$ & $0$ & $\ii\tau^0,\tau^{1,2,3}$ & $0$ & $J'''$\\
23 & Z011110(1) & $0$ & $\ii\tau^0$ & $\ii\tau^0$ & $\tau^{2}$ & $\tau^{1,3}$ & $\ii\tau^0,\tau^{2}$ & $\tau^{1,3}$ & $J'''$\\
24 & Z011111(1) & $0$ & $\ii\tau^0$ & $\ii\tau^0$ & $\ii\tau^0$ & $\tau^{1,3}$ & $\ii\tau^0,\tau^{2}$ & $\tau^{1,3}$ & $J'''$\\
25 & Z100010(1) & $0$ & $\tau^{1,3}$ & $\tau^{1,3}$ & $0$ & $\tau^{1,3}$ & $\tau^{1,3}$ & $\tau^{1,3}$ & $J'''$\\
26 & Z100100(1) & $\tau^{1,3}$ & $0$ & $0$ & $0$ & $\tau^{1,3}$ & $\tau^{1,3}$ & $\tau^{1,3}$ & $J'''$\\
27 & Z101000(0) & $\tau^{1,2,3}$ & $0$ & $0$ & $\ii\tau^0$ & $0$ & $\ii\tau^0,\tau^{1,2,3}$ & $0$ & $J'''$\\
28 & Z101001(0) & $\tau^{1,2,3}$ & $0$ & $0$ & $\tau^{1,2,3}$ & $0$ & $\ii\tau^0,\tau^{1,2,3}$ & $0$ & $J'''$\\
29 & Z101011(0) & $\ii\tau^0$ & $0$ & $0$ & $\tau^{1,2,3}$ & $0$ & $\ii\tau^0,\tau^{1,2,3}$ & $0$ & $J'''$\\
30 & Z101100(0) & $\tau^{1,2,3}$ & $0$ & $0$ & $\ii\tau^0$ & $0$ & $\ii\tau^0,\tau^{1,2,3}$ & $0$ & $J'''$\\
31 & Z101100(1) & $\tau^{2}$ & $0$ & $0$ & $\ii\tau^0$ & $\tau^{1,3}$ & $\ii\tau^0,\tau^{2}$ & $\tau^{1,3}$ & $J'''$\\
32 & Z101101(0) & $\tau^{1,2,3}$ & $0$ & $0$ & $\tau^{1,2,3}$ & $0$ & $\ii\tau^0,\tau^{1,2,3}$ & $0$ & $J'''$\\
33 & Z101101(1) & $\tau^{2}$ & $0$ & $0$ & $\tau^{2}$ & $\tau^{1,3}$ & $\ii\tau^0,\tau^{2}$ & $\tau^{1,3}$ & $J'''$\\
34 & Z101110(1) & $\ii\tau^0$ & $0$ & $0$ & $\ii\tau^0$ & $\tau^{1,3}$ & $\ii\tau^0,\tau^{2}$ & $\tau^{1,3}$ & $J'''$\\
35 & Z101111(0) & $\ii\tau^0$ & $0$ & $0$ & $\tau^{1,2,3}$ & $0$ & $\ii\tau^0,\tau^{1,2,3}$ & $0$ & $J'''$\\
36 & Z101111(1) & $\ii\tau^0$ & $0$ & $0$ & $\tau^{2}$ & $\tau^{1,3}$ & $\ii\tau^0,\tau^{2}$ & $\tau^{1,3}$ & $J'''$\\
37 & Z110000(1) & $\tau^{2}$ & $\tau^{2}$ & $\tau^{2}$ & $0$ & $\tau^{1,3}$ & $\tau^{1,3}$ & $\tau^{1,3}$ & $J'''$\\
38 & Z110010(1) & $\ii\tau^0$ & $\tau^{2}$ & $\tau^{2}$ & $0$ & $\tau^{1,3}$ & $\tau^{1,3}$ & $\tau^{1,3}$ & $J'''$\\
39 & Z110100(1) & $\tau^{2}$ & $\ii\tau^0$ & $\ii\tau^0$ & $0$ & $\tau^{1,3}$ & $\tau^{1,3}$ & $\tau^{1,3}$ & $J'''$\\
40 & Z110110(1) & $\ii\tau^0$ & $\ii\tau^0$ & $\ii\tau^0$ & $0$ & $\tau^{1,3}$ & $\tau^{1,3}$ & $\tau^{1,3}$ & $J'''$\\
41 & Z111000(0) & $0$ & $\tau^{1,2,3}$ & $\tau^{1,2,3}$ & $\ii\tau^0$ & $0$ & $\ii\tau^0,\tau^{1,2,3}$ & $0$ & $J'''$\\
42 & Z111010(0) & $0$ & $\tau^{1,2,3}$ & $\tau^{1,2,3}$ & $\ii\tau^0$ & $0$ & $\ii\tau^0,\tau^{1,2,3}$ & $0$ & $J'''$\\
43 & Z111010(1) & $0$ & $\tau^{2}$ & $\tau^{2}$ & $\ii\tau^0$ & $\tau^{1,3}$ & $\ii\tau^0,\tau^{2}$ & $\tau^{1,3}$ & $J'''$\\
44 & Z111011(1) & $0$ & $\tau^{2}$ & $\tau^{2}$ & $\tau^{2}$ & $\tau^{1,3}$ & $\ii\tau^0,\tau^{2}$ & $\tau^{1,3}$ & $J'''$\\
45 & Z111101(0) & $0$ & $\ii\tau^0$ & $\ii\tau^0$ & $\tau^{1,2,3}$ & $0$ & $\ii\tau^0,\tau^{1,2,3}$ & $0$ & $J'''$\\
46 & Z111110(1) & $0$ & $\ii\tau^0$ & $\ii\tau^0$ & $\ii\tau^0$ & $\tau^{1,3}$ & $\ii\tau^0,\tau^{2}$ & $\tau^{1,3}$ & $J'''$\\
47 & Z111111(0) & $0$ & $\ii\tau^0$ & $\ii\tau^0$ & $\tau^{1,2,3}$ & $0$ & $\ii\tau^0,\tau^{1,2,3}$ & $0$ & $J'''$\\
48 & Z111111(1) & $0$ & $\ii\tau^0$ & $\ii\tau^0$ & $\tau^{2}$ & $\tau^{1,3}$ & $\ii\tau^0,\tau^{2}$ & $\tau^{1,3}$ & $J'''$\\
\end{tabular}
\end{ruledtabular}
\end{table*}
\begin{table*}
\caption{$U(1)$ spin liquids realized with singlet amplitudes up to third neighbors ($J''$). A label U$w_{C_6}w_\sigma w_\T$-$n_{T_2}n_{\T\sigma}n_{\T C_6}n_\sigma n_{\sigma C_6}n_{C_6}$ gives the sector and the signs of the canonical form of Table~\ref{tab:u1} (digit $0$ for $+1$ and $1$ for $-1$); in the sector $(0,1,0)$ the continuous flux $\Phi$, here at a generic value, replaces $n_{T_2}$, and the $J'$ amplitude is the bond-rotated $\tau^3$, $e^{-\ii\Phi\tau^3/2}\tau^3$ (Table~S9 of the Supplemental Material~\cite{SM}). Columns as in Table~\ref{tab:z2ans}.}
\label{tab:u1ans}
\begin{ruledtabular}
\begin{tabular}{rlccccccc c}
No. & PSG & $J_b$ & $J_g$ & $J_r$ & $J'$ & $J''$ & $J'''$ & $a_\mu$ & range\\
\hline
1 & U010-$\Phi$10110 & $\ii\tau^0,\tau^{3}$ & $\ii\tau^0,\tau^{3}$ & $\ii\tau^0,\tau^{3}$ & $0$ & $0$ & $0$ & $0$ & $J_1$\\
2 & U110-010100 & $\tau^{3}$ & $\ii\tau^0,\tau^{3}$ & $\ii\tau^0,\tau^{3}$ & $0$ & $0$ & $0$ & $0$ & $J_1$\\
3 & U110-010101 & $\tau^{3}$ & $\ii\tau^0,\tau^{3}$ & $\ii\tau^0,\tau^{3}$ & $0$ & $0$ & $0$ & $0$ & $J_1$\\
4 & U110-110100 & $\tau^{3}$ & $\ii\tau^0,\tau^{3}$ & $\ii\tau^0,\tau^{3}$ & $0$ & $0$ & $0$ & $0$ & $J_1$\\
5 & U110-110101 & $\tau^{3}$ & $\ii\tau^0,\tau^{3}$ & $\ii\tau^0,\tau^{3}$ & $0$ & $0$ & $0$ & $0$ & $J_1$\\
6 & U001-000000 & $\tau^{3}$ & $\tau^{3}$ & $\tau^{3}$ & $\tau^{3}$ & $\tau^{3}$ & $\tau^{3}$ & $\tau^{3}$ & $J'$\\
7 & U001-100001 & $\tau^{3}$ & $\tau^{3}$ & $\tau^{3}$ & $\tau^{3}$ & $\tau^{3}$ & $\tau^{3}$ & $\tau^{3}$ & $J'$\\
8 & U010-$\Phi$11100 & $0$ & $\ii\tau^0,\tau^{3}$ & $\ii\tau^0,\tau^{3}$ & $e^{-\ii\Phi\tau^3/2}\tau^{3}$ & $0$ & $\ii\tau^0,\tau^{3}$ & $0$ & $J'$\\
9 & U010-$\Phi$11110 & $0$ & $\ii\tau^0,\tau^{3}$ & $\ii\tau^0,\tau^{3}$ & $e^{-\ii\Phi\tau^3/2}\tau^{3}$ & $0$ & $\ii\tau^0,\tau^{3}$ & $0$ & $J'$\\
10 & U011-000110 & $\tau^{3}$ & $\tau^{3}$ & $\tau^{3}$ & $\tau^{3}$ & $\tau^{3}$ & $\tau^{3}$ & $\tau^{3}$ & $J'$\\
11 & U011-100111 & $\tau^{3}$ & $\tau^{3}$ & $\tau^{3}$ & $\tau^{3}$ & $\tau^{3}$ & $\tau^{3}$ & $\tau^{3}$ & $J'$\\
12 & U101-000011 & $\tau^{3}$ & $\tau^{3}$ & $\tau^{3}$ & $\tau^{3}$ & $\tau^{3}$ & $\tau^{3}$ & $\tau^{3}$ & $J'$\\
13 & U101-100010 & $\tau^{3}$ & $\tau^{3}$ & $\tau^{3}$ & $\tau^{3}$ & $\tau^{3}$ & $\tau^{3}$ & $\tau^{3}$ & $J'$\\
14 & U110-011101 & $0$ & $\ii\tau^0,\tau^{3}$ & $\ii\tau^0,\tau^{3}$ & $\ii\tau^0,\tau^{3}$ & $0$ & $\ii\tau^0,\tau^{3}$ & $0$ & $J'$\\
15 & U110-111100 & $0$ & $\ii\tau^0,\tau^{3}$ & $\ii\tau^0,\tau^{3}$ & $\ii\tau^0,\tau^{3}$ & $0$ & $\ii\tau^0,\tau^{3}$ & $0$ & $J'$\\
16 & U111-000101 & $\tau^{3}$ & $\tau^{3}$ & $\tau^{3}$ & $\tau^{3}$ & $\tau^{3}$ & $\tau^{3}$ & $\tau^{3}$ & $J'$\\
17 & U111-100100 & $\tau^{3}$ & $\tau^{3}$ & $\tau^{3}$ & $\tau^{3}$ & $\tau^{3}$ & $\tau^{3}$ & $\tau^{3}$ & $J'$\\
18 & U001-000001 & $\tau^{3}$ & $\tau^{3}$ & $\tau^{3}$ & $0$ & $\tau^{3}$ & $\tau^{3}$ & $\tau^{3}$ & $J''$\\
19 & U001-000010 & $0$ & $\tau^{3}$ & $\tau^{3}$ & $\tau^{3}$ & $\tau^{3}$ & $\tau^{3}$ & $\tau^{3}$ & $J''$\\
20 & U001-000100 & $\tau^{3}$ & $0$ & $0$ & $\tau^{3}$ & $\tau^{3}$ & $\tau^{3}$ & $\tau^{3}$ & $J''$\\
21 & U001-100000 & $\tau^{3}$ & $\tau^{3}$ & $\tau^{3}$ & $0$ & $\tau^{3}$ & $\tau^{3}$ & $\tau^{3}$ & $J''$\\
22 & U001-100011 & $0$ & $\tau^{3}$ & $\tau^{3}$ & $\tau^{3}$ & $\tau^{3}$ & $\tau^{3}$ & $\tau^{3}$ & $J''$\\
23 & U001-100101 & $\tau^{3}$ & $0$ & $0$ & $\tau^{3}$ & $\tau^{3}$ & $\tau^{3}$ & $\tau^{3}$ & $J''$\\
24 & U011-000010 & $\tau^{3}$ & $0$ & $0$ & $\tau^{3}$ & $\tau^{3}$ & $\tau^{3}$ & $\tau^{3}$ & $J''$\\
25 & U011-000100 & $0$ & $\tau^{3}$ & $\tau^{3}$ & $\tau^{3}$ & $\tau^{3}$ & $\tau^{3}$ & $\tau^{3}$ & $J''$\\
26 & U011-000111 & $\tau^{3}$ & $\tau^{3}$ & $\tau^{3}$ & $0$ & $\tau^{3}$ & $\tau^{3}$ & $\tau^{3}$ & $J''$\\
27 & U011-100011 & $\tau^{3}$ & $0$ & $0$ & $\tau^{3}$ & $\tau^{3}$ & $\tau^{3}$ & $\tau^{3}$ & $J''$\\
28 & U011-100101 & $0$ & $\tau^{3}$ & $\tau^{3}$ & $\tau^{3}$ & $\tau^{3}$ & $\tau^{3}$ & $\tau^{3}$ & $J''$\\
29 & U011-100110 & $\tau^{3}$ & $\tau^{3}$ & $\tau^{3}$ & $0$ & $\tau^{3}$ & $\tau^{3}$ & $\tau^{3}$ & $J''$\\
30 & U101-000001 & $0$ & $\tau^{3}$ & $\tau^{3}$ & $\tau^{3}$ & $\tau^{3}$ & $\tau^{3}$ & $\tau^{3}$ & $J''$\\
31 & U101-000010 & $\tau^{3}$ & $\tau^{3}$ & $\tau^{3}$ & $0$ & $\tau^{3}$ & $\tau^{3}$ & $\tau^{3}$ & $J''$\\
32 & U101-000111 & $\tau^{3}$ & $0$ & $0$ & $\tau^{3}$ & $\tau^{3}$ & $\tau^{3}$ & $\tau^{3}$ & $J''$\\
33 & U101-100000 & $0$ & $\tau^{3}$ & $\tau^{3}$ & $\tau^{3}$ & $\tau^{3}$ & $\tau^{3}$ & $\tau^{3}$ & $J''$\\
34 & U101-100011 & $\tau^{3}$ & $\tau^{3}$ & $\tau^{3}$ & $0$ & $\tau^{3}$ & $\tau^{3}$ & $\tau^{3}$ & $J''$\\
35 & U101-100110 & $\tau^{3}$ & $0$ & $0$ & $\tau^{3}$ & $\tau^{3}$ & $\tau^{3}$ & $\tau^{3}$ & $J''$\\
36 & U111-000001 & $\tau^{3}$ & $0$ & $0$ & $\tau^{3}$ & $\tau^{3}$ & $\tau^{3}$ & $\tau^{3}$ & $J''$\\
37 & U111-000100 & $\tau^{3}$ & $\tau^{3}$ & $\tau^{3}$ & $0$ & $\tau^{3}$ & $\tau^{3}$ & $\tau^{3}$ & $J''$\\
38 & U111-000111 & $0$ & $\tau^{3}$ & $\tau^{3}$ & $\tau^{3}$ & $\tau^{3}$ & $\tau^{3}$ & $\tau^{3}$ & $J''$\\
39 & U111-100000 & $\tau^{3}$ & $0$ & $0$ & $\tau^{3}$ & $\tau^{3}$ & $\tau^{3}$ & $\tau^{3}$ & $J''$\\
40 & U111-100101 & $\tau^{3}$ & $\tau^{3}$ & $\tau^{3}$ & $0$ & $\tau^{3}$ & $\tau^{3}$ & $\tau^{3}$ & $J''$\\
41 & U111-100110 & $0$ & $\tau^{3}$ & $\tau^{3}$ & $\tau^{3}$ & $\tau^{3}$ & $\tau^{3}$ & $\tau^{3}$ & $J''$\\
\end{tabular}
\end{ruledtabular}
\end{table*}
\begin{table*}
\caption{$U(1)$ spin liquids that first appear when the fourth-neighbor bonds $J'''$ are included; labels as in Table~\ref{tab:u1ans} and columns as in Table~\ref{tab:z2ans}.}
\label{tab:u1ansb}
\begin{ruledtabular}
\begin{tabular}{rlccccccc c}
No. & PSG & $J_b$ & $J_g$ & $J_r$ & $J'$ & $J''$ & $J'''$ & $a_\mu$ & range\\
\hline
1 & U000-001000 & $\tau^{3}$ & $0$ & $0$ & $\tau^{3}$ & $0$ & $\ii\tau^0,\tau^{3}$ & $0$ & $J'''$\\
2 & U000-001001 & $\tau^{3}$ & $0$ & $0$ & $\ii\tau^0$ & $0$ & $\ii\tau^0,\tau^{3}$ & $0$ & $J'''$\\
3 & U000-011000 & $0$ & $\tau^{3}$ & $\tau^{3}$ & $\tau^{3}$ & $0$ & $\ii\tau^0,\tau^{3}$ & $0$ & $J'''$\\
4 & U000-011001 & $0$ & $\tau^{3}$ & $\tau^{3}$ & $\ii\tau^0$ & $0$ & $\ii\tau^0,\tau^{3}$ & $0$ & $J'''$\\
5 & U000-101000 & $\tau^{3}$ & $0$ & $0$ & $\ii\tau^0$ & $0$ & $\ii\tau^0,\tau^{3}$ & $0$ & $J'''$\\
6 & U000-101001 & $\tau^{3}$ & $0$ & $0$ & $\tau^{3}$ & $0$ & $\ii\tau^0,\tau^{3}$ & $0$ & $J'''$\\
7 & U000-111000 & $0$ & $\tau^{3}$ & $\tau^{3}$ & $\ii\tau^0$ & $0$ & $\ii\tau^0,\tau^{3}$ & $0$ & $J'''$\\
8 & U000-111001 & $0$ & $\tau^{3}$ & $\tau^{3}$ & $\tau^{3}$ & $0$ & $\ii\tau^0,\tau^{3}$ & $0$ & $J'''$\\
9 & U001-000011 & $0$ & $\tau^{3}$ & $\tau^{3}$ & $0$ & $\tau^{3}$ & $\tau^{3}$ & $\tau^{3}$ & $J'''$\\
10 & U001-000101 & $\tau^{3}$ & $0$ & $0$ & $0$ & $\tau^{3}$ & $\tau^{3}$ & $\tau^{3}$ & $J'''$\\
11 & U001-100010 & $0$ & $\tau^{3}$ & $\tau^{3}$ & $0$ & $\tau^{3}$ & $\tau^{3}$ & $\tau^{3}$ & $J'''$\\
12 & U001-100100 & $\tau^{3}$ & $0$ & $0$ & $0$ & $\tau^{3}$ & $\tau^{3}$ & $\tau^{3}$ & $J'''$\\
13 & U010-$\Phi$01010 & $\ii\tau^0,\tau^{3}$ & $0$ & $0$ & $e^{-\ii\Phi\tau^3/2}\tau^{3}$ & $0$ & $\ii\tau^0,\tau^{3}$ & $0$ & $J'''$\\
14 & U010-$\Phi$01110 & $\ii\tau^0,\tau^{3}$ & $0$ & $0$ & $e^{-\ii\Phi\tau^3/2}\tau^{3}$ & $0$ & $\ii\tau^0,\tau^{3}$ & $0$ & $J'''$\\
15 & U011-000011 & $\tau^{3}$ & $0$ & $0$ & $0$ & $\tau^{3}$ & $\tau^{3}$ & $\tau^{3}$ & $J'''$\\
16 & U011-000101 & $0$ & $\tau^{3}$ & $\tau^{3}$ & $0$ & $\tau^{3}$ & $\tau^{3}$ & $\tau^{3}$ & $J'''$\\
17 & U011-100010 & $\tau^{3}$ & $0$ & $0$ & $0$ & $\tau^{3}$ & $\tau^{3}$ & $\tau^{3}$ & $J'''$\\
18 & U011-100100 & $0$ & $\tau^{3}$ & $\tau^{3}$ & $0$ & $\tau^{3}$ & $\tau^{3}$ & $\tau^{3}$ & $J'''$\\
19 & U100-001011 & $\ii\tau^0,\tau^{3}$ & $0$ & $0$ & $\ii\tau^0,\tau^{3}$ & $0$ & $\ii\tau^0,\tau^{3}$ & $0$ & $J'''$\\
20 & U100-011001 & $0$ & $\tau^{3}$ & $\tau^{3}$ & $\ii\tau^0,\tau^{3}$ & $0$ & $\ii\tau^0,\tau^{3}$ & $0$ & $J'''$\\
21 & U100-011011 & $0$ & $\tau^{3}$ & $\tau^{3}$ & $\ii\tau^0,\tau^{3}$ & $0$ & $\ii\tau^0,\tau^{3}$ & $0$ & $J'''$\\
22 & U100-101010 & $\ii\tau^0,\tau^{3}$ & $0$ & $0$ & $\ii\tau^0,\tau^{3}$ & $0$ & $\ii\tau^0,\tau^{3}$ & $0$ & $J'''$\\
23 & U100-111000 & $0$ & $\tau^{3}$ & $\tau^{3}$ & $\ii\tau^0,\tau^{3}$ & $0$ & $\ii\tau^0,\tau^{3}$ & $0$ & $J'''$\\
24 & U100-111010 & $0$ & $\tau^{3}$ & $\tau^{3}$ & $\ii\tau^0,\tau^{3}$ & $0$ & $\ii\tau^0,\tau^{3}$ & $0$ & $J'''$\\
25 & U101-000000 & $0$ & $\tau^{3}$ & $\tau^{3}$ & $0$ & $\tau^{3}$ & $\tau^{3}$ & $\tau^{3}$ & $J'''$\\
26 & U101-000110 & $\tau^{3}$ & $0$ & $0$ & $0$ & $\tau^{3}$ & $\tau^{3}$ & $\tau^{3}$ & $J'''$\\
27 & U101-100001 & $0$ & $\tau^{3}$ & $\tau^{3}$ & $0$ & $\tau^{3}$ & $\tau^{3}$ & $\tau^{3}$ & $J'''$\\
28 & U101-100111 & $\tau^{3}$ & $0$ & $0$ & $0$ & $\tau^{3}$ & $\tau^{3}$ & $\tau^{3}$ & $J'''$\\
29 & U110-001001 & $\tau^{3}$ & $0$ & $0$ & $\ii\tau^0,\tau^{3}$ & $0$ & $\ii\tau^0,\tau^{3}$ & $0$ & $J'''$\\
30 & U110-001101 & $\tau^{3}$ & $0$ & $0$ & $\ii\tau^0,\tau^{3}$ & $0$ & $\ii\tau^0,\tau^{3}$ & $0$ & $J'''$\\
31 & U110-011100 & $0$ & $\ii\tau^0,\tau^{3}$ & $\ii\tau^0,\tau^{3}$ & $0$ & $0$ & $\ii\tau^0,\tau^{3}$ & $0$ & $J'''$\\
32 & U110-101000 & $\tau^{3}$ & $0$ & $0$ & $\ii\tau^0,\tau^{3}$ & $0$ & $\ii\tau^0,\tau^{3}$ & $0$ & $J'''$\\
33 & U110-101100 & $\tau^{3}$ & $0$ & $0$ & $\ii\tau^0,\tau^{3}$ & $0$ & $\ii\tau^0,\tau^{3}$ & $0$ & $J'''$\\
34 & U110-111101 & $0$ & $\ii\tau^0,\tau^{3}$ & $\ii\tau^0,\tau^{3}$ & $0$ & $0$ & $\ii\tau^0,\tau^{3}$ & $0$ & $J'''$\\
35 & U111-000000 & $\tau^{3}$ & $0$ & $0$ & $0$ & $\tau^{3}$ & $\tau^{3}$ & $\tau^{3}$ & $J'''$\\
36 & U111-000110 & $0$ & $\tau^{3}$ & $\tau^{3}$ & $0$ & $\tau^{3}$ & $\tau^{3}$ & $\tau^{3}$ & $J'''$\\
37 & U111-100001 & $\tau^{3}$ & $0$ & $0$ & $0$ & $\tau^{3}$ & $\tau^{3}$ & $\tau^{3}$ & $J'''$\\
38 & U111-100111 & $0$ & $\tau^{3}$ & $\tau^{3}$ & $0$ & $\tau^{3}$ & $\tau^{3}$ & $\tau^{3}$ & $J'''$\\
\end{tabular}
\end{ruledtabular}
\end{table*}
\begin{table*}
\caption{$SU(2)$ spin liquids realized up to fourth neighbors, with the singlet link $u=\ii\chi\tau^0$ allowed on the bonds marked $\ii\tau^0$; columns as in Table~\ref{tab:z2ans}.}
\label{tab:su2ans}
\begin{ruledtabular}
\begin{tabular}{rlccccccc c}
No. & PSG & $J_b$ & $J_g$ & $J_r$ & $J'$ & $J''$ & $J'''$ & $a_\mu$ & range\\
\hline
1 & SU010110 & $\ii\tau^0$ & $\ii\tau^0$ & $\ii\tau^0$ & $0$ & $0$ & $0$ & $0$ & $J_1$\\
2 & SU010111 & $\ii\tau^0$ & $\ii\tau^0$ & $\ii\tau^0$ & $0$ & $0$ & $0$ & $0$ & $J_1$\\
3 & SU110110 & $\ii\tau^0$ & $\ii\tau^0$ & $\ii\tau^0$ & $0$ & $0$ & $0$ & $0$ & $J_1$\\
4 & SU110111 & $\ii\tau^0$ & $\ii\tau^0$ & $\ii\tau^0$ & $0$ & $0$ & $0$ & $0$ & $J_1$\\
5 & SU011101 & $0$ & $\ii\tau^0$ & $\ii\tau^0$ & $\ii\tau^0$ & $0$ & $\ii\tau^0$ & $0$ & $J'$\\
6 & SU011111 & $0$ & $\ii\tau^0$ & $\ii\tau^0$ & $\ii\tau^0$ & $0$ & $\ii\tau^0$ & $0$ & $J'$\\
7 & SU111100 & $0$ & $\ii\tau^0$ & $\ii\tau^0$ & $\ii\tau^0$ & $0$ & $\ii\tau^0$ & $0$ & $J'$\\
8 & SU111110 & $0$ & $\ii\tau^0$ & $\ii\tau^0$ & $\ii\tau^0$ & $0$ & $\ii\tau^0$ & $0$ & $J'$\\
9 & SU001011 & $\ii\tau^0$ & $0$ & $0$ & $\ii\tau^0$ & $0$ & $\ii\tau^0$ & $0$ & $J'''$\\
10 & SU001111 & $\ii\tau^0$ & $0$ & $0$ & $\ii\tau^0$ & $0$ & $\ii\tau^0$ & $0$ & $J'''$\\
11 & SU011100 & $0$ & $\ii\tau^0$ & $\ii\tau^0$ & $0$ & $0$ & $\ii\tau^0$ & $0$ & $J'''$\\
12 & SU011110 & $0$ & $\ii\tau^0$ & $\ii\tau^0$ & $0$ & $0$ & $\ii\tau^0$ & $0$ & $J'''$\\
13 & SU101010 & $\ii\tau^0$ & $0$ & $0$ & $\ii\tau^0$ & $0$ & $\ii\tau^0$ & $0$ & $J'''$\\
14 & SU101110 & $\ii\tau^0$ & $0$ & $0$ & $\ii\tau^0$ & $0$ & $\ii\tau^0$ & $0$ & $J'''$\\
15 & SU111101 & $0$ & $\ii\tau^0$ & $\ii\tau^0$ & $0$ & $0$ & $\ii\tau^0$ & $0$ & $J'''$\\
16 & SU111111 & $0$ & $\ii\tau^0$ & $\ii\tau^0$ & $0$ & $0$ & $\ii\tau^0$ & $0$ & $J'''$\\
\end{tabular}
\end{ruledtabular}
\end{table*}

Table~\ref{tab:realized} gives the number of PSGs that are realized as spin liquids at each bond range. The small numbers at nearest-neighbor range reflect the bipartite structure of these bonds. Let $\epsilon_i=+1$ on sublattice $A$ and $-1$ on sublattice $B$. The gauge transformation $\epsilon_i\tau^0$ reverses the sign of every nearest-neighbor link, and by Eq.~\eqref{eq:psg-T} so does $G_\T(i)$. Hence $\epsilon_iG_\T(i)$ belongs to the IGG of every time-reversal-symmetric nearest-neighbor ansatz. For a $\Z_2$ PSG with $g_\T=\ii\tau^2$ ($t=1$) this element squares to $-1$ and generates a $U(1)$ subgroup; for a $U(1)$ PSG with $w_\T=1$ it enlarges the IGG to $SU(2)$. The on-site terms allowed by these PSGs anticommute with $G_\T$ and vanish identically at self-consistency. Without them, no $\Z_2$ PSG with $t=1$ is realized with nearest-neighbor bonds: the nearest-neighbor $\Z_2$ spin liquids all have $t=0$, with singlet as with triplet amplitudes, and the singlet $U(1)$ spin liquids all have $w_\T=0$. Each bond beyond nearest neighbors connects sites of the same sublattice, which removes this sublattice structure and introduces new loops. The number of realized PSGs then grows quickly with the range; at fourth neighbors 88 of the 112 realizable $\Z_2$ PSGs are realized with singlet amplitudes and all 112 with triplet amplitudes. The remaining sixteen $\Z_2$ PSGs, those with $g_\T=\tau^0$ and $\eta_{\T C_6}=\eta_{\T\sigma}=+1$, admit no amplitude at all, because $G_\T=\tau^0$ on every site turns Eq.~\eqref{eq:psg-T} into $u=-u$. The $\Z_2$ spin liquids realized up to third neighbors are listed in Table~\ref{tab:z2ans} and those that first appear with $J'''$ in Table~\ref{tab:z2ansb}; the $U(1)$ spin liquids are listed in the same way in Tables~\ref{tab:u1ans} and~\ref{tab:u1ansb}, and the $SU(2)$ spin liquids up to fourth neighbors in Table~\ref{tab:su2ans}. Each PSG is listed with the shortest range at which it is realized.

Several features of the tables follow directly from the PSG. On nearest-neighbor bonds the sublattice signs $h(\mu)$ of time reversal decide which components survive (Table~S9 of the Supplemental Material~\cite{SM}). For $SU(2)$ spin liquids, with singlet links along $\ii\tau^0$, this requires $h(\mu_i)h(\mu_j)=-1$ on every nonzero bond, so that the bond graph of a time-reversal-symmetric $SU(2)$ state must be bipartite with respect to $h$. In the sector $\eta_{\T\sigma}=-1$, $\eta_{\T C_6}=+1$ the function $h$ equals the sublattice parity. Only the nearest-neighbor bonds survive there, and the four connected states are the $SU(2)$ states with all three nearest-neighbor amplitudes nonzero. In the other three sectors $h$ differs from the sublattice parity: $J'$ or $J'''$ bonds become allowed, while some nearest-neighbor bonds are forbidden. In none of the four sectors does a time-reversal-symmetric $SU(2)$ state contain $J''$ bonds. A nonzero $J''$ amplitude puts all three bonds of the inner $J''$ triangle of a hexagon into the bond graph, and this odd cycle cannot be bipartite with respect to $h$.

\subsection{Hierarchy}
\label{sec:hierarchy}

A state $X$ descends from a parent $P$ if the PSG of $X$ is a subgroup of the PSG of $P$, so that the parent ansatz is a special point of the amplitude space of $X$, and if generic amplitudes of $X$ lower the IGG. For an $SU(2)$ parent with sign structure $\epsilon_O(i)$, Sec.~\ref{sec:alg-su2}, the descendants follow from choosing site-independent matrices $g_O\in SU(2)$ and forming $G'_O(i)=\epsilon_O(i)g_O$. A $U(1)$ descendant requires the $g_O$ to normalize a common $U(1)$ subgroup, and a $\Z_2$ descendant requires them to satisfy the relations up to signs. Up to conjugation the images of $p6mm\times\Z_2^\T$ in $SO(3)$ that arise in this way are subgroups of the dihedral group $D_6$ or of the octahedral group, so the enumeration is finite. Each connected $SU(2)$ parent realized up to $J'''$ (Table~\ref{tab:su2ans}) has one $U(1)$ descendant in each of the eight sectors $(w_{C_6},w_\sigma,w_\T)$ and sixteen algebraic $\Z_2$ descendants. One of the $\Z_2$ descendants is reached through no $U(1)$ intermediate. Its gauge matrices form an octahedral group, translations act projectively, and the IGG drops from $SU(2)$ directly to $\Z_2$ when hopping and pairing are added together. Every step of the hierarchy is effected by bond amplitudes: the $SU(2)\to U(1)$ steps by real hopping $\chi^3$ on specific bonds, and the $U(1)\to\Z_2$ steps by pairing on specific bonds. We call a descendant genuine at a given range if generic bond amplitudes up to that range have exactly the IGG of its PSG. The minimal range of a descent is the shortest range at which the parent is connected and the descendant is genuine. Because $\epsilon_iG_\T(i)$ belongs to the IGG of every time-reversal-symmetric nearest-neighbor ansatz, the descents to $\Z_2$ states with $t=1$ and to $U(1)$ states with $w_\T=1$ first occur once $J'$ or longer bonds are included, and with $J_1$ amplitudes no chiral descendant is genuine at half filling (Sec.~\ref{sec:chiral}).

\begin{figure*}
\centering
\includegraphics[width=0.97\textwidth]{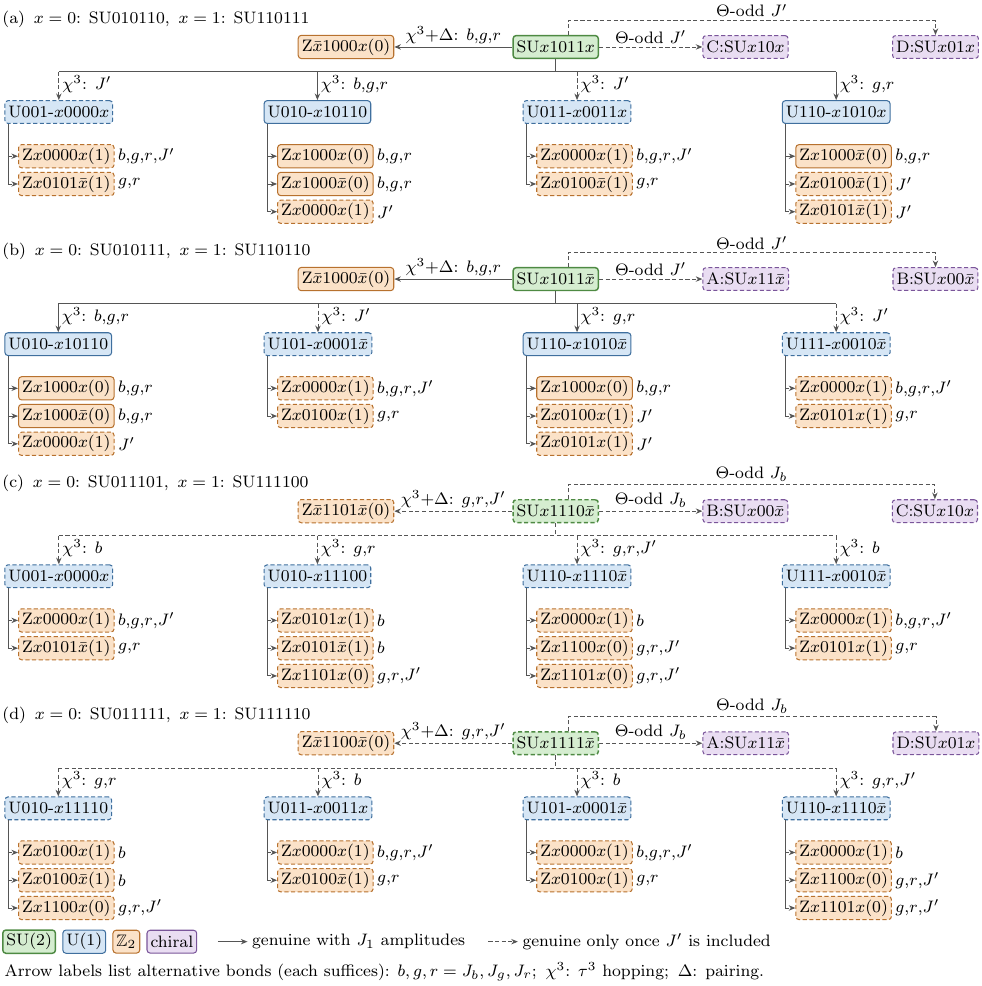}
\caption{Hierarchy of the symmetric parton mean-field states of the SHD lattice realized with bond amplitudes up to second neighbors ($J'$); labels as in Sec.~\ref{sec:hierarchy}. Each panel (a)--(d) shows two connected $SU(2)$ parents, related by flipping $n_{T_2}$ and $n_{C_6}$, whose identical hierarchies are labeled jointly with the digit $x$ ($x=0$ for the first parent in the panel heading, $x=1$ for the second, and $\bar{x}=1-x$). Box colors: green, $SU(2)$; blue, $U(1)$; orange, $\Z_2$; purple, chiral states with IGG $SU(2)$ and broken time reversal. Arrows point from parent to descendant. The arrow label names the bonds whose amplitudes, added to a generic parent ansatz, lower the IGG: $\chi^3$ ($\tau^3$ hopping) for $SU(2)\to U(1)$, and pairing $\Delta$ for $U(1)\to\Z_2$ (vertical lists). Comma-separated entries are alternatives, each sufficient by itself, with $b,g,r$ for $J_b,J_g,J_r$. The octahedral $\Z_2$ descendant (left of each parent) needs $\chi^3+\Delta$. Chiral descendants (right of each parent) require the listed $\Theta$-odd hopping. A $\Z_2$ state descending from several $U(1)$ states is listed under each of them. Solid boxes and arrows: the descendant is already genuine with nearest-neighbor ($J_1$) amplitudes; dashed: it becomes genuine only once $J'$ amplitudes are included (in panels (c) and (d) the parents themselves require $J'$). Descendants that first become genuine with $J''$ or $J'''$ amplitudes, the parents connected only at fourth neighbors, and the chiral descendants of the $U(1)$ and $\Z_2$ states are tabulated in Sec.~SIX of the Supplemental Material~\cite{SM}.}
\label{fig:hierarchy}
\end{figure*}

Figure~\ref{fig:hierarchy} shows the resulting hierarchy for the eight $SU(2)$ parents realized up to $J'$, which come in four pairs related by reversing $n_{T_2}$ and $n_{C_6}$ together. With $J_1$ amplitudes each of the four parents realized there has two $U(1)$ descendants, in the sectors $(0,1,0)$ and $(1,1,0)$, and three $\Z_2$ descendants, all with $t=0$, one of them octahedral; up to $J'$ these numbers grow to four and six. The descendants not shown in the figure, including those of the parents connected only at fourth neighbors, are listed in Sec.~SIX of the Supplemental Material~\cite{SM}. In the figure and in the Supplemental Material, SU and Z labels are those of Sec.~\ref{sec:algebraic} and U labels those of Table~\ref{tab:u1ans}; chiral labels give the class (A to D) and $n_{T_2}\,n_{\sigma}\,n_{\sigma C_6}\,n_{C_6}$ of the chiral PSG.

\section{Chiral Spin Liquids}
\label{sec:chiral}

\subsection{Realization and the Role of the Lattice}

A chiral PSG is realized only if its generic ansatz breaks time reversal; otherwise the state is an ordinary symmetric spin liquid. On the SHD lattice every nearest-neighbor ansatz is time-reversal symmetric at half filling. The transformation $W_i=\epsilon_i\tau^0$ of Sec.~\ref{sec:ansatz} reverses every nearest-neighbor link, singlet or triplet, $W_iu^{(\alpha)}_{ij}W_j^\dagger=-u^{(\alpha)}_{ij}$, so every nearest-neighbor bond ansatz satisfies Eq.~\eqref{eq:psg-T} with $G_\T(i)=\epsilon_i\tau^0$ for every PSG. On-site Lagrange multipliers are invariant under $W_i$ and can break time reversal algebraically, but at half filling they vanish self-consistently for a nearest-neighbor ansatz. At $a_i=0$, time reversal with $G_\T(i)=\epsilon_i\tau^0$ is an antiunitary symmetry of the mean-field Hamiltonian. An on-site term obeys Eq.~\eqref{eq:psg-T} with $j=i$, so this operation maps $a_i\to-\epsilon_i^2a_i=-a_i$, and the constraint operators $\psi_i^\dagger\tau^k\psi_i$ are odd under it. Their thermal expectation values therefore vanish at $a_i=0$, which thus satisfies $\langle\psi_i^\dagger\tau^k\psi_i\rangle=0$, and by the strict concavity of the free energy in the multipliers at $T>0$ (Sec.~\ref{sec:ansatz}) it is the only solution. At $T=0$ the statement holds in the limit $T\to0^+$. A chiral mean-field state on the SHD lattice therefore requires same-sublattice bonds, $J'$, $J''$, or $J'''$, each of which closes a triangle with nearest-neighbor bonds (Sec.~\ref{sec:lattice}).

Table~\ref{tab:chiralcounts} gives, for each IGG and range, the number of chiral PSGs whose generic ansatz with singlet bond amplitudes breaks time reversal on a connected bond graph. Chiral ans\"atze with triplet amplitudes obey the rules of Table~S9 of the Supplemental Material~\cite{SM} but are not scanned here; all chiral results below refer to singlet amplitudes. The counts coincide in all four chiral classes, although different labels appear in each; this coincidence is an observation of the enumeration, for which no bijection between the classes is derived here. In the $U(1)$ sector $(w_{C_6},w_\sigma)=(0,1)$ the counts at generic flux $\Phi$ equal those at $\Phi=0$. The PSGs not counted at $J''$ and $J'''$ also break time reversal, but only through bonds inside the hexagons, which form disconnected clusters.

\begin{table}
\caption{Number of chiral PSGs, in each of the classes A--D, whose generic singlet bond amplitudes break time reversal on a connected bond graph, as a function of the longest bond included; in parentheses the number whose generic bond amplitudes have exactly the IGG of the PSG. The last row gives the number of PSGs in each column; the $U(1)$ sectors are labeled by $(w_{C_6},w_\sigma)$, and in the sector $(0,1)$ the four PSGs are counted at $\Phi=0$ and at a generic flux $\Phi$. With nearest-neighbor bonds alone no PSG is chiral.}
\label{tab:chiralcounts}
\begin{ruledtabular}
\begin{tabular}{lcccccc}
 & $\Z_2$ & $SU(2)$ & \multicolumn{4}{c}{$U(1)$}\\
\cline{4-7}
range & & & $(0,0)$ & $(0,1)$ & $(1,0)$ & $(1,1)$\\
\hline
$+J'$ & 10 (6) & 2 (2) & 2 (0) & 2 (2) & 2 (2) & 2 (2)\\
$+J''$ & 16 (14) & 8 (8) & 4 (4) & 6 (6) & 6 (6) & 6 (6)\\
$+J'''$ & 16 (16) & 12 (12) & 4 (4) & 6 (6) & 8 (8) & 8 (8)\\
\hline
total & 16 & 16 & 4 & $4+4$ & 8 & 8
\end{tabular}
\end{ruledtabular}
\end{table}

\subsection{Time Reversal Broken Alone}
\label{sec:pt}

Class-A $U(1)$ spin liquids are excluded on many lattices by the argument of Ref.~\cite{BieriLhuillierMessio2016} [Eq.~(35) and Sec.~III~C~3 there], which extends to $SU(2)$. The flux $\Phi_C$ through an odd loop $C$ is reversed by time reversal, and if a mirror maps $C$ onto itself with reversed orientation, mirror symmetry also reverses $\Phi_C$, so that a mirror-symmetric state has $\Phi_C\in\{0,\pi\}$ and is time-reversal symmetric. The argument requires an odd loop mapped onto itself by a mirror. On the SHD lattice no site lies on a mirror line, so every loop mapped onto itself by a mirror crosses the mirror at two bond midpoints and has even length. In particular, no triangle built from $J'$, $J''$, or $J'''$ bonds is self-mirrored, and the mirrors map a triangle with flux $\Phi$ onto a different triangle, which carries $-\Phi$. Mirror symmetry is then compatible with a pattern of triangle fluxes $\pm\Phi$, $\Phi\neq0,\pi$, whose sign alternates under the mirrors, and an ansatz with such a pattern breaks time reversal. In this section, ``staggered'' refers to such sign patterns of time-reversal-odd quantities under point-group operations, not to translations combined with $\T$. This happens for $\Z_2$, $U(1)$, and $SU(2)$ spin liquids alike. For an $SU(2)$ ansatz every triangle carries $\Phi=\pm\pi/2$ (Sec.~\ref{sec:flux}), and class-A $SU(2)$ states break time reversal as soon as their bond graph contains an odd loop, e.g., a $J_b$--$J_r$--$J'$ triangle.

Table~\ref{tab:pt} lists the $\Z_2$ PSGs whose class-A ansatz breaks time reversal, together with the range at which this first happens. Four of them break time reversal already at $J'$ with an enlarged $U(1)$ or $SU(2)$ IGG and become $\Z_2$ spin liquids at $J''$ or $J'''$. With $J''$ bonds the generic singlet ansatz of every class-A $\Z_2$ PSG breaks time reversal. All these states have zero net chirality, spinon Chern number, and thermal Hall conductivity $\kappa_{xy}$, because the unitary mirrors stagger the triangle fluxes and reverse the Berry curvature.

\begin{table}
\caption{$\Z_2$ PSGs of chiral class A, in which all elements of $p6mm$ are unitary symmetries, whose generic singlet ansatz breaks time reversal. Labels are $n_{T_2}n_\sigma n_{\sigma C_6}n_{C_6}$ (digit 0 for $+1$ and 1 for $-1$). The second column gives the shortest range at which the ansatz breaks time reversal and has $\Z_2$ IGG; the third, the triangles whose flux is odd under time reversal.}
\label{tab:pt}
\begin{ruledtabular}
\begin{tabular}{lcc}
PSGs & range & odd triangles\\
\hline
0000, 0001, 0010, 1000, 1001, 1011 & $J'$ & $J_b$--$J_r$--$J'$\\
0100, 1101 & $J''$ & $J_b$--$J_r$--$J'$\\
0111, 1110 & $J'''$ & $J_b$--$J_r$--$J'$\\
0011, 0101, 0110, 1010, 1100, 1111 & $J''$ & $J_g$--$J_r$--$J''$
\end{tabular}
\end{ruledtabular}
\end{table}

\subsection{Chern Numbers, Edge States, and Thermal Hall Response}

The symmetry operations constrain the Berry curvature $F(\bm{k})$ of the occupied spinon bands and the Chern number $C=(2\pi)^{-1}\int_{\rm BZ}d^2k\,F(\bm{k})$. We use $F=\partial_{k_x}\mathcal{A}_y-\partial_{k_y}\mathcal{A}_x$ with $\bm{\mathcal{A}}=\ii\sum_n\langle u_n|\nabla_{\bm{k}}u_n\rangle$, summed over the occupied Bloch states $u_n$, in the right-handed axes of Fig.~\ref{fig:lattice}; with this sign, Eq.~\eqref{eq:kxy} below gives $\kappa_{xy}/T$ of the same sign as $C$. A unitary rotation maps $F(\bm{k})\to F(R\bm{k})$ and preserves $C$; a unitary mirror maps $F(\bm{k})\to-F(\sigma\bm{k})$ and reverses $C$, as does time reversal, $F(\bm{k})\to-F(-\bm{k})$. In class A the unitary mirror forces $C=0$. In classes C and D the antiunitary element $C_2\T=(C_6\T)^3$ leaves every momentum invariant and reverses the Berry curvature, so that $C=0$ again; in the periodic gauge it gives $F(\bm{k})=-F(\bm{k})$ pointwise. These constraints hold for every ansatz of the class, with singlet or triplet amplitudes, for either sign of $\eta_{12}$, and for degenerate bands. Class B, in which the rotations are unitary and the mirrors are combined with time reversal, is therefore the only one of the four chiral classes that admits a nonzero Chern number of the mean-field spinon bands, a quantized thermal Hall conductivity, and a uniform scalar spin chirality. The rule is that of Ref.~\cite{BieriLhuillierMessio2016} applied to $p6mm$. The same operations determine the symmetry of the dynamical structure factor. A unitary twofold rotation guarantees $S(\bm{q},\omega)=S(-\bm{q},\omega)$; it is present in classes A and B but not in classes C and D, where the structure factor is in general asymmetric under $\bm{q}\to-\bm{q}$.

For singlet ans\"atze the spinon problem reduces to the number-conserving block of $(f_{\uparrow},f^\dagger_{\downarrow})$, which has a particle-hole antisymmetry $\Xi H(\bm{k})^*\Xi^\dagger=-H(-\bm{k})$ with $\Xi\Xi^*=-1$, that of Altland--Zirnbauer class C~\cite{AltlandZirnbauer1997,Schnyder2008}. In two dimensions this class has even Chern numbers, so the Chern number $C_\psi$ of the block is even and that of the full Bogoliubov--de Gennes problem is $\nu=2C_\psi\in4\Z$. For $U(1)$ and $SU(2)$ ans\"atze we quote the Chern number $C$ per spin species, with $C_\psi=2C$.

\begin{figure*}
\centering
\includegraphics[width=\textwidth]{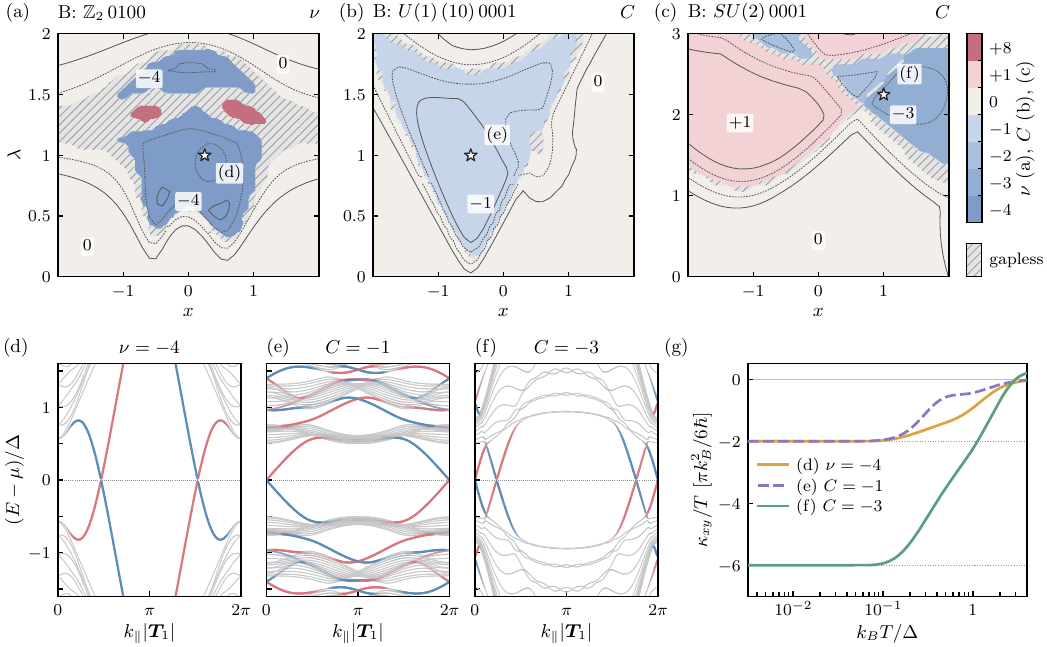}
\caption{Chern numbers, chiral edge modes, and thermal Hall response of class-B chiral spin liquids with singlet amplitudes up to $J'$. (a)--(c) Chern number on two-parameter slices of the ansatz space for (a) the $\Z_2$ PSG 0100, (b) the $U(1)$ PSG 0001 in the sector $(w_{C_6},w_\sigma)=(1,0)$, and (c) the $SU(2)$ PSG 0001 (labels $n_{T_2}n_\sigma n_{\sigma C_6}n_{C_6}$). The horizontal axis is the signed ratio $x$ of the $J_r$ and $J_b$ amplitudes, and the vertical axis the magnitude $\lambda$ of the $J'$ amplitude, the only time-reversal-odd part of the ansatz; the directions of the amplitudes in the allowed spaces are listed in Sec.~SX of the Supplemental Material~\cite{SM}. Colors give $\nu=2C_\psi$ in (a) and the Chern number $C$ per spin in (b) and (c); the small rose pockets in (a) have $\nu=+8$, and the narrow light-blue strips in (c) have $C=-2$. Hatched regions are gapless at half filling or have a gap below $5\times10^{-3}W$, with $W$ the spinon bandwidth; thin lines are contours of the spinon gap at $\Delta/W=0.02$ (dotted) and $0.05$ (solid). Stars mark the states of (d)--(g): $\nu=-4$ with $\Delta=0.016W$ (a); a Kalmeyer--Laughlin-type state with $C=-1$ and $\Delta=0.115W$ (b); and $C=-3$ with $\Delta=0.033W$ (c). (d)--(f) Spinon spectra of ribbons periodic along $\bm{T}_1$, fourteen unit cells wide, in units of $\Delta$ and measured from the chemical potential $\mu$, colored by the weight near the lower (red) and upper (blue) edge; bulk states are gray. Shown are the $(f_\uparrow,f^\dagger_\downarrow)$ block in (d), where each mode is a complex fermion, and the spin-up problem in (e) and (f). Each edge carries $|C_\psi|=2$ (d), $|C|=1$ (e), and $|C|=3$ (f) chiral modes. (g) Thermal Hall conductivity of the free spinons, Eq.~\eqref{eq:kxy}, in units of $\pi k_B^2/(6\hbar)$; dotted lines mark the free-spinon chiral central charges $\nu/2=-2$, $2C=-2$, and $2C=-6$. The plateaus are those of the mean-field spinons; the values expected after projection are $c=-2$ for the $\Z_2$ state, $c=-1$ for the Kalmeyer--Laughlin-type state, and $c=-21/5$ for the $SU(2)$ state with $C=-3$ (Sec.~\ref{sec:chiral}).}
\label{fig:chiral-topo}
\end{figure*}

Within class B, we compute the Chern numbers of gapped mean-field spinon bands with the Fukui--Hatsugai--Suzuki method~\cite{FukuiHatsugaiSuzuki2005}, on two-parameter slices of the ansatz space and on random samples of all class-B PSGs (Sec.~SX of the Supplemental Material~\cite{SM}). Figure~\ref{fig:chiral-topo}(a)--(c) shows representative maps. In each computed class-B map the time-reversal-symmetric bipartite ansatz at zero same-sublattice amplitude is a boundary of the chiral phases, and $\nu$ is odd under the sign of the same-sublattice amplitude. In the maps and samples we find $\nu\in\{0,\pm4,\pm8,-12,-16\}$ for $\Z_2$ states, $C\in\{0,\pm1,\pm2,+3,\pm4\}$ per spin for $U(1)$ states, and $C\in\{-4,-3,-2,-1,0,1,2\}$ for $SU(2)$ states. Gap closings at the phase boundaries occur at $\Gamma$, at the three $M$ points, and at $\pm K$. These momenta account for the steps of $\pm1$, $\pm2$, and $\pm3$ in the Chern number per spin of $U(1)$ and $SU(2)$ states between neighboring phases (and multiples of $4$ in $\nu$ for $\Z_2$ states).

For the three states marked in Fig.~\ref{fig:chiral-topo}(a)--(c), the number of chiral edge modes on ribbons equals the bulk invariant on both inequivalent terminations, with edges parallel to $\bm{T}_1$ and to $\bm{T}_1+\bm{T}_2$ [Fig.~\ref{fig:chiral-topo}(d)--(f)]. The thermal Hall conductivity of the free spinons is~\cite{SumiyoshiFujimoto2013}
\begin{equation}
\frac{\kappa_{xy}}{T}=d_s\,\frac{3}{\pi^2}\sum_n\int_{\rm BZ}\frac{d^2k}{2\pi}\,\Omega_n(\bm{k})\,g_2\!\left(\frac{\varepsilon_n(\bm{k})-\mu}{k_BT}\right),
\label{eq:kxy}
\end{equation}
in units of $\pi k_B^2/(6\hbar)$. Here $\Omega_n$ is the Berry curvature of band $n$, $g_2(x)=\int_x^\infty y^2(-\partial_yf)\,dy$ with $f$ the Fermi function, and $d_s=1$ for the $(f_\uparrow,f^\dagger_\downarrow)$ block of a $\Z_2$ state and $d_s=2$ for the two spin species of a $U(1)$ or $SU(2)$ state. For $k_BT\lesssim0.1\Delta$, where $\Delta$ is the spinon gap, $\kappa_{xy}/T$ approaches the plateau set by the free-spinon chiral central charge [Fig.~\ref{fig:chiral-topo}(g)]. Only for $\Z_2$ states is this the value expected after projection; for $U(1)$ and $SU(2)$ states the gauge field reduces it, as discussed next.

These Chern numbers are invariants of the mean-field spinon bands, and the topological order of the projected states is inferred from them. After projection, a $U(1)$ state with $C=\pm1$ per spin is a Kalmeyer--Laughlin spin liquid~\cite{KalmeyerLaughlin1987,WenWilczekZee1989} with semion topological order and chiral central charge $c=\pm1$. This is half the free-spinon value, because the emergent gauge field removes one chiral boson. A $\Z_2$ state with Bogoliubov--de Gennes Chern number $\nu$ realizes the sector $\nu\bmod16$ of Kitaev's sixteenfold way~\cite{Kitaev2006}, with $c=\nu/2$ already at mean-field level and vortices of topological spin $\theta=e^{\ii\pi\nu/8}$. Since $\nu\in4\Z$, the $\Z_2$ states found here realize four types of topological order: semion$\times$semion ($\theta=\ii$) for $\nu=4$ and $\nu=-12$ (with $c=2$ and $c=-6$); antisemion$\times$antisemion ($\theta=-\ii$, $c=-2$) for $\nu=-4$, the case of the state of Fig.~\ref{fig:chiral-topo}(d); three-fermion ($\theta=-1$) for $\nu=\pm8$; and toric-code anyons with $c=-8$ for $\nu=-16$. The $U(1)$ states with $|C|\geq2$ per spin are not analyzed here. Chiral $SU(2)$ states with $|C|>1$ are expected to be non-Abelian, since integrating out the spinons gives an $SU(2)$ Chern--Simons term at level $|C|$~\cite{Wen1991}. The expected topological order is $U(2|C|)_1/SU(2)_{|C|}$, with chiral central charge $|c|=2|C|-3|C|/(|C|+2)$, which reduces to the Kalmeyer--Laughlin value for $|C|=1$ and gives $c=-21/5$ for the state with $C=-3$ of Fig.~\ref{fig:chiral-topo}(c), compared with the free-spinon value $-6$. These states are not analyzed further.

\begin{figure}
\centering
\includegraphics[width=\columnwidth]{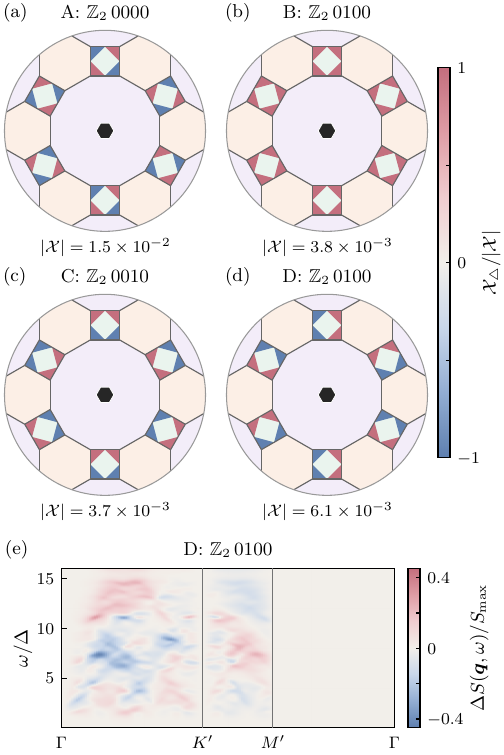}
\caption{Scalar spin chirality and structure-factor asymmetry of the four chiral classes. (a)--(d) Mean-field scalar spin chirality $\mathcal{X}_\triangle=\langle\bm{S}_i\cdot(\bm{S}_j\times\bm{S}_k)\rangle$, with $i,j,k$ ordered counterclockwise, on the triangles formed by two nearest-neighbor bonds and one $J'$ diagonal of a square, for one gapped $\Z_2$ spin liquid of each chiral class (ans\"atze in Sec.~SX of the Supplemental Material~\cite{SM}), on a patch centered on a dodecagon. Each square contains four such triangles, one at each corner; each triangle is drawn at half size, scaled about its corner site (the site opposite its $J'$ bond), and colored by $\mathcal{X}_\triangle/|\mathcal{X}|$. All these triangles are related by elements of $p6mm$, so $|\mathcal{X}_\triangle|=|\mathcal{X}|$ is the same on every triangle of a state (value below each patch), and the classes differ only in the sign pattern. (a) Class A, PSG 0000: $\mathcal{X}_\triangle$ is even under $C_6$ and odd under the mirrors, so that adjacent corners of every square carry opposite signs. (b) Class B, PSG 0100: uniform chirality. (c) Class C, PSG 0010, and (d) class D, PSG 0100: $\mathcal{X}_\triangle$ is odd under $C_6$, so that opposite corners of every square carry opposite signs, and odd (C) or even (D) under $\sigma$. PSG labels are $n_{T_2}n_\sigma n_{\sigma C_6}n_{C_6}$. (e) Asymmetry $\Delta S(\bm{q},\omega)=S(\bm{q},\omega)-S(-\bm{q},\omega)$ of the mean-field dynamical structure factor of the class-D state of (d) along $\Gamma\to K'\to M'\to\Gamma$ in the extended Brillouin zone, normalized by the maximum $S_{\max}$ of $S(\bm{q},\omega)$, with $\omega$ in units of the spinon gap $\Delta$. It vanishes on the $\Gamma$--$M'$ line, which lies on the mirror line of $\sigma$, because the antiunitary symmetry $\sigma\T$ maps $\bm{q}$ on this line to $-\bm{q}$. Panel (e) uses the commensurate site positions of Appendix~\ref{app:bz}.}
\label{fig:chiral-patterns}
\end{figure}

For singlet ans\"atze in each of the four classes, overlapping bands of the number-conserving $(f_\uparrow,f^\dagger_\downarrow)$ problem produce compensated Fermi pockets, a chiral spinon Fermi surface, in open regions of parameter space. For point nodes the class matters. In classes A and B no antiunitary symmetry leaves a generic momentum invariant, so a band crossing at the Fermi level has codimension three, and nodes occur only on the phase boundaries of the Chern-number maps. One exception is not explained by the symmetry group: for the class-B $\Z_2$ PSGs 0000, 0001, 1000, and 1001 with amplitudes up to $J''$, all bands of the $(f_\uparrow,f^\dagger_\downarrow)$ problem are twofold degenerate at every momentum. For B:0000, no operation that combines a space-group element with an on-site gauge transformation, unitary or antiunitary, leaves the ansatz invariant beyond those of the PSG, so the degeneracy has another origin, which is not identified here. It is lifted by $J'''$ amplitudes, and the statements on point nodes refer to the generic case. In classes C and D, $C_2\T$ acts at fixed momentum, and its square depends on the PSG. For the $\Z_2$ PSGs with $\eta_{12}=-\eta_{C_6}$, labeled 0001, 0011, 0101, 0111, 1000, 1010, 1100, and 1110 in both classes, $(C_2\T)^2=+1$ on the Bogoliubov--de Gennes problem and $C_2\T$ acts as an effective reality condition. Crossings then have codimension two, and Dirac nodes with Berry phase $\pm\pi$ are stable in open regions. For the other eight $\Z_2$ PSGs of each class, with $\eta_{12}=\eta_{C_6}$, $(C_2\T)^2=-1$ and every band is twofold degenerate at every momentum, as random ans\"atze with amplitudes up to $J'''$ show for all sixteen PSGs of each class. A crossing of two such doublets has codimension five, so these PSGs have no generic Dirac nodes. At mean-field level, gapless Dirac-type chiral $\Z_2$ spin liquids are therefore generic for the first eight $\Z_2$ PSGs of classes C and D and fine-tuned otherwise. In the per-spin hopping problem of the $U(1)$ and $SU(2)$ PSGs of classes C and D the pattern differs: the bands are twofold degenerate at every momentum for the $SU(2)$ PSGs and the $U(1)$ PSGs with $w_{C_6}=1$ that have $\eta_{12}=-\eta_{C_6}$, and nondegenerate for the others.

The scalar spin chirality $\mathcal{X}_\triangle=\langle\bm{S}_i\cdot(\bm{S}_j\times\bm{S}_k)\rangle$ on the elementary triangles distinguishes the classes in real space [Fig.~\ref{fig:chiral-patterns}(a)--(d)]. It is uniform in class B, staggered under the mirrors in class A, and staggered under $C_6$ in classes C and D, which differ in their behavior under $\sigma$. The net chirality vanishes in all classes but B. In momentum space, the dynamical structure factor of classes C and D is asymmetric under $\bm{q}\to-\bm{q}$, with differences reaching $0.42$ of the maximum of $S(\bm{q},\omega)$ for the state of Fig.~\ref{fig:chiral-patterns}(e), while the static structure factor is even in all classes. An asymmetry of the inelastic neutron-scattering intensity under $\bm{q}\to-\bm{q}$ can separate the rotation-breaking classes C and D from classes A and B, and thermal Hall measurements can single out class B. The same symmetry reasoning applies to bosonic partons. In Schwinger-boson states on the honeycomb lattice, a projective $C_6\T$ symmetry forces the zero-field thermal Hall response of a chiral flux state to vanish, and the sign and temperature dependence of $\kappa_{xy}$ distinguish parton flux sectors at fixed couplings~\cite{Sasamoto2026}.
\section{Spinon Spectra and Spin Correlations}
\label{sec:spectra}

At half filling, with one fermion per site, a spin liquid is gapped, has nodes at the Fermi level, or has Fermi pockets. Which of these occurs is decided by the indirect gap between bands $N_b/2$ and $N_b/2+1$ of the $N_b$-band problem per spin. The spinon gap quoted in this paper is this indirect band gap, the threshold of the two-spinon continuum; the quasiparticle gap $\min|E|$ of a Bogoliubov--de Gennes spectrum and the gap $\Delta$ of Sec.~SVII of the Supplemental Material~\cite{SM} are half of it. We denote by $\chi_o$ the amplitude on the reference bond of orbit $o$ (Table~\ref{tab:bonds}), with $\chi_{12}$, $\chi_{23}$, and $\chi_{27}$ on $J_b$, $J_g$, and $J_r$, and $\chi_{28}$, $\chi_{26}$, and $\chi_{24}$ on $J'$, $J''$, and $J'''$. For $SU(2)$ states $u_{ij}=\ii\chi_o\tau^0$, and for $U(1)$ states with $w_\T=1$, $u_{ij}=\chi_o\tau^3$. For $U(1)$ states the on-site multiplier $a^3_i\tau^3$ is determined by the local constraint $\langle n_i\rangle=1$. In the sectors with $w_\sigma=1$ or $w_{C_6}=1$ the PSG allows only a staggered multiplier with zero average (alternating between the sublattices $A$ and $B$ for $w_\sigma=1$, $w_{C_6}=0$, and with period four in $\mu$ for $w_{C_6}=1$), which is not a chemical potential. This multiplier must be nonzero as soon as same-sublattice bonds are present; without it the local occupation deviates from one by up to about $20\%$.

\subsection{$SU(2)$ Spin Liquids: Closed-Form Spectra and Nodal Lines}

\begin{figure*}
\centering
\includegraphics[width=\textwidth]{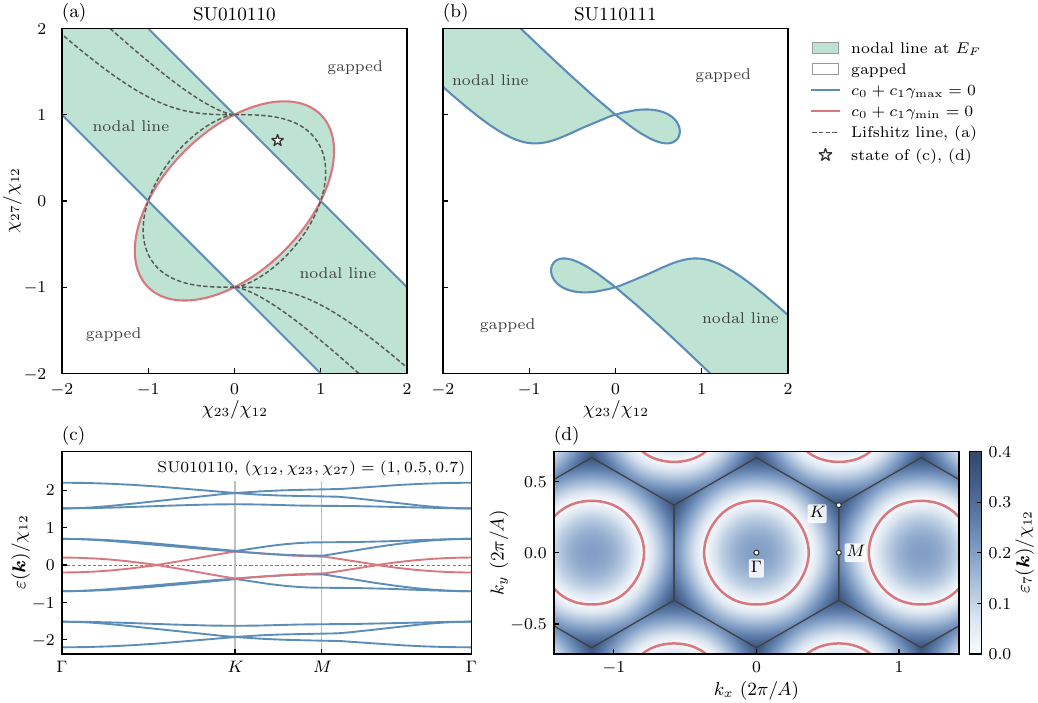}
\caption{Spinon spectra of the nearest-neighbor $SU(2)$ states with gapless regions, with singlet hoppings $u_{ij}=\ii\chi_o\tau^0$ on the $J_b$, $J_g$, and $J_r$ bonds and $E_F=0$ by sublattice symmetry. (a), (b) Closed-form phase diagrams of SU010110 and SU110111 (doubled cell) at half filling. With $\det B(\bm{k})=c_0+c_1\gamma(\bm{k})$, Eq.~\eqref{eq:detB}, the gapless phases (green) are nodal-line phases bounded by $c_0+c_1\gamma_{\max}=0$ (blue) and $c_0+c_1\gamma_{\min}=0$ (rose). For SU010110 these are the lines $|\chi_{23}+\chi_{27}|=\chi_{12}$ and the ellipse $\chi_{23}^2-\chi_{23}\chi_{27}+\chi_{27}^2=\chi_{12}^2$, Eq.~\eqref{eq:su2region}; on the dashed Lifshitz line the nodal line passes through $M$. Both diagrams are symmetric under the gauge equivalence $(\chi_{23},\chi_{27})\to-(\chi_{23},\chi_{27})$. (c) Bands of SU010110 at $(\chi_{12},\chi_{23},\chi_{27})=(1,0.5,0.7)$ [star in (a)], in units of $\chi_{12}$; the sixth and seventh bands (rose) cross $E_F$ where the path meets the nodal line. (d) Lowest positive band of the same state with the nodal line $\gamma(\bm{k})=-c_0/c_1$ (rose), a ring around $\Gamma$; momenta in units of $2\pi/A$.}
\label{fig:spec-su2}
\end{figure*}

For a nearest-neighbor $SU(2)$ ansatz the hopping is bipartite, $h(\bm{k})=\begin{pmatrix}0&B(\bm{k})\\B^\dagger(\bm{k})&0\end{pmatrix}$, the spectrum is $\pm$ the singular values of the $6\times6$ block $B$, and zero modes occur where $\det B(\bm{k})=0$. For the four connected nearest-neighbor $SU(2)$ states of Table~\ref{tab:su2ans} the determinant takes the form
\begin{equation}
\det B(\bm{k})=c_0+c_1\,\gamma(\bm{k}),
\label{eq:detB}
\end{equation}
with real coefficients $c_0$ and $c_1$ that are polynomials in $\chi_{12}$, $\chi_{23}$, and $\chi_{27}$, and a single real harmonic, $\gamma(\bm{k})=\cos k_1+\cos k_2+\cos(k_1-k_2)$ with $k_i=\bm{k}\cdot\bm{T}_i$ for $\eta_{12}=+1$, and $\gamma(2k_1+\pi,2k_2+\pi)$ in the doubled cell for $\eta_{12}=-1$. Since $\det B$ is real, its zero set is generically a curve, so the gapless states among these four are nodal-line states, with a line of zero-energy touchings along the contour $\gamma(\bm{k})=-c_0/c_1$ and a finite density of states at the Fermi level. For SU010110 the coefficients are $c_0=(\chi_{27}^2-\chi_{12}^2)^3+\chi_{23}^3(\chi_{23}^3+2\chi_{27}^3)$ and $c_1=2\chi_{12}^2\chi_{23}^3\chi_{27}$. The state is gapless when $-c_0/c_1$ lies in the range of $\gamma$, which gives the region
\begin{align}
&(\chi_{12}-\chi_{23}-\chi_{27})(\chi_{12}+\chi_{23}+\chi_{27})\nonumber\\
&\quad\times\left(\chi_{12}^2-\chi_{23}^2+\chi_{23}\chi_{27}-\chi_{27}^2\right)\le0,
\label{eq:su2region}
\end{align}
bounded in the quadrant $\chi_{23},\chi_{27}>0$ by the line $\chi_{27}=\chi_{12}-\chi_{23}$ and the ellipse $\chi_{23}^2-\chi_{23}\chi_{27}+\chi_{27}^2=\chi_{12}^2$ (Fig.~\ref{fig:spec-su2}). On these boundaries the nodal line shrinks to a quadratic band touching at $\Gamma$ or at $\pm K$, and inside the region the line changes from a ring around $\Gamma$ to rings around $\pm K$ at a Lifshitz transition where it passes through $M$. SU110110 is gapped whenever all three amplitudes are nonzero; its gap closes only in the disconnected limits $\chi_{27}\to0$ with $|\chi_{12}|=|\chi_{23}|$ and $\chi_{12}\to0$ with $\chi_{23}=\chi_{27}$. SU010111 is gapped except on the single ray $|\chi_{12}|:|\chi_{23}|:|\chi_{27}|=\sqrt3:2:1$ with $\chi_{23}\chi_{27}<0$, where $c_0+3c_1=0$ and a twofold-degenerate Dirac cone appears at $\Gamma$ (the singlet counterpart of the gapless Kitaev point, Sec.~\ref{sec:kitaev}). SU110111 is gapless in a lobe near $(\chi_{23},\chi_{27})=(0,\chi_{12})$, which is the tip of a gapless region extending over much of the quadrants with $\chi_{23}\chi_{27}<0$.

\begin{table}[h]
\caption{Fluxes of the nearest-neighbor $SU(2)$ states of Table~\ref{tab:su2ans} through a square ($\Phi_s$), a hexagon ($\Phi_h$), and a dodecagon ($\Phi_d$), $P_C\propto e^{\ii\Phi_C}\tau^0$, in the sectors $s=\operatorname{sgn}(\chi_{23}\chi_{27})$, with the spinon spectrum at half filling. Every row satisfies Eq.~\eqref{eq:fluxsum}, with $\Phi=0$ for $\eta_{12}=+1$ and $\Phi=\pi$ for $\eta_{12}=-1$. SU010111 with $s=-$ is the Lieb-flux state (Sec.~\ref{sec:kitaev}); its gap closes only on the ray $|\chi_{12}|:|\chi_{23}|:|\chi_{27}|=\sqrt3:2:1$. The gapless lobe of SU110111 near $(\chi_{23},\chi_{27})=(0,\chi_{12})$ extends into both sectors.}
\label{tab:su2flux}
\begin{ruledtabular}
\begin{tabular}{lcccl}
PSG & $s$ & $(\Phi_s,\Phi_h,\Phi_d)$ & $\Phi$ & spectrum\\
\hline
SU010110 & $+$ & $(0,0,0)$ & $0$ & nodal lines, Eq.~\eqref{eq:su2region}\\
 & $-$ & $(0,\pi,0)$ & $0$ & nodal lines, Eq.~\eqref{eq:su2region}\\
SU010111 & $+$ & $(\pi,\pi,\pi)$ & $0$ & gapped\\
 & $-$ & $(\pi,0,\pi)$ & $0$ & gapped except on one ray\\
SU110110 & $+$ & $(\pi,\pi,0)$ & $\pi$ & gapped\\
 & $-$ & $(\pi,0,0)$ & $\pi$ & gapped\\
SU110111 & $+$ & $(0,0,\pi)$ & $\pi$ & nodal lines in the lobe\\
 & $-$ & $(0,\pi,\pi)$ & $\pi$ & nodal lines in a large region
\end{tabular}
\end{ruledtabular}
\end{table}

The PSG alone does not fix the fluxes of these states. For $u_{ij}=\ii\chi_{ij}\tau^0$ the flux operator of a face $C$ of length $L$ is $P_C=\ii^L\prod_C\chi_{ij}\,\tau^0\propto e^{\ii\Phi_C}\tau^0$, where $\Phi_C$ is the flux of the spinon hopping through $C$. Reversing the sign of one orbit's amplitude shifts $\Phi_C$ by $\pi$ on every face that contains an odd number of its bonds. A square contains two $J_b$ and two $J_r$ bonds, a hexagon three $J_g$ and three $J_r$ bonds, and a dodecagon six $J_b$ and six $J_g$ bonds. The PSG therefore fixes $\Phi_s$ and $\Phi_d$, with $\Phi_d=\pi$ exactly when $\eta_{C_6}=-1$, whereas $\Phi_h$, and with it the flux sector, is fixed by $\operatorname{sgn}(\chi_{23}\chi_{27})$; the sign of $\chi_{12}$ changes no flux. Table~\ref{tab:su2flux} lists both sectors of each state. Since $2\Phi_h=0\pmod{2\pi}$, both sectors satisfy Eq.~\eqref{eq:fluxsum}.

The nodal lines are protected by two symmetries. The sublattice structure of the ansatz, which follows from the time-reversal signs $\eta_{\T\sigma}=-1$, $\eta_{\T C_6}=+1$, makes $h(\bm{k})$ off-diagonal, and the antiunitary symmetry $C_2\T$, which acts at fixed momentum, makes $\det B$ real. The sign of $\det B$ is then a $\Z_2$ invariant at every momentum off the line. It changes across the line, so the line can disappear only by shrinking to a point. Since these PSGs allow no singlet amplitude beyond nearest neighbors, the lines are stable within their $SU(2)$ singlet sector. Triplet amplitudes with the spin axis perpendicular to the plane gap them linearly, whereas in-plane triplet amplitudes only split them. For SU010110 at $(\chi_{12},\chi_{23},\chi_{27})=(1,0.5,0.9)$, the PSG-allowed out-of-plane triplet amplitudes open a quasiparticle gap $\min|E|\simeq0.28\,t$ for $t\le0.04$, with $t$ the norm of the triplet amplitudes on the three reference bonds, while with in-plane amplitudes of the same size zero modes persist and the number of zero crossings along straight cuts through the zone roughly doubles.

The time-reversal sector $\eta_{\T\sigma}=\eta_{\T C_6}=-1$, in which $J'$ and $J'''$ bonds are allowed (Sec.~\ref{sec:ansatz}), has point nodes instead. For SU011101, whose bonds are bipartite with respect to $h(\mu)$ but not to the sublattices, the block $B$ between the sites with $h=+1$ and $h=-1$ has a complex determinant, whose zeros are isolated points with winding $\pm1$. A large part of its parameter space is a Dirac phase with six nodes on the $\Gamma$--$M$ lines. The boundaries of this phase include the curves on which $\det B(M)$ vanishes, $F_1F_2=0$ with two cubic polynomials $F_{1,2}$ of the amplitudes. On these curves pairs of nodes of opposite winding merge at $M$ into semi-Dirac points of an $SU(2)$ state.

\subsection{$U(1)$ Spin Liquids}

Whether point nodes are stable in a $U(1)$ spin liquid is decided by the symmetries that act at fixed momentum. In the sectors $(w_{C_6},w_\sigma)=(0,0)$ and $(0,1)$ the antiunitary $C_2\T$ makes $h(\bm{k})$ real in a suitable gauge, Dirac nodes carry a Berry phase $\pi$, and they are stable in open regions of parameter space. In the sectors with $w_{C_6}=1$ the twofold rotation is realized with $\ii\tau^1$ and the Berry curvature is nonzero. Within these sectors, for $w_\T=0$ the gap can close only on surfaces of codimension one in parameter space. For $w_\T=1$ the twofold rotation itself acts at fixed momentum as an antiunitary operation that reverses the sign of $h(\bm{k})$, whereas $C_2\T$ maps $\bm{k}$ to $-\bm{k}$ (Sec.~\ref{sec:qsh}). The $U(1)$ states of these sectors that we studied with nearest-neighbor and $J'$ amplitudes are gapped except on critical surfaces, on which the gap closes at isolated Dirac points. With $J''$ bonds, however, U101-000001 has a Dirac phase with six nodes on the $\Gamma$--$M$ lines (Table~S10 of the Supplemental Material~\cite{SM}). In U101-100010 the spectrum contains exactly flat zero-energy bands when $\chi_{28}=\chi_{12}$ and $\chi_{27}=\chi_{23}$.

\begin{figure*}
\centering
\includegraphics[width=\textwidth]{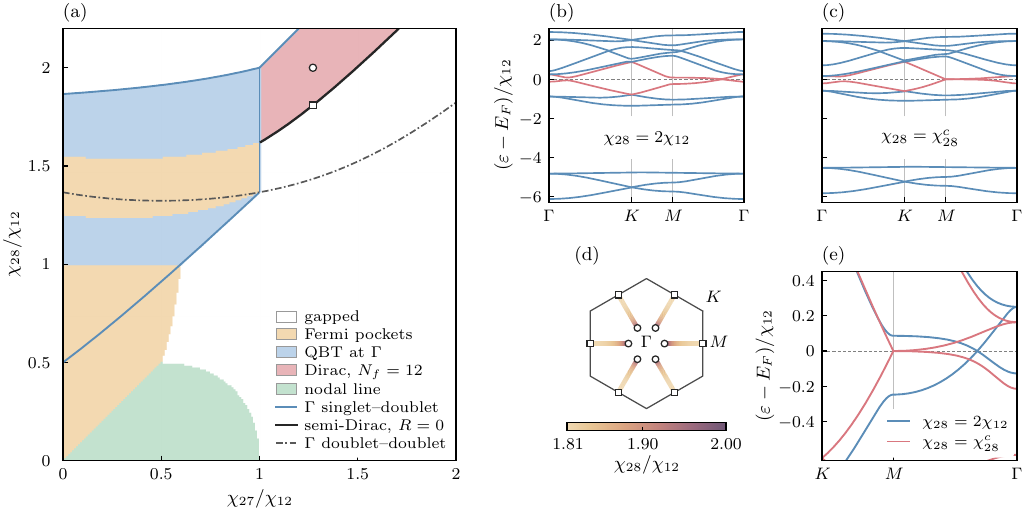}
\caption{Spinon spectra of the $U(1)$ state U001-000000 with singlet hoppings $\chi_o\tau^3$ on the $J_b$, $J_g$, $J_r$, and $J'$ bonds, $\chi_{12}=\chi_{23}=1$, and the chemical potential fixed by half filling. (a) Phase diagram: gapped, Fermi pockets, quadratic band touching at $\Gamma$ pinned by a $C_{6v}$ doublet, Dirac phase with six nodes on the $\Gamma$--$M$ lines ($N_f=12$), and a nodal line at constant energy (small $\chi_{28}$). Lines are closed-form: crossings of a singlet and a doublet (blue) or of two doublets (dash-dotted) at $\Gamma$ at the Fermi level, and the semi-Dirac line $R=0$ (black). The circle marks $(\chi_{27},\chi_{28})=(1.272,2)$ and the square the semi-Dirac point $(1.272,\chi_{28}^c)$, $\chi_{28}^c=1.8088$. (b), (c) Bands at these two points, measured from $E_F$; the sixth and seventh bands are rose. (d) Positions of the six Dirac nodes for $\chi_{28}^c<\chi_{28}\le2$ at $\chi_{27}=1.272$ (color scale; circles at $\chi_{28}=2$); the nodes $\bm{k}_D$ and $\bm{G}-\bm{k}_D$ meet at the $M$ points (squares) at $\chi_{28}^c$; the node positions follow the closed form of Sec.~SVI of the Supplemental Material~\cite{SM}. (e) Bands near $E_F$ along $K\to M\to\Gamma$: a linear crossing on $M\Gamma$ at $\chi_{28}=2$, and at $\chi_{28}^c$ a touching at $M$ that is linear along $KM$ and quadratic along $M\Gamma$.}
\label{fig:spec-u1}
\end{figure*}

We discuss in detail U001-000000, the $U(1)$ state without projective signs and with $w_\T=1$. Its levels at $\Gamma$ are known in closed form (in units in which the hopping on orbit $o$ is $-\chi_o$): four singlets $-(\chi_{12}+\chi_{23}+\chi_{27}+\chi_{28})$, $\chi_{12}-\chi_{23}-\chi_{27}+\chi_{28}$, $\chi_{12}+\chi_{23}+\chi_{27}-\chi_{28}$, and $-\chi_{12}+\chi_{23}+\chi_{27}+\chi_{28}$, and four doublets $\chi_{28}\pm q_1$ and $-\chi_{28}\pm q_2$, with $q_{1,2}=(\chi_{12}^2\pm\chi_{12}\chi_{23}\mp2\chi_{12}\chi_{27}+\chi_{23}^2-\chi_{23}\chi_{27}+\chi_{27}^2)^{1/2}$. Figure~\ref{fig:spec-u1} shows the phase diagram at $\chi_{12}=\chi_{23}=1$. Where a doublet sits at the Fermi level, the twofold representation of $C_{6v}$ enforces a quadratic band touching at $\Gamma$; this phase occupies large regions of the diagram. For $\chi_{27}>\chi_{12}$ a singlet drops below the doublet, and the band inversion between levels of opposite parity under the $\Gamma$--$M$ mirrors creates six Dirac nodes, one on each $\Gamma$--$M$ mirror line, all at the Fermi level. These six cones with two spin components each form a Dirac spin liquid with $N_f=12$ two-component Dirac fermions. On decreasing $\chi_{28}$ the nodes move outward along the mirror lines and annihilate pairwise at the three $M$ points. This happens on the curve $R(\chi_{12},\chi_{23},\chi_{27},\chi_{28})=0$, given by the resultant of the two cubic polynomials that govern the levels at $M$; beyond it the state is gapped. At small $\chi_{28}$ the nodal line of the $SU(2)$ parent survives the breaking of the sublattice symmetry by $J'$: it remains a contour of $\gamma(\bm{k})$, now at a constant energy that coincides with the Fermi level. The surviving line is a property of the bond graph rather than of the PSG, and $J'''$ bonds break it into pockets. Two further lines, also known in closed form, bound the Dirac phase. On $\chi_{27}=\chi_{12}$ a singlet and a doublet are degenerate at $\Gamma$ and form a threefold touching. On $2\chi_{28}=q_1+q_2$ the two doublets cross at $\Gamma$ and form a fourfold-degenerate linear touching between two gapped phases.

The energy and momentum of the six nodes follow in closed form from the mirror-block decomposition of $h(\bm{k})$ on the $\Gamma$--$M$ lines, $\bm{k}=s\bm{M}$ (Sec.~SVI of the Supplemental Material~\cite{SM}). Which root is the node at half filling is established empirically. Among $4500$ amplitude sets drawn uniformly from $[-1.5,1.5]^4$, all $325$ Dirac states have $|\chi_{28}|>|\chi_{12}|$, and in each of them the node at half filling is the root closest to zero of a quadratic polynomial $Q_1(E)$, so that the Fermi energy is
\begin{align}
E_D&=\frac{\chi_{27}\chi_{28}-\operatorname{sgn}(\chi_{27}\chi_{28})\,D_1^{1/2}}{\chi_{12}},\nonumber\\
D_1&=\chi_{27}^2\chi_{28}^2+\chi_{12}^2\left(\chi_{12}^2+\chi_{23}^2-\chi_{27}^2-\chi_{28}^2\right).
\label{eq:ED}
\end{align}
The node position $\cos(\pi s_D)$ is given in closed form by Eq.~(S74) of the Supplemental Material~\cite{SM}; the nodes reach $M$ exactly on the semi-Dirac line $R=0$, and $\Gamma$ at $\chi_{27}=\chi_{12}$ and at $\chi_{28}=\chi_{23}+\chi_{27}$, with square-root laws at both ends; near $M$ the node separation agrees with $2k_0=2(-m/a)^{1/2}$ from Eq.~\eqref{eq:semidirac}. At $(\chi_{27},\chi_{28})=(1.272,2)$, $E_D=0.854636$ and $|\bm{k}_D|=0.320599|\bm{M}|$. All $694$ states of the same sample with a constant-energy nodal line at the Fermi level have $|\chi_{28}|<|\chi_{12}|$, and a root of $Q_2$ gives the energy and contour of the line.

\subsection{Semi-Dirac Critical Spin Liquid}

On the line $R=0$ the spinon spectrum has semi-Dirac points at the three $M$ points: the dispersion is linear along the zone edge and quadratic along $\Gamma$--$M$~\cite{Montambaux2009}, as in the semi-Dirac spin liquid of the trellis lattice~\cite{Chatterjee2026}. The little group $C_{2v}$ of $M$ fixes the orientation of the semi-Dirac point~\cite{Chatterjee2026}. The two levels that cross belong to the representations $A_1$ and $B_1$, where $B_1$ is odd under the mirror through $\Gamma M$ and even under the mirror along the zone edge. Their product transforms like the momentum $q_\parallel$ along the zone edge, which allows a coupling linear in $q_\parallel$ and forbids one linear in the momentum $q_\perp$ along $\Gamma$--$M$; $C_2\T$ forbids the third Pauli matrix. To leading order in the anisotropic scaling $q_\parallel\sim E$, $q_\perp\sim E^{1/2}$, the low-energy Hamiltonian is
\begin{equation}
H(\bm{q})=t_0\,q_\perp^2+\left(m+a\,q_\perp^2\right)\rho_x+v\,q_\parallel\rho_y,
\label{eq:semidirac}
\end{equation}
with Pauli matrices $\rho$ in the space of the two levels, $m=0$ on the critical line and $m/a<0$ on the Dirac side, where the nodes sit at $q_\perp=\pm(-m/a)^{1/2}$. The symmetry-allowed terms $t_\parallel q_\parallel^2$ and $q_\parallel^2\rho_x$ are of higher order in this scaling and are omitted. At $\chi_{27}=1.272$ the critical value is $\chi_{28}^c=1.8088$, and $v=1.877$, $a=-0.197$, and $t_0=0.093$ in units of the nearest-neighbor bond length and of the amplitudes, in the basis in which $m>0$ on the Dirac side. The two merging nodes carry Berry phases $\pm\pi$, and the semi-Dirac point carries Berry phase zero.

The semi-Dirac state is critical for every PSG of the SHD lattice: within $p6mm$ and its projective realizations, no symmetry enforces a twofold band touching that is linear along one direction and quadratic along the perpendicular one, and enumerating all projective representations of the $M$-point symmetry with antiunitary and particle--hole operations confirms that every enforced twofold degeneracy is a Dirac point, part of a nodal line, or momentum independent. The operations that leave an $M$ point invariant, including time reversal, the particle--hole and chiral operations that the PSG combines with lattice symmetries, and their projective realizations, act on $\bm{q}$ by reflections and by $\bm{q}\to-\bm{q}$, so that $q_\parallel^2$ and $q_\perp^2$ are invariant under all of them. A splitting term that depends on $\bm{q}$ only through $q_\parallel^2$ and $q_\perp^2$ therefore transforms like the constant splitting at the point, the mass. In Eq.~\eqref{eq:semidirac} the mass $m$ and the curvature $a$ multiply the same matrix, and any symmetry that forbids $m$ also forbids $a$. If a twofold degeneracy at such a point is enforced by these symmetries, all splitting terms of this form vanish, and the splitting along each invariant axis is linear, cubic, or zero, never quadratic. Terms proportional to the unit matrix, such as $t_0q_\perp^2$, do not split the levels, and mixed terms such as $q_\parallel q_\perp\rho_y$ can be allowed. On an invariant axis along which no linear splitting is allowed, however, every allowed splitting vanishes, since a function of the momentum along this axis transforms either like the mass or like the forbidden linear term; the degeneracy then extends along the axis as a nodal line through the point. At points with a threefold rotation a single linear direction is excluded altogether. Symmetry-enforced nodal lines through high-symmetry points of the layer groups are analyzed in Ref.~\cite{Damljanovic2023}. Semi-Dirac fermions can be symmetry protected in two dimensions by a different mechanism, as fourfold degeneracies of doubly degenerate bands enforced by nonsymmorphic symmetries together with inversion and time reversal~\cite{Meng2023}, which lies outside the twofold touchings considered here. Bands that are twofold degenerate at every momentum occur in $75$ of the $183$ realized states ($8$ $SU(2)$, $29$ $U(1)$, and $38$ $\Z_2$ states). For all $183$ states, with random singlet amplitudes up to $J'''$ and random allowed on-site terms in two independent draws ($\Phi=2\pi/3$ in the sector $(0,1,0)$), we computed the levels at the momenta $(i\bm{b}_1+j\bm{b}_2)/12$, $0\le i,j\le11$, which include $\Gamma$, the $M$ and $K$ points, and the quarter points $(i\bm{b}_1+j\bm{b}_2)/4$. Fourfold degeneracies occur at some of these momenta in $42$ of the $75$ states with doubled bands, but in none of the $183$ states is a touching between the highest occupied and the lowest empty level enforced at any of them at half filling. Fourfold semi-Dirac points of this type are therefore not pinned to the Fermi level of these spin liquids. A survey of all $\Z_2$, $U(1)$, and $SU(2)$ PSGs and of the four chiral classes, with singlet and triplet amplitudes up to $J'''$, finds semi-Dirac points only where two Dirac nodes annihilate; under random symmetric perturbations they split into a Dirac pair or open a gap. The semi-Dirac points of U001-000000 occupy a hypersurface of codimension one in the space of PSG-allowed amplitudes, the boundary between the $N_f=12$ Dirac phase and a gapped $U(1)$ state, given by $R=0$ for the nearest-neighbor and $J'$ amplitudes. Both nearest- and further-neighbor amplitudes control the mass: at the representative point $\partial m/\partial\chi_{28}=0.85$, $\partial m/\partial\chi_{24}=0.47$, and $\partial m/\partial\chi_{26}=-0.26$, compared with $-0.11$, $-0.55$, and $-0.68$ for $\chi_{12}$, $\chi_{23}$, and $\chi_{27}$. The same mechanism produces semi-Dirac points on the boundaries of the $N_f=24$ phases, where pairs of nodes annihilate at $\Gamma$ or at the quarter points $\bm{b}_1/4$ and $\bm{b}_2/4$ (Sec.~\ref{sec:spectra-manyflavor}). The low-energy laws below describe the quantum critical regime of this transition.

\begin{figure}
\centering
\includegraphics[width=\columnwidth]{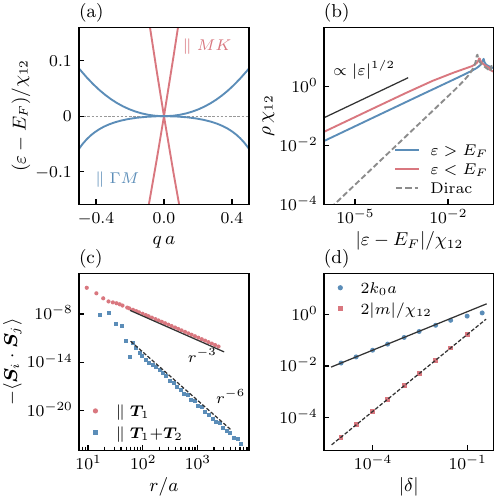}
\caption{Semi-Dirac spin liquid of U001-000000 at $(\chi_{12},\chi_{23},\chi_{27},\chi_{28})=(1,1,1.272,\chi_{28}^c)$; lengths in units of the nearest-neighbor bond length, energies in units of $\chi_{12}$. (a) Bands near $M$ along the zone edge $MK$ (rose, linear) and along $\Gamma M$ (blue, quadratic), with $q$ the distance from $M$. (b) Density of states per unit cell above (blue) and below (rose) $E_F$, with fitted exponents $0.4999(2)$ and $0.498(3)$; the solid line has slope $1/2$, and the dashed line is the Dirac phase at $\chi_{28}=2$. (c) Equal-time spin correlations between sites separated by $n\bm{T}_1$ (circles) and $n(\bm{T}_1+\bm{T}_2)$ (squares), with fitted exponents $2.98(1)$ and $6.00(3)$; lines are $r^{-3}$ and $r^{-6}$. (d) Separation $2k_0$ of the two Dirac nodes that merge at $M$ for $\delta=\chi_{28}-\chi_{28}^c>0$ (circles), and gap $2|m|$ at $M$ for $\delta<0$ (squares); the lines are the predictions of Eq.~\eqref{eq:semidirac}, $2k_0\propto\delta^{1/2}$ and $2|m|\propto|\delta|$.}
\label{fig:spec-sd}
\end{figure}

At the critical point the anisotropic dispersion gives distinct low-energy laws (Fig.~\ref{fig:spec-sd}). In units with $\chi_{12}=k_B=1$, the density of states is $\rho(E)\propto|E|^{1/2}$, with prefactors $14.3$ above and $29.4$ below the Fermi level per unit cell (both spins); the asymmetry comes from the tilt $t_0$. The specific heat at fixed particle number is $C_V=117\,T^{3/2}$ per unit cell. Equal-time spin correlations, $\langle\bm{S}_i\cdot\bm{S}_j\rangle=-\frac32|G_{ij}|^2$ at mean-field level, decay as $r^{-3}$ along the three bond-row directions $\bm{T}_1$, $\bm{T}_2$, and $\bm{T}_2-\bm{T}_1$, parallel to the linear axes of the nodes at the three $M$ points, and as $r^{-6}$ along all others, including $\bm{T}_1+\bm{T}_2$. For comparison, the Dirac phase has $\rho\propto|E|$, $C_V\propto T^2$, and $r^{-4}$ correlations in all directions. The fitted exponents agree with these values to within about $1\%$ [Fig.~\ref{fig:spec-sd}(b) and (c)]. For the correlations, $G_{ij}$ is evaluated as a Bloch integral with Gauss--Legendre quadrature graded toward the nodes rather than on a finite torus, which avoids image contributions and resolves the values $|G_{ij}|^2\approx10^{-24}$ reached at the largest distances. The exponents are least-squares fits of $\ln|G_{ij}|^2$ against $\ln r$ for $r_{\max}/4\le r\le r_{\max}$, with $r_{\max}=2423$ along $\bm{T}_1$ and $6295$ along $\bm{T}_1+\bm{T}_2$ in units of the bond length, and the uncertainties are the variations over shifted fit windows. The crossovers between the Dirac and semi-Dirac laws occur at energies of order $|m|$. The mean-field static structure factor is nearly identical in the semi-Dirac, Dirac, and gapped states. The dynamical structure factor distinguishes them: at momentum transfers within one node, $S(\bm{q},\omega)\propto\omega^{1/2}$ at low frequency, and its lower edge is linear along the zone edge and quadratic along $\Gamma$--$M$.

\subsection{Dirac Spin Liquids with Many Flavors}
\label{sec:spectra-manyflavor}

\begin{figure*}
\centering
\includegraphics[width=\textwidth]{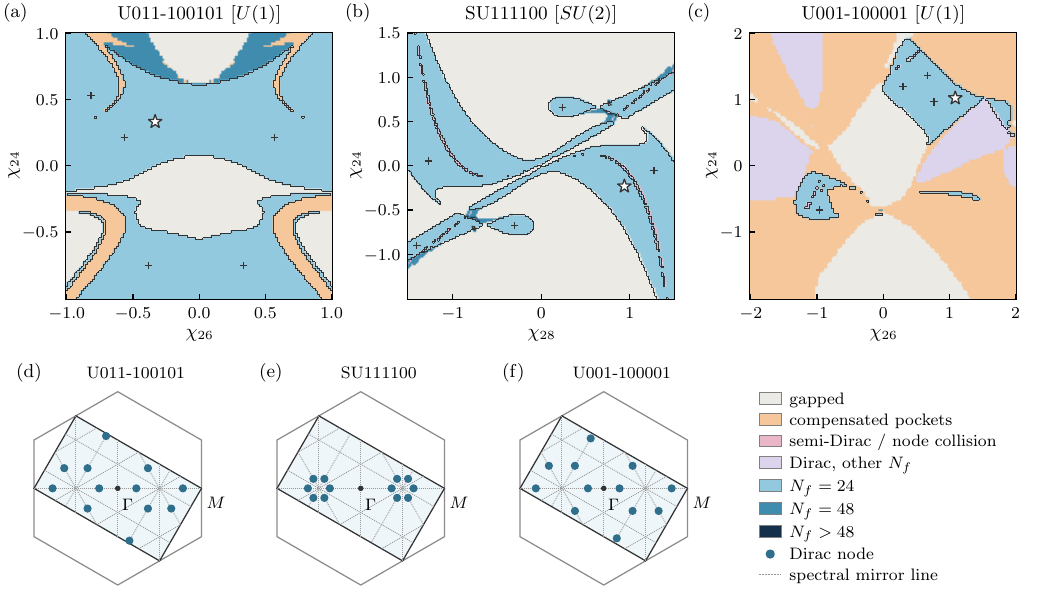}
\caption{Mean-field phase diagrams at half filling of three classes with 24-site cells ($n_{T_2}=-1$), with singlet bond amplitudes only and the on-site multiplier fixed by $\langle n_i\rangle=1$. (a) U011-100101 in the $(\chi_{26},\chi_{24})$ plane at $(\chi_{23},\chi_{27},\chi_{28})=(-0.2396,-0.2513,0.569)$. (b) SU111100 in the $(\chi_{28},\chi_{24})$ plane at $(\chi_{23},\chi_{27})=(-0.95,-0.92)$. (c) U001-100001 in the $(\chi_{26},\chi_{24})$ plane at $(\chi_{12},\chi_{23},\chi_{27},\chi_{28})=(0.6912,0.9889,1.4105,1.1267)$; ``Dirac, other $N_f$'' is $N_f=8$. $N_f=2\times$(nodes per spin in the reduced zone); every node is linear with a refined gap below $10^{-10}$ and Berry phase $\pm\pi$ (U(1)) or winding $\pm1$ (SU(2)). In (b), 16 grid cells near $(\chi_{28},\chi_{24})=\pm(0.75,0.65)$, where the two $N_f=24$ bands cross, carry $36$ or $48$ linear nodes ($N_f=72$ or $96$); such cells are absent from (a). Pink: semi-Dirac points, at which two nodes annihilate at $\Gamma$, $\bm{b}_1/4$, or $\bm{b}_2/4$, and collisions of six nodes at the $C_6$ centers $\pm M/2$ or other node collisions not resolved on the grid. Crosses and stars in (a)--(c): 10 (star 20) random 5\% perturbations of all allowed amplitudes each; 167 of 170 keep 12 nodes, the exceptions at the two crosses of (b) nearest a boundary. (d)--(f) Nodes of the starred states in the reduced zone (shaded); hexagon: zone of the 12-site cell. The spectrum is periodic under $\bm b_{1,2}/2$, and its mirror lines (dotted) cross at the $C_6$ centers $\pm M/2$; all twelve nodes lie on them. Parameters are listed in Sec.~SX of the Supplemental Material~\cite{SM}.}
\label{fig:nf24}
\end{figure*}

\begin{figure*}
\centering
\includegraphics[width=\textwidth]{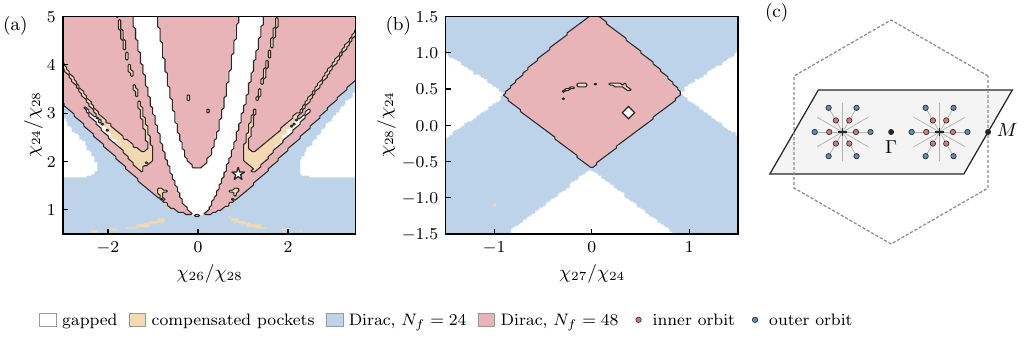}
\caption{Extended $N_f=48$ Dirac phase of the U(1) state U011-100101 at half filling, with the staggered Lagrange multiplier fixed by $\langle n_i\rangle=1$ at every point. (a) Phase diagram in the $(\chi_{26}/\chi_{28},\chi_{24}/\chi_{28})$ plane at $\chi_{23}/\chi_{28}=-0.247$, $\chi_{27}/\chi_{28}=0.692$; the star marks the survey point. (b) Phase diagram in the $(\chi_{27}/\chi_{24},\chi_{28}/\chi_{24})$ plane at $\chi_{23}/\chi_{24}=-0.023$, $\chi_{26}/\chi_{24}=0.530$ through the centered point (diamond). Colors distinguish Dirac states with $N_f=24$ and $N_f=48$ two-component Dirac fermions (one and two node orbits on the mirror lines), gapped states, and compensated pockets, that is, electron and hole spinon Fermi pockets of equal size at half filling; the black line encloses $N_f=48$. (c) Reduced Brillouin zone of the 24-site magnetic cell (shaded; dashed: zone of the 12-site cell) with the 24 Dirac nodes per spin at the centered point. They form two orbits of six (inner and outer) on the mirror lines through the $C_6$ centers $\pm M/2$ (crosses). The pocket regions are thinner than one grid cell in places and are drawn at the grid resolution. Parameters are listed in Sec.~SX of the Supplemental Material~\cite{SM}.}
\label{fig:nf48}
\end{figure*}

A Dirac spin liquid with $N_f$ flavors of two-component fermions is described at low energies by QED$_3$ with a compact $U(1)$ gauge field~\cite{Hermele2004}. We count $N_f$ as twice the number of nodes per spin in the reduced Brillouin zone of the magnetic unit cell, so that $N_f$ includes the spin. The phase is stable if the PSG allows no relevant fermion bilinear and if the symmetry-allowed monopole operators are irrelevant. To leading orders in $1/N_f$ the scaling dimension of the elementary monopole is $0.265N_f-0.038$~\cite{Borokhov2002,Pufu2014}, that is, $3.1$, $6.3$, and $12.7$ for $N_f=12$, $24$, and $48$, compared with the threshold $3$ of relevance, while the flavor-adjoint fermion bilinears have scaling dimensions close to $2$ and are relevant at all these $N_f$~\cite{Hermele2004}. Which monopoles are allowed is fixed by their lattice quantum numbers~\cite{Song2019,Song2020}; on the maple-leaf lattice an $N_f=12$ Dirac spin liquid has symmetry-allowed charge-one monopoles with a scaling dimension close to the threshold of relevance~\cite{Zhang2026}. For $N_f\ge24$ the stability is therefore decided by the fermion bilinears. The Dirac spin liquid of U001-000000 has $N_f=12$, and its PSG allows no spin-singlet fermion bilinear that gaps the cones. Its stability as a QED$_3$ phase then depends on its monopoles, which are not addressed here (Sec.~\ref{sec:conclusion}). In the frame in which $C_2\T$ is complex conjugation, the only single-valley mass is odd under $C_2\T$. Intervalley bilinears carry the incommensurate momentum difference of two nodes and are forbidden by translations. Triplet masses are forbidden by spin-rotation symmetry; with spin--orbit coupling they gap the nodes (Sec.~\ref{sec:qsh}). The $p6mm$ mirrors keep the six nodes on the $\Gamma$--$M$ lines for every symmetry-allowed singlet perturbation up to $J'''$, so that they can be removed only by annihilating at $M$ or at $\Gamma$. A survey of all $U(1)$ PSGs with $\Phi\in\{0,\pi\}$ and all $SU(2)$ PSGs, with amplitudes up to $J'''$, finds Dirac phases with $N_f=24$ in the $U(1)$ classes U000-011001, U100-011001, U001-100001, and U011-100101. Their nodes lie on two orbits of mirror lines, on a generic twelvefold orbit, or in the reduced zone of the doubled cell. The same survey finds extended phases with $N_f=48$ in U011-100101 (Fig.~\ref{fig:nf48}) and in U000-011001, discussed below, and U001-100001 also has a Dirac phase with $N_f=8$ [Fig.~\ref{fig:nf24}(c)]. At the generic amplitude point of Sec.~SVII of the Supplemental Material~\cite{SM}, whose ans\"atze contain the orbits up to the minimal range of each PSG, U010-$\Phi$11100 at $\Phi=\pi/2$ has $N_f=48$, with its $24$ nodes per spin in close pairs so that the linear law of the density of states holds only below $|E|\approx10^{-5}$, and further $N_f=12$ Dirac spin liquids occur in U011-000100, U101-000001, and Z011011(1). In the $N_f=24$ and $N_f=48$ phases of the survey the nodes are protected by $C_2\T$ (U001-100001, U011-100101) or by the sublattice (chiral) symmetry that time reversal induces for $w_\T=0$ (U000-011001, U100-011001). When all orbits allowed up to $J'''$ are added at this generic point, U011-000100 and U101-000001 keep their nodes, whereas U010-$\Phi$11100, U011-100101, and SU111100 become gapped, and at the second amplitude point of the Supplemental Material~\cite{SM}, U101-100000, U100-111000, and Z111010(1) are Dirac spin liquids with $N_f=24$ and Z100010(1) one with $N_f=48$. The Dirac phases found here, like the other spectral types, therefore occupy regions of amplitude space and are not properties of the PSGs alone. The absence of a symmetry-allowed singlet mass is established above for U001-000000 only. For the other $N_f=12$ states and for all states with $N_f\ge24$ we have not classified the fermion bilinears, including, in the doubled-cell states, the intervalley bilinears between nodes at $\bm{k}$ and $\bm{k}+\bm{b}_{1,2}/2$. Their stability rests on the observation that the nodes survive random symmetric perturbations of the singlet amplitudes in the samples tested (numbers below), a statement about the mean-field spectra rather than a classification of the symmetry-allowed masses. In the sublattice-symmetric states SU011101, U000-011001, and U100-011001, whose $h$ is off-diagonal, the nodes on the $\Gamma$--$M$ lines are roots of an explicit quadratic, and in the doubled-cell states U011-100101 and U001-100001 roots of polynomials of known degree (Sec.~SVI of the Supplemental Material); nodes at generic momenta remain numerical. The $SU(2)$ Dirac states belong to the sector $\eta_{\T\sigma}=\eta_{\T C_6}=-1$: SU011101 has $N_f=12$, $24$, and $36$, and SU111100 has $N_f=24$ and, in thin regions, $72$ and $96$ [Fig.~\ref{fig:nf24}(b)]. In SU011101 the nodes are the zeros of the complex $\det B$, and the three values occur in $25\%$, $3\%$, and $1\%$ of amplitude sets drawn uniformly from $[-1.5,1.5]^4$; these fractions refer to this measure. At the $N_f=36$ point listed in Sec.~SX of the Supplemental Material the 18 linear nodes per spin form one orbit of six on the $\Gamma$--$M$ lines and one generic orbit of twelve, and they survive $44$ of $48$ random symmetric perturbations of $5\%$ of the amplitude norm. Altogether, the Dirac phases analyzed in this section and in Sec.~SVII of the Supplemental Material have $N_f=8$, $12$, $24$, $36$, and $48$, and $N_f=72$ and $96$ occur only in thin regions of SU111100. The $SU(2)$ Dirac states are described by quantum chromodynamics in $2+1$ dimensions (QCD$_3$) with gauge group $SU(2)$ and $N_f/2$ fundamental Dirac doublets.

The $N_f=24$ states are extended phases as well. Figure~\ref{fig:nf24} shows phase diagrams of the three classes with a doubled unit cell that host twelve nodes per spin, U011-100101, SU111100, and U001-100001, in planes spanned by further-neighbor amplitudes, with the on-site multiplier fixed by the local constraint at every point. The $N_f=24$ phase covers 63\% and 33\% of the planes shown for U011-100101 and SU111100, and 10\% for U001-100001, where compensated electron--hole pockets dominate; these phase diagrams are cuts through the space of ansatz amplitudes, and the area fractions depend on the chosen planes and windows. The twelve nodes are linear, carry Berry phase $\pi$ or winding number $\pm1$, and sit at the Fermi level. Because the translations act projectively, the spectrum is periodic under half reciprocal-lattice vectors and its mirror lines are shifted to pass through sixfold rotation centers at $\pm M/2$. At the starred points of Fig.~\ref{fig:nf24} all twelve nodes lie on these spectral mirror lines. In the spectrum, $\Gamma$ and the quarter points $\bm{b}_1/4$ and $\bm{b}_2/4$, two of the midpoints of the $\Gamma$--$M$ lines, are related by the sixfold rotation about $M/2$ and have only a twofold rotation, whereas the third midpoint, $\pm M/2$, is the sixfold rotation center. The phase ends by pairwise annihilation of nodes at $\Gamma$, $\bm{b}_1/4$, or $\bm{b}_2/4$, through semi-Dirac points, by a Lifshitz transition to compensated pockets in the $U(1)$ classes, or by the appearance of twelve further nodes, $N_f=48$. At $\pm M/2$ the six nodes of an orbit can meet only all at once, in a touching of higher order; such collisions occur on lines inside the phase in Fig.~\ref{fig:nf24}(b) and leave $N_f=24$ on both sides. In SU111100 small regions with $N_f=72$ and $96$ also occur.

In U011-100101 three properties of the PSG act together. Its projective translations carry flux $\pi$ per unit cell, so that the spectrum is periodic under half reciprocal-lattice vectors and the sixfold rotation acts about $\pm M/2$; every mirror is accompanied by a particle--hole gauge transformation, $E(\bm{k})=-E(\sigma\bm{k})$; and $C_2\T$ gives each node a Berry phase $\pi$. On a mirror line the spectrum is therefore symmetric about $E=0$, and a node there sits at the Fermi level. A node off the mirror lines at energy $\varepsilon$ has a mirror partner at $-\varepsilon$. Both are at the Fermi level only if $\varepsilon=0$, a condition of codimension one, and otherwise the state has compensated pockets. Protected nodes thus lie on the mirror lines, in orbits of six around the rotation centers, and two orbits around each of $\pm M/2$ give $N_f=48$ [Fig.~\ref{fig:nf48}(c)]. The node count changes only when two orbits annihilate on a line, when an orbit passes through a high-symmetry point, or when compensated pockets appear. The same mechanism bounds the flavor number. On each spectral mirror line an involution $\Lambda$, unique and independent of the amplitudes, anticommutes with $h(\bm{k})$, and $h$ is off-diagonal in its eigenbasis. On the $\Gamma$--$M$ lines the determinant of the off-diagonal block is real up to a constant phase, and for amplitudes up to $J'''$ it is a quadratic polynomial in $\cos(2\pi s)$, Eq.~(S100) of the Supplemental Material~\cite{SM}; this form is established numerically, not analytically. With at most two roots, at most two orbits of six nodes surround each rotation center, so that $N_f\leq48$ in this class with amplitudes up to $J'''$. On the other mirror lines the determinant is complex at all sampled amplitudes, and the levels show only avoided crossings.

The $N_f=48$ phase of U011-100101 requires a fourth-neighbor amplitude $|\chi_{24}|\gtrsim|\chi_{28}|$ together with a finite $\chi_{26}$, while the $J_b$ amplitude $\chi_{12}$ is forbidden by the PSG; at the centered point of Fig.~\ref{fig:nf48}(b), $\chi_{24}$ is the largest amplitude and exceeds $\chi_{28}$ by a factor of $5.7$. Such an ansatz is unusual for exchange models with dominant nearest-neighbor couplings, and whether it is energetically favorable for any spin Hamiltonian is not addressed here. The phase covers about $44\%$ of the window of Fig.~\ref{fig:nf48}(a) and about $6\%$ of amplitude sets with all five amplitudes drawn uniformly from $[-1.5,1.5]$; like the other area fractions quoted here, these numbers depend on the chosen coordinates and measure in the space of ans\"atze. At the centered point, with the staggered multiplier re-solved, $N_f=48$ survives symmetric perturbations of $19\%$ of the amplitude norm in all of $48$ random directions perpendicular to the amplitude vector, and perturbations of $30\%$ in $33$ of them. The boundaries of the phase are transitions to the $N_f=24$ phase, in which the outer orbit has annihilated at $M$, to gapped states, and to compensated pockets. The two node orbits are not related by symmetry; at the centered point they lie at $0.137|\bm{M}|$ and $0.288|\bm{M}|$ from $M/2$ and have different velocities. When the nodes form inequivalent orbits, as here, the symmetry of the continuum theory is the subgroup implemented by the PSG, much smaller than $SU(N_f)$.

Classes with a sublattice (chiral) symmetry have no such bound. In U000-011001 time reversal makes $h(\bm{k})$ off-diagonal (Sec.~SVI of the Supplemental Material~\cite{SM}), so that every zero of the complex determinant $\det B(\bm{k})$ is a node at $E_F=0$ with winding $\pm1$, and nodes at generic momenta are as stable as nodes on the mirror lines. Besides the $N_f=24$ phase, this class has an open region with $N_f=48$. At the point listed in Sec.~SX of the Supplemental Material the 24 linear nodes per spin form two orbits of six on the $\Gamma$--$M$ lines and one generic orbit of twelve, and they survive $46$ of $48$ random symmetric perturbations of $5\%$ of the amplitude norm, $26$ of $48$ at $10\%$, and $13$ of $48$ at $20\%$. Among $3000$ amplitude sets drawn uniformly from $[-1.5,1.5]^5$, about $2\%$ have $N_f=48$, and none has a larger flavor number.

\subsection{Classical Orders}
\label{sec:classical}

\begin{figure*}
\centering
\includegraphics[width=\textwidth]{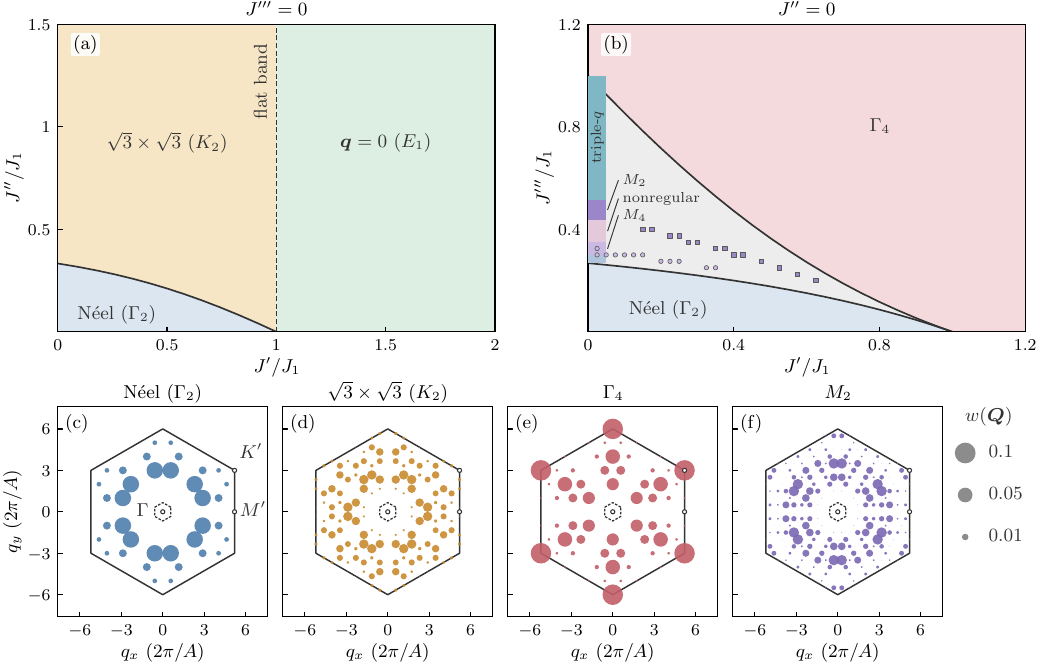}
\caption{Classical ground states of the Heisenberg model on the SHD lattice with $J_b=J_g=J_r=J_1=1$. (a) Phase diagram in the $(J',J'')$ plane at $J'''=0$, where a regular state reaches the Luttinger--Tisza (LT) bound everywhere; on the dashed line $J'=1$ the lowest LT band is flat (a second flat-band line, $J'=-1$, lies outside the range shown). (b) Phase diagram in the $(J',J''')$ plane at $J''=0$. In the gray wedge the LT bound is generically not reached; the bar along $J'=0$ gives the lowest states found by simulated annealing and local minimization on tori of up to 1728 spins: a modulated N\'eel state for $J'''\lesssim0.30$, the triple-$\bm{q}$ order $M_4$ up to $J'''\approx0.351$, nonregular noncoplanar states with mixed wave vectors for $0.351\lesssim J'''\lesssim0.438$, the triple-$\bm{q}$ order $M_2$ for $0.438\lesssim J'''\lesssim0.516$, and incommensurate noncoplanar triple-$\bm{q}$ states for $0.516\lesssim J'''<1$. Symbols mark points where $M_4$ (circles) or $M_2$ (squares) reaches the LT bound. (c)--(f) Bragg weights $w(\bm{Q})$ computed with the ideal site positions, in the hexagon with corners $K'=9K$ (solid; dashed: first Brillouin zone), normalized to unit total weight in the hexagon (peaks on its edges and corners counted with weights $1/2$ and $1/3$), for (c) N\'eel order, (d) $K_2$ at $J''=1$, (e) $\Gamma_4$ at $J'''=1.5$, and (f) $M_2$ at $J'''=0.5$, with the other further-neighbor couplings zero; disc areas are proportional to $w$; momenta as in Fig.~\ref{fig:spec-sq}. In $K_2$, $M_2$, and $M_4$ the spin directions, and hence the relative weights, vary with the couplings.}
\label{fig:classical}
\end{figure*}

We compare the structure factors of the spin liquids with those of the classical orders. Regular magnetic orders, in which every lattice operation is undone by a global $O(3)$ spin transformation, are classified by the mappings of $p6mm$ into $O(3)$ of Ref.~\cite{MessioLhuillierMisguich2011}. There are $40$ of them: $28$ invariant under translations, $8$ with the wave vector $K$, and $4$ with the three wave vectors $M$. Because $p6mm$ acts freely on the SHD sites, the site constraint of Ref.~\cite{MessioLhuillierMisguich2011} is empty, and the direction of the reference spin within a multidimensional representation is not fixed by symmetry. As a result, the canting angles of these orders vary with the couplings. The energies of the regular orders reproduce all Luttinger--Tisza eigenvalues~\cite{LuttingerTisza1946,KaplanMenyuk2007} at $\Gamma$ and $M$ and part of those at $K$, so a regular order is the exact classical ground state whenever the Luttinger--Tisza minimum is one of these eigenvalues.

For antiferromagnetic nearest-neighbor couplings the bipartite lattice is unfrustrated and the N\'eel state is the exact classical ground state for any $J_b,J_g,J_r>0$, with energy $-(J_b+J_g+J_r)/2$ per site (unit spins). Further neighbors frustrate it (Fig.~\ref{fig:classical}). With $J_b=J_g=J_r=1$ and $J'''=0$, for $|J'|<1$ the N\'eel state gives way at $J''=\frac23[1-J'/2-\frac12(J'^2-J'+1)^{1/2}]$ to the coplanar $\sqrt3\times\sqrt3$ state $K_2$. For $J'>1$ the coplanar $q=0$ state $E_1$ is the ground state for all $J''>0$. The states $K_2$ and $E_1$ meet on the line $J'=1$, where the lowest Luttinger--Tisza band is flat: every square with equal edge and diagonal couplings becomes a complete graph, and its energy depends only on its total spin. The line stays flat for $J''<0$, where it separates N\'eel from $\Gamma_4$. There a N\'eel pattern on each hexagon, antiparallel on the $J_g$ and $J_r$ edges and parallel on the $J''$ bonds, leaves every square with zero total spin for any relative orientation of the hexagons, so N\'eel, $\Gamma_4$, and all states built from independent hexagon N\'eel vectors are degenerate (Sec.~SVIII of the Supplemental Material~\cite{SM}). For $J'<-1$ the N\'eel state gives way instead at $J''=1-1/\sqrt3$ to the coplanar $q=0$ state $E_2$. The lowest Luttinger--Tisza band is flat also on the line $J'=-1$ for $J''>1-1/\sqrt3$, where $E_2$, $K_2$, and $M_3$ are degenerate. With $J''=0$ and $J'''>0$, N\'eel order is stable up to $J'''=2-\sqrt3$ at $J'=0$, and the collinear state $\Gamma_4$, with ferromagnetic $J_b$ pairs and antiferromagnetic hexagons, is the exact classical ground state for $J'''\geq1$. In the wedge between them the Luttinger--Tisza bound is generically not reached (except on narrow strips where $M_4$ or $M_2$ saturates it). Along $J'=0$ we searched for the lowest states by simulated annealing followed by local minimization on tori of $L\times L$ unit cells with $L=4$, $6$, $8$, and $12$. Above a modulated N\'eel state for $J'''\lesssim0.30$, the triple-$\bm{q}$ order $M_4$ is the lowest state found up to $J'''\approx0.351$, somewhat beyond the strip $0.301<J'''<0.332$ on which it saturates the Luttinger--Tisza bound. For $0.351\lesssim J'''\lesssim0.438$ nonregular noncoplanar states lie lower, by up to $3\times10^{-3}$ per site. Up to $J'''\approx0.39$ they combine the wave vectors $M$ and $K$ (at $J'''=0.36$ with weights $0.47$ and $0.53$), and above it one $M$ point with the pair $(3\bm{b}_1+\bm{b}_2)/4$, $(\bm{b}_1+3\bm{b}_2)/4$. The triple-$\bm{q}$ order $M_2$ is the lowest state found for $0.438\lesssim J'''\lesssim0.516$, and incommensurate noncoplanar triple-$\bm{q}$ states lie lower for $0.516\lesssim J'''<1$. Since larger tori can only add competing states, these windows bound the extent of $M_4$ and $M_2$ from above.

\begin{table*}
\caption{Irreducible regular orders of the Heisenberg model on the SHD lattice. $\bm{Q}$ is the ordering wave vector, $N_m$ the number of sites in the magnetic unit cell, and $n_d$ the number of distinct spin directions for generic couplings. Energies are per site for unit spins. For the coplanar orders $E/N=a-(c^2+s^2)^{1/2}$ with $a$, $c$, $s$ of Eq.~(S104) of the Supplemental Material~\cite{SM}, and for the $M$ orders $E/N$ is half the lowest eigenvalue of the matrices of Eq.~(S109) there, given in closed form by Eq.~(S110). The chiral class of the noncoplanar orders is $(\det G_{C_6},\det G_\sigma)$ (Sec.~SVIII of the Supplemental Material). The last column gives coupling regions or representative points, with $J_1=J_b=J_g=J_r$, at which the order saturates the Luttinger--Tisza bound and is therefore an exact classical ground state; regions related by the sublattice-sign duality $J_{b,g,r}\to-J_{b,g,r}$ are listed once. NN denotes nearest-neighbor couplings alone, with arbitrary magnitudes; all other entries are for $J_1=1$ with the remaining further-neighbor couplings zero unless stated. $J''_{NK}(J')=\frac23[1-\frac12J'-\frac12(J'^2-J'+1)^{1/2}]$, and $J'''_N$, $J'''_G$ are the instabilities of the N\'eel and $\Gamma_4$ states toward incommensurate order.}
\label{tab:regular}
\begin{ruledtabular}
\begin{tabular}{lcclll}
order & $\bm{Q}$ & $N_m$ & spins ($n_d$) & $E/N$ & exact ground state\\
\hline
$\Gamma_1$ & $\Gamma$ & 12 & collinear, F (1) & $\frac12(J_b+J_g+J_r+J')+J''+J'''$ & \parbox[t]{6.0cm}{\raggedright NN: $J_b,J_g,J_r<0$; dual of $\Gamma_2$\strut}\\
$\Gamma_2$ & $\Gamma$ & 12 & collinear, N\'eel (2) & $-\frac12(J_b+J_g+J_r)+\frac12J'+J''+J'''$ & \parbox[t]{6.0cm}{\raggedright NN: $J_b,J_g,J_r>0$; $J''<J''_{NK}(J')$ for $|J'|<1$ and $J''<1-1/\sqrt3$ for $J'<-1$ ($J'''=0$); $J'''<J'''_N(J')$, $J'''_N(0)=2-\sqrt3$ ($J''=0$)\strut}\\
$\Gamma_3$ & $\Gamma$ & 12 & collinear (2) & $\frac12(-J_b+J_g+J_r-J')+J''-J'''$ & \parbox[t]{6.0cm}{\raggedright NN: $J_b>0>J_g,J_r$; dual of $\Gamma_4$\strut}\\
$\Gamma_4$ & $\Gamma$ & 12 & collinear (2) & $\frac12(J_b-J_g-J_r-J')+J''-J'''$ & \parbox[t]{6.0cm}{\raggedright NN: $J_b<0<J_g,J_r$; $J'>1$, $J''<0$ ($J'''=0$); $J'''>J'''_G(J')$, $J'''_G(0)=1$ ($J''=0$)\strut}\\
$\Gamma_5=E_1$ & $\Gamma$ & 12 & coplanar (12) & $a-\rho$ & \parbox[t]{6.0cm}{\raggedright $J'>1$, $J''>0$ ($J'''=0$); $J'>1-2J'''+3J'''^2$, $J'''<0$ ($J''=0$)\strut}\\
$\Gamma_6=E_2$ & $\Gamma$ & 12 & coplanar (6) & $a-\rho$ & \parbox[t]{6.0cm}{\raggedright $J'<-1$, $J''>1-1/\sqrt3$ ($J'''=0$); $J'''_E(J'')<J'''<1$ for $J'=0$, $J''\gtrsim0.17$, with $J'''_E$ decreasing from $0.29$ ($J''=0.17$) to $2/\sqrt3-1$ ($J''\approx1$)\strut}\\
$K_1$ & $K$ & 36 & coplanar (6) & $a-\rho$ & \parbox[t]{6.0cm}{\raggedright NN: $J_b<0$, $J_gJ_r<0$; dual of $K_2$\strut}\\
$K_2$ & $K$ & 36 & coplanar (12) & $a-\rho$ & \parbox[t]{6.0cm}{\raggedright NN: $J_b>0$, $J_gJ_r<0$; $J''>J''_{NK}(J')$ for $|J'|<1$ ($J'''=0$); $1<J'<1-2J'''+3J'''^2$, $J'''<0$ ($J''=0$)\strut}\\
$M_1$ & $M$ & 48 & noncoplanar, B (24) & $\frac12\lambda_{\min}(2M_{M_1})$ & \parbox[t]{6.0cm}{\raggedright dual of $M_2$\strut}\\
$M_2$ & $M$ & 48 & noncoplanar, A (24) & $\frac12\lambda_{\min}(2M_{M_2})$ & \parbox[t]{6.0cm}{\raggedright strips, e.g., $0.376<J'''<0.394$ at $J'=0.2$, $J''=0$; $0.496<J'''<0.570$ at $J'=0$, $J''=-0.5$\strut}\\
$M_3$ & $M$ & 48 & noncoplanar, D (48) & $\frac12\lambda_{\min}(2M_{M_3})$ & \parbox[t]{6.0cm}{\raggedright dual of $M_4$\strut}\\
$M_4$ & $M$ & 48 & noncoplanar, C (48) & $\frac12\lambda_{\min}(2M_{M_4})$ & \parbox[t]{6.0cm}{\raggedright strips, e.g., $0.301<J'''<0.332$ at $J'=J''=0$; $0.354<J'''<0.457$ at $J'=-1$, $J''=0$\strut}
\end{tabular}
\end{ruledtabular}
\end{table*}

\begin{figure*}
\centering
\includegraphics[width=\textwidth]{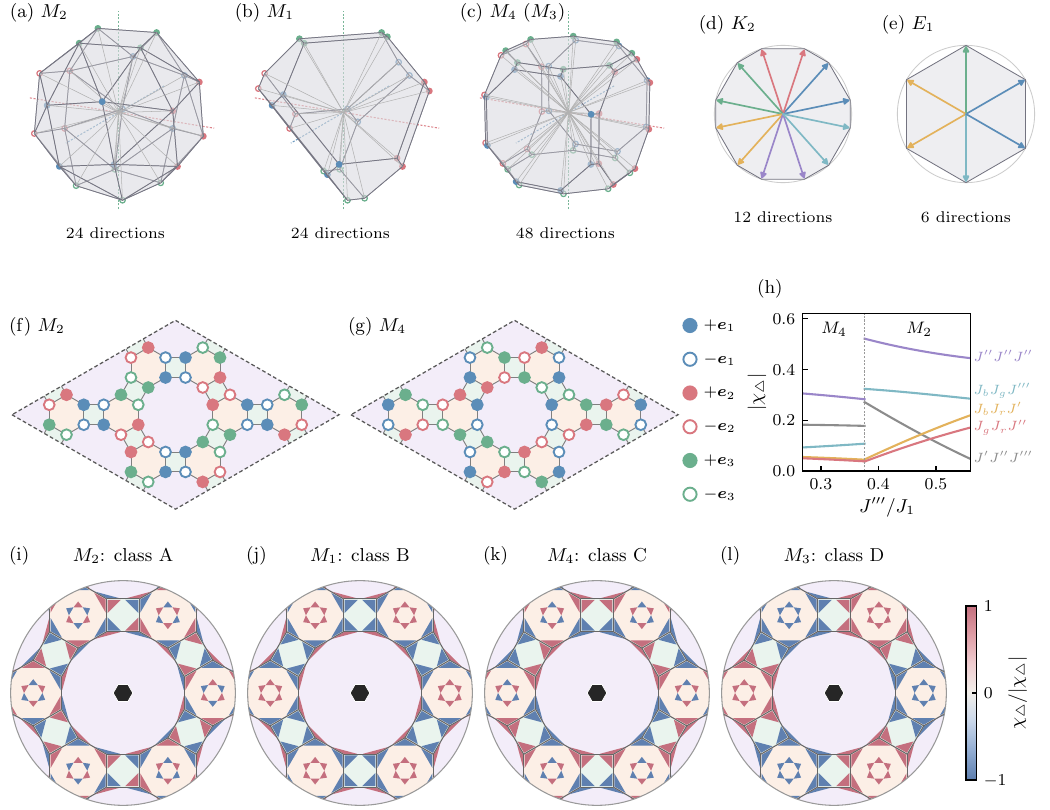}
\caption{Noncoplanar regular orders of the SHD Heisenberg model. (a)--(e) Common-origin plots of all distinct spin directions of the magnetic unit cell, with the convex polyhedron of their tips, for (a) $M_2$ at $J'''=0.5$, (b) $M_1$ at the dual point $J_1=-1$, $J'''=0.5$, (c) $M_4$ at $(J',J''')=(0.1,0.3)$ ($M_3$ at the dual point has the same directions), and the coplanar (d) $K_2$ at $J'=J''=0.6$ and (e) $E_1$ at $J'=1.5$, $J''=1$; other couplings zero and $J_1=1$ unless stated. Dashed lines are the axes $\bm{e}_a$, along which the spin components are modulated with the three $M$ wave vectors. (f), (g) Spin structure of $M_2$ and $M_4$ on one magnetic unit cell of $48$ sites (dashed); each site is colored by the axis $\bm{e}_a$ of its largest spin component, filled for a positive and open for a negative component. (h) Magnitude $|\chi_\triangle|$ of the scalar chirality on the five triangle types along $J'=J''=0$, for $M_4$ below and $M_2$ above their energy crossing at $J'''=0.376$. (i)--(l) Sign of $\chi_\triangle=\bm{S}_i\cdot(\bm{S}_j\times\bm{S}_k)$ (counterclockwise) on the triangles with two nearest-neighbor bonds, drawn at half size about their common site in the squares ($J_bJ_rJ'$), hexagons ($J_gJ_rJ''$), and dodecagons ($J_bJ_gJ'''$), and on the $J''J''J''$ triangles (star points in the hexagons), for the states of (a)--(c). $M_2$, $M_1$, $M_4$, and $M_3$ realize the chirality patterns of classes A, B, C, and D; only $M_1$ has a nonzero net chirality.}
\label{fig:regular}
\end{figure*}

The energy of every regular order is fixed by those of twelve irreducible ones: the collinear orders $\Gamma_1$--$\Gamma_4$, the coplanar orders $E_1$ and $E_2$ (wave vector $\Gamma$) and $K_1$ and $K_2$ (36-site magnetic cells), and the noncoplanar triple-$\bm{q}$ orders $M_1$--$M_4$ with 48-site magnetic cells, in which the spin component along the axis $\bm{e}_a$ is modulated with the wave vector of the $a$th $M$ point. Table~\ref{tab:regular} gives their energies for all six couplings, and Sec.~SVIII of the Supplemental Material~\cite{SM} their spin structures. For the $E$, $K$, and $M$ orders the spin directions rotate with the couplings [Fig.~\ref{fig:regular}(a)--(e)]. With nearest-neighbor couplings alone a regular order is the exact ground state for all signs and magnitudes. Only the hexagons can then be frustrated, when $J_gJ_r<0$, and in that case $K_2$ ($J_b>0$) or $K_1$ ($J_b<0$) reaches the energy of decoupled hexagons and $J_b$ dimers. With further neighbors each of the twelve orders is the exact ground state in some region (Table~\ref{tab:regular}). The orders $M_4$ and $M_2$ saturate the Luttinger--Tisza bound on strips of finite width (Table~\ref{tab:regular}), which widen when $J'$ or $J''$ is ferromagnetic. The $M$ orders carry a scalar chirality $\chi_\triangle=\bm{S}_i\cdot(\bm{S}_j\times\bm{S}_k)$ on the five types of triangles formed by $J_bJ_rJ'$, $J_gJ_rJ''$, $J_bJ_gJ'''$, $J''J''J''$, and $J'J''J'''$ bonds, and the chirality pattern of a regular order is that of the chiral class $(\chi(C_6),\chi(\sigma))=(\det G_{C_6},\det G_\sigma)$ of Sec.~\ref{sec:chiral} (Sec.~SVIII of the Supplemental Material). The orders $M_2$, $M_1$, $M_4$, and $M_3$ realize the chirality patterns of classes A, B, C, and D, respectively [Fig.~\ref{fig:regular}(i)--(l)], and of these four only $M_1$ carries a net chirality. Melting these orders while retaining their chirality pattern is a natural route to the chiral spin liquids of the corresponding class~\cite{MessioLhuillierMisguich2013}.

All classical orders considered here place little or no Bragg weight inside the first Brillouin zone. With the ideal site positions the N\'eel state, for example, carries $76\%$ of the weight in the hexagon with corners $K'=9K$ (Appendix~\ref{app:bz}) on the twelve reciprocal-lattice vectors of length $\sqrt7|\bm{b}_1|$. The strongest individual peaks of $\Gamma_4$ lie at the six corners $K'$, whereas the six vectors of length $4\times2\pi/A$ carry the largest total weight ($29\%$). The suppression is a form-factor effect of the twelve-site basis. For this reason we display the structure factors in this hexagon, nine times larger than the first Brillouin zone in linear size. The commensurate realization of Appendix~\ref{app:bz} redistributes the weight among the outer shells; for the N\'eel state it gives $71\%$ on the $\sqrt7$ shell and moves weight from the $\sqrt{13}$ shell ($18\%$ in the ideal geometry, $9\%$) to the $\sqrt{19}$ shell ($6\%$, $20\%$).

\subsection{Spin Structure Factors}

\begin{figure*}
\centering
\includegraphics[width=\textwidth]{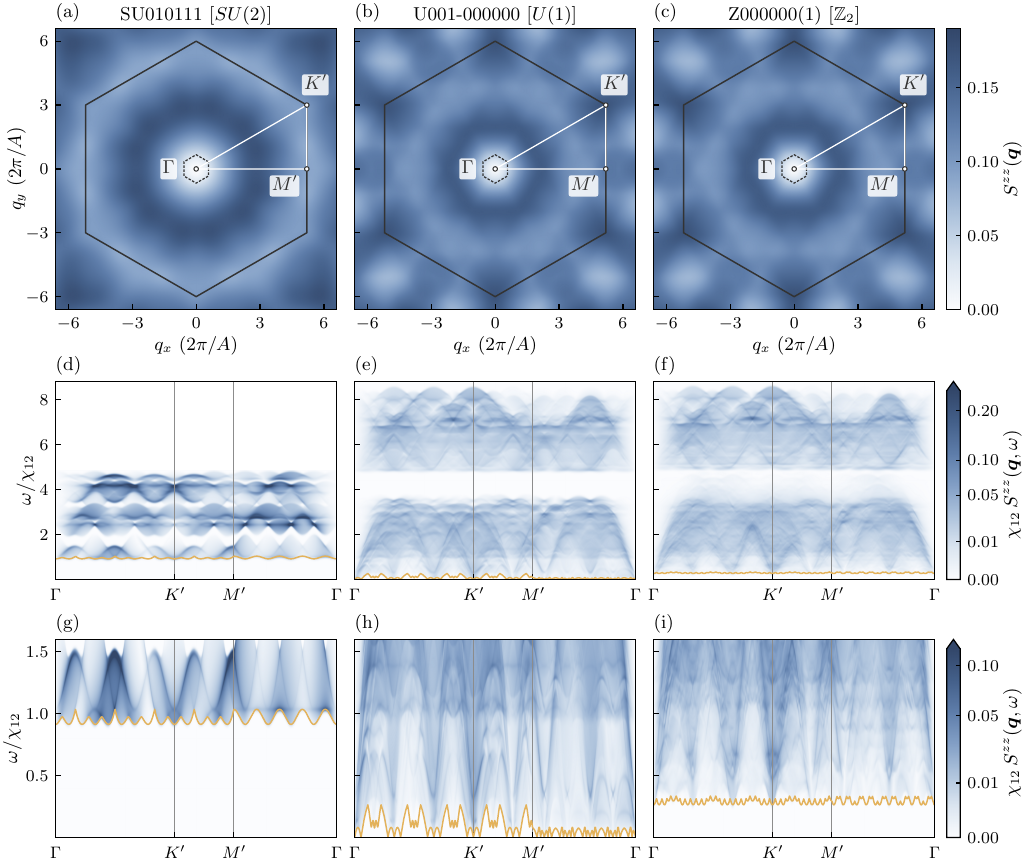}
\caption{Mean-field spin structure factors $S^{zz}$, Eq.~\eqref{eq:sqw}, of (a), (d), (g) the gapped $SU(2)$ state SU010111 at $\chi_{12}=\chi_{23}=\chi_{27}=1$, with narrow twofold-degenerate bands, fluxes $\pi$ through all faces, and spinon gap $0.914$ (single-spinon gap $0.457$); the Lieb-flux state of the same PSG ($\chi_{27}=-1$, the singlet part of the Kitaev ansatz of Sec.~\ref{sec:kitaev}) has the same kind of narrow twofold-degenerate bands, spinon gap $1.035$, and $S(K')=0.121$, $S(M')=0.069$; (b), (e), (h) the $U(1)$ Dirac state U001-000000 at $(\chi_{12},\chi_{23},\chi_{27},\chi_{28})=(1,1,1.272,2)$; and (c), (f), (i) its $\Z_2$ descendant in class Z000000(1), with the same hoppings, singlet pairing $0.4\tau^1$ on the $J_r$ bonds, and self-consistent on-site multipliers; its quasiparticle gap is $\min|E|=0.132$ (two-spinon threshold $0.263$). Energies are in units of $\chi_{12}$. All structure factors use the site positions of the ideal lattice. (a)--(c) $S^{zz}(\bm{q})$ in and around the hexagon with corners $K'=9K$ and edge midpoints $M'=9M$ (solid; the extended Brillouin zone of the commensurate realization), with the first Brillouin zone (dashed) at its center and the path of (d)--(i) (white); momenta are in units of $2\pi/A$. (d)--(f) $S^{zz}(\bm{q},\omega)$ along $\Gamma\to K'\to M'\to\Gamma$ over the full two-spinon range, and (g)--(i) its low-energy part; the amber line is the lower edge of the two-spinon continuum. Spectra are broadened by Gaussians of width $0.03$ in (d)--(f) and $0.01$ in (g)--(i), on a $120\times120$ momentum mesh; color scales follow a square-root law.}
\label{fig:spec-sq}
\end{figure*}

The spin structure factors are computed from the mean-field two-spinon bubble with the site positions of the ideal lattice, including all normal and anomalous coherence factors~\cite{Chauhan2023}, as described in Appendix~\ref{app:sf}. In the ideal geometry $S(\bm{q})$ is defined at every momentum but is not periodic (Appendix~\ref{app:bz}). We show it in the hexagon with corners $K'=9K$, the extended Brillouin zone of a commensurate realization of the tiling, and plot $S(\bm{q},\omega)$ along the path $\Gamma\to K'\to M'\to\Gamma$ (Fig.~\ref{fig:spec-sq}). The commensurate realization gives the same qualitative structure factors; for the states of Fig.~\ref{fig:spec-sq} the two differ by at most $9\%$ (SU010111) and $15\%$ (U001-000000 and its $\Z_2$ descendant) of the maximum of $S(\bm{q})$ in the hexagon (Appendix~\ref{app:bz}). For $\eta_{12}=-1$ the bands are shown in the reduced zone of the doubled cell~\cite{Maity2023}, whereas the spin correlations, being gauge invariant, are periodic with the original Bravais lattice.

Figure~\ref{fig:spec-sq} compares a gapped $SU(2)$ state, the $N_f=12$ Dirac state, and a gapped $\Z_2$ descendant of it. In all three, $S(\bm{q})$ vanishes at $\Gamma$ because the total spin is conserved. It has broad maxima in the outer part of the hexagon, near the shells that carry the Bragg weight of the N\'eel state. The Dirac state and its $\Z_2$ descendant have nearly the same $S(\bm{q})$, with $S(K')=0.173$ in both and $S(M')=0.131$ and $0.127$, and maxima near the corners $K'$; they differ by at most $3\%$ of the maximum. The gapped $SU(2)$ state differs markedly: its maxima lie on a ring at $|\bm{q}|\approx2.9\times2\pi/A$, and $S(K')=0.102$ and $S(M')=0.090$. The dynamical structure factor reflects the spinon bands directly. The narrow bands of SU010111 produce a sequence of nearly dispersionless intensity bands above the spinon gap. In the Dirac state the lower edge of the continuum vanishes at the momenta connecting two nodes, including the reciprocal-lattice vectors, which connect a node with itself; at these momenta spectral weight extends down to zero frequency. Pairing gaps these soft points in the $\Z_2$ descendant, and the lower edge of its continuum follows twice the quasiparticle gap with a weak dispersion. The structure factors of all realized states are collected in the Supplemental Material~\cite{SM}.
\section{Spin--Orbit Coupling}
\label{sec:soc}

\subsection{Projective Moriya Rules}

In the presence of spin--orbit coupling the triplet links of Eq.~\eqref{eq:links} are allowed, and their spin index rotates with the lattice according to Eq.~\eqref{eq:psg-spatial}. By Eq.~\eqref{eq:TP}, the component $t^{a3}_{ij}$ multiplies the imaginary spin-dependent hopping $\ii f^\dagger_i\sigma^af_j$, which is even under time reversal and odd under exchange of $i$ and $j$. With these two properties, the vector $\bm{\lambda}_{ij}=(t^{x3}_{ij},t^{y3}_{ij},t^{z3}_{ij})$ is the parton counterpart of a Dzyaloshinskii--Moriya vector. The component $t^{a0}$ gives real spin-dependent hopping, odd under time reversal, and $t^{a1}$, $t^{a2}$ give triplet pairing.

For a PSG in which all symmetries act without projective signs, the allowed directions of $\bm{\lambda}_{ij}$ reproduce Moriya's rules~\cite{Moriya1960} for the bond stabilizers of Table~\ref{tab:bonds}. On the nearest-neighbor bonds, which are reversed by a mirror perpendicular to the bond through its midpoint, $\bm{\lambda}$ lies in that mirror plane. It has an out-of-plane component and an in-plane component perpendicular to the bond, the latter along the line joining the dodecagon and square centers on $J_b$, the dodecagon and hexagon centers on $J_g$, and the square and hexagon centers on $J_r$. On the $J'$ diagonals, reversed by the twofold rotation about the square center, $\bm{\lambda}$ lies in the plane. On $J''$ and $J'''$, which have no stabilizer, all three components are allowed on the reference bond and the other bonds of the orbit follow by symmetry. If the lattice plane is in addition a mirror plane, which is not part of $p6mm$, only out-of-plane components survive on the nearest-neighbor, $J''$, and $J'''$ bonds. On $J'$ the vector $\bm{\lambda}$ then vanishes, since the mirror combined with the twofold rotation is an inversion center.

A projective symmetry modifies these rules. Let $\bar O$ be the element that reverses a bond. The triplet link satisfies $R_{\bar O}\bm{\lambda}=\epsilon\,\rho\,\bm{\lambda}$, where $\rho=+1$ for the $\tau^0$ channel and $\rho=-1$ for the $\ii\tau^{1,2,3}$ channels, and $\epsilon$ is the sign with which the PSG realizes $\bar O$ on the bond: $\epsilon=\eta_{\sigma C_6}$ on $J_b$, $\epsilon=\eta_\sigma$ on $J_g$ and $J_r$, and $\epsilon=\eta_{12}\eta_{C_6}$ on $J'$. In a chiral class, $\epsilon$ changes sign when $\bar O$ is combined with time reversal, which happens for $J_b$, $J_g$, and $J_r$ in class B, for $J_b$ and $J'$ in class C, and for $J_g$, $J_r$, and $J'$ in class D. When $\epsilon\rho=-1$ the Moriya rule above holds. When $\epsilon\rho=+1$ it is inverted: on the nearest-neighbor bonds the Dzyaloshinskii--Moriya vector must point along the bond, and on $J'$ it must point out of the plane. Since the geometry of a physical Dzyaloshinskii--Moriya interaction always follows Moriya's rules, the sign $\epsilon$ determines the gauge channel in which it can enter the ansatz: the $\ii\tau^3$ channel when $\epsilon=+1$ and the $\tau^0$ channel when $\epsilon=-1$, provided that time reversal admits this channel on the bond, which requires $\zeta=-\theta_k$ in the notation of Sec.~SV\,B of the Supplemental Material~\cite{SM}. Otherwise a coupling with Moriya's geometry has no counterpart on that bond, which can then carry only the inverted direction, in the other channel if time reversal admits it. Time reversal does not alter the allowed directions. Through the product $h(\mu_i)h(\mu_j)$ of its sublattice signs it only selects which gauge channels survive. This product equals $\eta_{\T\sigma}\eta_{\T C_6}$ on $J_b$, $\eta_{\T\sigma}$ on $J_g$ and $J_r$, $\eta_{\T C_6}$ on $J'$ and $J'''$, and $+1$ on $J''$. Because it is $+1$ on $J''$, these bonds carry no amplitude at all, singlet or triplet, in any $\Z_2$ PSG with $g_\T=\tau^0$, in any $U(1)$ PSG with $w_\T=0$, and in any time-reversal-symmetric $SU(2)$ spin liquid. The complete rules for all channels and chiral classes are collected in Sec.~SV\,B and Table~S9 of the Supplemental Material. They agree with a direct solution of the symmetry conditions in all PSG classes without a continuous flux.

\subsection{Spinon Quantum Spin Hall States}
\label{sec:qsh}

Triplet amplitudes of the form $\ii\tau^3$ describe spin-dependent imaginary hopping, the parton counterpart of the Kane--Mele spin--orbit coupling~\cite{KaneMele2005a,KaneMele2005b}. In a $U(1)$ spin liquid they can turn the spinon bands into a band insulator with a nontrivial $\Z_2$ invariant, in analogy with the topological Mott insulator~\cite{PesinBalents2010} and with the fractionalized quantum spin Hall states of Refs.~\cite{YoungLeeKallin2008,RachelLeHur2010}. Which PSGs permit such a state is decided first by how time reversal acts on the spinons. Requiring $\Tr[\sigma^\alpha\psi_iu^{(\alpha)}_{ij}\psi_j^\dagger]$ to transform correctly under the physical antiunitary operation $\psi_i\to\ii\sigma^y\psi_iB_i$ fixes $B_i=G_\T(i)\,\ii\tau^2$. For $w_\T=1$, $G_\T\propto\ii\tau^1$, time reversal acts on the spinons as $f_i\to\pm\ii\sigma^yf_i$ and squares to $-1$. The number-conserving spinon problem is then in class AII, which admits a $\Z_2$ invariant in two dimensions~\cite{AltlandZirnbauer1997,Schnyder2008}. For $w_\T=0$, time reversal maps $f_{i\sigma}\to\pm f^\dagger_{i\sigma}$ and acts as a sublattice-type chiral symmetry (class AIII). The Chern number and the spin Chern number then vanish, and no spinon quantum spin Hall state is possible in these classes.

Two further constraints apply to $S^z$-conserving, out-of-plane spin--orbit coupling, for which the $\Z_2$ invariant equals $C_\uparrow\bmod2$, with $C_\uparrow=-C_\downarrow$ the Chern number of the spin-up bands. For $w_{C_6}=1$ the twofold rotation $C_2=C_6^3$ carries the gauge factor $\ii\tau^1$ and maps $f_{i\uparrow}$ to $f^\dagger_{C_2(i)\downarrow}$ up to a phase. On the Bloch Hamiltonian $h(\bm{k})$ it acts as an antiunitary operation $P'$ with $P'h(\bm{k})P'^{-1}=-h(\bm{k})$ and $P'^2=\eta_{C_6}$, which commutes with time reversal, maps spin-up onto spin-down states, and leaves the momentum invariant (for $\eta_{12}=-1$, up to half a reciprocal vector of the doubled cell). The product $C_2\T$ maps $\bm{k}$ to $-\bm{k}$ and is not a chiral symmetry of $h(\bm{k})$, so these classes remain in class AII. With $S^z$ conserved, $P'$ maps the occupied spin-up bands onto the empty spin-down bands and reverses the Berry curvature, so that $C_\downarrow=C_\uparrow$. Together with $C_\downarrow=-C_\uparrow$ this gives $C_\uparrow=0$ in all 32 classes with $w_{C_6}=1$. For $\eta_{12}=-1$ the projective translations anticommute on the spinons, which therefore see a flux $\pi$ per unit cell. A gapped set of spin-up bands that holds six spinons per unit cell then has an even Chern number, by the Diophantine relation for magnetic translations~\cite{TKNN1982,DanaAvronZak1985}, so that $C_\uparrow$ is even in all 32 classes with $\eta_{12}=-1$. Spin--orbit couplings that break the $S^z$ conservation are not covered by these two arguments.

The remaining 16 classes, with $w_{C_6}=0$ and $\eta_{12}=+1$, admit a parity criterion. Let $\xi$ be the product, over the four time-reversal-invariant momenta, of the eigenvalues of the twofold rotation (including its PSG gauge transformation) for the occupied spin-up bands. With $S^z$-conserving spin--orbit coupling, $(-1)^{C_\uparrow}=\xi$~\cite{FuKane2007,HughesProdanBernevig2011,FangGilbertBernevig2012}. For $\eta_{12}=-1$ the twofold rotation exchanges the time-reversal-invariant momenta of the doubled cell in pairs, and the criterion does not apply. A gapped singlet ansatz has $C_\uparrow=0$, so a singlet parent with $\xi=-1$ is necessarily gapless at half filling. An $S^z$-conserving spinon quantum spin Hall state can then arise in two ways. Out-of-plane spin--orbit coupling that gaps a parent with $\xi=-1$ without inverting bands at the time-reversal-invariant momenta keeps $\xi=-1$ and produces an odd $C_\uparrow$. Alternatively, out-of-plane spin--orbit coupling that inverts two spin-up levels of opposite twofold-rotation eigenvalue at a time-reversal-invariant momentum changes $\xi$ from $+1$ to $-1$ and turns a gapped trivial parent into a state with odd $C_\uparrow$. These statements and the exclusions above are analytic.

Which classes realize these mechanisms, and whether couplings that break the $S^z$ conservation produce $\Z_2$-odd states elsewhere, was determined numerically for all 64 $U(1)$ classes with $w_\T=1$. For each class we drew random singlet parents with amplitudes up to $J'''$, 40 per class for $\eta_{12}=+1$ and 16 per class for the 32 classes with $\eta_{12}=-1$ (40 for U001-100000), whose unit cell is doubled, with nearest-neighbor amplitudes of order one and further-neighbor amplitudes reduced by a random factor between 0 and 0.8. Each parent received four descendants, with out-of-plane or with generic symmetry-allowed spin--orbit coupling, all allowed coefficients drawn as $0.1$ (weak) or $0.4$ (strong) times a standard normal variable. At every sample the Lagrange multiplier was solved for $\langle n_i\rangle=1$ on all sites, the gap was obtained by refining the maximum of band $N_b/2$ and the minimum of band $N_b/2+1$ separately, together with dense scans of the mirror lines, and the $\Z_2$ invariant of every gapped sample was computed from the Wannier-center flow~\cite{SoluyanovVanderbilt2011} and with the Fukui--Hatsugai method~\cite{FukuiHatsugai2007}, odd values being confirmed on finer grids and by a strengthened gap search. Of the $3632$ descendants with out-of-plane couplings, $2637$ are gapped, and of the $3632$ with generic couplings $2744$ are gapped. With out-of-plane couplings, $\Z_2$-odd descendants occur only in U001-000000 ($14$ of $66$ gapped descendants) and U001-000001 ($9$ of $61$). Every gapped $S^z$-conserving descendant in the 16 classes with $w_{C_6}=0$ and $\eta_{12}=+1$ obeys $(-1)^{C_\uparrow}=\xi$, $C_\uparrow=0$ in every class with $w_{C_6}=1$, and $C_\uparrow$ is even in every class with $\eta_{12}=-1$, as the arguments above require. Couplings that break the $S^z$ conservation give $\Z_2$-odd descendants in ten classes: U001-000000 ($11$ of $39$ gapped descendants), U001-000001 ($7$ of $64$), U011-000000 ($3$ of $20$), U011-000011 ($23$ of $80$), U011-000101 ($13$ of $80$), U011-000110 ($8$ of $69$), U101-000001 ($15$ of $68$), U101-000011 ($20$ of $80$), U111-000001 ($14$ of $58$), and U111-000101 ($11$ of $80$). In the other 54 classes every gapped descendant is $\Z_2$ trivial, apart from three with gaps below $0.007$ whose invariant is not resolved by our grids; no $\Z_2$-odd descendant was found in the 32 classes with $\eta_{12}=-1$. Within these samples, mean-field spinon quantum spin Hall states with conserved $S^z$ thus occur in U001-000000 and U001-000001, and with generic spin--orbit coupling in at least ten classes, among them classes with $w_{C_6}=1$.

Singlet parents with $\xi=-1$ occur in these samples only in U001-000000, U011-000010, and U011-000100, as in a separate set of 80 parents per class. In the two U011 classes the mirrors act as particle--hole transformations, so that the spin-up spectrum is symmetric under $E\to-E$ on every mirror line of the Brillouin zone and nodes on these lines are pinned to the Fermi level. A survey of $4000$ spin--orbit-coupled descendants per class, restricted to U011-000010, U011-000100, and U001-000000, with out-of-plane or generic couplings and the multiplier solved at every sample, finds that every $S^z$-conserving descendant with $\xi=-1$ in the two U011 classes remains gapless and that all their gapped descendants are $\Z_2$ trivial, whereas about $30\%$ of the gapped descendants of U001-000000 are $\Z_2$ odd. On the diamond lattice the nonprojective class contains the spinon Fu--Kane--Mele model~\cite{FuKaneMele2007,Chauhan2023}; on the SHD lattice the nonprojective class hosts spinon quantum spin Hall states as well, but they are not confined to it.

\begin{figure*}
\centering
\includegraphics[width=\textwidth]{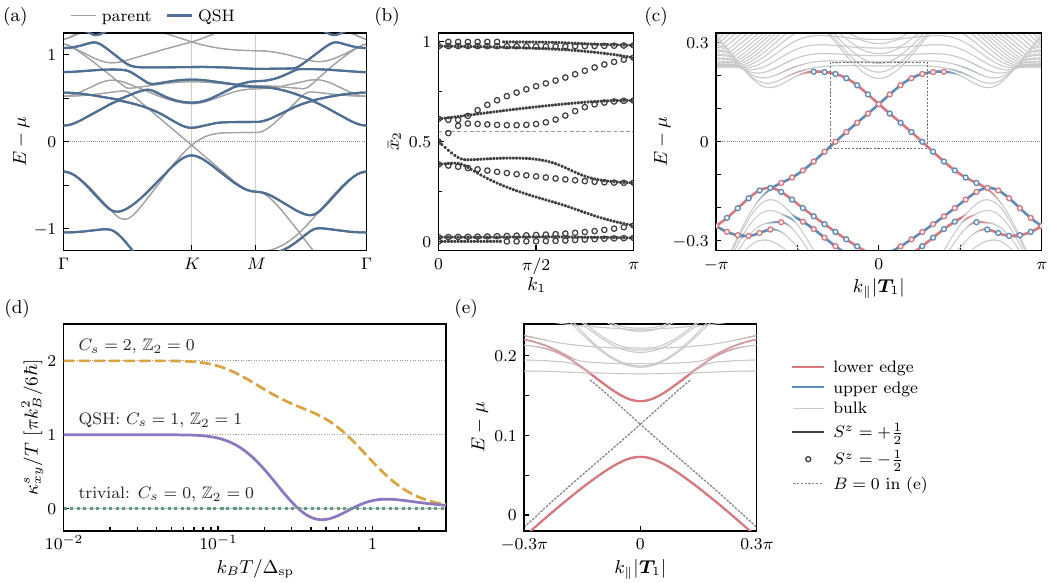}
\caption{Spinon quantum spin Hall (QSH) state in the $U(1)$ class U001-000000, with $(w_{C_6},w_\sigma,w_\T)=(0,0,1)$ and no projective signs, and singlet and out-of-plane triplet amplitudes up to $J'''$ (parameters in Sec.~SX of the Supplemental Material~\cite{SM}). Energies are in units of the amplitudes. (a) Spinon bands of the QSH state (blue) and of its singlet parent (gray), measured from the chemical potential $\mu$; the parent has Dirac nodes at $\pm K$, and the triplet term opens the gap $\Delta_{\rm sp}=0.317$. All bands are twofold degenerate. (b) Wannier-center flow of the occupied bands over half the Brillouin zone, with $k_{1,2}=\bm{k}\cdot\bm{T}_{1,2}$; spin-up (filled) and spin-down (open) centers flow in opposite directions, and every horizontal line is crossed an odd number of times, so that the $\Z_2$ invariant is $1$. The spin-resolved Chern numbers are $C_\uparrow=-C_\downarrow=1$. (c) Spectrum of a ribbon periodic along $\bm{T}_1$, twenty-four unit cells wide; lines (spin up) and circles (spin down) are colored by their weight near the lower (red) and upper (blue) edge; bulk states are gray. Each edge carries one Kramers pair of counterpropagating modes of opposite spin, which meet at $k_\parallel=0$. (d) Spin thermal Hall conductivity $\kappa^s_{xy}=(\kappa^\uparrow_{xy}-\kappa^\downarrow_{xy})/2$ in units of $\pi k_B^2T/(6\hbar)$, for the QSH state, a $\Z_2$-trivial state with spin Chern number $C_s=2$ (class U001-000100), and a trivial state ($C_s=0$); temperatures are in units of the gap of each state. At low temperature $\kappa^s_{xy}/T=C_s$ (dotted lines). (e) Lower-edge branches of the ribbon of (c) near the Kramers point [window marked in (c)] in a Zeeman field $B=0.2\Delta_{\rm sp}$ along $x$, which opens a gap $0.070\approx1.1B$; dashed lines show $B=0$.}
\label{fig:qsh}
\end{figure*}

In U001-000000 the quantum spin Hall states arise by the first mechanism, and their parent singlet states require substantial further-neighbor amplitudes. With $J''$ and $J'''$ the class contains Dirac semimetals with nodes at $\pm K$ and $\xi=-1$. Out-of-plane spin--orbit coupling turns them into spinon quantum spin Hall insulators with $C_\uparrow=-C_\downarrow=\pm1$ and $\Z_2$ invariant $1$ (Fig.~\ref{fig:qsh}). For the parent of Fig.~\ref{fig:qsh}, a coupling on any single one of the orbits that admit it in this class, $J_b$, $J_g$, $J_r$, $J''$, or $J'''$, opens a gap linear in the coupling. The $\Z_2$ invariants are obtained from the Wannier-center flow~\cite{SoluyanovVanderbilt2011} and with the Fukui--Hatsugai method~\cite{FukuiHatsugai2007}, and the spin-resolved Chern numbers with the Fukui--Hatsugai--Suzuki method~\cite{FukuiHatsugaiSuzuki2005}. Ribbons with edges parallel to $\bm{T}_1$ and to $\bm{T}_1+\bm{T}_2$ carry one helical Kramers pair of spinon edge modes per edge. The pair survives random time-reversal-symmetric edge perturbations and is gapped linearly by a Zeeman field. In-plane Dzyaloshinskii--Moriya components break the $S^z$ conservation but leave the $\Z_2$ invariant unchanged as long as the gap remains open. The six-node Dirac spin liquid of Sec.~\ref{sec:spectra} lies in the same class and has $\xi=-1$. Out-of-plane spin--orbit coupling on the $J_b$ or $J_r$ bonds gaps its nodes at first order and yields a spinon quantum spin Hall insulator with spin Chern number $3$, whereas the same coupling on the $J_g$ bonds alone leaves the nodes gapless. On the semi-Dirac line, where $\xi=+1$, the same coupling opens only a trivial gap, of second order in the coupling. Ribbons of this insulator with edges parallel to $\bm{T}_1$ and to $\bm{T}_1+\bm{T}_2$ carry a net three helical pairs per edge for every straight termination. Perturbations that break the $S^z$ conservation but preserve time reversal can remove two of them, leaving the single pair protected by the $\Z_2$ invariant. A state with spin Chern number $2$ and $\Z_2$ invariant $0$, which descends from the quadratic band touching of U001-000100, has edge modes protected only by $S^z$ conservation.

In U001-000001 the second mechanism operates. Its only projective sign, $\eta_{C_6}=-1$, gives $\epsilon=\eta_{12}\eta_{C_6}=-1$ on the $J'$ diagonals, which forbids the singlet hopping there and, by the inverted Moriya rule, allows an out-of-plane Dzyaloshinskii--Moriya vector, absent in U001-000000 (Sec.~SV\,B of the Supplemental Material~\cite{SM}). All sampled singlet parents of this class have $\xi=+1$. The out-of-plane coupling on $J'$ splits the levels at $\Gamma$ at first order and can invert two spin-up levels of opposite twofold-rotation eigenvalue. For the gapped parent specified in Sec.~SX of the Supplemental Material, with gap $1.149$, this coupling alone closes the gap at $\Gamma$ at $t^{z3}_{J'}=0.348$ and reopens it with $\xi=-1$. At $t^{z3}_{J'}=0.6$ the state is a spinon quantum spin Hall insulator with gap $0.979$, $C_\uparrow=-C_\downarrow=1$, and $\Z_2$ invariant $1$, and ribbons with edges parallel to $\bm{T}_1$ and to $\bm{T}_1+\bm{T}_2$ carry one helical Kramers pair per edge. With amplitudes on all orbits, weak out-of-plane couplings can suffice: in the survey, $2$ of the $34$ gapped descendants of U001-000001 with weak out-of-plane couplings are $\Z_2$ odd, with gaps $0.099$ and $0.075$ and $C_\uparrow=\pm1$.

Integrating out the gapped spinons of a state with $S^z$ conservation and spin Chern number $C_s$ gives no Chern--Simons term for the emergent gauge field $a$, because time reversal forces $C_\uparrow+C_\downarrow=0$, but a mutual term $(C_s/2\pi)\,a\,dA^s$ with a probe field $A^s$ that couples to $S^z$. Each $2\pi$ flux of $a$ therefore carries spin $S^z=C_s$. The response of the free spinons to a temperature gradient is a spin thermal Hall effect, with $\kappa^s_{xy}/T\to C_s\,\pi k_B^2/(6\hbar)$ at low temperature, while the thermal Hall conductivity vanishes. The structure-factor tensor acquires the uniaxial anisotropy $S^{xx}=S^{yy}\neq S^{zz}$ of the spin--orbit coupling, and its low-frequency weight at the momenta connecting the parent nodes is removed below twice the spinon gap. These are properties of the mean-field state, in which the gauge field is not compact. Beyond mean-field theory, the mutual term implies that monopoles of the compact $U(1)$ gauge field carry $S^z=C_s$, in analogy with the spin bound to $\pi$ fluxes in electronic quantum spin Hall insulators~\cite{RanVishwanathLee2008,QiZhang2008}. With exact $S^z$ conservation, monopole operators therefore cannot appear in the effective action. The gauge field does not confine, but its photon becomes, through the mutual term, the Goldstone mode of the spontaneously broken spin rotations about $\hat{\bm{z}}$, and the physical state is ordered, with an order parameter that carries $\Delta S^z=C_s$. For $C_s=1$ this is in-plane dipolar magnetic order; for $C_s=2$ it is spin-nematic order, with $\langle S^+_iS^+_j\rangle\neq0$ and no dipole moment; and for $C_s=3$ it is an order parameter of the form $(S^+)^3$, which leaves rotations by $2\pi/3$ about $\hat{\bm{z}}$ unbroken. With generic spin--orbit coupling $S^z$ is not conserved, monopoles are allowed, and the compact $U(1)$ gauge field confines in two dimensions~\cite{Polyakov1977}. In both cases the spinon quantum spin Hall insulator is a mean-field state whose physical fate in two dimensions is a magnetically ordered or confined phase rather than a stable spin liquid. This differs from the diamond lattice~\cite{Chauhan2023}, where a compact $U(1)$ gauge field in three dimensions has a stable Coulomb phase~\cite{Polyakov1977}. A stable descendant requires the gauge field to be Higgsed to $\Z_2$ by spinon pairing. In the $\Z_2$ descendants with $g_\T=\ii\tau^2$ the spinon Bogoliubov--de Gennes Hamiltonian is in class DIII. Spinon pairing added to the quantum spin Hall insulator of Fig.~\ref{fig:qsh} gaps its helical edge. Topological spinon superconductors with $\Z_2$ invariant $1$ and a helical pair of Majorana edge modes occur in several of these $\Z_2$ classes, among them Z000000(1), Z010110(1), and Z011100(1).
\section{Kitaev Spin Liquids}
\label{sec:kitaev}

The Kitaev model~\cite{Kitaev2006} on the tricoordinated SHD lattice is exactly solvable~\cite{Barari2026}, and its projected parton ansatz, an exact eigenstate up to a global spin rotation, is a reference point for the classification.

\subsection{Model and Symmetry}

The Kitaev Hamiltonian reads
\begin{equation}
H_K=\sum_{a=x,y,z}K_a\sum_{\langle ij\rangle_a}S^a_iS^a_j,
\label{eq:HK}
\end{equation}
where $\langle ij\rangle_a$ runs over the bonds of color $a$. In the coloring of Ref.~\cite{Barari2026} the $x$, $y$, and $z$ bonds are the square-hexagon ($J_r$), hexagon-dodecagon ($J_g$), and square-dodecagon ($J_b$) edges, respectively. Each color is a single orbit of $p6mm$, so every lattice operation maps each bond onto a bond of the same color, and no spin rotation needs to accompany a lattice operation. The symmetry group of $H_K$ is
\begin{equation}
G_K=p6mm\times D_2\times\Z_2^\T,
\label{eq:GK}
\end{equation}
where $p6mm$ acts on the sites only, $R_O=\one$ in Eq.~\eqref{eq:psg-spatial}, and $D_2=\{\one,R_x,R_y,R_z\}$ consists of the global $\pi$ rotations of all spins about the three axes. Time reversal and the $\pi$ rotations commute with all lattice operations. On the honeycomb lattice, by contrast, the threefold rotation permutes the bond colors and is a symmetry of the Kitaev model only together with a spin rotation that permutes $x$, $y$, and $z$; this is the setting of the parton mean-field descriptions of the honeycomb Kitaev spin liquid~\cite{BurnellNayak2011,YouKimchiVishwanath2012,SchafferBhattacharjeeKim2012} and of the use of its PSG to expand the spin operator in Majorana fermions beyond the exactly solvable limit~\cite{SongYouBalents2016}. Majorana spin liquids with a projectively realized $SU(2)$ spin symmetry were studied in Ref.~\cite{ChenEssinHermele2012}.

The coupling of spin and lattice in $G_K$ differs from the spin--orbit-coupled action of Sec.~\ref{sec:parton}, in which $C_6$ and $\sigma$ rotate the spin. In a spin--orbit-coupled magnet with symmetry $p6mm$, $C_6$ rotates the spins by $60^\circ$ about $\hat{z}$ and a mirror rotates them by $\pi$ about the in-plane normal of its mirror line. Couplings $K_aS^a_iS^a_j$ with one fixed spin axis for all bonds of an orbit are not invariant under these operations, because the Ising axis of a bond would have to rotate together with the bond. Equation~\eqref{eq:HK} with $R_O=\one$ is therefore an idealized model, not the effective Hamiltonian of a spin--orbit-coupled $p6mm$ magnet, and the results of this section are statements about this model and its symmetry group $G_K$. The algebraic classification of Sec.~\ref{sec:algebraic} nevertheless applies: the relations~\eqref{eq:relations} and the PSG conditions on the gauge parts $G_O(i)$ do not involve $R_O$, which enters only the constraints on triplet amplitudes. The PSGs of $p6mm\times\Z_2^\T$ obtained above therefore classify the Kitaev spin liquids as well, supplemented by the projective action of $D_2$, while the triplet links of the Kitaev ansatz below satisfy the constraints with $R_O=\one$ and not those of Sec.~\ref{sec:parton}.

\begin{figure*}
\centering
\includegraphics[width=\textwidth]{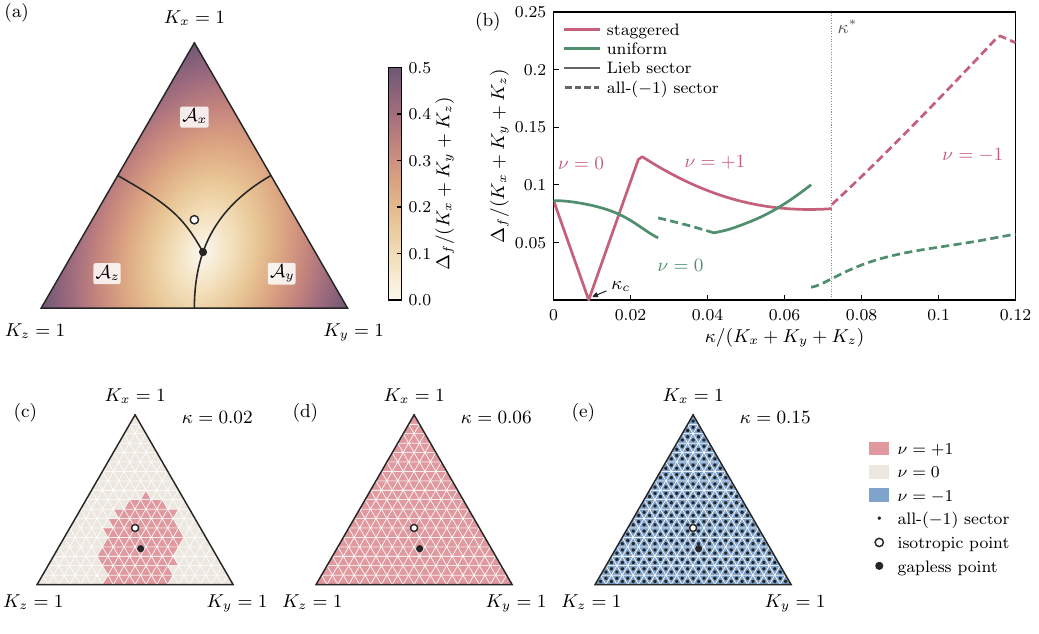}
\caption{Majorana spectrum of the Kitaev model $H_K=\sum_{a=x,y,z}K_a\sum_{\langle ij\rangle_a}S^a_iS^a_j$ ($\bm{S}=\bm{\sigma}/2$) on the SHD lattice, with $K_x+K_y+K_z=1$ setting the unit of energy. The bond colors are $x$ (square-hexagon edges, $J_r$), $y$ (hexagon-dodecagon edges, $J_g$), and $z$ (square-dodecagon edges, $J_b$). (a) Majorana fermion gap $\Delta_f=\min_{\bm{k},n}\varepsilon_n(\bm{k})$ over the triangle $K_a\geq0$, in the ground-state (Lieb) flux sector, $W_p=-1$ on squares and dodecagons and $W_p=+1$ on hexagons. $\Delta_f$ was computed on a grid of spacing $1/72$ in each $K_a$. Open circle: isotropic point $K_x=K_y=K_z$, where $\Delta_f=\sqrt{2-\sqrt3}/6=0.0863$. Filled circle: the only gapless point, $K_x=K_y/2=K_z/\sqrt3=1/(3+\sqrt3)$. Apart from the gapless point, the triangle is a single toric-code phase with Chern number $\nu=0$. Thin lines separate the regions $\mathcal{A}_x$, $\mathcal{A}_y$, and $\mathcal{A}_z$ of Ref.~\cite{Barari2026}; the boundary of $\mathcal{A}_y$ is the curve $K_y^2=K_x^2+K_z^2$, and the three lines meet at the gapless point and end at the midpoints of the edges. (b) $\Delta_f$ at the isotropic point as a function of the uniform or sublattice-staggered three-spin coupling $\kappa$ of Sec.~\ref{sec:kitaev-kappa}, with equal $\kappa$ on square, hexagon, and dodecagon paths. Each curve shows $\Delta_f$ in the flux sector of lowest energy among the 32 sectors realizable with a twelve-site-periodic gauge, which is either the Lieb sector (solid) or the sector with $W_p=-1$ on every plaquette (all-$(-1)$, dashed). For the staggered term, the gap closes at $\kappa_c=0.009$, where $\nu$ changes from $0$ to $+1$, and the ground state moves to the all-$(-1)$ sector, with $\nu=-1$, at $\kappa^\ast=0.072$ (dotted line); including all sectors of a doubled unit cell and 256 sectors of a $\sqrt3\times\sqrt3$ cell leaves $\kappa^\ast$ unchanged. For the uniform term the two sectors shown differ by less than $4\times10^{-4}$ per site for $0.02\leq\kappa\leq0.07$, and sectors with a doubled unit cell are degenerate with them to within about $10^{-5}$ per site, so the ground sector is not resolved (Sec.~\ref{sec:kitaev-kappa}); all these sectors have $\nu=0$ and an open gap. (c)--(e) Chern number $\nu$ of the negative-energy Majorana bands in the ground flux sector for the staggered term with $\kappa=0.02$, $0.06$, and $0.15$. Each small triangle of an $18\times18$ subdivision of the triangle is colored by $\nu$ at its centroid. A dot marks tiles whose ground state lies in the all-$(-1)$ sector; all other tiles are in the Lieb sector. Corners and special points are labeled as in (a). The sign convention for $\nu$ is given in Sec.~\ref{sec:kitaev-kappa}.}
\label{fig:kitaev-phase}
\end{figure*}

For any signs of the $K_a$, the Kitaev model reduces to free Majorana fermions $c_i$ hopping in a static $\Z_2$ gauge field $\mathsf{u}_{ij}=\pm1$~\cite{Kitaev2006}, $H_K=\frac{\ii}{4}\sum_{jk}A_{jk}c_jc_k$ with $A_{ij}=-(K_a/2)\mathsf{u}_{ij}$ on an $a$ bond. We define the plaquette variable $W_p=\prod_{\langle ij\rangle\in p}\mathsf{u}_{ij}$, with $i$ on sublattice A and $j$ on sublattice B for every edge of $p$, which does not depend on the orientation of the plaquette. The flux $\Phi_p$ of the Hermitian hopping matrix $\ii A$ through $p$, the phase of the cyclic product of its matrix elements around $p$, is then given by $e^{\ii\Phi_p}=W_p\prod_{\langle ij\rangle_a\in p}\mathrm{sgn}\,K_a$. The SHD lattice is planar and bipartite, all its faces are even, and the mirror lines of $p6mm$ contain no sites, bisect every plaquette, and preserve the bond colors. Lieb's theorem~\cite{Lieb1994} for half-filled hopping problems that are symmetric under reflections in such lines therefore places the ground state in the sector with $\Phi_p=\pi$ through squares and dodecagons and $\Phi_p=0$ through hexagons~\cite{Barari2026}, that is, $W_p=-1$ on squares and dodecagons and $W_p=\mathrm{sgn}(K_xK_y)$ on hexagons [Fig.~\ref{fig:kitaev-phase}(a)]. In units with $K_x+K_y+K_z=1$, the Majorana spectrum is gapped in the whole triangle except at the single point $K_x=K_y/2=K_z/\sqrt3$. This follows from the block $M(\bm{k})$ of $A$ between the two sublattices, since $\ii A(\bm{k})$ has a zero eigenvalue only where $\det M(\bm{k})=0$. For a twelve-site-periodic gauge in this sector, up to an overall sign and a shift of $\bm{k}$ that depend on the gauge,
\begin{align}
-2^6\det M(\bm{k})={}&(K_x^2+K_z^2)^3+K_y^6\nonumber\\
&+2K_xK_y^3\left[K_x^2+K_z^2\gamma(\bm{k})\right],
\label{eq:kitaev-detM}
\end{align}
with $\gamma(\bm{k})=\cos k_1-\cos k_2+\cos(k_1-k_2)\in[-3,\frac32]$ and $k_a=\bm{k}\cdot\bm{T}_a$. Writing $K_x+\ii K_z=re^{\ii\theta}$, for $K_a\geq0$ the right-hand side is bounded from below by its value at $\gamma=-3$, $(r^3-K_y^3)^2+2K_y^3r^3(1+\cos3\theta)\geq0$. The bound vanishes only for $r=K_y$ and $\theta=\pi/3$, which is the point $K_x=K_y/2=K_z/\sqrt3$, and there only at the single momentum with $\gamma=-3$. At this point a twofold-degenerate Dirac cone appears, at an $M$ point in the Bloch gauge of Ref.~\cite{Barari2026} and at $\Gamma$ in the canonical gauge of Eq.~\eqref{eq:z2psg}, in which translations and $C_6$ act without momentum shift. All bands are twofold degenerate (Sec.~\ref{sec:kitaev-psg}), and the Chern number vanishes. The gapped phase is a single $\Z_2$ toric-code phase.

The toric-code phase contains three regions $\mathcal{A}_x$, $\mathcal{A}_y$, and $\mathcal{A}_z$, which differ in which plaquettes host the $e$ and the $m$ anyons and which are separated by gap closings in two-vortex sectors only~\cite{Barari2026,ChernMoessnerCastelnovo2026}. Across their boundaries the ground state stays in the Lieb sector with an open Majorana gap, so no phase transition separates them: they belong to one toric-code phase and differ only in which lattice excitation, a vortex on a given plaquette class with or without a bound fermion, is labeled $e$ and which $m$. We locate the boundaries with the physical-parity criterion of Ref.~\cite{ChernMoessnerCastelnovo2026}. On $L\times L$ tori with up to $L=48$, a reference vortex and a second vortex on a square, hexagon, or dodecagon at a distance of order $L$ are inserted into the ground-state flux sector. Two plaquette classes host the same anyon when the corresponding two-vortex sectors have the same physical fermion parity; the parity follows from the sign of $\det M\prod\mathsf{u}_{ij}$. On all tori studied, the assignment is independent of the plaquette chosen within a class and of the Wilson-loop sector. It reproduces the assignments of Ref.~\cite{Barari2026} at the three strong-coupling corners, with the isotropic point in $\mathcal{A}_x$. The boundary between $\mathcal{A}_x$ and $\mathcal{A}_z$, a zero crossing of a level bound to a square vortex, runs from the gapless point to the midpoint of the edge $K_y=0$ and crosses the line $K_y=K_z$ at $K_x=0.266$. The boundary of $\mathcal{A}_y$ converges with increasing $L$ to the curve $K_y^2=K_x^2+K_z^2$, that is, $r=K_y$ in the notation of Eq.~\eqref{eq:kitaev-detM}; the size of the finite-$L$ deviation depends on which pair of vortex species is used as a probe. The curve passes through the gapless point and ends at the midpoints of the edges $K_z=0$ and $K_x=0$; on it a level bound to the hexagon and dodecagon vortices crosses zero. The parity criterion and the zero crossing of a vortex-bound Majorana level, with the vortex-free gap open, give the same boundaries [Fig.~\ref{fig:kitaev-phase}(a)], which converge exponentially in $L$ away from the gapless point and meet there up to a finite-size offset proportional to $1/L$.

\subsection{Parton Ansatz}

An exact parton description of the Kitaev model follows from the correspondence between Abrikosov fermions and the four Majorana fermions $b^x_i,b^y_i,b^z_i,c_i$ of Kitaev's representation established by Burnell and Nayak~\cite{BurnellNayak2011}. We use their identification [Eq.~(A1) of Ref.~\cite{BurnellNayak2011}]
\begin{align}
b^x&=\ii\left(f^\dagger_\uparrow-f_\uparrow\right),& b^y&=f^\dagger_\uparrow+f_\uparrow,\nonumber\\
b^z&=f^\dagger_\downarrow+f_\downarrow,& c&=\ii\left(f^\dagger_\downarrow-f_\downarrow\right).
\label{eq:bn}
\end{align}
On the four-dimensional Fock space of one site, $D_i=b^x_ib^y_ib^z_ic_i=(2n_{i\uparrow}-1)(2n_{i\downarrow}-1)$. With this identification the physical, singly occupied states have $D_i=-1$, and $(1-D_i)/2=n_{i\uparrow}+n_{i\downarrow}-2n_{i\uparrow}n_{i\downarrow}$ is the projector onto single occupancy. On these states $\ii b^x_ic_i$, $\ii b^y_ic_i$, and $\ii b^z_ic_i$ correspond to $-f^\dagger_i\sigma^yf_i$, $f^\dagger_i\sigma^xf_i$, and $-f^\dagger_i\sigma^zf_i$, respectively, as in Eq.~(A5) of Ref.~\cite{BurnellNayak2011}. In compact form, $\frac{1-D_i}{2}\,\ii b^a_ic_i=-Uf^\dagger_i\sigma^af_iU^\dagger$, where $U=\exp(-\ii\frac{\pi}{2}\sum_jS^z_j)$ rotates all spins by $\pi/2$ about $\hat{z}$, $\sigma^x\to\sigma^y$ and $\sigma^y\to-\sigma^x$. Since $H_K$ is quadratic in the spins, the overall sign drops out of every bond term, and Kitaev's Majorana Hamiltonian written with Eq.~\eqref{eq:bn} reduces on the physical states to $UH_KU^\dagger$, which is $H_K$ with $S^xS^x$ and $S^yS^y$ interchanged on all bonds. Gutzwiller projection of the ansatz below therefore yields Kitaev's exact eigenstates up to this global spin rotation: for a projected state $|\Psi\rangle$, $U^\dagger|\Psi\rangle$ is the corresponding eigenstate of $H_K$. The rotation acts on the spins only, so it commutes with all lattice operations, which do not rotate the spins in this idealized model, and with time reversal. It therefore leaves the PSG unchanged, and we apply it when spin correlation functions are computed from the parton state.

Writing the $a$-bond term of Kitaev's Majorana Hamiltonian as $(\ii b^a_ic_i)(\ii b^a_jc_j)=-(\ii b^a_ib^a_j)(\ii c_ic_j)$, decoupling it, and inserting Eq.~\eqref{eq:bn} gives, in the conventions of Eq.~\eqref{eq:links}, the link matrices
\begin{align}
u^{(0)}_{ij}&=-\frac{\mathsf{u}_{ij}}{4}\left(\frac{K_a}{4}+\lambda_a\right)\ii\tau^0,\nonumber\\
u^{(b)}_{ij}&=\frac{\mathsf{u}_{ij}}{4}\,p_b\left(\frac{K_a}{4}\pm\lambda_a\right)\ii\tau^b,
\label{eq:kitaev-ansatz}
\end{align}
for $i\in A$, $j\in B$, with $p_b=+1$, $-1$, and $+1$ for $b=x$, $y$, and $z$, the upper sign for $b=\bar{a}$ and the lower sign for $b\neq\bar{a}$, where $\bar{x}=y$, $\bar{y}=x$, and $\bar{z}=z$ label the spin components that represent the bond colors after the rotation $U$. Here $\mathsf{u}_{ij}$ is the gauge field of the Lieb sector, and $\lambda_a>0$ is the amplitude of the $b^a$ dimer on the bond. The projected state does not depend on $\lambda_a$, and the mean-field decoupling fixes it self-consistently to $\lambda_a=(K_a/4)|\langle\ii c_ic_j\rangle|$. The singlet link is pure hopping, and in each triplet channel the spin index $b$ is locked to the gauge direction $\tau^b$. For $\lambda_a=K_a/4$ only $u^{(0)}$ and the triplet $u^{(\bar{a})}$ survive. The Bogoliubov--de Gennes spectrum of the ansatz consists of the exact Majorana spectrum of the Lieb sector together with flat bands at $\pm2\lambda_a$ from the $b$ dimers, at every point of the phase diagram.

\subsection{Projective Symmetry Group}
\label{sec:kitaev-psg}

The IGG of the ansatz~\eqref{eq:kitaev-ansatz} is $\Z_2$ at every point of the phase diagram: the solutions of $W_iu^{(\alpha)}_{ij}=u^{(\alpha)}_{ij}W_j$ for all four link matrices are $W_i=\pm\tau^0$. Solving Eqs.~\eqref{eq:psg-spatial} and~\eqref{eq:psg-T} with $R_O=\one$ for each generator yields a unique gauge transformation up to the IGG, and evaluating the relations gives
\begin{equation}
(\eta_{12},\eta_{C_6},\eta_\sigma,\eta_{\sigma C_6},\eta_{\T C_6},\eta_{\T\sigma},t)=(+,-,-,-,+,-,0),
\label{eq:kitaev-labels}
\end{equation}
which is the class Z010111(0) of Sec.~\ref{sec:alg-z2}. A sign gauge transformation $W(x,y,\mu)=(-1)^xw_\mu$ brings the solution to the canonical form~\eqref{eq:z2psg}: $G_{T_1}=G_{T_2}=\tau^0$, $G_{C_6}$ and $G_\sigma$ equal to the sublattice signs of Table~\ref{tab:z2}, and $G_\T(\mu)=(-1)^{\mu+1}\tau^0$. The labels are the same in the three regions $\mathcal{A}_x$, $\mathcal{A}_y$, and $\mathcal{A}_z$, at the gapless point, and for every choice of the signs of the $K_a$. This PSG accounts for two features of the exact solution of Ref.~\cite{Barari2026}. That work traced the twofold degeneracy of every Majorana band to $(C_2\T)^2=-1$: since $C_2\T$ maps $\bm{k}$ onto itself, it acts at fixed momentum as an antiunitary operator on the $c$ Majoranas, and $(C_2\T)^2=-1$ there enforces the degeneracy. The gauge part of $(C_2\T)^2$, with $C_2=C_6^3$, is $\eta_{C_6}\eta_{\T C_6}^3\eta_{\T}=\eta_{C_6}=-1$, and $\T$ alone squares to $+1$ on the $c$ Majoranas. The sign $\eta_{C_6}$ has a direct origin in the gauge field: for the $c$ Majoranas the gauge part of $C_6^6$ equals the flux variable $W_D$ of the dodecagon at the rotation center, as we have checked for $C_6$-symmetric flux configurations on open clusters. The $\pi$ flux through the dodecagon thus gives the $c$ Majoranas half-integer angular momentum about its center, the analog of the double-cover point-group PSG found for Kitaev's honeycomb model on spherical clusters~\cite{MelladoPetrovaTchernyshyov2015}. In a sector with $W_D=+1$ on every dodecagon and Lieb values elsewhere the flux per unit cell is $\pi$, so that $\eta_{12}=-1$, while $\eta_{C_6}=+1$ and $(C_2\T)^2=+1$. The momentum shift that accompanies $C_3$ and the mirror in the Bloch representation of Ref.~\cite{Barari2026} is a gauge artifact. It corresponds to the sign $n_{C_6T_1}$ of Sec.~\ref{sec:alg-z2}, which is removed by a sign gauge transformation that preserves the twelve-site unit cell, and the translations carry no projective sign, $\eta_{12}=+1$. The distinction between $\mathcal{A}_x$, $\mathcal{A}_y$, and $\mathcal{A}_z$ concerns which lattice excitations carry the labels $e$ and $m$, which the fermionic PSG does not encode, and the gapless point is a property of the band structure within one PSG class.

The $\pi$ rotations are realized as $G_{R_a}=\ii\tau^a$ on every site, with gauge parts $R_a^2=(R_xR_y)^2=-1$ and trivial commutators with all elements of $p6mm\times\Z_2^\T$. The gauge parts $\ii\tau^a$ realize the same projective representation of $D_2$ as the spin lifts $-\ii\sigma^a$, so that the combined action on the Majoranas is linear: the Kitaev spin liquid locks the global spin rotations to gauge rotations, as expressed by the triplet links of Eq.~\eqref{eq:kitaev-ansatz}. The relation values depend on how $C_6$ and $\sigma$ are lifted. Lifting them together with a $\pi$ rotation, which is also a symmetry of $H_K$, changes $\eta_{C_6}$, $\eta_\sigma$, and $\eta_{\sigma C_6}$, while $\eta_{12}$, $\eta_{\T C_6}$, $\eta_{\T\sigma}$, and $t$ are independent of the lift. The $\pi$ rotations act on the $c$ Majoranas at most by a global sign, so $(C_2\T)^2=-1$ on the $c$ Majoranas, and with it the degeneracy, does not depend on the lift. Equation~\eqref{eq:kitaev-labels} refers to the pure lattice operations of Eq.~\eqref{eq:action}.

In the hierarchy of Sec.~\ref{sec:hierarchy}, Z010111(0) descends from an $SU(2)$ state. Dropping the triplet links of Eq.~\eqref{eq:kitaev-ansatz} leaves nearest-neighbor hopping $\ii\tau^0$ with the Lieb fluxes, which is the $SU(2)$ state SU010111 of Table~\ref{tab:su2ans} with $\chi_{23}\chi_{27}<0$ in the conventions of Table~\ref{tab:bonds}. With singlet amplitudes $\chi\propto K_a$ its hopping problem is the Majorana problem of the Lieb sector, and the gapless point $K_x=K_y/2=K_z/\sqrt3$ is the ray of SU010111 on which a Dirac cone appears at $\Gamma$ (Sec.~\ref{sec:spectra}). By Lieb's theorem~\cite{Lieb1994} its fluxes are those that minimize the energy of reflection-symmetric nearest-neighbor hopping at half filling. For hopping magnitudes that respect the lattice reflections, it is thus the lowest-energy flux pattern for a nearest-neighbor $SU(2)$ mean-field state of the Heisenberg model on this lattice. The Kitaev spin liquid is the $\Z_2$ descendant of this $SU(2)$ state in which the bond-dependent triplet links, locked to the bond color, break the gauge group down to $\Z_2$ without changing any projective sign.

\subsection{Projected Wave Functions}

Projecting the mean-field ground state of the ansatz~\eqref{eq:kitaev-ansatz} onto the singly occupied subspace removes the $b$ dimers and leaves, up to the rotation $U$, Kitaev's exact eigenstate of the corresponding gauge sector, provided that the Majorana vacuum of that sector has the physical fermion parity. Otherwise the projected mean-field ground state vanishes, and the lowest physical state of the sector is the projection of the state with one $c$ quasiparticle in the lowest level. We confirm this numerically. On the 12-site torus, at six points of the phase diagram and in the four Wilson-loop sectors, each projected state, rotated by $U^\dagger$, is an eigenstate of $H_K$ with the free-Majorana energy of its gauge sector, independent of $\lambda_a$, with the Lieb plaquette values and the two-spin correlations of exact diagonalization (ED). At $(0.6,0.25,0.15)$, in two-vortex sectors with the unphysical vacuum parity, the projected one-quasiparticle state is the eigenstate. On 24-site tori the rotated projected states are again exact eigenstates of $H_K$, and Lanczos diagonalization at the isotropic point gives a twofold-degenerate ground state of energy $-1.6334$, equal to that of the projected states in the two lowest Wilson-loop sectors.

\begin{table}
\caption{Nearest-neighbor spin correlations $\langle S^a_iS^a_j\rangle$ on bonds of color $a$, in the infinite system (Majorana solution) and on a 24-site torus (exact ground states, equal to the Gutzwiller-projected parton states rotated by $U^\dagger$; averaged over the inequivalent bonds of each color), at the isotropic point and at points inside the regions $\mathcal{A}_x$ and $\mathcal{A}_z$, with $K_x+K_y+K_z=1$. All other two-spin correlations of the projected states vanish.}
\label{tab:kitaev-corr}
\begin{ruledtabular}
\begin{tabular}{lccc}
$(K_x,K_y,K_z)$ & $a$ & infinite & 24 sites\\
\hline
$(\frac13,\frac13,\frac13)$ & $x$ & $-0.1539$ & $-0.1489$\\
 & $y$ & $-0.1091$ & $-0.1174$\\
 & $z$ & $-0.1413$ & $-0.1420$\\
$(0.6,0.2,0.2)$ & $x$ & $-0.2299$ & $-0.2296$\\
 & $y$ & $-0.0475$ & $-0.0483$\\
 & $z$ & $-0.0758$ & $-0.0762$\\
$(0.2,0.2,0.6)$ & $x$ & $-0.0773$ & $-0.0777$\\
 & $y$ & $-0.0405$ & $-0.0430$\\
 & $z$ & $-0.2310$ & $-0.2303$
\end{tabular}
\end{ruledtabular}
\end{table}

A spin operator $\sigma^a_j$ changes the fluxes of the two plaquettes adjacent to the $a$ bond at $j$, so $\langle S^a_iS^b_j\rangle$ vanishes unless $a=b$ and either $i=j$ or $i$ and $j$ are joined by an $a$ bond~\cite{BaskaranMandalShankar2007}. Table~\ref{tab:kitaev-corr} lists the surviving correlations; their strong dependence on the bond color reflects the absence of a symmetry permuting $x$, $y$, and $z$. On the 12-site torus the ground state at some couplings lies in the sector with $W_p=-1$ on every plaquette. This is a finite-size effect: at $(K_x,K_y,K_z)=(0.15,0.15,0.7)$ the energy per site of this sector relative to the Lieb sector is $-3.9\times10^{-5}$ on 12 sites and $+1.9\times10^{-6}$ on 24 sites, and the Lieb sector is the ground sector for all larger systems studied (up to 768 sites).

\subsection{Spin Structure Factors of the Kitaev Model}

\begin{figure*}
\centering
\includegraphics[width=\textwidth]{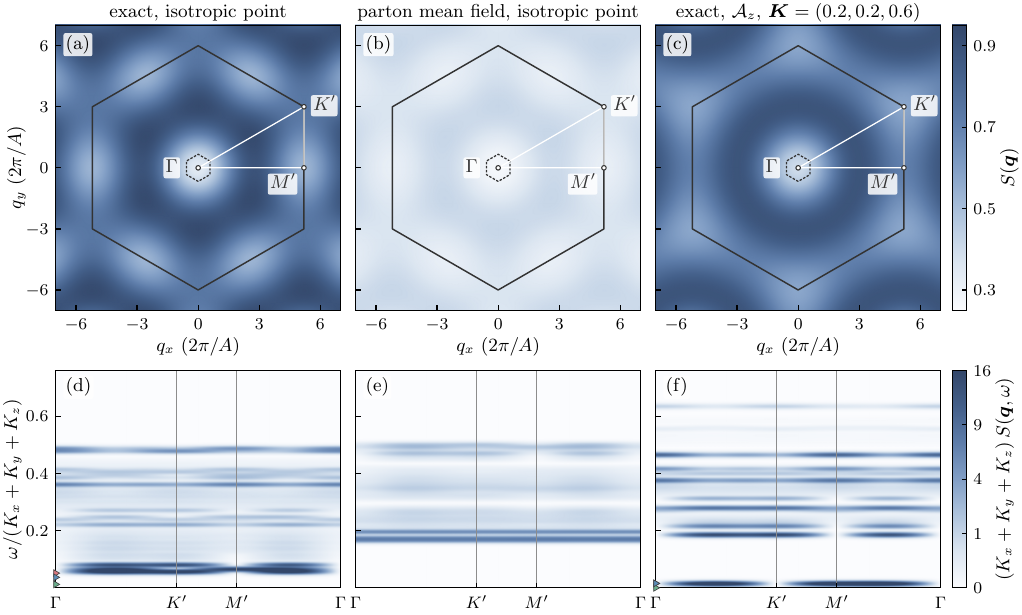}
\caption{Spin structure factors of the SHD Kitaev model. The isotropic point $K_x=K_y=K_z$ is shown in the exact solution [(a), (d)] and in the parton mean field [(b), (e)]; the toric-code region $\mathcal{A}_z$ at $(K_x,K_y,K_z)=(0.2,0.2,0.6)$ is shown in the exact solution [(c), (f)]. Energies are in units of $K_x+K_y+K_z=1$. All panels are computed with the site positions of the commensurate realization of Appendix~\ref{app:bz}, not with the ideal positions of Fig.~\ref{fig:spec-sq}. (a)--(c) Static structure factor $S(\bm{q})=\sum_{a}S^{aa}(\bm{q})$, with $S^{aa}(\bm{q})$ and its mean-field counterpart given by Eqs.~\eqref{eq:kitaev-sq} and~\eqref{eq:kitaev-sqmf} and $C_a$ from Table~\ref{tab:kitaev-corr}; zones, path (white), and momentum units as in Fig.~\ref{fig:spec-sq}. (d)--(f) Dynamical structure factor $S(\bm{q},\omega)=\sum_aN^{-1}\sum_{i,j}e^{\mathrm{i}\bm{q}\cdot(\bm{r}_i-\bm{r}_j)}\int\frac{dt}{2\pi}e^{\mathrm{i}\omega t}\langle S^a_i(t)S^a_j(0)\rangle$ along $\Gamma\to K'\to M'\to\Gamma$. Each Lehmann line is broadened by a Gaussian of standard deviation $0.005$. Each row has a single color scale, which is linear in (a)--(c) and follows a square-root law in (d)--(f). Results for the exact solution are computed with the local-quench method of Ref.~\cite{Knolle2014} on a $10\times10$-cell torus (1200 sites) in the Lieb flux sector, keeping the one-particle sector of the Lehmann sum. This sector carries $98.5$--$99.3\%$ (isotropic) and $99.5$--$99.7\%$ ($\mathcal{A}_z$) of the static weight along the path. Triangles on the left axis mark the two-flux gaps $\Delta_a$ created by flipping one $a$ bond, colored by bond type ($x$ red, $y$ green, $z$ blue). In $\mathcal{A}_z$, $\Delta_{x,y,z}=0.0155$, $0.0002$, and $0.0156$; the line at $\omega=\Delta_z$ in (f) is the zero-particle $\delta$ peak of $S^{zz}$, which carries about $95\%$ of its on-site weight. The mean field (e) uses the ansatz of Eq.~\eqref{eq:kitaev-ansatz} with the identification of Eq.~\eqref{eq:bn}, the spin rotation $U$ applied, and self-consistent $\lambda_a=(K_a/4)|\langle\mathrm{i}c_ic_j\rangle_a|=0.0513$, $0.0364$, and $0.0471$ for $a=x,y,z$. Its spectrum includes $\bm{q}$-independent lines of weight $1/16$ at $2(\lambda_b+\lambda_c)$, $\{a,b,c\}=\{x,y,z\}$.}
\label{fig:kitaev-sq}
\end{figure*}

Because only on-site and nearest-neighbor correlations of matching color survive, the static structure factor is
\begin{equation}
S^{aa}(\bm{q})=\frac14+\frac16\sum_{\bm{d}\in a}C_a\cos(\bm{q}\cdot\bm{d}),
\label{eq:kitaev-sq}
\end{equation}
where $C_a=\langle S^a_iS^a_j\rangle$ on $a$ bonds and the sum runs over the six $a$ bonds of a unit cell with bond vectors $\bm{d}$. The unprojected mean-field state obeys the same selection rule, since the $b$ Majoranas are strictly dimerized, but with incorrect weights. A site is singly occupied with probability $1/2$, so $\langle S^a_iS^a_i\rangle_{\rm MF}=1/8$, and the nearest-neighbor correlations are reduced by a factor of four, giving
\begin{equation}
S^{aa}_{\rm MF}(\bm{q})=\frac18+\frac14\left[S^{aa}(\bm{q})-\frac14\right].
\label{eq:kitaev-sqmf}
\end{equation}
At mean-field level the static structure factor has the correct momentum dependence, and projection restores its weights; see Fig.~\ref{fig:kitaev-sq}.

The dynamical structure factor differs qualitatively between the two. In the exact solution, $\sigma^a_j$ inserts a pair of fluxes adjacent to an $a$ bond, and $S^{aa}(\bm{q},\omega)$ is the response of the Majoranas to this local quench~\cite{Knolle2014}. Its spectral weight starts at the two-flux gap $\Delta_a$, which is small and depends strongly on the bond color: at the isotropic point $\Delta_x=0.051$, $\Delta_y=0.012$, and $\Delta_z=0.037$, the hexagon-dodecagon pair being the cheapest. Each component has a sharp bound state above $\Delta_a$, at $\omega=0.057$, $0.081$, and $0.066$ for $a=x$, $y$, and $z$, respectively, carrying $66\%$, $44\%$, and $61\%$ of the weight, followed by a continuum. Inside the region $\mathcal{A}_a$ the component $S^{aa}$ is dominated by a single $\bm{q}$-modulated $\delta$ function at $\omega=\Delta_a$ with about $95\%$ of the weight: on a strong bond the spin flip leaves the dimer in its ground state and creates only the flux pair. In the mean-field state, by contrast, the spin operator creates a $c$ Majorana together with a $b$ Majorana on a dimer, so $S^{aa}_{\rm MF}(\bm{q},\omega)$ is the free Majorana spectral function shifted by $2\lambda_a$. It has no flux gap and no bound states; its continuum starts at $2\lambda_a+\Delta_f$, equal to $0.19$, $0.16$, and $0.18$ at the isotropic point for the self-consistent $\lambda_a$, and for $a=x$ and $z$ the $\bm{q}$-independent lines at $2(\lambda_b+\lambda_c)$, at $0.167$ and $0.175$, lie below it. These are artifacts of the unprojected mean-field state, not of the PSG; the projected state, rotated by $U^\dagger$, is the exact eigenstate and has the exact dynamics.

\subsection{Three-Spin Term and Chiral Descendants}
\label{sec:kitaev-kappa}

For any couplings $K_a$, the SHD Kitaev model has no chiral phase: all faces are even, the ground state lies in the real flux sector, and $C_2\T$ forces the Berry curvature of the Majorana bands to vanish pointwise, so that the thermal Hall conductivity $\kappa_{xy}$ vanishes identically. As an exactly solvable deformation we add the three-spin term $H_\kappa=\kappa\sum\epsilon_k\sigma^\alpha_j\sigma^\gamma_k\sigma^\beta_l$, summed over paths $j$--$k$--$l$ with $jk$ an $\alpha$ bond, $kl$ a $\beta$ bond, and $\gamma$ the third color, each unordered path counted once. The uniform term has $\epsilon_k=1$; the staggered term has the sublattice sign $\epsilon_k=\pm1$ of Sec.~\ref{sec:ansatz}. On the SHD lattice the colors $x\to y\to z$ run counterclockwise around the sites with $\epsilon_k=+1$ and clockwise around those with $\epsilon_k=-1$. $H_\kappa$ commutes with all $W_p$, and in the physical subspace it is a second-neighbor Majorana hopping, $\ii c_jc_l$, on the $J'$, $J''$, and $J'''$ bonds, which are the square, hexagon, and dodecagon second neighbors. These three path types are not related by symmetry, and taking the same $\kappa$ on all of them is a choice of model; we use equal weights unless stated otherwise. We define $\nu$ as the Chern number of the negative-energy bands of $\ii A(\bm{k})$, $\nu=(2\pi)^{-1}\sum_{n}\int d^2k\,\Omega_n$ with $\Omega_n=\ii(\langle\partial_{k_1}u_n|\partial_{k_2}u_n\rangle-\langle\partial_{k_2}u_n|\partial_{k_1}u_n\rangle)$; with this convention the chiral edge modes of the phase with $\nu=+1$ propagate clockwise around the sample.

In Kitaev's perturbative argument~\cite{Kitaev2006}, a magnetic field $\bm{h}$ coupled to $\sum_i\bm{h}\cdot\bm{\sigma}_i$ generates the uniform term at third order, with $\kappa\sim h_xh_yh_z/\Delta^2$, where $\Delta$ is the energy of the intermediate flux excitations. A field $\epsilon_i\bm{h}$ along $[111]$ with opposite signs on the two sublattices generates the staggered term instead, since the product $\epsilon_j\epsilon_k\epsilon_l$ of the three field factors along a path equals $\epsilon_k$. This link holds only for $h_a$ small compared with the two-flux gaps, which at the isotropic point are $\Delta_x=0.051$, $\Delta_y=0.012$, and $\Delta_z=0.037$. With $\Delta$ of this order, the values $\kappa\gtrsim\kappa_c$ studied below correspond to field components comparable to or larger than $\Delta_y$, outside the perturbative regime, and near the corners of the triangle the two-flux gaps are smaller still. Like the pure lattice action of $p6mm$, a sublattice-staggered field is an idealization. We therefore treat $\kappa$ as a parameter of the exactly solvable model and do not relate its values to a field strength.

Since the lattice mirrors preserve the colors, the uniform three-spin term is invariant under all of $p6mm$ and breaks only time reversal. In the Lieb and all-$(-1)$ sectors its PSG is the class-A descendant of Eq.~\eqref{eq:kitaev-labels}, the chiral $\Z_2$ PSG A:0111 (labels $n_{T_2}n_\sigma n_{\sigma C_6}n_{C_6}$, Eq.~(S112) of the Supplemental Material~\cite{SM}). As in Sec.~\ref{sec:chiral}, the unitary mirrors of class A and the antiunitary $C_2\T$ of classes C and D force $\nu=0$ for the Majorana bands, so the uniform term gives $\nu=0$. A stronger statement holds for the spin model: $H_K+\sum_i\bm{h}\cdot\bm{\sigma}_i$ is invariant under the pure lattice mirrors for every direction of $\bm{h}$, and a unitary orientation-reversing symmetry forces the chiral central charge and $\kappa_{xy}$ to vanish. Within this model a uniform field therefore cannot produce a phase with nonzero chiral central charge, such as the non-Abelian Ising phase, unless the mirrors are broken spontaneously. With the uniform term the ground flux sector is not resolved. At the isotropic point the lowest of the 2048 sectors realizable with a doubled (24-site) unit cell is the all-$(-1)$ sector for $\kappa=0.04$, with six sectors within $2\times10^{-5}$ per site, and for $\kappa=0.05$ and $0.06$ the sector with $W_D=+1$ on every dodecagon and Lieb values elsewhere, $1.2\times10^{-5}$ and $1.1\times10^{-6}$ per site below the Lieb sector. That sector has $\eta_{12}=-1$ and $\eta_{C_6}=+1$ (Sec.~\ref{sec:kitaev-psg}), so its PSG is not a descendant of Eq.~\eqref{eq:kitaev-labels}. Every doubled-cell sector within $4\times10^{-5}$ per site of the lowest is gapped with $\nu=0$, in line with the mirror argument. At the gapless point the uniform term opens a trivial gap, $0.031$ for $\kappa=0.05$ in the all-$(-1)$ sector, which is the lowest doubled-cell sector there.

\begin{figure}
\centering
\includegraphics[width=\columnwidth]{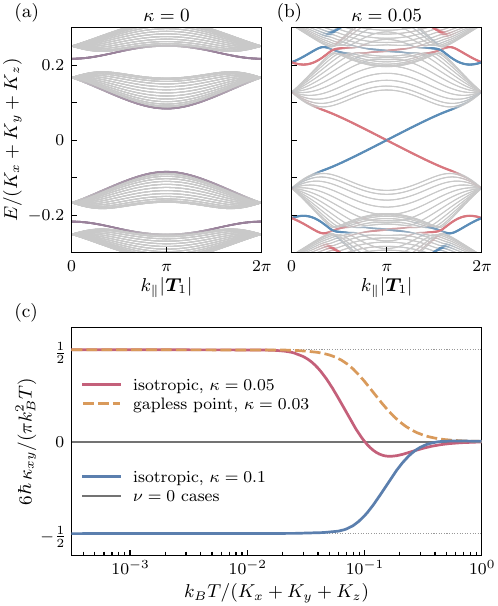}
\caption{Chiral Majorana edge modes and thermal Hall response of the SHD Kitaev model, with energies in units of $K_x+K_y+K_z=1$. (a), (b) Majorana spectrum of a ribbon periodic along $\bm{T}_1$, with momentum $k_\parallel$ along the edge, at the isotropic point $K_x=K_y=K_z$ in the Lieb flux sector. Each state is colored by its weight on the outer 1.5 rows of the lower (red) and upper (blue) edge; bulk states are gray. For exactly degenerate levels the weights are averaged over the degenerate subspace, so a degenerate pair localized on opposite edges appears in the mixed color. (a) $\kappa=0$: no state crosses the bulk gap ($0.0863$). The only states below the bulk band edge are a degenerate pair of nonchiral states, one on each edge, which lies $0.002$ below the band edge near $k_\parallel|\bm{T}_1|=\pi$. (b) Staggered three-spin term $\kappa=0.05$, defined as in Fig.~\ref{fig:kitaev-phase}(b), for which $\nu=+1$ and the bulk gap is $0.085$. One chiral Majorana mode crosses the gap on each edge, with opposite velocities on the two edges, and the two modes cross $E=0$ at $k_\parallel|\bm{T}_1|=\pi$. (c) Thermal Hall conductivity $\kappa_{xy}/T$ in units of $\pi k_B^2/(6\hbar)$ of the free Majorana fermions in the ground flux sector, given by Eq.~\eqref{eq:kxy} with $d_s=1/2$, which accounts for the Majorana doubling, and $\mu=0$. The curves are the isotropic point with staggered $\kappa=0.05$ (Lieb sector, $\nu=+1$, $\Delta_f=0.085$), the gapless point $K_x=K_y/2=K_z/\sqrt3$ with staggered $\kappa=0.03$ (Lieb, $\nu=+1$, $\Delta_f=0.105$), and the isotropic point with staggered $\kappa=0.1$ (all-$(-1)$ sector, $\nu=-1$, $\Delta_f=0.174$). For $k_BT\ll\Delta_f$ they approach the half-quantized plateau $\nu/2$ (dotted lines), which corresponds to chiral central charge $c=\nu/2$. The gray line shows the cases with $\nu=0$, where $\kappa_{xy}$ vanishes identically: the isotropic and gapless points at $\kappa=0$, and the gapless point with uniform $\kappa=0.1$ (all-$(-1)$ sector). The free-Majorana result neglects thermally excited fluxes and is quantitative only for $k_BT$ well below the two-flux gaps, which are $0.058$--$0.125$ at the isotropic point with $\kappa=0.05$.}
\label{fig:kitaev-edge}
\end{figure}

The staggered term is the chiral counterpart of Kitaev's honeycomb term on the SHD lattice. It is odd under $\sigma$ and under $\T$ but invariant under $\sigma\T$, and its PSG is the class-B descendant B:0001, obtained from Eq.~\eqref{eq:kitaev-labels} through $(\sigma\T)^2=\eta_\sigma\eta_{\T\sigma}\eta_{\T}$. In both cases the IGG remains $\Z_2$, and the projected ansatz with second-neighbor $c$ hopping is again an exact eigenstate up to the rotation $U$ (ED on 12 sites). At the isotropic point the staggered term closes the Majorana gap at $\kappa_c=0.009$ through a single nondegenerate Dirac cone at a $C_6$-invariant momentum, beyond which $\nu=+1$. This first transition, from $\nu=0$ to $\nu=+1$ at a single cone, persists for unequal weights of the three path types, with $\kappa_c$ between $0.015$ and $0.043$ for the weights we tried. Toward the gapless point $\kappa_c$ decreases linearly to zero ($\kappa_c=0.0046$ halfway between the isotropic and gapless points), and at the gapless point itself $\nu=+1$ already for $\kappa=0.001$, with a gap of $0.105$ at $\kappa=0.03$. At $\kappa=0.06$ the bands carry $\nu=+1$ at every sampled point of the triangle [Fig.~\ref{fig:kitaev-phase}(d)]; on the tiles at the corners, where the gap is about $0.09$, $\kappa$ exceeds the two weak couplings and the three-spin term is not a small perturbation of $H_K$. The resulting phase is the non-Abelian Ising phase of Kitaev's sixteenfold way~\cite{Kitaev2006}, with one chiral Majorana mode per edge and a thermal Hall conductivity $\kappa_{xy}/T=\frac12\,\pi k_B^2/(6\hbar)$ for $T$ below the Majorana and flux gaps~\cite{SumiyoshiFujimoto2013,NasuYoshitakeMotome2017} [Fig.~\ref{fig:kitaev-edge}; the ribbon geometry is specified in Sec.~SX of the Supplemental Material~\cite{SM}]. Lieb's theorem does not apply for $\kappa\neq0$. At the isotropic point the sector with $W_p=-1$ on every plaquette becomes the lowest in energy among the 32 sectors of Fig.~\ref{fig:kitaev-phase}(b) at $\kappa^\ast=0.072$, and there $\nu=-1$ [Fig.~\ref{fig:kitaev-phase}(e)]. Including all 2048 doubled-cell sectors and 256 sectors of a $\sqrt3\times\sqrt3$ cell does not change this: at $\kappa=0.07$, $0.0723$ (doubled cell only), and $0.075$ all other sectors lie at least $4\times10^{-4}$ per site above the lower of the two. The value of $\kappa^\ast$, the reversal of $\nu$ there, and $\nu=-1$ at $\kappa=0.15$ depend, however, on the equal weights of the three path types. For weights $0.3:1:0.3$ (square:hexagon:dodecagon) the Lieb sector remains the lowest of the 32 sectors up to $\kappa=0.15$, and $\nu$ changes to $-1$ within it; for $0.3:0.3:1$ the ground state keeps $\nu=+1$ up to $\kappa=0.1$; and with dodecagon paths only, $\nu=-2$ at $\kappa=0.1$. The uniform term, in contrast, only opens a trivial gap. Within the chiral descendants of Eq.~\eqref{eq:kitaev-labels} in classes A--D, a non-Abelian Kitaev phase on the SHD lattice requires a time-reversal-odd perturbation that breaks the site mirrors $\sigma$ but preserves $\sigma\T$. The staggered three-spin term is such a perturbation; for fields small compared with the two-flux gaps, a $[111]$ field of opposite sign on the two sublattices generates it at third order.
\section{Conclusions and Outlook}
\label{sec:conclusion}

Given that the space group acts freely on the sites of the square-hexagon-dodecagon lattice and only time reversal fixes them, its $\Z_2$ PSGs are the classes of $H^2(p6mm\times\Z_2^\T,\Z_2)$ themselves, whereas for the IGG $U(1)$ the on-site condition of time reversal removes $64$ of the $208$ discrete cohomology classes (Sec.~\ref{sec:alg-u1}). For every lattice whose sites form a single orbit of $p6mm$, the $\Z_2$ PSGs follow by restricting each SHD class to the stabilizer of a site and counting its on-site realizations. On the triangular, honeycomb, kagome, star, and ruby lattices this agrees with the earlier classifications where they overlap, adds eight PSGs with $g_\T=\ii\tau^3$ to the honeycomb list, and shows that PSGs with identical relation values can differ in their on-site representations and hence in the mean-field states they allow (Sec.~\ref{sec:parent}). The ans\"atze were constructed up to fourth neighbors, the pairs at graph distance two, with singlet and triplet amplitudes for symmetric states and singlet amplitudes for chiral states.

For the Kitaev model, an idealized model in which the lattice operations do not rotate the spins, the classification is tested against an exact solution. A single $\Z_2$ class, Z010111(0), descending from the Lieb-flux $SU(2)$ state SU010111, covers the whole anisotropic phase diagram. Its sign $\eta_{C_6}=-1$ equals the flux variable of the dodecagon at the rotation center, gives $(C_2\T)^2=-1$ on the Majorana fermions, and enforces the twofold degeneracy of all Majorana bands. Because the mirrors preserve the bond colors, the model in a uniform field of any direction keeps a unitary mirror symmetry and has no phase with a nonzero chiral central charge unless the mirrors break spontaneously, whereas a sublattice-staggered three-spin term, odd under $\sigma$ but even under $\sigma\T$, drives the non-Abelian Ising phase. The relation of this term to a sublattice-staggered field holds only for fields small compared with the two-flux gaps, and the reversal of the Chern number at larger couplings depends on the weights of the three types of paths. By Lieb's theorem the Lieb-flux pattern also minimizes the mean-field energy of nearest-neighbor $SU(2)$ hopping with reflection-symmetric magnitudes, which makes SU010111 and its descendants natural starting points for Heisenberg models on this lattice.

The gapless and topological states found away from the Kitaev point are mean-field states, and their status beyond mean-field theory differs. Gapped $\Z_2$ states and chiral states with nonzero Chern number, which occur only in chiral class B, can be stable topological phases; their topological orders are inferred from mean-field Chern numbers, and their free-spinon thermal Hall plateaus coincide with the physical ones only for $\Z_2$ states. Gapped $U(1)$ and $SU(2)$ states without a Chern--Simons term confine in two dimensions~\cite{Polyakov1977}. This applies to the mean-field spinon quantum spin Hall insulators, which occur with conserved $S^z$ in U001-000000 and U001-000001 and are excluded for conserved $S^z$ in the classes with $w_{C_6}=1$ or $\eta_{12}=-1$, and which occur in at least ten classes with generic spin--orbit coupling: with conserved $S^z$ their monopoles carry spin and the physical state orders magnetically, and with generic spin--orbit coupling the gauge field confines. Spinon pairing Higgses the gauge field to $\Z_2$, and topological spinon superconductors of class DIII~\cite{AltlandZirnbauer1997,Kitaev2009} occur in several $\Z_2$ classes. The Dirac spin liquids, with $N_f$ between $8$ and $48$ and in thin regions $72$ and $96$, occupy regions of amplitude space, and several of them are gapped at other amplitude points (Sec.~SVII of the Supplemental Material~\cite{SM}). For the $N_f=12$ state of U001-000000 the PSG forbids every spin-singlet mass, so its stability depends on its monopoles. For $N_f\ge24$ the monopoles are irrelevant at large $N_f$, and the stability is decided by fermion bilinears that we have not classified. The $N_f=48$ phase of U011-100101 requires a fourth-neighbor amplitude at least comparable to the second-neighbor one, together with a third-neighbor amplitude, in a class that forbids $J_b$ amplitudes, and whether such an ansatz is favorable for any spin Hamiltonian is open. The semi-Dirac state is critical, a hypersurface of codimension one in amplitude space, and its exponents are mean-field values. The nodal lines are protected only within the $SU(2)$ singlet sector.

The approach has several limitations. The PSG constrains the mean-field amplitudes but does not decide which ansatz has the lowest energy for a given spin Hamiltonian, and none of the states was optimized for one. Spectral types, phase diagrams, and area fractions refer to chosen points, planes, and measures in amplitude space. The flavor numbers of the Dirac phases, the classes that host spinon quantum spin Hall states with generic spin--orbit coupling, and the absence of other semi-Dirac mechanisms rest on random samples. Fermion bilinears were classified only for one Dirac state, chiral ans\"atze with triplet amplitudes were not scanned, and except at the Kitaev point no projected wave function was evaluated, so that structure factors, edge states, and thermal Hall responses are those of the mean-field spinons. The classical analysis combines the Luttinger--Tisza method with simulated annealing on finite tori along one cut and contains no estimate of quantum fluctuations.

For the many-flavor Dirac phases the open questions are the symmetry-allowed fermion bilinears, including the intervalley bilinears of the doubled-cell states, and the lattice quantum numbers of the monopoles~\cite{Song2019,Song2020}. For the $N_f=12$ state, following the analysis of the maple-leaf lattice~\cite{Zhang2026}, this requires the action of the PSG on the Dirac valleys, including the mirror $\sigma$ that is absent on the maple-leaf lattice, and the Berry phases of the monopoles under $C_6$, $\sigma$, and translations. Since the SHD unit cell contains an even number of spins, no Lieb--Schultz--Mattis constraint forbids symmetry-allowed charge-one monopoles. Whether such monopoles exist, and what their scaling dimensions are, decides whether the SHD lattice is more favorable for a stable QED$_3$ phase than the lattices studied so far. Whether the nodes of the gapless $\Z_2$ states are enforced by their symmetry-fractionalization classes, as for the symmetry-protected gapless $\Z_2$ spin liquids of Ref.~\cite{Lu2018}, is also open. The $SU(2)$ Dirac states are mean-field realizations of QCD$_3$ with large flavor numbers; they have no conserved gauge flux, and the question of their stability takes a different form. The energetics of frustrated SHD Heisenberg models can be addressed by variational Monte Carlo~\cite{Gros1989,BeccaSorella2017} with Gutzwiller-projected ans\"atze, Pfaffian wave functions for the $\Z_2$ states~\cite{Bajdich2006,Bajdich2008,TaharaImada2008}, and the insertion of monopoles into Dirac states~\cite{Budaraju2023,Budaraju2025}, as done for the kagome and triangular lattices~\cite{Iqbal2014,Iqbal2016}. Such a study of the $J_1$--$J'$--$J''$--$J'''$ model would compare the projected states with the cluster-singlet products and with magnetic order near the flat-band lines and in the classically frustrated wedge. The contraction argument could also organize the bosonic PSGs of $p6mm$ lattices. A Schwinger-boson classification~\cite{WangVishwanath2006,YangWang2016,JinZhou2020,LiuWang2024} would extend the present analysis to spin liquids proximate to magnetic order, whose bosonic spinons give flux-sector-dependent thermal Hall signatures~\cite{Sasamoto2026}.

\begin{acknowledgments}
Y.I. thanks Chunxiao Liu, Julien Vidal and Atanu Maity for helpful discussions.
H.R. acknowledges the Council of Scientific and Industrial Research (CSIR), Government of India, for financial support through a Senior Research Fellowship (File No.~09/0084(16360)/2023-EMR-I).
The work of Y.I. was performed, in part, at the Aspen Center for Physics, which is supported by a grant from the Simons Foundation (1161654, Troyer). This research was also supported in part by grant NSF PHY-2309135 to the Kavli Institute for Theoretical Physics (KITP). Y.I. acknowledges support from the Abdus Salam International Centre for Theoretical Physics (ICTP) through the Associates Programme, from the Simons Foundation through Grant No.~284558FY19, from IIT Madras through the Institute of Eminence (IoE) program for establishing QuCenDiEM (Project No.~SP22231244CPETWOQCDHOC), and thanks the International Centre for Theoretical Sciences (ICTS), Bengaluru, for participation in the Discussion Meeting ``Fractionalized Quantum Matter'' (code: ICTS/DMFQM2025/07).  We acknowledge the use of the AI tools Claude Opus 5.5 and Fable 5.1 (Anthropic) for assistance with code debugging, verifying some of the calculations, and preparing parts of the manuscript. The authors reviewed all results and changes and take full responsibility for the content.
\end{acknowledgments}

\section*{Data availability}
The data that support the findings of this article are available from the authors upon reasonable request.

\makeatletter\let\SM@hangfrom@section\@hangfrom@section\let\SM@sectioncntformat\@sectioncntformat\makeatother
\appendix
\section{Fermion Bilinears of the Link Matrices}
\label{app:bilinears}

This appendix expresses the mean-field Hamiltonian~\eqref{eq:HMF} in terms of the fermions $f_{i\alpha}$. Write $f_i=(f_{i\uparrow},f_{i\downarrow})^{\rm T}$ and $f^{\dagger\rm T}_j=(f^\dagger_{j\uparrow},f^\dagger_{j\downarrow})^{\rm T}$. Inserting the matrix~\eqref{eq:psi} into $\Tr[\sigma^\alpha\psi_iu^{(\alpha)}_{ij}\psi_j^\dagger]$ and multiplying out the $2\times2$ products gives a sum of terms $f^\dagger_{i\alpha}f_{j\beta}$, $f_{i\alpha}f^\dagger_{j\beta}$, $f^\dagger_{i\alpha}f^\dagger_{j\beta}$, and $f_{i\alpha}f_{j\beta}$. Adding the term with $i$ and $j$ exchanged, which carries $u^{(\alpha)}_{ji}=u^{(\alpha)\dagger}_{ij}$, and anticommuting all operators into the order $f^\dagger_i(\cdots)f_j$ and $f^\dagger_i(\cdots)f^{\dagger\rm T}_j$ yields, for each bond,
\begin{align}
&\sum_{\alpha}\Tr\left[\sigma^\alpha\psi_iu^{(\alpha)}_{ij}\psi_j^\dagger\right]+(i\leftrightarrow j)\nonumber\\
&\qquad=f_i^\dagger\,\mathcal{T}_{ij}\,f_j+f_i^\dagger\,\mathcal{P}_{ij}\,f^{\dagger\rm T}_j+\mathrm{h.c.},
\label{eq:bilinear}
\end{align}
with the spin-space hopping and pairing matrices
\begin{align}
\mathcal{T}_{ij}&=2\left(\ii\chi^0_{ij}-\chi^3_{ij}\right)\sigma^0\nonumber\\
&\quad+2\sum_{a}\left(-t^{a0}_{ij}+\ii t^{a3}_{ij}\right)\sigma^a,\nonumber\\
\mathcal{P}_{ij}&=-2\left(\Delta^1_{ij}+\ii\Delta^2_{ij}\right)\ii\sigma^y\nonumber\\
&\quad+2\sum_{a}\left(\ii t^{a1}_{ij}-t^{a2}_{ij}\right)\sigma^a\,\ii\sigma^y,
\label{eq:TP}
\end{align}
where the coefficients are those of Eq.~\eqref{eq:links}. The triplet link matrices produce spin-dependent hopping $\propto\sigma^a$ and triplet pairing $\propto\sigma^a\ii\sigma^y$, whose vector of coefficients is the usual $\bm{d}$-vector. The on-site term gives
\begin{align}
\Tr\left[\psi_ia_i\psi_i^\dagger\right]=&-2a^3_i\left(n_i-1\right)\nonumber\\
&-2\left[\left(a^1_i+\ii a^2_i\right)f^\dagger_{i\uparrow}f^\dagger_{i\downarrow}+\mathrm{h.c.}\right],
\label{eq:onsite}
\end{align}
with $n_i=f^\dagger_{i\alpha}f_{i\alpha}$: $a^3_i$ is a chemical potential for the fermion number and $a^{1,2}_i$ an on-site singlet pairing field, which together impose the three components of the constraint on average.

\section{Algebraic Projective Symmetry Groups: $\Z_2$ Classification}
\label{app:psg}

This appendix summarizes the derivation of the $\Z_2$ PSGs of the SHD lattice. The complete step-by-step derivation is given in Sec.~SIII of the Supplemental Material~\cite{SM}, and the $SU(2)$, $U(1)$, and space-group PSGs are derived in Sec.~SIV there.

\subsection{Conventions}
\label{app:psg-conv}

A site is labeled $i=(x,y,\mu)$, where $x,y\in\Z$ label the unit cell at $x\bm{T}_1+y\bm{T}_2$ and $\mu=1,\dots,12$ labels the sublattice, with $[k]\equiv((k-1)\bmod 12)+1$. By Eq.~\eqref{eq:action},
\begin{align}
C_6&:\ (x,y,\mu)\mapsto(-y,\,x+y,\,[\mu-2]),\nonumber\\
\sigma&:\ (x,y,\mu)\mapsto(y,\,x,\,[9-\mu]),\nonumber\\
C_6^{-1}&:\ (x,y,\mu)\mapsto(x+y,\,-x,\,[\mu+2]),\qquad \sigma^{-1}=\sigma ,
\label{eq:psg-inverses}
\end{align}
and time reversal $\T$ leaves every site fixed. By Eq.~\eqref{eq:compose}, a word $O_1O_2\cdots O_n$ is accompanied by
\begin{align}
G_{O_1\cdots O_n}(i)&=G_{O_1}(i_0)\,G_{O_2}(i_1)\cdots G_{O_n}(i_{n-1}),\nonumber\\
i_0&=i,\qquad i_k=O_k^{-1}(i_{k-1}),
\label{eq:psg-word}
\end{align}
where a letter $O_k=O^{-1}$ contributes $G_{O^{-1}}(i_{k-1})=[G_O(O(i_{k-1}))]^{-1}$; time reversal, which acts linearly on the amplitudes, composes by the same rule. In the canonical gauge the IGG consists of site-independent matrices, $\{\pm\tau^0\}$ for a $\Z_2$ spin liquid. If a word equals the identity of the symmetry group, $G_{O_1\cdots O_n}(i)$ belongs to the IGG and does not depend on $i$; we call it the value of the relation. In the $\Z_2$ case the values of the relations~\eqref{eq:relations} are denoted by $\eta$: $\eta_{12}$ for $T_2T_1T_2^{-1}T_1^{-1}$, $\eta_{T_1C_6}$ for $T_1C_6T_1^{-1}T_2C_6^{-1}$, $\eta_{T_2C_6}$ for $T_2C_6T_1^{-1}C_6^{-1}$, $\eta_{T_2\sigma}$ and $\eta_{T_1\sigma}$ for $T_2\sigma T_1^{-1}\sigma^{-1}$ and $T_1\sigma T_2^{-1}\sigma^{-1}$, $\eta_{\sigma C_6}$, $\eta_{C_6}$, $\eta_\sigma$, and $\eta_\T$ for $(\sigma C_6)^2$, $C_6^6$, $\sigma^2$, and $\T^2$, and $\eta_{T_a\T}$, $\eta_{\T C_6}$, and $\eta_{\T\sigma}$ for the commutators of $\T$ with $T_a$, $C_6$, and $\sigma$. Each $\eta$ equals $\pm\tau^0$ and is identified with the sign $\pm1$. Two sets of matrices $\{G_O\}$ describe the same PSG if they are related by a gauge transformation $W_i\in SU(2)$,
\begin{equation}
G_O(i)\ \to\ W_i\,G_O(i)\,W^\dagger_{O^{-1}(i)},
\label{eq:psg-gauge}
\end{equation}
or by a re-lifting $G_O\to\epsilon_OG_O$ with $\epsilon_O$ in the IGG.

\paragraph*{Lemma (invariance of relation values).}
Under Eq.~\eqref{eq:psg-gauge} the inner factors of Eq.~\eqref{eq:psg-word} cancel pairwise, and since a relation returns to its starting site its value becomes $W_i\,\eta\,W_i^\dagger=\eta$: relation values are gauge invariant. A re-lifting $G_O\to-G_O$ multiplies the value of a relation by $(-1)^{N_O}$, where $N_O$ is the number of letters $O$ or $O^{-1}$ in the word. The value of a relation is therefore a label of the $\Z_2$ PSG if and only if every generator occurs an even number of times in it. This holds for $T_2T_1T_2^{-1}T_1^{-1}$, $(\sigma C_6)^2$, $C_6^6$, $\sigma^2$, $\T^2$, and the four commutators with $\T$, and fails for the four mixed relations of the translations with $C_6$ and $\sigma$. Of the nine candidate labels, $\eta_{T_1\T}=\eta_{T_2\T}=+1$ are forced (Sec.~\ref{app:psg-z2}), so that seven remain.

\subsection{Free Action and the Spanning-Tree Gauge}
\label{app:psg-tree}

By Eq.~\eqref{eq:action}, $C_6^a$ maps the sublattice $\mu=1$ onto $[1-2a]$ and $C_6^a\sigma$ maps it onto $[8-2a]$, so that the twelve elements of $C_{6v}$ map $\mu=1$ onto twelve different sublattices. Combined with the translations, every site $i$ is the image $g_i(i_0)$ of the reference site $i_0=(0,0,1)$ under exactly one element $g_i\in p6mm$: the spins occupy the general Wyckoff position $12f$, on which $p6mm$ acts freely and transitively. Attach the matrix $G_O(i)$ of each spatial generator to the edge joining $O^{-1}(i)$ to $i$. By Eq.~\eqref{eq:psg-gauge} these matrices transform like link variables of a lattice gauge field. Since the action is free, a path that reads the letters of a word is closed precisely when the word is the identity of $p6mm$, that is, a product of conjugated relators, and the ordered product of the matrices along any closed path is a product of relation values. Choosing the gauge such that every edge of a spanning tree carries $\tau^0$, every other edge carries such a product. Therefore, in this tree gauge, every matrix $G_O(i)$ of a $\Z_2$ PSG equals $\pm\tau^0$ with a sign fixed by the relation values, and two PSGs with the same relation values are gauge equivalent. We use the tree formed by all $T_1$ edges, the $T_2$ edges in the column $x=0$, the $C_6$ edges $G_{C_6}(0,0,\mu)$ joining $[\mu+2]$ to $\mu$ for $\mu\in\{1,3,5,7,9\}$ and $\mu\in\{8,10,12,2,4\}$, and the $\sigma$ edge $G_\sigma(0,0,1)$ joining $8$ to $1$. The time-reversal matrices $G_\T(i)$ are on-site, and the only constraint specific to them is $G_\T(i)^2=\eta_\T$.

\subsection{Derivation of the $\Z_2$ PSG}
\label{app:psg-z2}

The derivation follows the tree in eleven steps, which are given in full in Sec.~SIII\,C of the Supplemental Material~\cite{SM}.

\paragraph*{Steps 1--4: translations and the $x,y$ dependence.}
Fixing the gauge along the $T_1$ and $T_2$ edges of the tree and evaluating $T_2T_1T_2^{-1}T_1^{-1}$ with Eq.~\eqref{eq:psg-word} gives
\begin{equation}
G_{T_1}(x,y,\mu)=\tau^0,\qquad G_{T_2}(x,y,\mu)=\eta_{12}^{\,x}\,\tau^0 .
\label{eq:z2-trans}
\end{equation}
The mixed relations of the translations with $C_6$ and $\sigma$ then become recursions in $x$ and $y$, whose solutions are
\begin{align}
G_{C_6}(x,y,\mu)&=\eta_{T_1C_6}^{\,x}\,\eta_{T_2C_6}^{\,y}\,\eta_{12}^{\,x(x+1)/2+xy}\;g_{C_6}(\mu),\nonumber\\
G_\sigma(x,y,\mu)&=\eta_{T_1\sigma}^{\,x}\,\eta_{T_2\sigma}^{\,y}\,\eta_{12}^{\,xy}\;g_\sigma(\mu),
\label{eq:z2-xy}
\end{align}
with the sublattice matrices $g_O(\mu)\equiv G_O(0,0,\mu)$.

\paragraph*{Steps 5--8: point-group relations and residual gauge.}
The relation $\sigma^2$ is independent of $x+y$ only if $\eta_{T_1\sigma}=\eta_{T_2\sigma}\equiv\eta_{T\sigma}$, and it then reduces to $g_\sigma(\mu)\,g_\sigma([9-\mu])=\eta_\sigma$. The gauge transformation $W_{(x,y,\mu)}=\nu^{\,y}\tau^0$ with $\nu=\pm1$, combined with a re-lifting of $T_2$, multiplies $\eta_{T_1C_6}$, $\eta_{T_2C_6}$, and $\eta_{T\sigma}$ by $\nu$, and we set $\eta_{T\sigma}=+1$. The relation $(\sigma C_6)^2$ then requires $\eta_{T_2C_6}=\eta_{12}$. Finally, $W_{(x,y,\mu)}=\nu^{\,x+y}\tau^0$, together with re-liftings of the translations, changes only $\eta_{T_1C_6}\to\nu\,\eta_{T_1C_6}$, so that $\eta_{T_1C_6}$ is pure gauge and is set to $+1$ (Step~8). Hence
\begin{align}
G_\sigma(x,y,\mu)&=\eta_{12}^{\,xy}\,g_\sigma(\mu),\nonumber\\
G_{C_6}(x,y,\mu)&=\eta_{12}^{\,x(x+1)/2+xy+y}\;g_{C_6}(\mu).
\label{eq:z2-final}
\end{align}

\paragraph*{Step 9: the relation $C_6^6$.}
The exponent of $\eta_{12}$ in Eq.~\eqref{eq:z2-final}, summed over the orbit of $(x,y)$ under $C_6^{-1}$, vanishes identically, so that $C_6^6$ is a condition on the sublattice matrices alone, $g_{C_6}(\mu)\,g_{C_6}([\mu+2])\cdots g_{C_6}([\mu+10])=\eta_{C_6}$.

\paragraph*{Step 10: the sublattice matrices.}
The remaining gauge freedom, $W_{(x,y,\mu)}=W_\mu$, fixed along the tree, gives $g_{C_6}(\mu)=\tau^0$ for $\mu\neq6,11$ and $g_\sigma(1)=\tau^0$. The relation $C_6^6$ then gives $g_{C_6}(6)=g_{C_6}(11)=\eta_{C_6}$, and alternating $\sigma^2$ and $(\sigma C_6)^2$ around the sublattices determines the other $g_\sigma(\mu)$ as products of $\eta_\sigma$, $\eta_{\sigma C_6}$, and $\eta_{C_6}$, listed in Table~\ref{tab:z2}. All other instances of the three relations are then satisfied identically, and the only remaining freedom is a global $W\in SU(2)$.

\paragraph*{Step 11: time reversal.}
The commutators of $\T$ with $T_1$ and $T_2$ force $\eta_{T_1\T}=\eta_{T_2\T}=+1$, and those with $C_6$ and $\sigma$ give $g_\T([\mu+2])=\eta_{\T C_6}\,g_\T(\mu)$ and $g_\T([9-\mu])=\eta_{\T\sigma}\,g_\T(\mu)$, so that $G_\T(x,y,\mu)=h(\mu)\,g_\T$ with
\begin{equation}
h(\mu)=\begin{cases}\eta_{\T C_6}^{\,(\mu-1)/2}, & \mu\ \text{odd},\\[2pt] \eta_{\T\sigma}\,\eta_{\T C_6}^{\,\mu/2}, & \mu\ \text{even}.\end{cases}
\label{eq:z2-h}
\end{equation}
The sign function $h(i)=\chi(g_i)$ is a character $\chi$ of $p6mm$, which is trivial on the translations (Sec.~\ref{sec:parton}). The relation $\T^2$ gives $g_\T^2=\eta_\T$; a re-lifting of $G_\T$ and a global rotation then give $g_\T=\tau^0$ for $\eta_\T=+1$ and $g_\T=\ii\tau^2$ for $\eta_\T=-1$.

\paragraph*{Result.}
Every $\Z_2$ PSG of the SHD lattice is gauge equivalent to the form~\eqref{eq:z2psg}, with $g_{C_6}(\mu)$, $g_\sigma(\mu)$, and $h(\mu)$ as listed in Table~\ref{tab:z2} and $g_\T\in\{\tau^0,\ii\tau^2\}$, and is labeled as in Sec.~\ref{sec:alg-z2}. Substituting Eq.~\eqref{eq:z2psg} into the thirteen relations~\eqref{eq:relations} reproduces the label values for all $2^7$ choices of the labels; the non-invariant relation values are $\eta_{T_1C_6}=\eta_{T_2C_6}\eta_{12}=\eta_{T_1\sigma}=\eta_{T_2\sigma}=+1$, and $\eta_{T_1\T}=\eta_{T_2\T}=+1$.

\subsection{Gauge Invariance, Independence, and Cohomology}
\label{app:psg-z2inv}

By the Lemma of Sec.~\ref{app:psg-conv} the seven labels are invariant; by Steps 1 to 11 they fix the PSG; and since Eq.~\eqref{eq:z2psg} satisfies all relations for each of the $2^7$ choices, they are independent, so that $N_{\Z_2}=2^6\times2=128$.

For a $\Z_2$ IGG the relation values define a central extension of the symmetry group by $\Z_2$, and inequivalent consistent assignments are classified by $H^2(p6mm\times\Z_2^\T,\Z_2)$. On the SHD lattice each cohomology class is realized by exactly one gauge class: uniqueness follows from the free spatial action, and existence from the solvability of the only on-site constraint, $G_\T(i)^2=\eta_\T$, in $SU(2)$ for both signs. With coefficients in $\mathbb{F}_2=\Z_2$, the K\"unneth formula gives
\begin{align}
H^2(p6mm\times\Z_2^\T)={}&H^2(p6mm)\nonumber\\
&\oplus\big[H^1(p6mm)\otimes H^1(\Z_2^\T)\big]\nonumber\\
&\oplus H^2(\Z_2^\T),
\label{eq:coh-kunneth}
\end{align}
where $H^2(\Z_2^\T)=\Z_2$ is detected by $\eta_\T$, and $H^1(p6mm,\Z_2)=\mathrm{Hom}(p6mm,\Z_2)=\Z_2^2$ by $\eta_{\T C_6}$ and $\eta_{\T\sigma}$. For $H^2(p6mm,\Z_2)$ the Lyndon--Hochschild--Serre spectral sequence of $1\to\Z^2\to p6mm\to C_{6v}\to1$ has $E_2^{2,0}=H^2(C_{6v},\Z_2)=\Z_2^3$, detected by $\eta_{C_6}$, $\eta_\sigma$, and $\eta_{\sigma C_6}$; $E_2^{1,1}=H^1(C_{6v},M)=0$, because the module $M=\mathrm{Hom}(\Z^2,\Z_2)$ is the two-dimensional Steinberg module of $GL(2,\mathbb{F}_2)\cong S_3$, which is projective; and $E_2^{0,2}=H^2(\Z^2,\Z_2)^{C_{6v}}=\Z_2$, detected by $\eta_{12}$. Since $p6mm$ is symmorphic, the inflation from $C_{6v}$ is injective, and the differentials that could act on these terms vanish. Hence $H^2(p6mm,\Z_2)=\Z_2^4$ and
\begin{equation}
H^2(p6mm\times\Z_2^\T,\Z_2)=\Z_2^4\oplus\Z_2^2\oplus\Z_2=\Z_2^7 ,
\end{equation}
whose seven generators correspond one to one to the seven labels $\eta_{12}$, $\eta_{C_6}$, $\eta_\sigma$, $\eta_{\sigma C_6}$, $\eta_{\T C_6}$, $\eta_{\T\sigma}$, $\eta_\T$. The vanishing of $E_2^{1,1}$ is the cohomological counterpart of Steps 6 to 8: the mixed relation values $\eta_{T_1C_6}$, $\eta_{T_2C_6}$, and $\eta_{T\sigma}$ are either gauge or fixed by $\eta_{12}$ and carry no independent label. The spectral sequence is evaluated in detail in Sec.~SIII\,D of the Supplemental Material~\cite{SM}.
\section{Projective Symmetry Groups of Lattices with Nontrivial Site Symmetry}
\label{app:parent}

\subsection{Reduction to the Site-Symmetry Group}

Let the spins occupy a single orbit of $p6mm$, and let $S$ be the site-symmetry group of a reference site $i_0$; time reversal fixes every site, so the stabilizer of $i_0$ in the full symmetry group is $S\times\Z_2^\T$. A $\Z_2$ PSG assigns to every element $g$ of the symmetry group a gauge transformation $G_g(i)$, defined up to the IGG, such that $G_{g_1}G_{g_2}=\eta(g_1,g_2)\,G_{g_1g_2}$ with $\eta(g_1,g_2)=\pm\tau^0$ uniform in space. The function $\eta$ is a two-cocycle, and re-lifting $G_g\to\epsilon(g)G_g$ changes it by a coboundary, so a PSG determines a class $[\eta]\in H^2(p6mm\times\Z_2^\T,\Z_2)$. Its seven components are the relation values of Sec.~\ref{sec:alg-z2}. Equivalently, the PSG operators $\hat g=(G_g,g)$ and the IGG generate a central extension $\tilde G$ of $p6mm\times\Z_2^\T$ by $\Z_2$, whose isomorphism class is $[\eta]$, and the PSG is a representation of $\tilde G$ on the fermions of all sites in which the central element acts as $-\tau^0$.

For an element $s$ of the stabilizer, $G_s(i_0)$ is an $SU(2)$ matrix acting at the single site $i_0$. These matrices satisfy $G_s(i_0)G_t(i_0)=\eta(s,t)\,G_{st}(i_0)$ for $s,t\in S\times\Z_2^\T$. They form a linear representation $\rho$ of the preimage $\tilde S\subset\tilde G$ of $S\times\Z_2^\T$, the central extension of $S\times\Z_2^\T$ by $\Z_2$ defined by the restriction of $\eta$, in which the nontrivial central element is represented by $-\tau^0$. The statement that the pair $([\eta],[\rho])$ classifies the PSGs, Eq.~\eqref{eq:wyckoff}, is the induction theorem for representations induced from a subgroup~\cite{Mackey1952,Brown1982}. It follows in three steps.

(a) Existence. Given a class $[\eta]$ and a representation $\rho$ of $\tilde S$ with $\rho(-1)=-\tau^0$, the induced representation $\mathrm{Ind}_{\tilde S}^{\tilde G}\rho$ acts on $\bigoplus_i\mathbb{C}^2$, one copy for each coset of $\tilde S$, that is, for each site. Each element of $\tilde G$ maps the copy of site $i$ onto that of $g(i)$, so its matrix is a set of $SU(2)$ blocks $G_g(g(i))$, and since $\mathrm{Ind}\,\rho$ is a representation of $\tilde G$, these blocks satisfy every relation~\eqref{eq:relations} with the value prescribed by $\eta$. Every pair $([\eta],[\rho])$ is therefore realized by a PSG.

(b) Uniqueness. Choose for every site $i$ an element $r_i\in p6mm$ with $r_i(i_0)=i$ and fix the gauge at $i$ by $G_{r_i}(i)=\tau^0$. The remaining values of $G_g(i)$ are then fixed by the cocycle and by $\rho$, because $r_{g(i)}^{-1}g\,r_i$ belongs to the stabilizer. Two PSGs with the same class and conjugate on-site representations are therefore gauge equivalent, provided that the two PSGs are brought to the same normal form of all thirteen relation values (Sec.~\ref{sec:alg-z2}). The re-liftings that preserve this normal form are exactly the characters of $p6mm\times\Z_2^\T$: the relation $T_1C_6T_1^{-1}T_2C_6^{-1}$ contains $T_2$ once and $T_2C_6T_1^{-1}C_6^{-1}$ contains $T_1$ and $T_2$ once each, so a re-lifting $\epsilon$ that preserves their values has $\epsilon(T_1)=\epsilon(T_2)=1$, while $\epsilon(C_6)$, $\epsilon(\sigma)$, and $\epsilon(\T)$ are free. A gauge transformation $W_{i_0}$ conjugates $\rho$, and a character $\lambda$ multiplies $\rho$ by its restriction to $\tilde S$.

(c) Counting. The number of algebraic PSGs is therefore $N_S=\sum_{[\eta]}n_S(\eta)$, Eq.~\eqref{eq:wyckoff}, with $n_S(\eta)$ the number of two-dimensional unit-determinant linear representations of $\tilde S$ that map the central element to $-\tau^0$, counted up to conjugation and up to twisting by the restrictions of characters of $p6mm\times\Z_2^\T$.

\subsection{Evaluation}

The extension $\tilde S$ can be constructed explicitly from any SHD representative of the class $\eta$. Collapsing a cluster of the SHD lattice (a dodecagon, hexagon, square, or edge) to a point maps its stabilizer onto $S$, and $S$ permutes the sites of the cluster freely. The SHD PSG matrices of the elements of $S$, restricted to the cluster, are signed permutation matrices; together with $G_\T$ and the IGG they generate a finite matrix group isomorphic to $\tilde S$. Equivalently, $\tilde S$ is fixed by the values in the SHD PSG of a complete set of relators of $S\times\Z_2^\T$, written in the generators of $p6mm\times\Z_2^\T$. For the honeycomb lattice, for instance, $S$ is generated by the threefold rotation $T_1C_6^2$ about a site and the mirror $\sigma$, and the relators are $(T_1C_6^2)^3$, $\sigma^2$, $(\sigma T_1C_6^2)^2$, $\T^2$, and the commutators of $\T$ with the two generators. The image of a representation $\rho$ is a finite subgroup of $SU(2)$ whose quotient by $\{\pm\tau^0\}$ is a quotient of $S\times\Z_2^\T$. For the groups that occur here ($6mm\times\Z_2$, $3m\times\Z_2$, $2mm\times\Z_2$, $m\times\Z_2$) every such quotient is supersolvable, which excludes the binary polyhedral groups; the image is therefore cyclic or binary dihedral, and after a global rotation it lies in the binary dihedral group of order 24 generated by $e^{\ii\pi\tau^3/6}$ and $\ii\tau^1$. Enumerating the images of the generators of $\tilde S$ in this group, discarding assignments that violate any relation of $\tilde S$ or that do not map the central element to $-\tau^0$, and identifying representations with equal characters up to twisting yields $n_S(\eta)$. The totals are listed in Table~\ref{tab:parent}. On all lattices of Table~\ref{tab:parent} $n_S(\eta)\le2$; classes with $n_S(\eta)=2$ occur only on the triangular lattice (eight classes) and on the honeycomb lattice (32 classes).

\subsection{Correspondence with Earlier Classifications}

\paragraph*{Honeycomb lattice.} The conventions of Ref.~\cite{LuRan2011} differ from ours in the choice of mirror. Its mirror $\sigma_{\rm LR}$ exchanges the two sublattices, whereas the SHD mirror is $\sigma=C_6\sigma_{\rm LR}$, which maps each honeycomb sublattice onto itself. Relation values quoted in the conventions of Ref.~\cite{LuRan2011} translate into SHD labels as $\eta_\sigma=[(\sigma_{\rm LR}C_6)^2]$, $\eta_{\sigma C_6}=[\sigma_{\rm LR}^2]$, and $\eta_{\T\sigma}=[\sigma_{\rm LR}\T\sigma_{\rm LR}^{-1}\T^{-1}]\,\eta_{\T C_6}$, where $[w]$ denotes the value of the relation $w$, while $\eta_{12}$, $\eta_{C_6}$, $\eta_{\T C_6}$, and $\eta_\T$ keep their meaning. To avoid a clash of notation we write $\epsilon_\sigma$ for the parameter called $\eta_\sigma$ in Ref.~\cite{LuRan2011}; it equals $\eta_{\sigma C_6}$ in SHD labels.

With these conventions the 112 algebraic PSGs correspond to the 160 entries listed in Appendix~B of Ref.~\cite{LuRan2011} as follows; we evaluated all 160 entries in the SHD labels and in the trace of the on-site matrix of the threefold rotation about a site. Entries with $\psi_3=-2\pi/3$ in the families (I)(B), (II)(i)(B)($\beta$), and (II)(ii)(B)($\alpha$) are related to those with $\psi_3=+2\pi/3$ by the global gauge transformation $\ii\tau^1$ and represent the same classes, which accounts for 32 entries. The entries (II)(ii)(B)($\beta$) with $\psi_3=0$ are gauge equivalent to the corresponding entries (II)(ii)(A), which accounts for 8 more. For the 16 entries (II)(ii)(B)($\beta$) with $\psi_3=\pm2\pi/3$ the relation $(\sigma_{\rm LR}C_6)^2$ does not take a value in $\{\pm\tau^0\}$. For example, for $\eta_{12}=\eta_{C_6}=\epsilon_\sigma=+1$ and $\psi_3=2\pi/3$, Eq.~(B30) of Ref.~\cite{LuRan2011} gives $g_{C_6}=(\ii\tau^1,-\ii\tau^1e^{\ii\psi_3\tau^3})$ and $g_\sigma=(\ii\tau^1,-\ii\tau^1)$ on the two sublattices, and the product of the four PSG matrices along $(\sigma_{\rm LR}C_6)^2$ equals $e^{-2\pi\ii\tau^3/3}$ at every site of one sublattice and $\tau^0$ at every site of the other. These entries therefore do not correspond to $\Z_2$ PSGs and are not counted here. The remaining 104 entries are pairwise inequivalent. The PSGs with $\psi_3=0$ and $\psi_3=2\pi/3$ in the families (I)(B), (II)(i)(B)($\beta$), and (II)(ii)(B)($\alpha$) have the same relation values and are inequivalent; the trace of the on-site matrix of the threefold rotation about a site, $2$ for one and $-1$ for the other, distinguishes them. They are the 32 classes with $n_S(\eta)=2$. Together with a family of eight PSGs with $g_\T=\ii\tau^3$, $g_{C_6}=(\ii\tau^1,-\ii\eta_{C_6}\eta_{12}\tau^1)$, and $g_\sigma=(\ii\tau^2,-\ii\epsilon_\sigma\tau^2)$ on the two sublattices, the 104 classes give the 112 PSGs. In the conventions of Ref.~\cite{LuRan2011} this family has $(\sigma_{\rm LR}C_6)^2=-\tau^0$ and $\sigma_{\rm LR}$ and $C_6$ both anticommuting with $\T$ in the PSG; in SHD labels it consists of the eight classes Z$n_{T_2}$011$n_{\sigma C_6}n_{C_6}$(1), with $\eta_\sigma=\eta_{\T C_6}=\eta_\T=-1$, $\eta_{\T\sigma}=+1$, and $\eta_{12}$, $\eta_{\sigma C_6}=\epsilon_\sigma$, $\eta_{C_6}$ free. Each of these classes has $n_S(\eta)=1$, and none of the 160 entries has these relation values. All eight PSGs admit nonzero singlet amplitudes, four of them on nearest-neighbor bonds and the others on third- or fourth-neighbor bonds. Of the 112 PSGs, 24 force all mean-field amplitudes to vanish, so that 88 are realizable.

\paragraph*{Triangular lattice.} Enumerating the admissible time-reversal matrices $g_\T\in\{\ii\tau^1,\ii\tau^2,\ii\tau^3\}$ for the seven point-group representations of Table~II of Ref.~\cite{BieriLhuillierMessio2016} and each value of the translation sign gives $1,2,2,2,3,1,1$ inequivalent choices, hence 24 noncentral classes. These are the classes of Table~I of Ref.~\cite{Lu2016} together with the additional pair of solutions noted there, taken for both values of $\eta_{12}$. The classes $(\mathrm{No.}\,2,\ii\tau^1)$ and $(\mathrm{No.}\,6,\ii\tau^3)$, and $(\mathrm{No.}\,5,\ii\tau^3)$ and $(\mathrm{No.}\,7,\ii\tau^3)$, have identical relation values for each translation sign and differ in the trace of the on-site matrix of the threefold rotation, which takes the values $2$ and $-1$ in the first pair and $-2$ and $1$ in the second. Labeled by relation values alone, as in Ref.~\cite{ZhengMeiQi2015}, these classes form 20 distinct sets of symmetry-fractionalization labels. The same two pairs of point-group representations, combined with a central time-reversal matrix, give the four central classes with $n_S(\eta)=2$, Z000110(0), Z000111(0), Z100110(0), and Z100111(0); since a central $G_\T$ forces every amplitude to vanish by Eq.~\eqref{eq:psg-T}, these pairs are not realizable. The four noncentral PSGs without amplitudes all have $\eta_{12}=+1$. For them we tested the on-site term and every bond with lattice displacement $(a,b)$, $|a|,|b|\le12$, which covers all residues modulo four of the cell coordinates, the period of the sign factors in the PSG.
\section{Brillouin Zones}
\label{app:bz}

We use the first Brillouin zone (FBZ), the reduced Brillouin zone (RBZ) of the doubled cell ($n_{T_2}=-1$), and the extended Brillouin zone (EBZ) of the structure factors.

\subsection{Reciprocal Lattice and First Brillouin Zone}

The Bravais vectors $\bm{T}_{1,2}$ are those of Eq.~\eqref{eq:T}, with lattice constant $A=|\bm{T}_{1,2}|$. The reciprocal vectors are defined by $\bm{b}_i\cdot\bm{T}_j=2\pi\delta_{ij}$. They are
\begin{equation}
\bm{b}_{1,2}=\frac{2\pi}{A}\left(\frac{1}{\sqrt3},\mp1\right),\qquad
|\bm{b}_{1,2}|=\frac{4\pi}{\sqrt3 A}.
\label{eq:recip}
\end{equation}
The vectors $\bm{b}_1$ and $\bm{b}_2$ enclose an angle of $120^\circ$, so $\bm{b}_1+\bm{b}_2=(4\pi/\sqrt3 A,0)$ has the same length. The six shortest reciprocal-lattice vectors are therefore $\pm\bm{b}_1$, $\pm\bm{b}_2$, and $\pm(\bm{b}_1+\bm{b}_2)$, pointing along the directions $0^\circ,\pm60^\circ,\pm120^\circ,180^\circ$.

The FBZ is the Wigner--Seitz cell of the reciprocal lattice: the set of momenta closer to $\bm{0}$ than to any other reciprocal-lattice vector. Its boundary consists of segments of the perpendicular bisectors of the six shortest vectors. The midpoints of these segments are the $M$ points,
\begin{equation}
M=\tfrac12(\bm{b}_1+\bm{b}_2)=\frac{2\pi}{A}\left(\frac{1}{\sqrt3},0\right),
\label{eq:Mpt}
\end{equation}
and its five images under $C_6$. Two neighboring bisectors, for instance those of $\bm{b}_1+\bm{b}_2$ and $\bm{b}_2$, intersect at a vertex $K$ that is equidistant from $\bm{0}$, $\bm{b}_2$, and $\bm{b}_1+\bm{b}_2$. Solving $|K|=|K-\bm{b}_2|=|K-\bm{b}_1-\bm{b}_2|$ gives
\begin{equation}
K=\tfrac13(\bm{b}_1+2\bm{b}_2)=\frac{2\pi}{A}\left(\frac{1}{\sqrt3},\frac13\right),\qquad |K|=\frac{4\pi}{3A},
\label{eq:Kpt}
\end{equation}
at polar angle $30^\circ$. The FBZ is a regular hexagon with vertices $C_6^n K$ ($n=0,\dots,5$), oriented with its vertices along $30^\circ+60^\circ n$ and its edge midpoints along $60^\circ n$. Band structures are plotted along $\Gamma\to K\to M\to\Gamma$ with $\Gamma=\bm{0}$.

\subsection{Reduced Brillouin Zone of the Doubled Cell}

For $n_{T_2}=-1$ every projective symmetry group realizes translations as $G_{T_1}=\tau^0$, $G_{T_2}(x,y,\mu)=(-1)^x\tau^0$ (Appendix~\ref{app:psg}), so a gauge-fixed mean-field ansatz is invariant under $T_1$ and $2T_2$ but not under $T_2$. The spinon Hamiltonian is then periodic on the doubled cell spanned by $\bm{T}_1$ and $2\bm{T}_2$, with reciprocal vectors $\bm{B}_1=\bm{b}_1$ and $\bm{B}_2=\bm{b}_2/2$. Since $\bm{B}_2\cdot(\bm{B}_1+\bm{B}_2)=0$, this reciprocal lattice is rectangular, with orthogonal primitive vectors $\bm{B}_2=(2\pi/A)(1/(2\sqrt3),1/2)$ and $\bm{B}_1+\bm{B}_2=(2\pi/A)(\sqrt3/2,-1/2)$ of lengths $2\pi/(\sqrt3A)$ and $2\pi/A$. Its Wigner--Seitz cell, the RBZ, is a rectangle of half the FBZ area whose corners are
\begin{equation}
\pm\frac{2\pi}{A}\left(\frac{1}{\sqrt3},0\right),\qquad
\pm\frac{2\pi}{A}\left(\frac{1}{2\sqrt3},-\frac12\right).
\label{eq:rbz}
\end{equation}
The spectrum is in addition invariant under $\bm{k}\to\bm{k}+\bm{b}_1/2$, because the $T_2$ translation of the ansatz is a gauge transformation, and the spin structure factors do not depend on the doubling.

\subsection{Extended Brillouin Zone}

The spin structure factors
\begin{equation}
S(\bm{q})=\frac{1}{N}\sum_{i,j}e^{\mathrm{i}\bm{q}\cdot(\bm{r}_i-\bm{r}_j)}\langle \bm{S}_i\cdot\bm{S}_j\rangle
\label{eq:sq-bz}
\end{equation}
and $S(\bm{q},\omega)$, as well as their individual spin components $S^{ab}$, depend on $\bm{q}$ only through the phases $e^{\mathrm{i}\bm{q}\cdot(\bm{r}_i-\bm{r}_j)}$, where $\bm{r}_i$ are the actual site positions. A momentum $\bm{G}$ is a period of $S(\bm{q})$ for every set of correlations if and only if $\bm{G}\cdot\bm{d}\in2\pi\mathbb{Z}$ for every difference $\bm{d}=\bm{r}_i-\bm{r}_j$. The periods therefore form the dual of the additive group $D$ generated by all site differences. If $D$ is a two-dimensional lattice, its reciprocal lattice is the period lattice and the EBZ is its Wigner--Seitz cell. If $D$ is dense in the plane, the only period is $\bm{G}=\bm{0}$: $\bm{G}\cdot\bm{d}$ is a continuous function of $\bm{d}$ that takes values in the discrete set $2\pi\mathbb{Z}$ on a dense set, hence vanishes identically.

In the ideal Archimedean geometry all edges have equal length and the bonds point along the twelve directions $30^\circ n$. Differences of site positions then involve both $1$ and $\sqrt3$ with independent integer coefficients, $D$ has rank four and is dense, and $S(\bm{q})$ has no period beyond those of the individual Fourier components. It is nevertheless well defined at every momentum: the two-spinon bubble of Appendix~\ref{app:sf} holds at every lift $\bm{q}+\bm{G}$ of a mesh momentum, and the Bragg weight of a classical order can be evaluated at every $\bm{q}$. It is only not periodic on any finite extended zone. Unless stated otherwise, the structure factors in this paper are computed with the ideal site positions, and we show them in the hexagon with corners $K'=9K$ and edge midpoints $M'=9M$ defined below, which is the extended Brillouin zone of a commensurate realization of the tiling; in the ideal geometry it is a display window, not a period cell.

In the commensurate realization of the lattice the twelve sites of the reference cell lie on a circle of radius $\sqrt{13}$ about the dodecagon center and $A=9$ (all lengths in units of the spacing of the fine lattice introduced below). In this realization the $J_b$ bonds have length $\sqrt3$, the $J_g$ and $J_r$ bonds have length $2$, the hexagons are regular, the dodecagons are equiangular, and the squares become $2\times\sqrt3$ rectangles; the space group is $p6mm$ as for the ideal lattice. With the fine primitive vectors
\begin{equation}
\bm{a}_1=\bm{T}_1/9,\qquad \bm{a}_2=\bm{T}_2/9,\qquad |\bm{a}_{1,2}|=1,
\label{eq:fine}
\end{equation}
every site of the reference cell has integer coordinates, $\bm{\epsilon}_\mu=n_1\bm{a}_1+n_2\bm{a}_2$:
\begin{center}
\begin{tabular}{c|cccccccccccc}
$\mu$ & 1 & 2 & 3 & 4 & 5 & 6 & 7 & 8 & 9 & 10 & 11 & 12\\\hline
$n_1$ & $-4$ & $-3$ & $-1$ & $1$ & $3$ & $4$ & $4$ & $3$ & $1$ & $-1$ & $-3$ & $-4$\\
$n_2$ & $3$ & $4$ & $4$ & $3$ & $1$ & $-1$ & $-3$ & $-4$ & $-4$ & $-3$ & $-1$ & $1$
\end{tabular}
\end{center}
In Cartesian coordinates, $\bm{\epsilon}_2=(\sqrt3/2,7/2)$, $\bm{\epsilon}_3=(3\sqrt3/2,5/2)$, $\bm{\epsilon}_4=(2\sqrt3,1)$, and the remaining sites follow from the mirror $y\to-y$ and the rotation $C_6$. Since $\bm{T}_{1,2}=9\bm{a}_{1,2}$, every site position is an integer combination of $\bm{a}_1,\bm{a}_2$. The site differences $\bm{\epsilon}_2-\bm{\epsilon}_1=\bm{a}_1+\bm{a}_2$ and $\bm{\epsilon}_3-\bm{\epsilon}_2=2\bm{a}_1$ together with $\bm{T}_1=9\bm{a}_1$ already generate $\bm{a}_1$ and $\bm{a}_2$ [indeed $\bm{a}_1=\bm{T}_1-4(\bm{\epsilon}_3-\bm{\epsilon}_2)$ and $\bm{a}_2=(\bm{\epsilon}_2-\bm{\epsilon}_1)-\bm{a}_1$], so $D$ is the triangular lattice spanned by $\bm{a}_1,\bm{a}_2$. Its reciprocal lattice is spanned by $9\bm{b}_1$ and $9\bm{b}_2$, and the EBZ is the hexagon obtained from the FBZ by a dilation by a factor of nine (eighty-one in area). Its high-symmetry points are
\begin{equation}
M'=9M=\left(\frac{2\pi}{\sqrt3},0\right),\qquad
K'=9K=\left(\frac{2\pi}{\sqrt3},\frac{2\pi}{3}\right),
\label{eq:ebz}
\end{equation}
in units of the inverse fine-lattice spacing; in the units $2\pi/A$ of the figures, $M'=(3\sqrt3,0)$ and $K'=(3\sqrt3,3)$. A structure factor computed on the commensurate lattice is periodic under $9\bm{b}_{1,2}$ and under no shorter reciprocal vector, and its average over the EBZ equals the on-site value of the correlation. Intensity maps are shown in this hexagon, with the FBZ drawn inside it, and dynamical structure factors are plotted along $\Gamma\to K'\to M'\to\Gamma$, also when they are computed with the ideal positions.

The commensurate realization distorts the squares into rectangles, with $J_b$ bonds shorter than $J_g$ and $J_r$ bonds by the factor $\sqrt3/2$, and changes $S(\bm{q})$ quantitatively but not qualitatively. For the three mean-field states of Fig.~\ref{fig:spec-sq}, computed on the same $120\times120$ mesh in both geometries, the largest difference in the hexagon is $9\%$ of the maximum of $S(\bm{q})$ for SU010111 and $15\%$ for U001-000000 and its $\Z_2$ descendant, with pattern correlation coefficients $0.98$ and $0.92$. The values at $K'$ change from $0.096$ (commensurate) to $0.102$ (ideal) for SU010111 and from $0.181$ to $0.173$ for U001-000000, while those at $M'$ change by at most $0.003$. In the ideal geometry the average of $S^{zz}(\bm{q})$ over the hexagon differs from $1/8$ by up to $1.4\%$ for these states, since the hexagon is not a period cell. For the classical orders the weight is redistributed among the outer shells of Bragg peaks: the N\'eel state carries $71\%$, $9\%$, and $20\%$ of its weight on the shells of length $\sqrt7$, $\sqrt{13}$, and $\sqrt{19}$ (in units of $|\bm{b}_1|$) in the commensurate realization and $76\%$, $18\%$, and $6\%$ in the ideal geometry.
\section{Mean-Field Spin Structure Factors}
\label{app:sf}

For an ansatz with singlet amplitudes only, the mean-field Hamiltonian conserves the number of the fermions $\Psi_{i1}=f_{i\uparrow}$ and $\Psi_{i2}=f^\dagger_{i\downarrow}$. Indeed, singlet hopping $f^\dagger_{i\uparrow}f_{j\uparrow}+f^\dagger_{i\downarrow}f_{j\downarrow}=\Psi^\dagger_{i1}\Psi_{j1}-\Psi^\dagger_{j2}\Psi_{i2}$ and singlet pairing $f^\dagger_{i\uparrow}f^\dagger_{j\downarrow}=\Psi^\dagger_{i1}\Psi_{j2}$ are both of the form $\Psi^\dagger\Psi$, so that
\begin{equation}
H_{\rm MF}=\sum_{i,j}\Psi_i^\dagger M_{ij}\Psi_j,
\end{equation}
with a Hermitian $2N\times2N$ matrix $M$. Because $\Psi_i^\dagger\Psi_i=n_{i\uparrow}-n_{i\downarrow}+1$, the spin component along $z$ is $S^z_i=\frac12(\Psi^\dagger_i\Psi_i-1)$, and its Fourier transform with the site positions $\bm{r}_i$ of the ideal lattice (Appendix~\ref{app:bz}) is, for $\bm{q}\neq0$,
\begin{equation}
S^z_{\bm{q}}=\sum_ie^{-\ii\bm{q}\cdot\bm{r}_i}S^z_i=\frac12\sum_{\bm{k}}\sum_{\mu}\Psi^\dagger_{\bm{k},\mu}\Psi_{\bm{k}+\bm{q},\mu},
\end{equation}
where $\Psi_{\bm{k},\mu}=N_c^{-1/2}\sum_{\bm{R}}e^{-\ii\bm{k}\cdot(\bm{R}+\bm{\epsilon}_\mu)}\Psi_{\bm{R},\mu}$ is defined with the phase of the actual site position (atomic gauge) and $N_c$ is the number of unit cells. In this gauge the Bloch matrix $M(\bm{k})$ is not periodic in the reciprocal lattice, but the operator $S^z_{\bm{q}}$ contains no form factors, and the expressions below hold at every momentum $\bm{q}$, not only in the first Brillouin zone. The structure factor is therefore defined at every $\bm{q}$; it is periodic only under those reciprocal vectors $\bm{G}$ with $\bm{G}\cdot(\bm{r}_i-\bm{r}_j)\in2\pi\mathbb{Z}$ for all pairs of sites, of which the ideal geometry has none besides $\bm{G}=0$ (Appendix~\ref{app:bz}).

Let $v_n(\bm{k})$ and $E_n(\bm{k})$ be the eigenvectors and eigenvalues of $M(\bm{k})$. The mean-field ground state fills all states with $E_n(\bm{k})<0$, and $S^z_{\bm{q}}$ creates a particle-hole pair. Inserting a complete set of states in $S^{zz}(\bm{q},\omega)=N^{-1}\sum_f|\langle f|S^z_{\bm{q}}|0\rangle|^2\delta(\omega-E_f+E_0)$ gives
\begin{align}
S^{zz}(\bm{q},\omega)=\frac{1}{4N}\sum_{\bm{k}}{\sum_{a,b}}'&\left|v_b^\dagger(\bm{k})\,v_a(\bm{k}+\bm{q})\right|^2\nonumber\\
&\times\delta\big(\omega-E_b(\bm{k})+E_a(\bm{k}+\bm{q})\big),
\label{eq:sqw}
\end{align}
where the primed sum runs over occupied states $a$ with $E_a(\bm{k}+\bm{q})<0$ and empty states $b$ with $E_b(\bm{k})>0$, and $N=12N_c$ is the number of sites. The overlaps contain the full coherence factors of the Bogoliubov quasiparticles of the original fermions, so Eq.~\eqref{eq:sqw} applies to $\Z_2$ as well as to $U(1)$ and $SU(2)$ ans\"atze~\cite{Chauhan2023}. Spin-rotation symmetry gives $S^{xx}=S^{yy}=S^{zz}$, and we plot $S^{zz}(\bm{q},\omega)$ and its frequency integral $S^{zz}(\bm{q})$; the trace over the spin components is three times larger. Its average over momentum is the on-site value $N^{-1}\sum_i\langle(S^z_i)^2\rangle_{\rm MF}$, exactly for the average over the extended zone of the commensurate realization (Appendix~\ref{app:bz}) and, in the ideal geometry, for averages over large momentum regions. With the $2\times2$ density matrix $\rho_i$ of $\Psi_i$, Wick's theorem gives $\langle(S^z_i)^2\rangle_{\rm MF}=\frac14\mathrm{Tr}[\rho_i(1-\rho_i)]=\frac18-\frac18|\bm{t}_i|^2$, where $t^k_i=\langle\Psi^\dagger_i\tau^k\Psi_i\rangle$ are the residuals of the on-site constraints and $\mathrm{Tr}\,\rho_i=1$ by spin-rotation symmetry. The value $1/8$, half the value $1/4$ of the physical spin because the mean-field state is singly occupied only with probability $1/2$, therefore holds when the constraints are satisfied on every site; for the residuals of the ans\"atze of the Supplemental Material (below $4\times10^{-4}$) the deviation is below $2\times10^{-8}$. For ans\"atze with triplet amplitudes the number of $\Psi$ fermions is not conserved, and the same calculation is carried out with the four-component Nambu spinor $(f_{i\uparrow},f_{i\downarrow},f^\dagger_{i\uparrow},f^\dagger_{i\downarrow})$, which yields the full tensor $S^{ab}(\bm{q},\omega)$.

\bibliography{references}

\clearpage
\onecolumngrid
\makeatletter\let\@hangfrom@section\SM@hangfrom@section\let\@sectioncntformat\SM@sectioncntformat\@removefromreset{equation}{section}\makeatother
\setcounter{figure}{0}\setcounter{table}{0}\setcounter{equation}{0}\setcounter{section}{0}
\renewcommand{\thefigure}{S\arabic{figure}}\renewcommand{\theHfigure}{S\arabic{figure}}
\renewcommand{\thetable}{S\arabic{table}}\renewcommand{\theHtable}{S\arabic{table}}
\renewcommand{\theequation}{S\arabic{equation}}\renewcommand{\theHequation}{S\arabic{equation}}
\renewcommand{\thesection}{S\Roman{section}}\renewcommand{\theHsection}{S\Roman{section}}
\renewcommand{\thesubsection}{\Alph{subsection}}
\begin{center}
{\large\bfseries Supplemental Material for ``Kitaev, semi-Dirac, and spinon quantum spin Hall liquids on the square-hexagon-dodecagon lattice: A parent projective symmetry group classification''}\\[1ex]
Harsh Raj and Yasir Iqbal\\[0.5ex]
{\small\itshape Department of Physics, Indian Institute of Technology Madras, Chennai 600036, India}
\end{center}
\phantomsection\addcontentsline{toc}{section}{Supplemental Material}
\vspace{1ex}
This Supplemental Material collects derivations, tables, and data that complement the main text. Sections~\ref{sm:sec:ansatz} and~\ref{sm:sec:sf} give the mean-field ans\"atze and spin structure factors of all symmetric spin liquids realized up to fourth neighbors. Section~\ref{sm:app:psg} gives the complete derivation of the $\Z_2$ projective symmetry groups summarized in Appendix~B of the main text, Sec.~\ref{sm:app:psg-u1} derives the $SU(2)$, $U(1)$, and space-group projective symmetry groups, and Sec.~\ref{sm:sec:bonds} gives the bond transformation rules and the rules for the symmetry-allowed singlet and triplet amplitudes on the reference bonds. Section~\ref{sm:app:nodes} derives closed forms for the Dirac nodes on the mirror lines, Sec.~\ref{sm:app:dos} gives the spinon densities of states of all realized symmetric spin liquids, Sec.~\ref{sm:app:regular} the spin structures of the regular magnetic orders, Sec.~\ref{sm:app:hierarchy} the descendants not shown in Fig.~4 of the main text and the chiral descendants, and Sec.~\ref{sm:app:params} the parameters of the ans\"atze in the figures of the main text. Equations, figures, tables, and sections numbered with the prefix S belong to this Supplemental Material; all other numbers refer to the main text.

\begin{figure*}[p]
\centering
\includegraphics{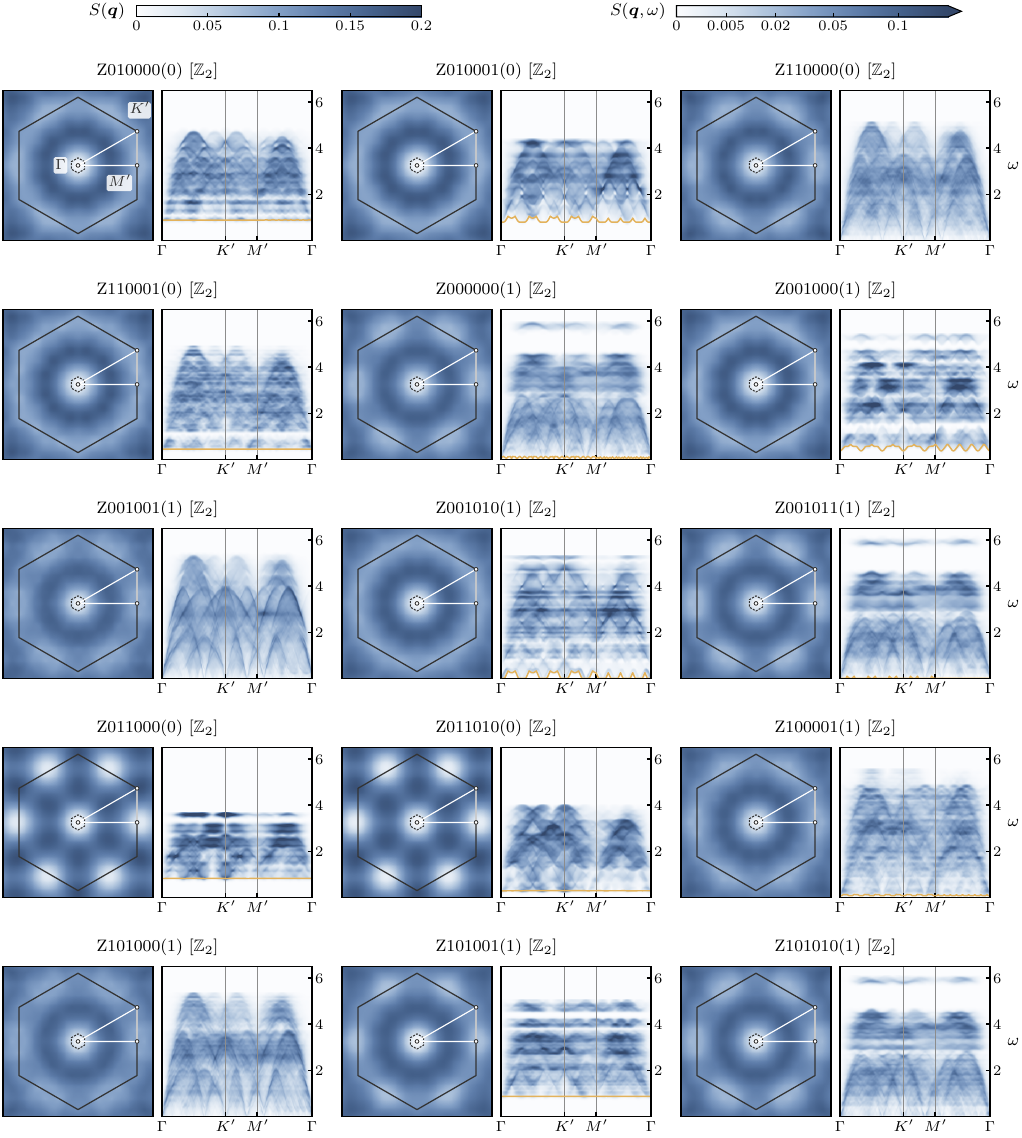}
\caption{Mean-field spin structure factors of the $\Z_2$ spin liquids Nos.~1--15, which are Nos.~1--15 of Table~\ref{tab:z2ans} of the main text. For each state the left panel shows $S(\bm{q})$, Eq.~\eqref{sm:eq:sq}, in the EBZ (solid hexagon), with the first Brillouin zone (dashed) and the path $\Gamma\to K'\to M'\to\Gamma$ (white line; the three points are labeled in the first panel), and the right panel shows $S(\bm{q},\omega)$ along this path, with $\omega$ in units of the amplitude $|u|$ on $J_b$, together with the lower edge of the two-spinon continuum (amber line). The title gives the PSG label and the IGG. The color scales are common to all panels of a figure, linear for $S(\bm{q})$ and following a square-root law for $S(\bm{q},\omega)$, which saturates above the end of its color bar. The amplitudes of these states are listed in Tables~\ref{sm:tab:z2} and \ref{sm:tab:z2b}.}
\label{sm:fig:first}
\end{figure*}

\begin{figure*}[p]
\centering
\includegraphics{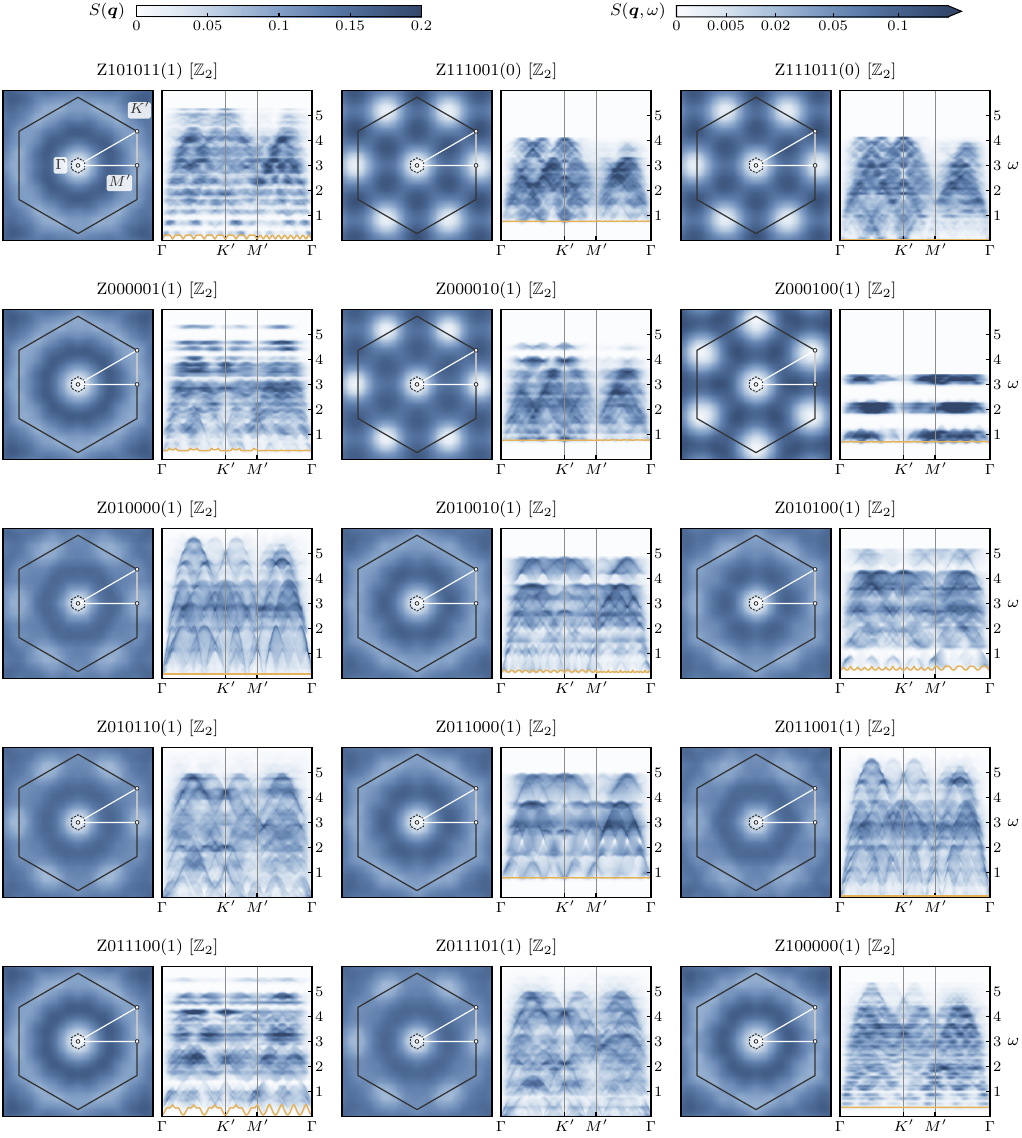}
\caption{As Fig.~\ref{sm:fig:first}, for the $\Z_2$ spin liquids Nos.~16--30, which are Nos.~16--30 of Table~\ref{tab:z2ans} of the main text; amplitudes and multipliers in Tables~\ref{sm:tab:z2} and \ref{sm:tab:z2b}.}
\label{sm:fig:p2}
\end{figure*}

\begin{figure*}[p]
\centering
\includegraphics{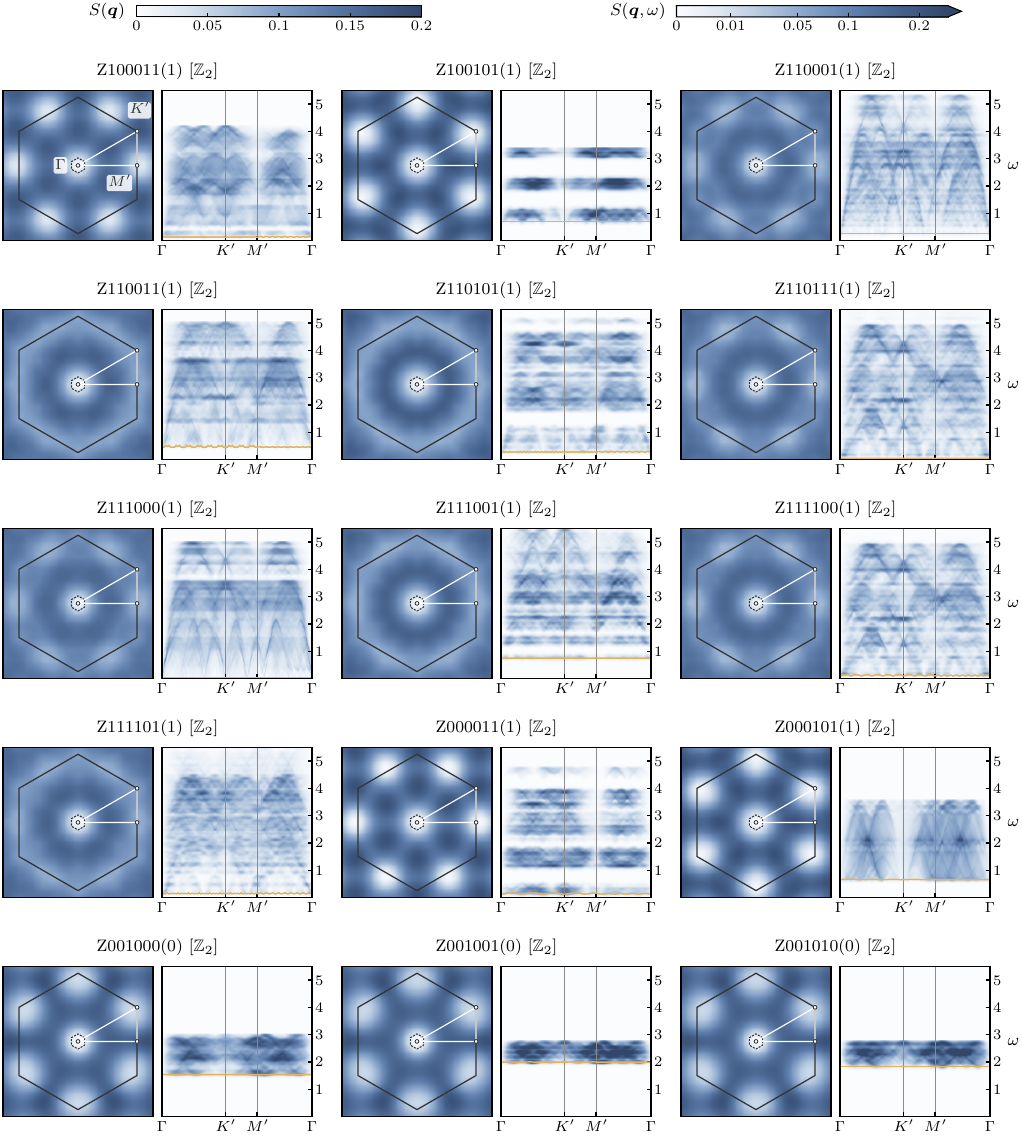}
\caption{As Fig.~\ref{sm:fig:first}, for the $\Z_2$ spin liquids Nos.~31--45, which are Nos.~31--40 of Table~\ref{tab:z2ans} and Nos.~1--5 of Table~\ref{tab:z2ansb} of the main text; amplitudes and multipliers in Tables~\ref{sm:tab:z2} and \ref{sm:tab:z2b}.}
\label{sm:fig:p3}
\end{figure*}

\begin{figure*}[p]
\centering
\includegraphics{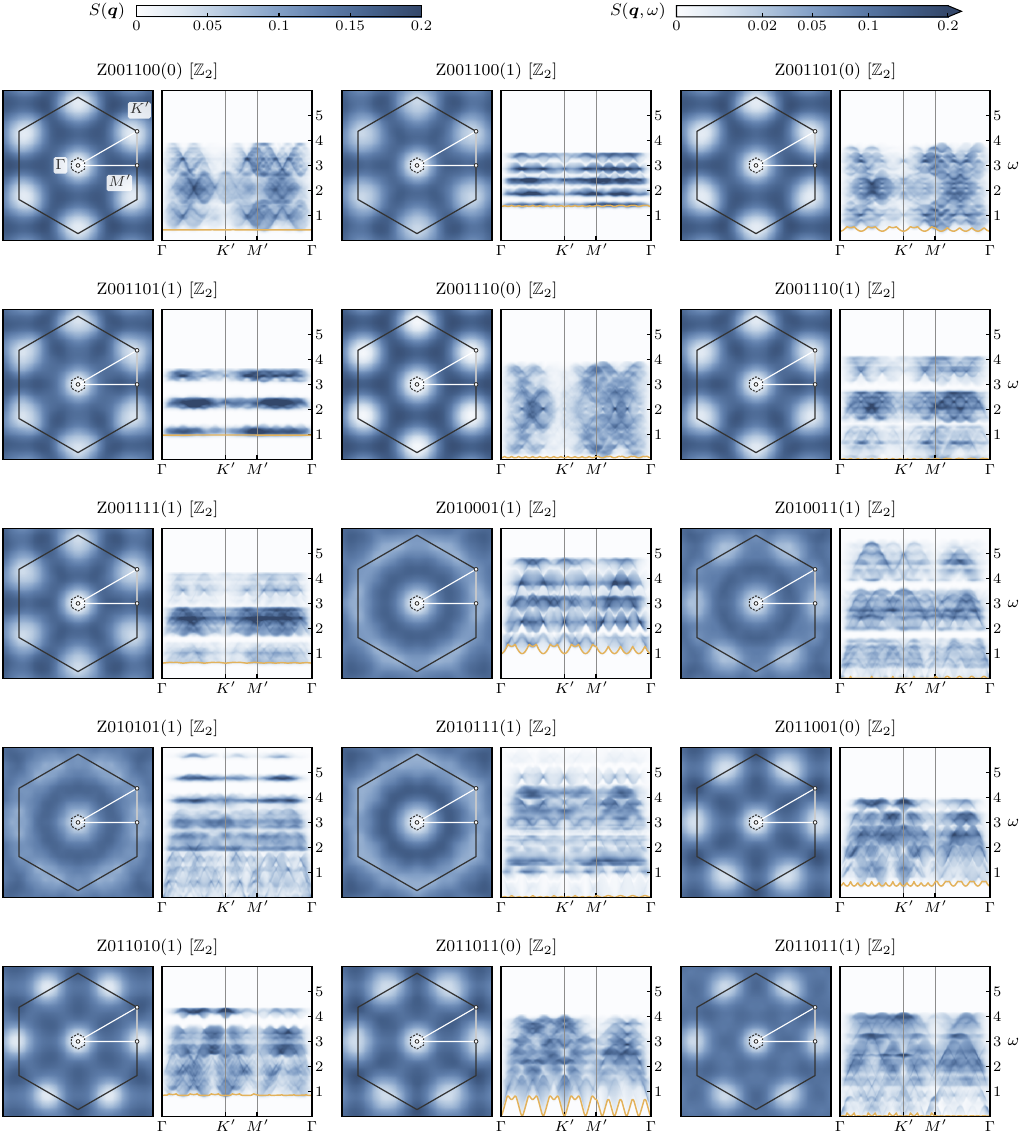}
\caption{As Fig.~\ref{sm:fig:first}, for the $\Z_2$ spin liquids Nos.~46--60, which are Nos.~6--20 of Table~\ref{tab:z2ansb} of the main text; amplitudes and multipliers in Tables~\ref{sm:tab:z2} and \ref{sm:tab:z2b}.}
\label{sm:fig:p4}
\end{figure*}

\begin{figure*}[p]
\centering
\includegraphics{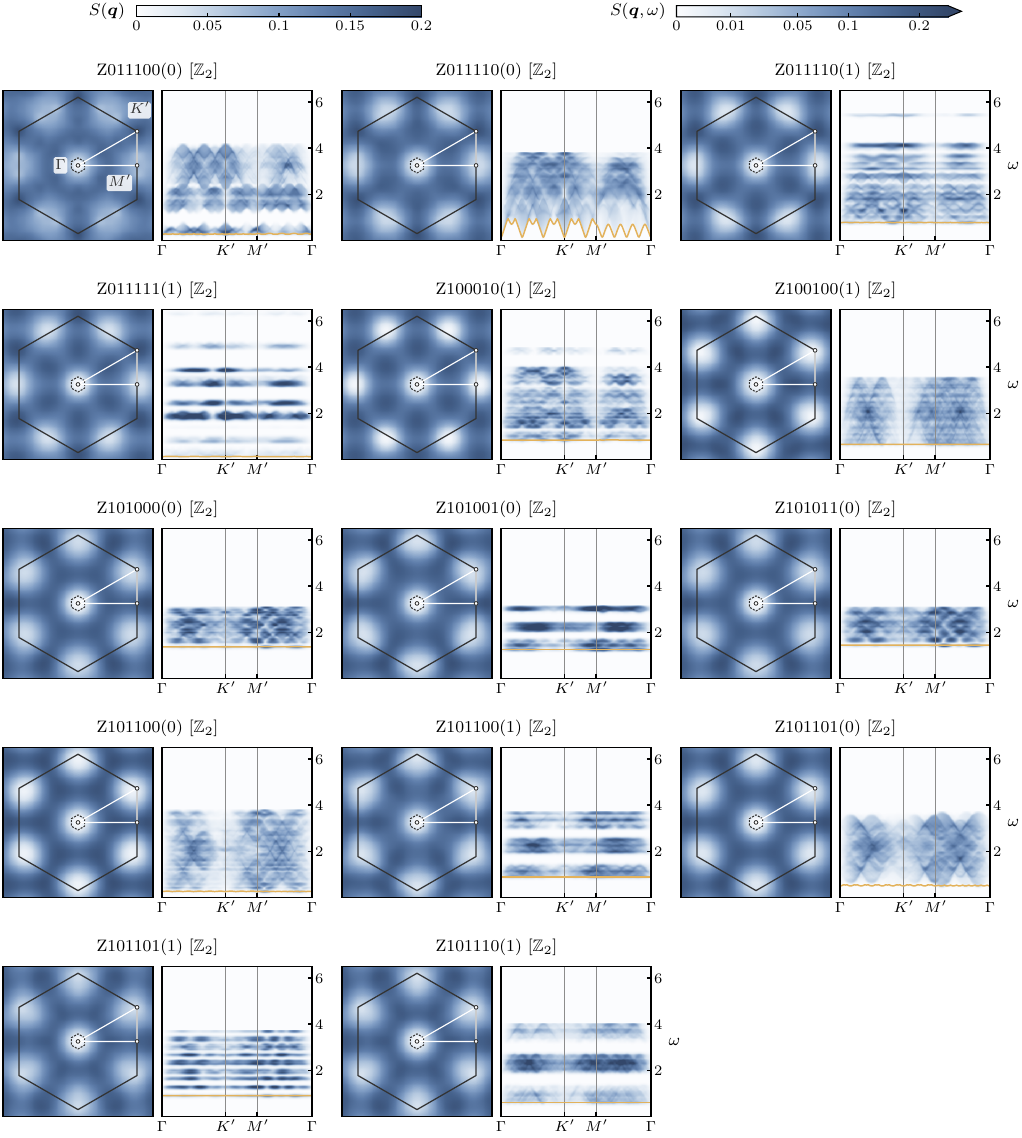}
\caption{As Fig.~\ref{sm:fig:first}, for the $\Z_2$ spin liquids Nos.~61--74, which are Nos.~21--34 of Table~\ref{tab:z2ansb} of the main text; amplitudes and multipliers in Tables~\ref{sm:tab:z2} and \ref{sm:tab:z2b}.}
\label{sm:fig:p5}
\end{figure*}

\begin{figure*}[p]
\centering
\includegraphics{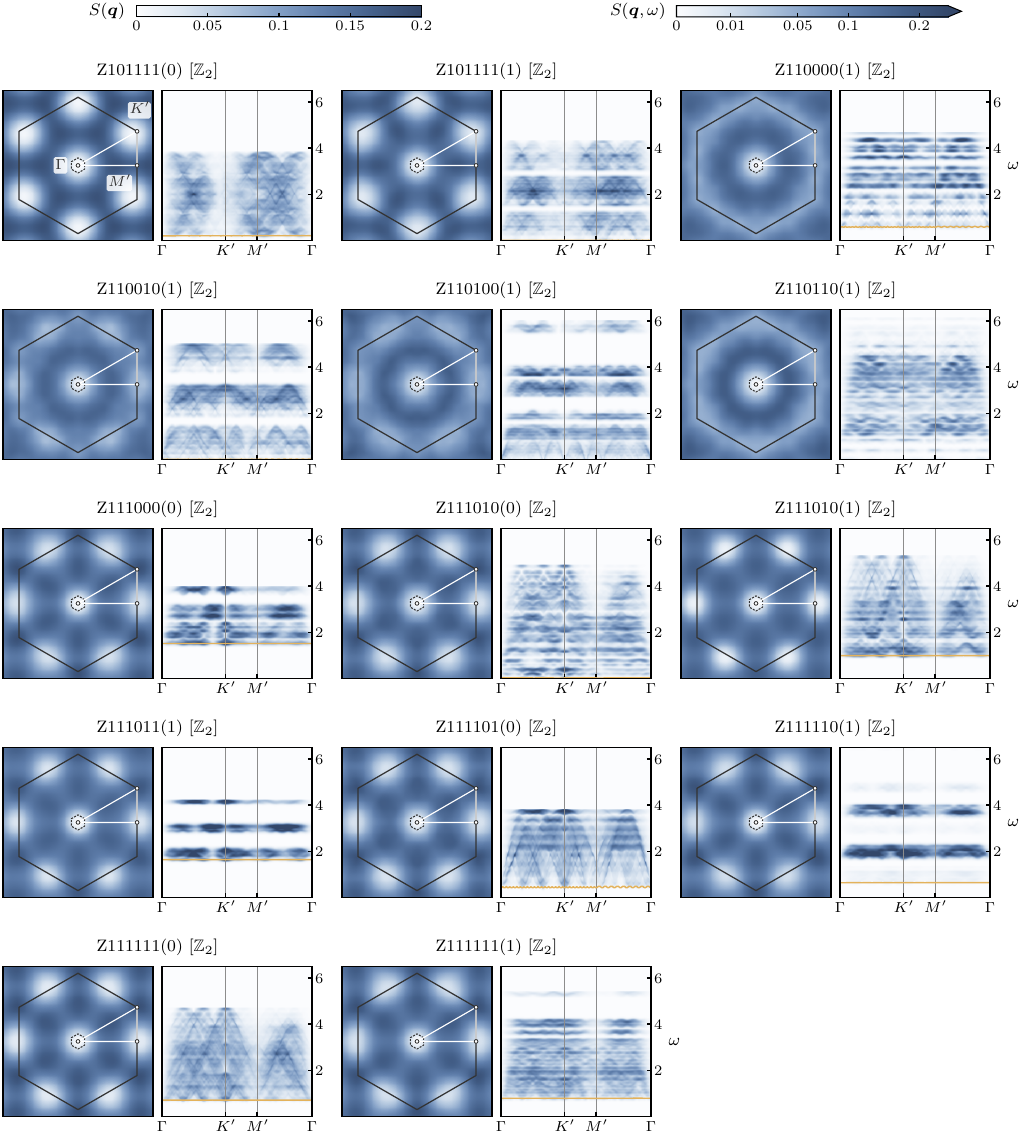}
\caption{As Fig.~\ref{sm:fig:first}, for the $\Z_2$ spin liquids Nos.~75--88, which are Nos.~35--48 of Table~\ref{tab:z2ansb} of the main text; amplitudes and multipliers in Tables~\ref{sm:tab:z2} and \ref{sm:tab:z2b}.}
\label{sm:fig:p6}
\end{figure*}

\begin{figure*}[p]
\centering
\includegraphics{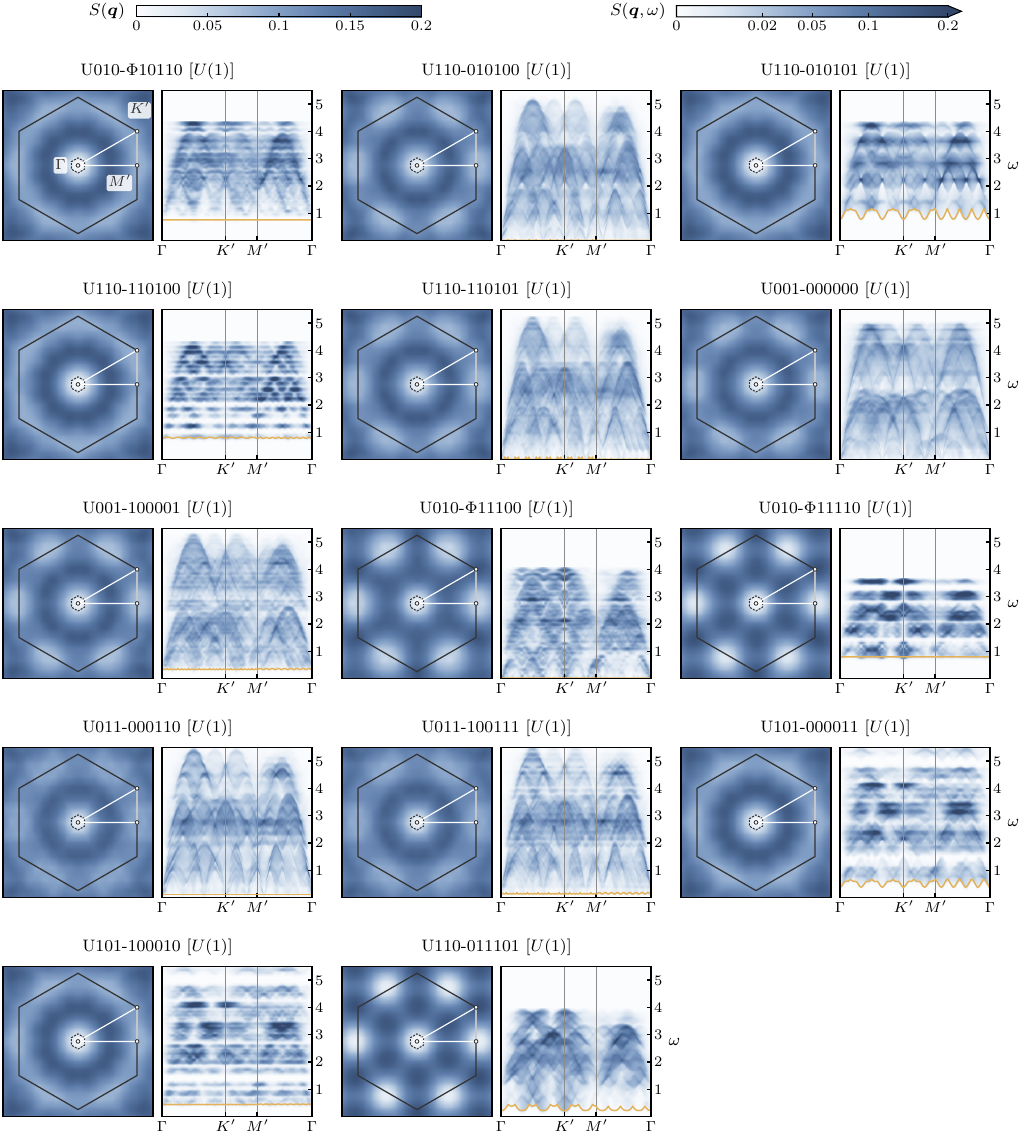}
\caption{As Fig.~\ref{sm:fig:first}, for the $U(1)$ spin liquids Nos.~1--14, which are Nos.~1--14 of Table~\ref{tab:u1ans} of the main text; amplitudes and multipliers in Tables~\ref{sm:tab:u1} and \ref{sm:tab:u1b}.}
\label{sm:fig:p7}
\end{figure*}

\begin{figure*}[p]
\centering
\includegraphics{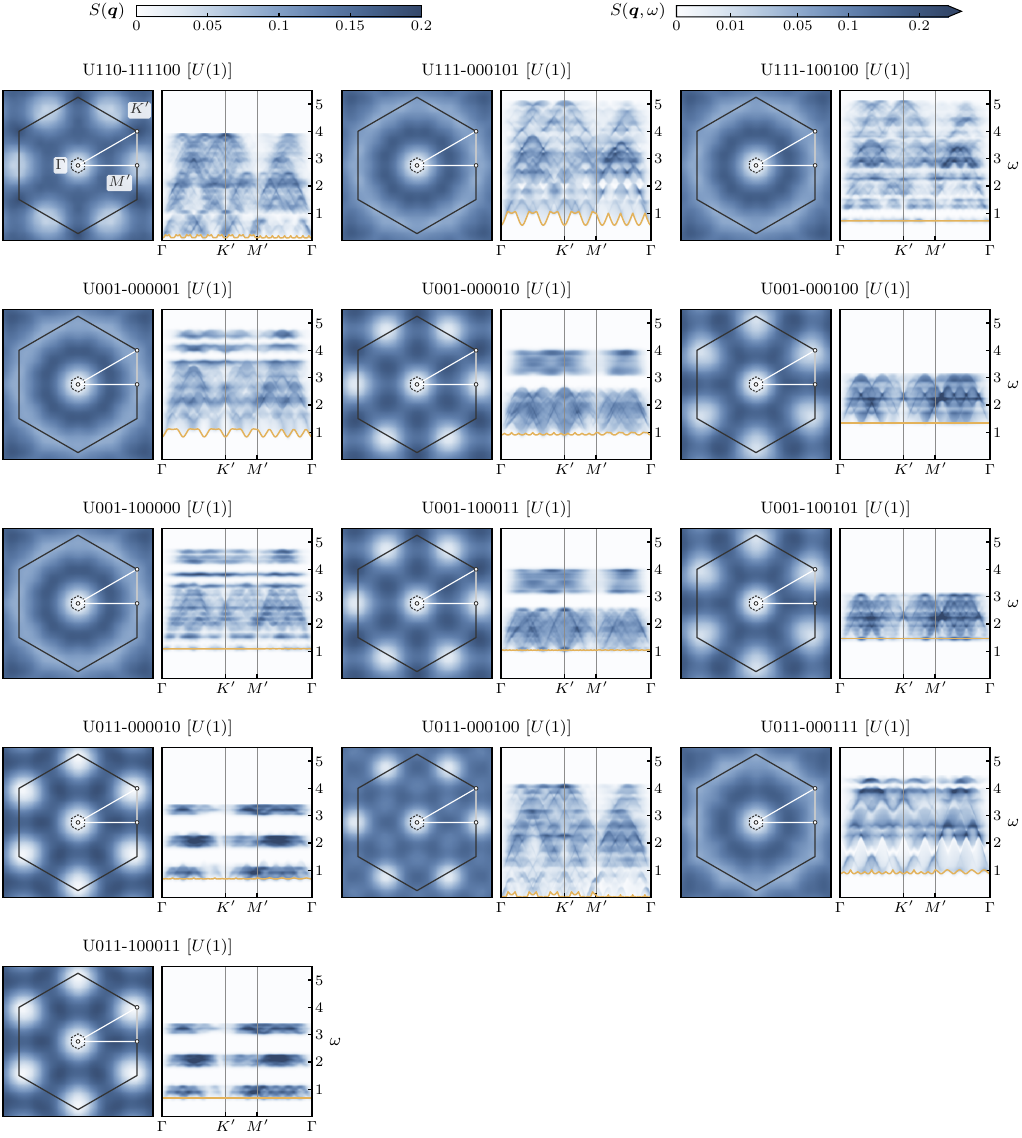}
\caption{As Fig.~\ref{sm:fig:first}, for the $U(1)$ spin liquids Nos.~15--27, which are Nos.~15--27 of Table~\ref{tab:u1ans} of the main text; amplitudes and multipliers in Tables~\ref{sm:tab:u1} and \ref{sm:tab:u1b}.}
\label{sm:fig:p8}
\end{figure*}

\begin{figure*}[p]
\centering
\includegraphics{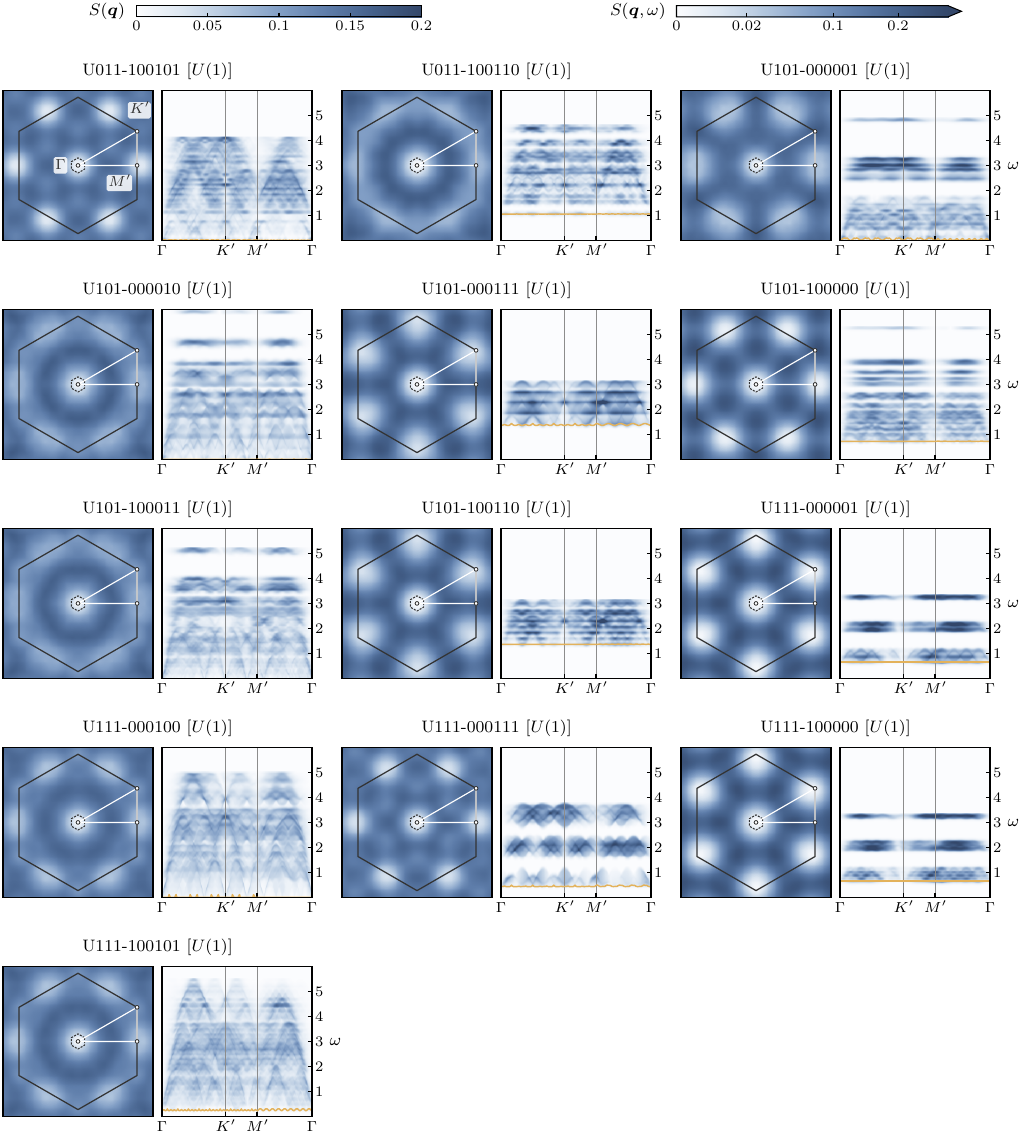}
\caption{As Fig.~\ref{sm:fig:first}, for the $U(1)$ spin liquids Nos.~28--40, which are Nos.~28--40 of Table~\ref{tab:u1ans} of the main text; amplitudes and multipliers in Tables~\ref{sm:tab:u1} and \ref{sm:tab:u1b}.}
\label{sm:fig:p9}
\end{figure*}

\begin{figure*}[p]
\centering
\includegraphics{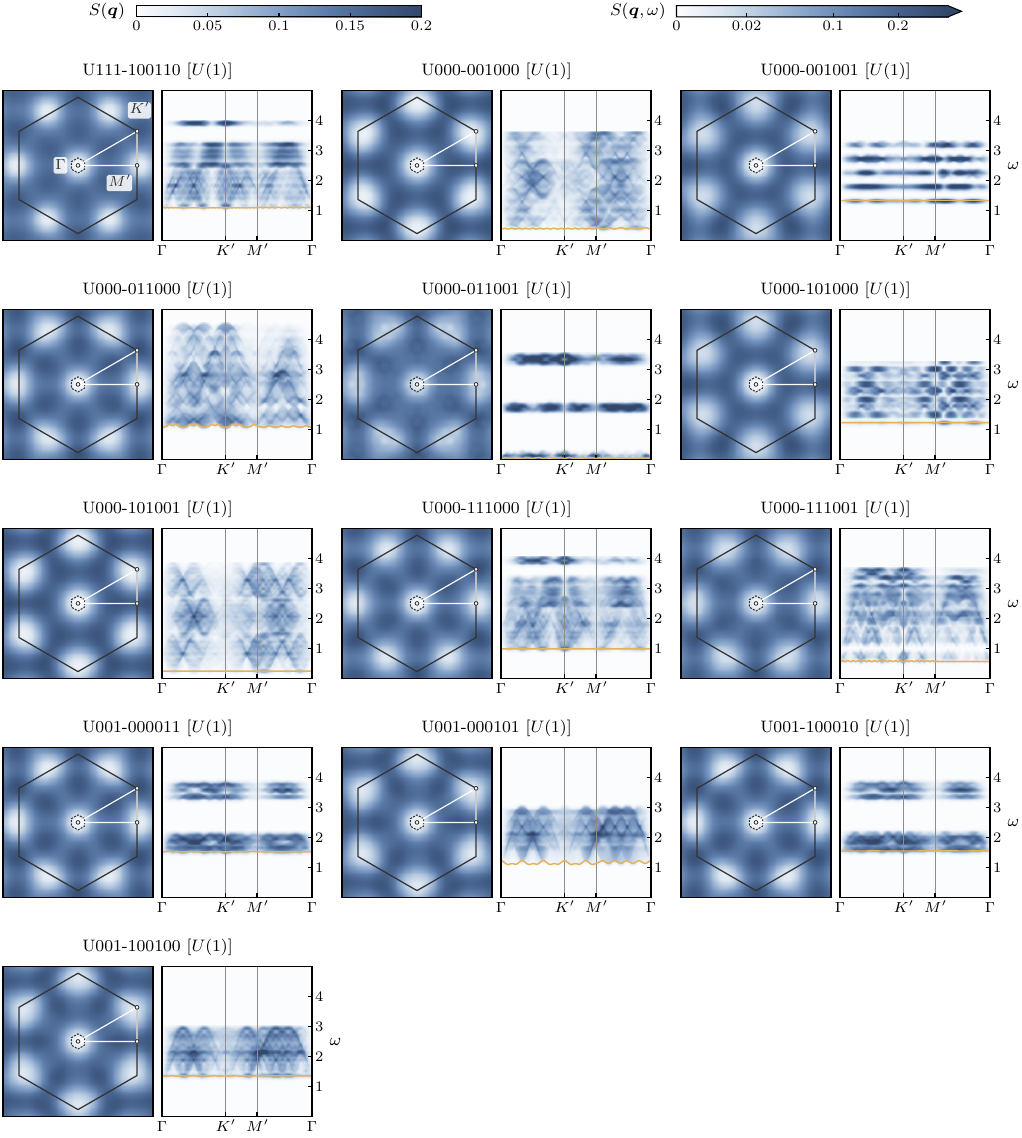}
\caption{As Fig.~\ref{sm:fig:first}, for the $U(1)$ spin liquids Nos.~41--53, which are No.~41 of Table~\ref{tab:u1ans} and Nos.~1--12 of Table~\ref{tab:u1ansb} of the main text; amplitudes and multipliers in Tables~\ref{sm:tab:u1} and \ref{sm:tab:u1b}.}
\label{sm:fig:p10}
\end{figure*}

\begin{figure*}[p]
\centering
\includegraphics{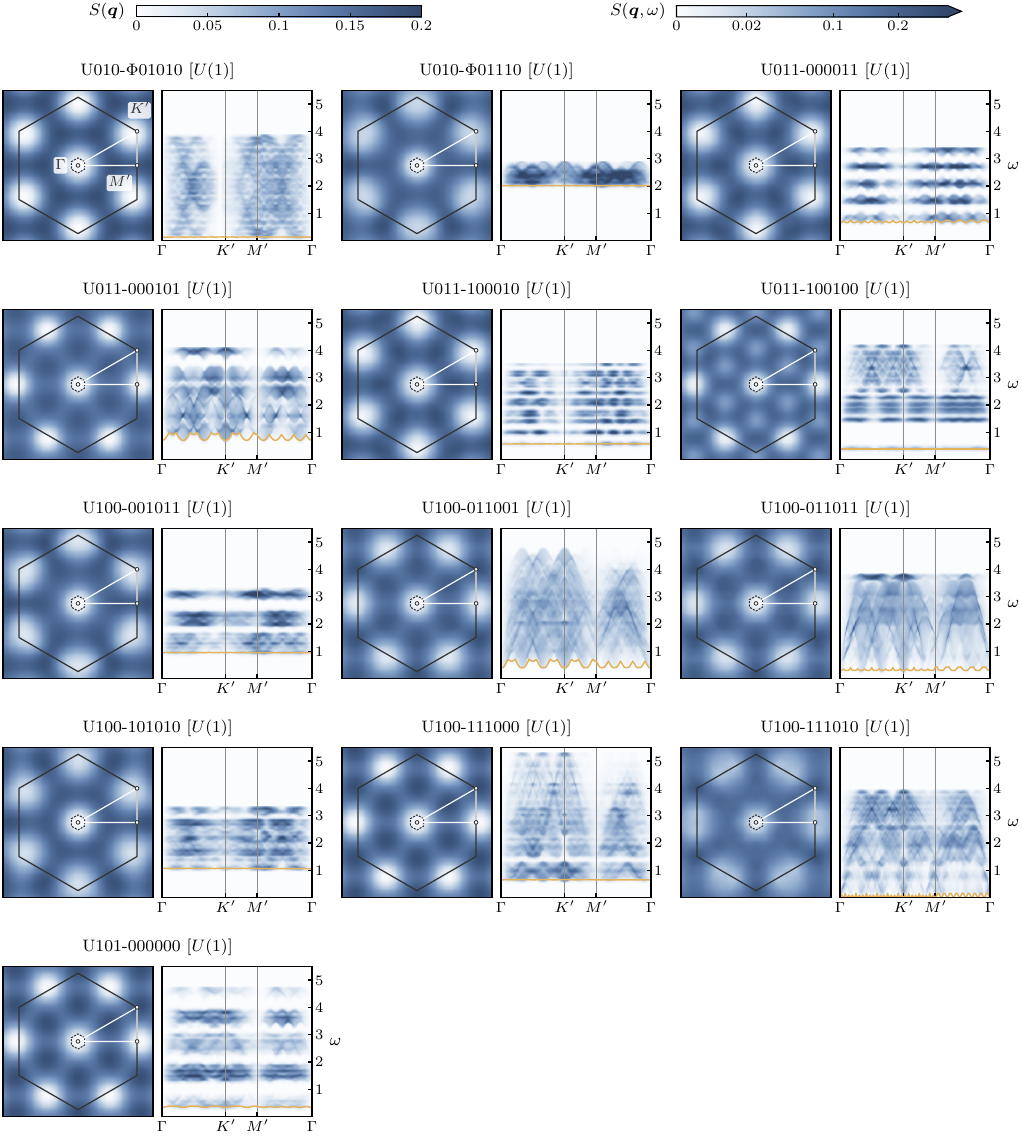}
\caption{As Fig.~\ref{sm:fig:first}, for the $U(1)$ spin liquids Nos.~54--66, which are Nos.~13--25 of Table~\ref{tab:u1ansb} of the main text; amplitudes and multipliers in Tables~\ref{sm:tab:u1} and \ref{sm:tab:u1b}.}
\label{sm:fig:p11}
\end{figure*}

\begin{figure*}[p]
\centering
\includegraphics{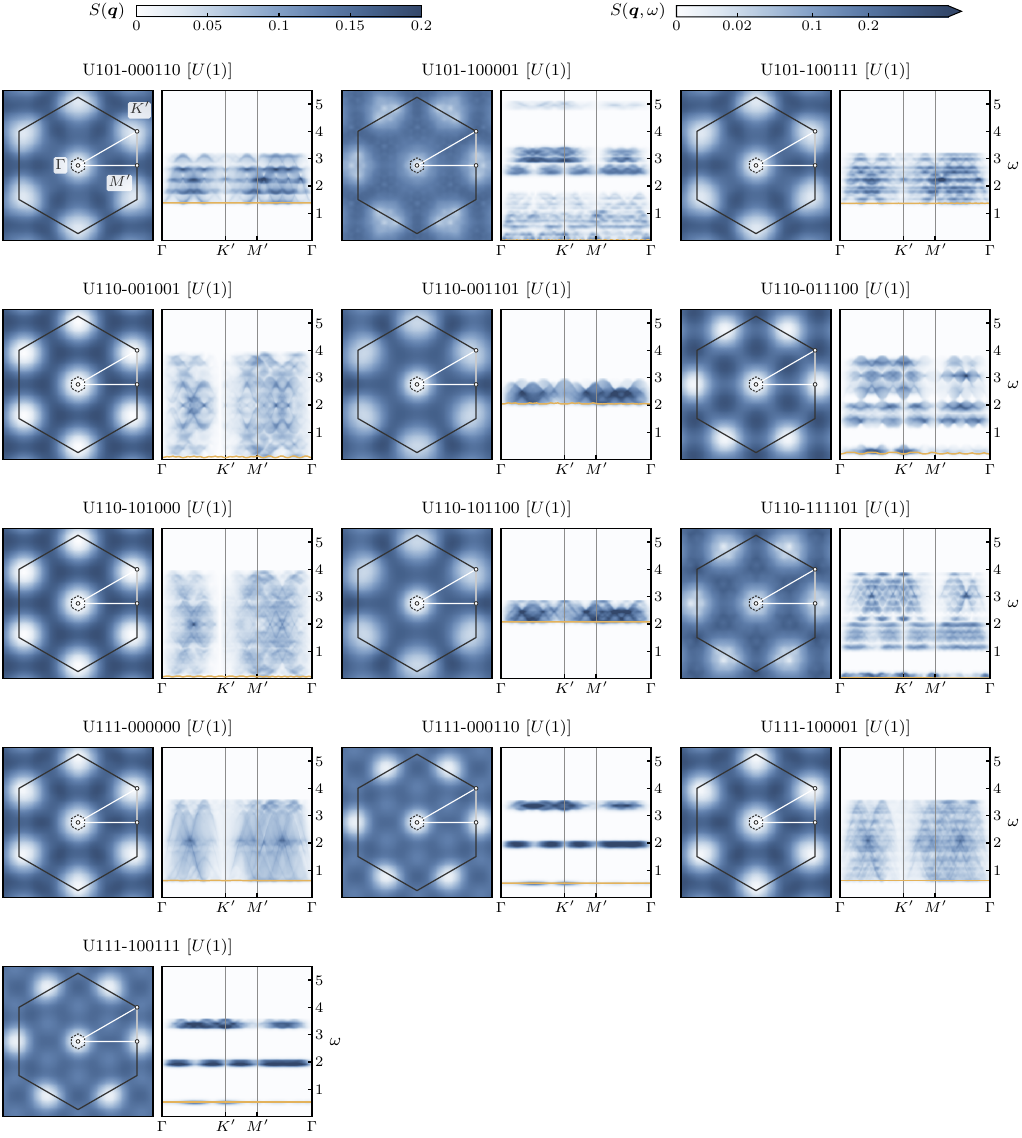}
\caption{As Fig.~\ref{sm:fig:first}, for the $U(1)$ spin liquids Nos.~67--79, which are Nos.~26--38 of Table~\ref{tab:u1ansb} of the main text; amplitudes and multipliers in Tables~\ref{sm:tab:u1} and \ref{sm:tab:u1b}.}
\label{sm:fig:p12}
\end{figure*}

\begin{figure*}[p]
\centering
\includegraphics{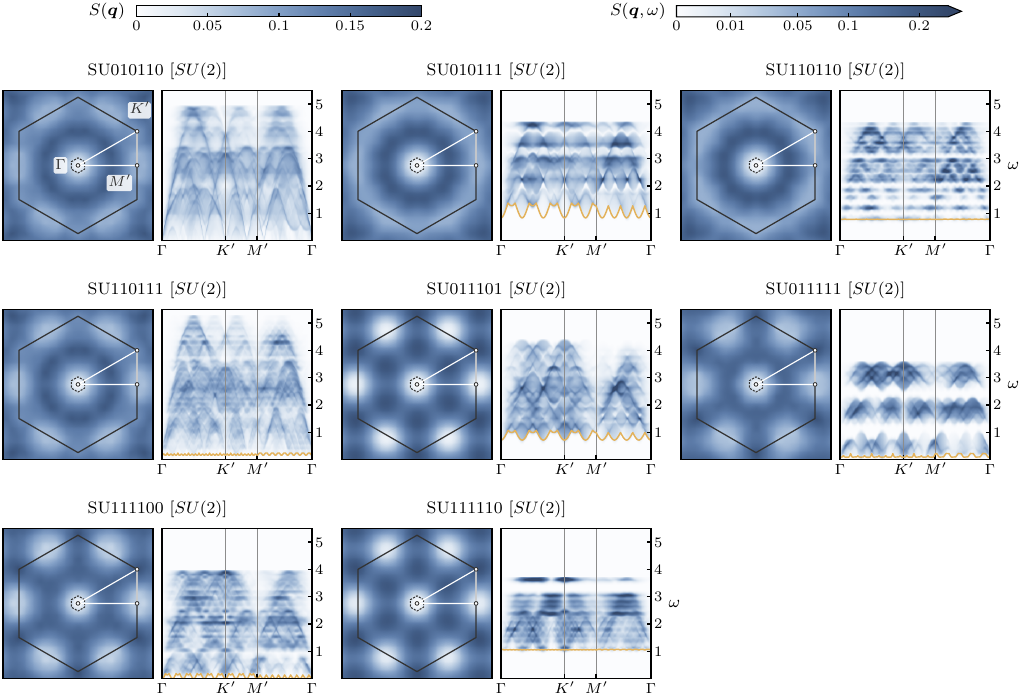}
\caption{As Fig.~\ref{sm:fig:first}, for the $SU(2)$ spin liquids Nos.~1--8, which are Nos.~1--8 of Table~\ref{tab:su2ans} of the main text; amplitudes and multipliers in Table~\ref{sm:tab:su2}.}
\label{sm:fig:p13}
\end{figure*}

\begin{figure*}[p]
\centering
\includegraphics{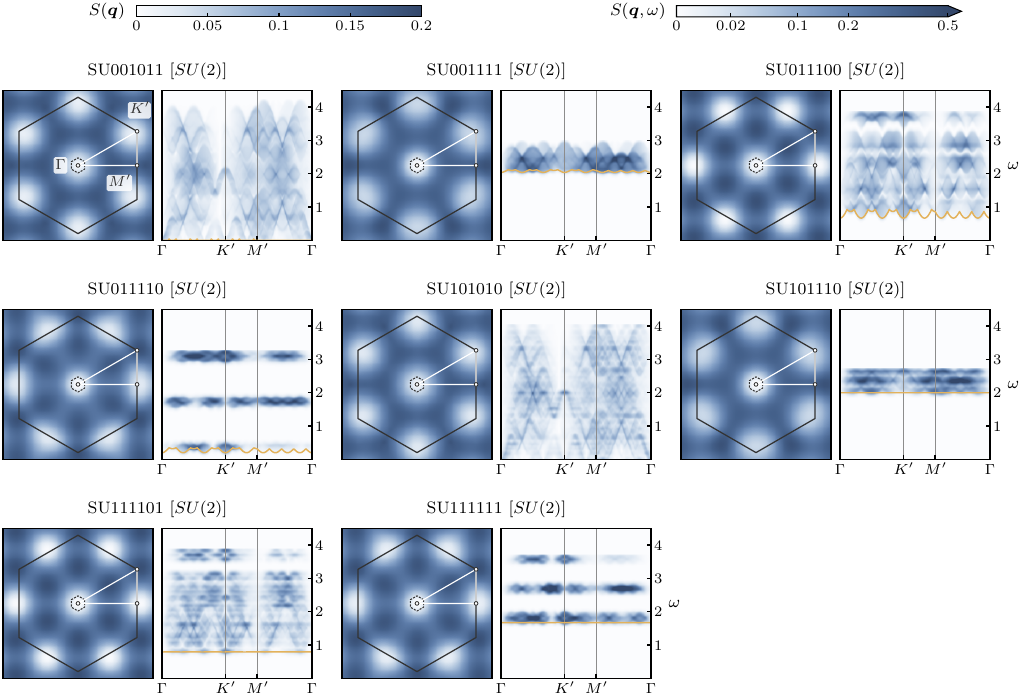}
\caption{As Fig.~\ref{sm:fig:first}, for the $SU(2)$ spin liquids Nos.~9--16, which are Nos.~9--16 of Table~\ref{tab:su2ans} of the main text; amplitudes and multipliers in Table~\ref{sm:tab:su2}.}
\label{sm:fig:last}
\end{figure*}

Sections~\ref{sm:sec:ansatz} and~\ref{sm:sec:sf} give the mean-field static and dynamical spin structure factors of all $183$ symmetric spin liquids realized with singlet amplitudes up to fourth neighbors in Tables~\ref{tab:z2ans}--\ref{tab:su2ans} of the main text: $88$ states with invariant gauge group (IGG) $\Z_2$ (Tables~\ref{tab:z2ans} and \ref{tab:z2ansb}), $79$ with IGG $U(1)$ (Tables~\ref{tab:u1ans} and \ref{tab:u1ansb}), among them five families with a continuous flux $\Phi$, and $16$ with IGG $SU(2)$ (Table~\ref{tab:su2ans}). Within each IGG the states are numbered consecutively in the order of these tables, so that the $\Z_2$ states Nos.~41--88 are Nos.~1--48 of Table~\ref{tab:z2ansb} and the $U(1)$ states Nos.~42--79 are Nos.~1--38 of Table~\ref{tab:u1ansb}. Figures~\ref{sm:fig:first}--\ref{sm:fig:last} show, for each state, the static structure factor $S(\bm{q})$ in the extended Brillouin zone (EBZ) and the dynamical structure factor $S(\bm{q},\omega)$ along $\Gamma\to K'\to M'\to\Gamma$. Tables~\ref{sm:tab:z2}--\ref{sm:tab:su2} list the amplitudes and on-site Lagrange multipliers of each ansatz, the type of its spinon spectrum, the threshold of its two-spinon continuum, and the values $S(K')$ and $S(M')$. All states are taken at the generic amplitude point of Sec.~\ref{sm:app:dos}, so that the spectrum types agree with the density-of-states Tables~\ref{sm:tab:dos1} and~\ref{sm:tab:dos2}. The type of the spectrum belongs to a region of amplitude space that contains this point, not to the PSG; Table~\ref{sm:tab:compare} lists the states whose type differs at a second amplitude point.

\section{Ans\"atze}
\label{sm:sec:ansatz}

Each ansatz is written in the gauge in which the projective symmetry group (PSG)~\SMcite{SM:Wen2002} takes the form of Eq.~\eqref{eq:z2psg} of the main text ($\Z_2$ and $SU(2)$), Eq.~\eqref{eq:u1psg} ($U(1)$), or Eq.~\eqref{eq:u1flux} ($U(1)$ sector $(w_{C_6},w_\sigma,w_\Theta)=(0,1,0)$ with flux $\Phi$). On the reference bond of each orbit (Table~\ref{tab:bonds} of the main text) the singlet amplitude
\begin{equation}
u=u^0\,\ii\tau^0+u^1\tau^1+u^2\tau^2+u^3\tau^3,
\label{sm:eq:u}
\end{equation}
with real coefficients $u^0=\chi^0$, $u^3=\chi^3$, and $u^{1,2}=\Delta^{1,2}$ in the notation of Eq.~\eqref{eq:links} of the main text~\SMcite{SM:LiuHalaszBalents2021}, lies in the subspace allowed by the PSG [the channels listed in Tables~\ref{tab:z2ans}--\ref{tab:su2ans}; $u^1=u^2=0$ for $U(1)$ and $u=u^0\ii\tau^0$ for $SU(2)$]. The amplitudes on all other bonds follow from the PSG. The ans\"atze are fixed as follows.
\begin{itemize}
\item \textit{Bond range.} As in Sec.~\ref{sec:ansatz} of the main text, a PSG is realized at a given range if generic bond amplitudes up to that range have exactly its IGG; the self-consistent multipliers commute with the IGG of the bond amplitudes and do not lower it. As in Sec.~\ref{sm:app:dos}, each ansatz contains all orbits allowed by its PSG up to $J'$ and, when the minimal range listed in Tables~\ref{tab:z2ans}--\ref{tab:su2ans} is $J''$ or $J'''$, up to that range, and no others.
\item \textit{Magnitudes.} The norm $|u|=[\sum_\mu(u^\mu)^2]^{1/2}$ is $1$, $0.9$, and $0.8$ on $J_b$, $J_g$, and $J_r$, $0.5$ on $J'$, and $0.3$ on $J''$ and $J'''$. The unequal nearest-neighbor magnitudes avoid fine-tuned points. Energies are in units of $|u|$ on $J_b$.
\item \textit{Directions.} The direction of $u$ is drawn at random from an isotropic distribution in the allowed space of the orbit. For a one-dimensional allowed space this fixes $u$ up to a random sign, which selects a flux sector; SU010111, for instance, is taken in the Lieb-flux sector, whereas Fig.~\ref{fig:spec-sq} of the main text shows its $\pi$-flux state. The tables give the coefficients rounded to three decimals, with the leading zero omitted in the amplitude columns.
\item \textit{Flux.} The five families of the $U(1)$ sector $(0,1,0)$, U010-$\Phi$10110, U010-$\Phi$11100, and U010-$\Phi$11110 of Table~\ref{tab:u1ans} and U010-$\Phi$01010 and U010-$\Phi$01110 of Table~\ref{tab:u1ansb}, are taken at $\Phi=\pi/2$. The ansatz is then periodic on the magnetic cell spanned by $\bm{T}_1$ and $4\bm{T}_2$ ($48$ sites), and the allowed spaces have the same dimensions as at a generic flux.
\item \textit{Unit cell.} For PSGs with $n_{T_2}=-1$ the bands are computed on the doubled cell spanned by $\bm{T}_1$ and $2\bm{T}_2$ (Appendix~D of the main text). The structure factors depend only on gauge-invariant correlations and are evaluated with the site positions of the ideal lattice, so they do not depend on the choice of cell.
\end{itemize}
For every ansatz the solutions of $W_iu_{ij}=u_{ij}W_j$ on the connected bond graph, with $W_i$ in the real span of $\tau^0$ and $\ii\tau^{1,2,3}$, form a space of dimension $4$, $2$, or $1$ for the $SU(2)$, $U(1)$, and $\Z_2$ states, respectively, with or without the condition $W_ia_i=a_iW_i$ for the on-site fields. Each ansatz therefore has exactly the IGG of its PSG.

\textit{Lagrange multipliers.} The on-site term $\sum_i\mathrm{Tr}[\psi_i a_i\psi_i^\dagger]$ of Eq.~\eqref{eq:HMF} of the main text, with $a_i$ in the space allowed by the PSG (column $a_\mu$ of Tables~\ref{tab:z2ans}--\ref{tab:su2ans}), is fixed by the constraints $\langle\Psi_i^\dagger\tau^k\Psi_i\rangle=0$ ($k=1,2,3$) on every site, where $\Psi_i=(f_{i\uparrow},f_{i\downarrow}^\dagger)^{T}$ is the Nambu spinor of the Abrikosov fermions~\SMcite{SM:Abrikosov1965,SM:AffleckZouHsuAnderson1988}. By symmetry the constraints reduce to one equation per allowed component of $a_i$, which is solved on the $60\times60$ momentum mesh used for the structure factors (Sec.~\ref{sm:sec:sf}), with levels at zero energy counted as half filled. For the $U(1)$ states whose on-site field is the uniform $a^3\tau^3$, which acts as a chemical potential, $a^3$ fixes half filling and lies at the center of the gap when the spinon spectrum is gapped. States without an allowed on-site field satisfy the constraints by symmetry. The constraints hold on every site to better than $10^{-8}$, except for $13$ gapless states and the $\Z_2$ state Z111000(1), whose gap $\Delta=0.023$ is comparable to the level spacing of the mesh; for these $14$ states the occupation on the mesh changes discontinuously with the multipliers, and the largest residual is $4\times10^{-4}$. By Wick's theorem the residuals $t_i$ change $\langle(S^z_i)^2\rangle$ from $1/8$ by $|t_i|^2/8<2\times10^{-8}$. Tables~\ref{sm:tab:z2b} and \ref{sm:tab:u1b} give the field $a_1$ on site~1 of the reference cell; the fields on the other sites follow from the PSG.

\textit{Spinon spectra.} The particle-hole pairs created by $S^z_{\bm{q}}$ involve the bands of $f_\uparrow$ for the hopping-only ($SU(2)$ and $U(1)$) ans\"atze and the bands $E_n(\bm{k})$ of the Bloch matrix of $\Psi$ for the $\Z_2$ ans\"atze. With $E_+(\bm{k})$ and $E_-(\bm{k})$ the lowest positive and the highest negative of these energies, the threshold of the two-spinon continuum is
\begin{equation}
\Delta=\min_{\bm{k}}E_+(\bm{k})-\max_{\bm{k}}E_-(\bm{k}),
\label{sm:eq:gap}
\end{equation}
which for the $\Z_2$ states equals twice the minimum quasiparticle energy. It is twice the single-spinon gap quoted in Sec.~\ref{sm:app:dos}. The spectrum is gapped when $\Delta>0$. Otherwise $E_+-E_-$ vanishes along lines (NL) or at $n$ isolated points with linear (D$n$) or quadratic (Q$n$) dispersion. To find the touchings, every local minimum of the direct gap on a mesh of at least $360\times360$ momenta of the zone of the cell used is refined; it is a touching if the refined gap at $E_F$ is below $10^{-10}$, and touchings that differ by a reciprocal-lattice vector of the cell are counted once. The dispersion is determined along $24$ directions at distances from $10^{-8}$ to $10^{-5}$ from each touching. All touchings of type D$n$ are linear down to the smallest distance, and $N_f=2n$ in the notation of Sec.~\ref{sm:app:dos}: in the $U(1)$ and $SU(2)$ states each node carries one two-component fermion per spin, and in the $\Z_2$ states each node of the Bogoliubov spectrum is fourfold degenerate, with two pairs of bands crossing at the same momentum. Nodal lines are distinguished from isolated touchings by minimizing the direct gap along $64$ lines of the zone in each of two directions and by following the zeros from a refined zero along the direction in which the gap does not grow. At point A a nodal line gives zeros on at least $20$ of the $64$ lines, isolated touchings on at most four. At point B (Table~\ref{sm:tab:compare}) the nodal lines of Z111100(1) and Z101111(1) are small closed loops, crossed by at most $14$ and by none of the lines, respectively, and are identified by following them. P marks the four states of type P of Sec.~\ref{sm:app:dos}, whose zero-energy quasiparticles at $E_F$ are pinned by the constraints; with the zero levels counted as half filled their constraints are satisfied only to $2\times10^{-4}$. The number $n$ refers to the Brillouin zone of the cell used: the first Brillouin zone (FBZ), the reduced zone of the doubled cell, or the magnetic zone. The number of filled bands is the same at every momentum, so no state has Fermi pockets; a nodal line at $E_F$ is nevertheless a Fermi surface of zero enclosed area, with a finite density of states at $E_F$. Of the $183$ states, $152$ are gapped, $21$ have nodal lines, $6$ are Dirac spin liquids, and $4$ are of type P, as in Sec.~\ref{sm:app:dos}.

\section{Structure Factors}
\label{sm:sec:sf}

The dynamical structure factor $S(\bm{q},\omega)$ is the two-spinon bubble of Eq.~(E3) of the main text~\SMcite{SM:Chauhan2023}, with the full Bogoliubov coherence factors and the site positions of the ideal lattice (unit edges). In this geometry $S(\bm{q})$ is defined at every momentum but is not periodic (Appendix~D of the main text); we show it in the hexagon with corners $K'=9K$, the extended Brillouin zone (EBZ) of the commensurate realization of the lattice constructed there. The static structure factor
\begin{equation}
S(\bm{q})=\frac1N\sum_{i,j}e^{\ii\bm{q}\cdot(\bm{r}_i-\bm{r}_j)}\langle S^z_iS^z_j\rangle=\int_0^\infty d\omega\,S(\bm{q},\omega)
\label{sm:eq:sq}
\end{equation}
is normalized as in Fig.~\ref{fig:spec-sq} of the main text: its average over large momentum regions equals the mean-field value $\langle(S^z_i)^2\rangle=1/8$ (over the EBZ it differs from $1/8$ by up to $0.007$, since the EBZ is not a period cell in the ideal geometry), and the trace $\sum_aS^{aa}$ is three times larger. Momenta are in units of $2\pi/A$. The EBZ is the hexagon with corners $K'=9K=(3\sqrt3,3)$ and edge midpoints $M'=9M=(3\sqrt3,0)$, nine times larger than the FBZ in linear size, and the FBZ lies at its center.

All quantities are computed on a $60\times60$ momentum mesh of the Brillouin zone of the cell used, which contains the $541$ momenta of the path. The maps of $S(\bm{q})$ assign to each point of a $241\times241$ raster the value at the nearest mesh momentum. The spectra are binned in steps of $0.005$ and broadened by a Gaussian of standard deviation $0.04$. The lower edge of the two-spinon continuum, $\omega_{\min}(\bm{q})=\min_{\bm{k}}[E_+(\bm{k})-E_-(\bm{k}+\bm{q})]$, is obtained on a $120\times120$ mesh and refined by continuous minimization over $\bm{k}$.

\begingroup
\setlength{\LTcapwidth}{\textwidth}\renewcommand{\arraystretch}{0.975}
\begingroup\scriptsize\setlength{\tabcolsep}{4pt}
\begin{longtable}{rlcccc}
\caption{$\Z_2$ spin liquids of Tables~\ref{tab:z2ans} (Nos.~1--40) and \ref{tab:z2ansb} (Nos.~41--88, listed there as Nos.~1--48) of the main text: amplitudes $u$, Eq.~\eqref{sm:eq:u}, on the reference bonds of $J_b$, $J_g$, $J_r$, and $J'$ (empty: orbit not included).}\label{sm:tab:z2}\\
\hline\hline
No. & PSG & $J_b$ & $J_g$ & $J_r$ & $J'$\\
\hline
\endfirsthead
\multicolumn{6}{l}{\textsc{Table}~\ref{sm:tab:z2} (\textit{continued})}\\
\hline\hline
No. & PSG & $J_b$ & $J_g$ & $J_r$ & $J'$\\
\hline
\endhead
\hline\hline
\endlastfoot
1 & Z010000(0) & $.341\tau^1{-}.702\tau^2{+}.625\tau^3$ & $-.656\tau^1{+}.563\tau^2{+}.251\tau^3$ & $.032\tau^1{-}.759\tau^2{-}.25\tau^3$ & \\
2 & Z010001(0) & $-.644\tau^1{+}.442\tau^2{+}.624\tau^3$ & $.89\tau^1{+}.101\tau^2{-}.088\tau^3$ & $.547\tau^1{-}.368\tau^2{-}.453\tau^3$ & \\
3 & Z110000(0) & $-.894\tau^1{+}.331\tau^2{+}.301\tau^3$ & $.235\tau^1{+}.868\tau^2{+}.048\tau^3$ & $-.454\tau^1{+}.265\tau^2{-}.603\tau^3$ & \\
4 & Z110001(0) & $-.457\tau^1{-}.871\tau^2{-}.177\tau^3$ & $-.157\tau^1{-}.641\tau^2{-}.612\tau^3$ & $-.24\tau^1{-}.318\tau^2{-}.694\tau^3$ & \\
5 & Z000000(1) & $-.858\tau^1{+}.514\tau^3$ & $.463\tau^1{+}.772\tau^3$ & $.727\tau^1{-}.335\tau^3$ & $.306\tau^1{-}.395\tau^3$\\
6 & Z001000(1) & $-\tau^2$ & $-.811\tau^1{+}.39\tau^3$ & $-.592\tau^1{+}.538\tau^3$ & $.5\tau^2$\\
7 & Z001001(1) & $-\tau^2$ & $-.894\tau^1{-}.103\tau^3$ & $.167\tau^1{-}.782\tau^3$ & $-.5\ii\tau^0$\\
8 & Z001010(1) & $-\ii\tau^0$ & $.794\tau^1{-}.424\tau^3$ & $.445\tau^1{-}.665\tau^3$ & $.5\tau^2$\\
9 & Z001011(1) & $-\ii\tau^0$ & $-.461\tau^1{+}.773\tau^3$ & $.763\tau^1{+}.24\tau^3$ & $-.5\ii\tau^0$\\
10 & Z011000(0) &  & $-.523\tau^1{+}.711\tau^2{+}.179\tau^3$ & $.257\tau^1{-}.185\tau^2{+}.735\tau^3$ & $.452\tau^1{+}.075\tau^2{-}.201\tau^3$\\
11 & Z011010(0) &  & $.262\tau^1{-}.016\tau^2{-}.861\tau^3$ & $.729\tau^1{+}.091\tau^2{+}.316\tau^3$ & $.492\tau^1{+}.082\tau^2{-}.03\tau^3$\\
12 & Z100001(1) & $.471\tau^1{+}.882\tau^3$ & $.048\tau^1{+}.899\tau^3$ & $-.791\tau^1{-}.121\tau^3$ & $.022\tau^1{+}.5\tau^3$\\
13 & Z101000(1) & $-\tau^2$ & $-.24\tau^1{+}.867\tau^3$ & $.571\tau^1{-}.56\tau^3$ & $.5\ii\tau^0$\\
14 & Z101001(1) & $\tau^2$ & $-.671\tau^1{+}.6\tau^3$ & $-.109\tau^1{-}.793\tau^3$ & $.5\tau^2$\\
15 & Z101010(1) & $-\ii\tau^0$ & $.59\tau^1{-}.68\tau^3$ & $.482\tau^1{+}.639\tau^3$ & $.5\ii\tau^0$\\
16 & Z101011(1) & $\ii\tau^0$ & $-.887\tau^1{-}.154\tau^3$ & $.273\tau^1{-}.752\tau^3$ & $.5\tau^2$\\
17 & Z111001(0) &  & $-.345\tau^1{-}.693\tau^2{-}.46\tau^3$ & $-.361\tau^1{+}.134\tau^2{+}.701\tau^3$ & $.215\tau^1{+}.055\tau^2{+}.448\tau^3$\\
18 & Z111011(0) &  & $-.013\tau^1{-}.073\tau^2{-}.897\tau^3$ & $-.283\tau^1{-}.416\tau^2{-}.622\tau^3$ & $-.451\tau^1{+}.158\tau^2{-}.146\tau^3$\\
19 & Z000001(1) & $.055\tau^1{+}.998\tau^3$ & $-.881\tau^1{-}.186\tau^3$ & $.42\tau^1{-}.681\tau^3$ & \\
20 & Z000010(1) &  & $-.846\tau^1{-}.307\tau^3$ & $-.72\tau^1{-}.348\tau^3$ & $.122\tau^1{+}.485\tau^3$\\
21 & Z000100(1) & $.638\tau^1{-}.77\tau^3$ &  &  & $-.327\tau^1{-}.378\tau^3$\\
22 & Z010000(1) & $-\tau^2$ & $-.9\tau^2$ & $-.8\tau^2$ & $-.132\tau^1{+}.482\tau^3$\\
23 & Z010010(1) & $-\ii\tau^0$ & $.9\tau^2$ & $-.8\tau^2$ & $.276\tau^1{+}.417\tau^3$\\
24 & Z010100(1) & $-\tau^2$ & $.9\ii\tau^0$ & $-.8\ii\tau^0$ & $-.178\tau^1{+}.467\tau^3$\\
25 & Z010110(1) & $\ii\tau^0$ & $-.9\ii\tau^0$ & $.8\ii\tau^0$ & $-.134\tau^1{-}.482\tau^3$\\
26 & Z011000(1) & $-\tau^1{+}.027\tau^3$ & $-.9\tau^2$ & $.8\tau^2$ & $-.5\tau^2$\\
27 & Z011001(1) & $-.998\tau^1{-}.067\tau^3$ & $.9\tau^2$ & $-.8\tau^2$ & $.5\ii\tau^0$\\
28 & Z011100(1) & $.581\tau^1{-}.814\tau^3$ & $.9\ii\tau^0$ & $.8\ii\tau^0$ & $-.5\tau^2$\\
29 & Z011101(1) & $-.905\tau^1{+}.425\tau^3$ & $-.9\ii\tau^0$ & $-.8\ii\tau^0$ & $.5\ii\tau^0$\\
30 & Z100000(1) & $-.522\tau^1{-}.853\tau^3$ & $-.457\tau^1{-}.775\tau^3$ & $.05\tau^1{-}.798\tau^3$ & \\
31 & Z100011(1) &  & $.9\tau^1{+}.007\tau^3$ & $-.394\tau^1{+}.696\tau^3$ & $.156\tau^1{+}.475\tau^3$\\
32 & Z100101(1) & $.918\tau^1{-}.396\tau^3$ &  &  & $.277\tau^1{+}.417\tau^3$\\
33 & Z110001(1) & $\tau^2$ & $-.9\tau^2$ & $.8\tau^2$ & $.256\tau^1{+}.43\tau^3$\\
34 & Z110011(1) & $\ii\tau^0$ & $.9\tau^2$ & $-.8\tau^2$ & $-.352\tau^1{+}.355\tau^3$\\
35 & Z110101(1) & $-\tau^2$ & $-.9\ii\tau^0$ & $-.8\ii\tau^0$ & $-.263\tau^1{+}.425\tau^3$\\
36 & Z110111(1) & $-\ii\tau^0$ & $-.9\ii\tau^0$ & $.8\ii\tau^0$ & $.287\tau^1{+}.409\tau^3$\\
37 & Z111000(1) & $-.402\tau^1{+}.916\tau^3$ & $-.9\tau^2$ & $-.8\tau^2$ & $.5\ii\tau^0$\\
38 & Z111001(1) & $.819\tau^1{+}.574\tau^3$ & $.9\tau^2$ & $.8\tau^2$ & $.5\tau^2$\\
39 & Z111100(1) & $-.819\tau^1{+}.573\tau^3$ & $.9\ii\tau^0$ & $.8\ii\tau^0$ & $-.5\ii\tau^0$\\
40 & Z111101(1) & $-.696\tau^1{-}.718\tau^3$ & $-.9\ii\tau^0$ & $.8\ii\tau^0$ & $-.5\tau^2$\\
41 & Z000011(1) &  & $-.897\tau^1{-}.075\tau^3$ & $.694\tau^1{-}.399\tau^3$ & \\
42 & Z000101(1) & $-.296\tau^1{+}.955\tau^3$ &  &  & \\
43 & Z001000(0) & $.072\tau^1{+}.96\tau^2{-}.271\tau^3$ &  &  & $.305\tau^1{-}.095\tau^2{+}.385\tau^3$\\
44 & Z001001(0) & $-.837\tau^1{-}.537\tau^2{-}.105\tau^3$ &  &  & $-.5\ii\tau^0$\\
45 & Z001010(0) & $\ii\tau^0$ &  &  & $.493\tau^1{-}.02\tau^2{+}.08\tau^3$\\
46 & Z001100(0) & $.731\tau^1{+}.681\tau^2{+}.044\tau^3$ &  &  & $.218\tau^1{-}.436\tau^2{-}.11\tau^3$\\
47 & Z001100(1) & $\tau^2$ &  &  & $-.5\tau^2$\\
48 & Z001101(0) & $.509\tau^1{+}.725\tau^2{-}.464\tau^3$ &  &  & $.5\ii\tau^0$\\
49 & Z001101(1) & $\tau^2$ &  &  & $.5\ii\tau^0$\\
50 & Z001110(0) & $\ii\tau^0$ &  &  & $-.152\tau^1{-}.034\tau^2{+}.475\tau^3$\\
51 & Z001110(1) & $\ii\tau^0$ &  &  & $.5\tau^2$\\
52 & Z001111(1) & $-\ii\tau^0$ &  &  & $.5\ii\tau^0$\\
53 & Z010001(1) & $-\tau^2$ & $-.9\tau^2$ & $.8\tau^2$ & \\
54 & Z010011(1) & $-\ii\tau^0$ & $.9\tau^2$ & $-.8\tau^2$ & \\
55 & Z010101(1) & $-\tau^2$ & $-.9\ii\tau^0$ & $.8\ii\tau^0$ & \\
56 & Z010111(1) & $\ii\tau^0$ & $.9\ii\tau^0$ & $.8\ii\tau^0$ & \\
57 & Z011001(0) &  & $.728\tau^1{-}.405\tau^2{-}.34\tau^3$ & $-.433\tau^1{+}.281\tau^2{+}.611\tau^3$ & $-.5\ii\tau^0$\\
58 & Z011010(1) &  & $-.9\tau^2$ & $-.8\tau^2$ & $-.5\tau^2$\\
59 & Z011011(0) &  & $.626\tau^1{+}.611\tau^2{-}.213\tau^3$ & $.77\tau^1{+}.062\tau^2{+}.209\tau^3$ & $-.5\ii\tau^0$\\
60 & Z011011(1) &  & $-.9\tau^2$ & $-.8\tau^2$ & $-.5\ii\tau^0$\\
61 & Z011100(0) &  & $-.9\ii\tau^0$ & $.8\ii\tau^0$ & $-.084\tau^1{+}.192\tau^2{-}.454\tau^3$\\
62 & Z011110(0) &  & $-.9\ii\tau^0$ & $-.8\ii\tau^0$ & $-.273\tau^1{+}.415\tau^2{-}.052\tau^3$\\
63 & Z011110(1) &  & $-.9\ii\tau^0$ & $-.8\ii\tau^0$ & $.5\tau^2$\\
64 & Z011111(1) &  & $.9\ii\tau^0$ & $-.8\ii\tau^0$ & $.5\ii\tau^0$\\
65 & Z100010(1) &  & $.847\tau^1{+}.305\tau^3$ & $-.745\tau^1{-}.293\tau^3$ & \\
66 & Z100100(1) & $.258\tau^1{-}.966\tau^3$ &  &  & \\
67 & Z101000(0) & $.407\tau^1{+}.75\tau^2{+}.521\tau^3$ &  &  & $-.5\ii\tau^0$\\
68 & Z101001(0) & $-.716\tau^1{+}.462\tau^2{-}.524\tau^3$ &  &  & $.34\tau^1{+}.04\tau^2{+}.365\tau^3$\\
69 & Z101011(0) & $\ii\tau^0$ &  &  & $-.172\tau^1{+}.451\tau^2{-}.132\tau^3$\\
70 & Z101100(0) & $-.669\tau^1{-}.631\tau^2{-}.394\tau^3$ &  &  & $-.5\ii\tau^0$\\
71 & Z101100(1) & $-\tau^2$ &  &  & $-.5\ii\tau^0$\\
72 & Z101101(0) & $.254\tau^1{-}.919\tau^2{+}.302\tau^3$ &  &  & $.024\tau^1{-}.441\tau^2{-}.234\tau^3$\\
73 & Z101101(1) & $\tau^2$ &  &  & $-.5\tau^2$\\
74 & Z101110(1) & $\ii\tau^0$ &  &  & $.5\ii\tau^0$\\
75 & Z101111(0) & $\ii\tau^0$ &  &  & $.423\tau^1{+}.246\tau^2{-}.103\tau^3$\\
76 & Z101111(1) & $\ii\tau^0$ &  &  & $-.5\tau^2$\\
77 & Z110000(1) & $\tau^2$ & $.9\tau^2$ & $.8\tau^2$ & \\
78 & Z110010(1) & $-\ii\tau^0$ & $-.9\tau^2$ & $-.8\tau^2$ & \\
79 & Z110100(1) & $\tau^2$ & $-.9\ii\tau^0$ & $-.8\ii\tau^0$ & \\
80 & Z110110(1) & $-\ii\tau^0$ & $.9\ii\tau^0$ & $.8\ii\tau^0$ & \\
81 & Z111000(0) &  & $.1\tau^1{-}.803\tau^2{-}.395\tau^3$ & $.487\tau^1{+}.03\tau^2{-}.634\tau^3$ & $-.5\ii\tau^0$\\
82 & Z111010(0) &  & $-.82\tau^1{+}.232\tau^2{-}.29\tau^3$ & $.158\tau^1{-}.569\tau^2{+}.54\tau^3$ & $-.5\ii\tau^0$\\
83 & Z111010(1) &  & $-.9\tau^2$ & $.8\tau^2$ & $-.5\ii\tau^0$\\
84 & Z111011(1) &  & $-.9\tau^2$ & $-.8\tau^2$ & $-.5\tau^2$\\
85 & Z111101(0) &  & $-.9\ii\tau^0$ & $-.8\ii\tau^0$ & $-.327\tau^1{+}.2\tau^2{-}.321\tau^3$\\
86 & Z111110(1) &  & $.9\ii\tau^0$ & $-.8\ii\tau^0$ & $.5\ii\tau^0$\\
87 & Z111111(0) &  & $-.9\ii\tau^0$ & $-.8\ii\tau^0$ & $.164\tau^1{+}.468\tau^2{+}.066\tau^3$\\
88 & Z111111(1) &  & $-.9\ii\tau^0$ & $-.8\ii\tau^0$ & $.5\tau^2$\\
\end{longtable}
\endgroup

\begingroup\scriptsize\setlength{\tabcolsep}{2pt}
\begin{longtable}{rlccclccc}
\caption{$\Z_2$ spin liquids of Table~\ref{sm:tab:z2}: amplitudes on the reference bonds of $J''$ and $J'''$ (empty: orbit not included), on-site multiplier $a_1$ on site~1 of the reference cell, type of the spinon spectrum (gapped; NL, nodal lines; D$n$ and Q$n$, $n$ isolated zero-energy band touchings with linear and quadratic dispersion, respectively; P, zero modes pinned at $E_F$ by the constraints), threshold $\Delta$ of the two-spinon continuum, Eq.~\eqref{sm:eq:gap}, and $S(\bm{q})$ at $K'$ and $M'$.}\label{sm:tab:z2b}\\
\hline\hline
No. & PSG & $J''$ & $J'''$ & $a_1$ & spectrum & $\Delta$ & $S(K')$ & $S(M')$\\
\hline
\endfirsthead
\multicolumn{9}{l}{\textsc{Table}~\ref{sm:tab:z2b} (\textit{continued})}\\
\hline\hline
No. & PSG & $J''$ & $J'''$ & $a_1$ & spectrum & $\Delta$ & $S(K')$ & $S(M')$\\
\hline
\endhead
\hline\hline
\endlastfoot
1 & Z010000(0) &  &  &  & gapped & 0.875 & 0.1068 & 0.0843\\
2 & Z010001(0) &  &  &  & gapped & 0.789 & 0.1043 & 0.0867\\
3 & Z110000(0) &  &  &  & NL & 0 & 0.1110 & 0.0836\\
4 & Z110001(0) &  &  &  & gapped & 0.446 & 0.0966 & 0.0983\\
5 & Z000000(1) &  &  & $-0.27066\tau^1{+}0.23575\tau^3$ & gapped & 0.047 & 0.1174 & 0.0817\\
6 & Z001000(1) &  &  & $0.24606\tau^1{-}0.21991\tau^3$ & gapped & 0.371 & 0.1060 & 0.0866\\
7 & Z001001(1) &  &  & $-0.11064\tau^1{-}0.02947\tau^3$ & NL & 0 & 0.1140 & 0.0941\\
8 & Z001010(1) &  &  & $-0.39803\tau^1{-}0.04546\tau^3$ & gapped & 0.002 & 0.1066 & 0.0959\\
9 & Z001011(1) &  &  & $-0.32907\tau^1{-}0.19671\tau^3$ & NL & 0 & 0.1180 & 0.0785\\
10 & Z011000(0) &  &  &  & gapped & 0.820 & 0.1831 & 0.0311\\
11 & Z011010(0) &  &  &  & gapped & 0.287 & 0.1834 & 0.0251\\
12 & Z100001(1) &  &  & $0.19742\tau^1{+}0.06364\tau^3$ & gapped & 0.075 & 0.1103 & 0.0892\\
13 & Z101000(1) &  &  & $-0.1477\tau^1{-}0.11558\tau^3$ & NL & 0 & 0.1144 & 0.0960\\
14 & Z101001(1) &  &  & $0.06619\tau^1{+}0.22777\tau^3$ & gapped & 0.869 & 0.1175 & 0.0711\\
15 & Z101010(1) &  &  & $-0.26588\tau^1{-}0.27454\tau^3$ & NL & 0 & 0.1166 & 0.0806\\
16 & Z101011(1) &  &  & $-0.21149\tau^1{-}0.10041\tau^3$ & gapped & 0.062 & 0.1061 & 0.0905\\
17 & Z111001(0) &  &  &  & gapped & 0.757 & 0.1834 & 0.0242\\
18 & Z111011(0) &  &  &  & gapped & 0.022 & 0.1836 & 0.0272\\
19 & Z000001(1) & $.228\tau^1{-}.195\tau^3$ &  & $-0.16403\tau^1{+}0.18319\tau^3$ & gapped & 0.351 & 0.0948 & 0.0976\\
20 & Z000010(1) & $.009\tau^1{-}.3\tau^3$ &  & $-0.01451\tau^1{-}0.12443\tau^3$ & gapped & 0.765 & 0.1712 & 0.0208\\
21 & Z000100(1) & $-.224\tau^1{-}.2\tau^3$ &  & $-0.00715\tau^1{-}0.0042\tau^3$ & gapped & 0.697 & 0.0172 & 0.1616\\
22 & Z010000(1) & $-.299\tau^1{-}.028\tau^3$ &  & $-0.15301\tau^1{+}0.19929\tau^3$ & gapped & 0.178 & 0.1141 & 0.0881\\
23 & Z010010(1) & $-.039\tau^1{-}.297\tau^3$ &  & $0.23367\tau^1{-}0.12163\tau^3$ & gapped & 0.244 & 0.0984 & 0.0978\\
24 & Z010100(1) & $-.075\tau^1{-}.291\tau^3$ &  & $0.04393\tau^1{+}0.3213\tau^3$ & gapped & 0.338 & 0.0973 & 0.0899\\
25 & Z010110(1) & $.249\tau^1{+}.168\tau^3$ &  & $0.0256\tau^1{+}0.17658\tau^3$ & NL & 0 & 0.1121 & 0.0802\\
26 & Z011000(1) & $-.003\tau^1{+}.3\tau^3$ &  & $0.26784\tau^1{-}0.13981\tau^3$ & gapped & 0.780 & 0.0994 & 0.0932\\
27 & Z011001(1) & $.283\tau^1{+}.101\tau^3$ &  & $-0.03264\tau^1{+}0.19474\tau^3$ & gapped & 0.058 & 0.1120 & 0.0928\\
28 & Z011100(1) & $-.263\tau^1{+}.144\tau^3$ &  & $0.15258\tau^1{+}0.21029\tau^3$ & gapped & 0.058 & 0.1038 & 0.0903\\
29 & Z011101(1) & $.296\tau^1{+}.051\tau^3$ &  & $-0.14046\tau^1{+}0.10472\tau^3$ & NL & 0 & 0.1133 & 0.0794\\
30 & Z100000(1) & $.062\tau^1{+}.293\tau^3$ &  & $-0.00219\tau^1{+}0.19063\tau^3$ & gapped & 0.360 & 0.0890 & 0.1034\\
31 & Z100011(1) & $.117\tau^1{+}.276\tau^3$ &  & $0.26378\tau^1{+}0.06494\tau^3$ & gapped & 0.117 & 0.1732 & 0.0359\\
32 & Z100101(1) & $.183\tau^1{+}.238\tau^3$ &  & $0.00411\tau^1{+}0.00684\tau^3$ & gapped & 0.700 & 0.0175 & 0.1617\\
33 & Z110001(1) & $.019\tau^1{+}.299\tau^3$ &  & $-0.11102\tau^1{-}0.26605\tau^3$ & gapped & 0.253 & 0.1126 & 0.0816\\
34 & Z110011(1) & $-.257\tau^1{+}.154\tau^3$ &  & $0.24303\tau^1{+}0.19577\tau^3$ & gapped & 0.450 & 0.0964 & 0.0959\\
35 & Z110101(1) & $-.109\tau^1{-}.28\tau^3$ &  & $0.21159\tau^1{+}0.02144\tau^3$ & gapped & 0.265 & 0.1034 & 0.0923\\
36 & Z110111(1) & $-.259\tau^1{-}.152\tau^3$ &  & $-0.16244\tau^1{-}0.19001\tau^3$ & gapped & 0.039 & 0.1114 & 0.0813\\
37 & Z111000(1) & $.294\tau^1{-}.061\tau^3$ &  & $-0.08915\tau^1{-}0.0691\tau^3$ & gapped & 0.023 & 0.1093 & 0.0854\\
38 & Z111001(1) & $-.246\tau^1{+}.172\tau^3$ &  & $-0.33039\tau^1{-}0.11489\tau^3$ & gapped & 0.740 & 0.1058 & 0.0881\\
39 & Z111100(1) & $-.234\tau^1{+}.187\tau^3$ &  & $0.14261\tau^1{-}0.09547\tau^3$ & gapped & 0.088 & 0.1116 & 0.0811\\
40 & Z111101(1) & $-.282\tau^1{-}.102\tau^3$ &  & $0.27818\tau^1{+}0.13173\tau^3$ & gapped & 0.129 & 0.0996 & 0.0910\\
41 & Z000011(1) & $.294\tau^1{-}.058\tau^3$ & $-.038\tau^1{-}.298\tau^3$ & $0.04467\tau^1{-}0.0079\tau^3$ & gapped & 0.123 & 0.1828 & 0.0131\\
42 & Z000101(1) & $.098\tau^1{-}.284\tau^3$ & $-.27\tau^1{-}.132\tau^3$ & $0.02275\tau^1{-}0.08802\tau^3$ & gapped & 0.652 & 0.0170 & 0.1619\\
43 & Z001000(0) &  & $.112\ii\tau^0{+}.26\tau^1{+}.097\tau^2{-}.018\tau^3$ &  & gapped & 1.532 & 0.0552 & 0.1692\\
44 & Z001001(0) &  & $-.169\ii\tau^0{-}.035\tau^1{+}.053\tau^2{+}.239\tau^3$ &  & gapped & 1.991 & 0.0580 & 0.1692\\
45 & Z001010(0) &  & $-.064\ii\tau^0{+}.188\tau^1{+}.192\tau^2{-}.116\tau^3$ &  & gapped & 1.836 & 0.0577 & 0.1692\\
46 & Z001100(0) &  & $.136\ii\tau^0{+}.033\tau^1{+}.228\tau^2{-}.135\tau^3$ &  & gapped & 0.421 & 0.0372 & 0.1691\\
47 & Z001100(1) & $-.281\tau^1{-}.105\tau^3$ & $.18\ii\tau^0{-}.24\tau^2$ & $-0.03546\tau^1{+}0.00687\tau^3$ & gapped & 1.358 & 0.0533 & 0.1636\\
48 & Z001101(0) &  & $-.034\ii\tau^0{+}.063\tau^1{-}.181\tau^2{+}.228\tau^3$ &  & gapped & 0.369 & 0.0366 & 0.1689\\
49 & Z001101(1) & $-.299\tau^1{+}.018\tau^3$ & $-.02\ii\tau^0{+}.299\tau^2$ & $0.01325\tau^1{+}0.04212\tau^3$ & gapped & 0.976 & 0.0427 & 0.1631\\
50 & Z001110(0) &  & $-.114\ii\tau^0{-}.193\tau^1{+}.196\tau^2{+}.038\tau^3$ &  & gapped & 0.066 & 0.0217 & 0.1701\\
51 & Z001110(1) & $-.258\tau^1{+}.154\tau^3$ & $-.255\ii\tau^0{-}.159\tau^2$ & $0.44293\tau^1{+}0.15242\tau^3$ & P & 0 & 0.0350 & 0.1615\\
52 & Z001111(1) & $.257\tau^1{-}.156\tau^3$ & $-.178\ii\tau^0{-}.242\tau^2$ & $0.54167\tau^1{+}0.11697\tau^3$ & gapped & 0.611 & 0.0496 & 0.1630\\
53 & Z010001(1) & $.257\tau^1{-}.154\tau^3$ & $.059\tau^1{-}.294\tau^3$ & $-0.05052\tau^1{-}0.10154\tau^3$ & gapped & 0.993 & 0.1105 & 0.0879\\
54 & Z010011(1) & $-.296\tau^1{-}.05\tau^3$ & $-.268\tau^1{+}.134\tau^3$ & $-0.05124\tau^1{-}0.20694\tau^3$ & NL & 0 & 0.1144 & 0.0731\\
55 & Z010101(1) & $.229\tau^1{+}.194\tau^3$ & $-.014\tau^1{-}.3\tau^3$ & $0.04229\tau^1{-}0.15578\tau^3$ & NL & 0 & 0.1086 & 0.0943\\
56 & Z010111(1) & $.206\tau^1{+}.218\tau^3$ & $-.268\tau^1{+}.135\tau^3$ & $0.15946\tau^1{+}0.07957\tau^3$ & P & 0 & 0.0905 & 0.1016\\
57 & Z011001(0) &  & $.045\ii\tau^0{+}.082\tau^1{-}.218\tau^2{+}.184\tau^3$ &  & gapped & 0.455 & 0.1835 & 0.0423\\
58 & Z011010(1) & $.235\tau^1{+}.187\tau^3$ & $-.284\ii\tau^0{-}.098\tau^2$ & $0.08982\tau^1{+}0.09587\tau^3$ & gapped & 0.843 & 0.1708 & 0.0316\\
59 & Z011011(0) &  & $.004\ii\tau^0{+}.016\tau^1{-}.231\tau^2{-}.191\tau^3$ &  & gapped & 0.018 & 0.1836 & 0.0455\\
60 & Z011011(1) & $-.01\tau^1{-}.3\tau^3$ & $-.058\ii\tau^0{-}.294\tau^2$ & $-0.05925\tau^1{-}0.1909\tau^3$ & D6 & 0 & 0.1462 & 0.0488\\
61 & Z011100(0) &  & $-.268\ii\tau^0{+}.093\tau^1{+}.021\tau^2{+}.094\tau^3$ &  & gapped & 0.270 & 0.1824 & 0.0787\\
62 & Z011110(0) &  & $.042\ii\tau^0{+}.077\tau^1{-}.287\tau^2{-}.015\tau^3$ &  & gapped & 0.144 & 0.1836 & 0.0482\\
63 & Z011110(1) & $-.127\tau^1{-}.272\tau^3$ & $.296\ii\tau^0{+}.049\tau^2$ & $0.05922\tau^1{-}0.23203\tau^3$ & gapped & 0.766 & 0.1822 & 0.0367\\
64 & Z011111(1) & $.294\tau^1{-}.058\tau^3$ & $.248\ii\tau^0{+}.168\tau^2$ & $-0.54515\tau^1{-}0.66531\tau^3$ & gapped & 0.123 & 0.1811 & 0.0406\\
65 & Z100010(1) & $.141\tau^1{+}.265\tau^3$ & $.195\tau^1{-}.228\tau^3$ & $-0.00322\tau^1{-}0.10297\tau^3$ & gapped & 0.832 & 0.1746 & 0.0139\\
66 & Z100100(1) & $-.097\tau^1{+}.284\tau^3$ & $-.297\tau^1{-}.041\tau^3$ & $-0.02357\tau^1{+}0.08206\tau^3$ & gapped & 0.651 & 0.0166 & 0.1618\\
67 & Z101000(0) &  & $.2\ii\tau^0{+}.13\tau^1{+}.148\tau^2{+}.105\tau^3$ &  & gapped & 1.370 & 0.0544 & 0.1694\\
68 & Z101001(0) &  & $-.036\ii\tau^0{+}.154\tau^1{+}.234\tau^2{-}.102\tau^3$ &  & gapped & 1.242 & 0.0500 & 0.1692\\
69 & Z101011(0) &  & $-.203\ii\tau^0{-}.103\tau^1{-}.039\tau^2{-}.191\tau^3$ &  & gapped & 1.446 & 0.0554 & 0.1693\\
70 & Z101100(0) &  & $.175\ii\tau^0{+}.017\tau^1{-}.243\tau^2{+}.002\tau^3$ &  & gapped & 0.253 & 0.0276 & 0.1693\\
71 & Z101100(1) & $-.055\tau^1{+}.295\tau^3$ & $.071\ii\tau^0{+}.291\tau^2$ & $0.04565\tau^1{-}0.0031\tau^3$ & gapped & 0.892 & 0.0420 & 0.1630\\
72 & Z101101(0) &  & $.13\ii\tau^0{+}.087\tau^1{+}.184\tau^2{-}.178\tau^3$ &  & gapped & 0.498 & 0.0438 & 0.1694\\
73 & Z101101(1) & $-.299\tau^1{+}.018\tau^3$ & $-.3\ii\tau^0{-}.002\tau^2$ & $-0.00877\tau^1{+}0.04203\tau^3$ & gapped & 0.899 & 0.0497 & 0.1624\\
74 & Z101110(1) & $-.111\tau^1{+}.279\tau^3$ & $-.029\ii\tau^0{-}.299\tau^2$ & $0.40588\tau^1{+}0.24818\tau^3$ & gapped & 0.600 & 0.0452 & 0.1631\\
75 & Z101111(0) &  & $-.124\ii\tau^0{-}.196\tau^1{+}.157\tau^2{-}.108\tau^3$ &  & gapped & 0.200 & 0.0228 & 0.1700\\
76 & Z101111(1) & $.172\tau^1{+}.246\tau^3$ & $.241\ii\tau^0{+}.179\tau^2$ & $-0.1747\tau^1{+}0.32622\tau^3$ & P & 0 & 0.0349 & 0.1606\\
77 & Z110000(1) & $.284\tau^1{+}.095\tau^3$ & $-.066\tau^1{-}.293\tau^3$ & $0.11828\tau^1{-}0.12873\tau^3$ & gapped & 0.570 & 0.0917 & 0.1000\\
78 & Z110010(1) & $.15\tau^1{-}.26\tau^3$ & $-.258\tau^1{-}.153\tau^3$ & $-0.09085\tau^1{+}0.14986\tau^3$ & NL & 0 & 0.1042 & 0.0763\\
79 & Z110100(1) & $-.203\tau^1{-}.221\tau^3$ & $.086\tau^1{-}.288\tau^3$ & $-0.40495\tau^1{-}0.27993\tau^3$ & NL & 0 & 0.1090 & 0.0865\\
80 & Z110110(1) & $-.292\tau^1{+}.068\tau^3$ & $-.03\tau^1{+}.298\tau^3$ & $-0.16359\tau^1{+}0.19357\tau^3$ & P & 0 & 0.0905 & 0.1016\\
81 & Z111000(0) &  & $-.031\ii\tau^0{-}.046\tau^1{-}.213\tau^2{+}.204\tau^3$ &  & gapped & 1.514 & 0.1833 & 0.0411\\
82 & Z111010(0) &  & $.267\ii\tau^0{-}.122\tau^1{+}.056\tau^2{-}.022\tau^3$ &  & gapped & 0.018 & 0.1834 & 0.0418\\
83 & Z111010(1) & $.248\tau^1{+}.169\tau^3$ & $.199\ii\tau^0{-}.224\tau^2$ & $-0.12075\tau^1{-}0.0602\tau^3$ & gapped & 0.984 & 0.1683 & 0.0210\\
84 & Z111011(1) & $-.279\tau^1{-}.11\tau^3$ & $.072\ii\tau^0{+}.291\tau^2$ & $-0.10938\tau^1{-}0.04761\tau^3$ & gapped & 1.635 & 0.1732 & 0.0370\\
85 & Z111101(0) &  & $.036\ii\tau^0{+}.041\tau^1{+}.268\tau^2{-}.122\tau^3$ &  & gapped & 0.419 & 0.1833 & 0.0401\\
86 & Z111110(1) & $.136\tau^1{+}.268\tau^3$ & $-.297\ii\tau^0{+}.041\tau^2$ & $-0.53447\tau^1{-}0.58262\tau^3$ & gapped & 0.637 & 0.1810 & 0.0435\\
87 & Z111111(0) &  & $.166\ii\tau^0{-}.045\tau^1{-}.032\tau^2{+}.244\tau^3$ &  & gapped & 0.704 & 0.1839 & 0.0298\\
88 & Z111111(1) & $-.11\tau^1{-}.279\tau^3$ & $-.296\ii\tau^0{+}.046\tau^2$ & $0.02713\tau^1{-}0.22467\tau^3$ & gapped & 0.787 & 0.1821 & 0.0410\\
\end{longtable}
\endgroup

\begingroup\scriptsize\setlength{\tabcolsep}{4pt}
\begin{longtable}{rlcccccc}
\caption{$U(1)$ spin liquids of Tables~\ref{tab:u1ans} (Nos.~1--41) and \ref{tab:u1ansb} (Nos.~42--79, listed there as Nos.~1--38) of the main text: amplitudes $u$, Eq.~\eqref{sm:eq:u}, on the reference bonds (empty: orbit not included). The five families of the sector $(0,1,0)$ are taken at $\Phi=\pi/2$.}\label{sm:tab:u1}\\
\hline\hline
No. & PSG & $J_b$ & $J_g$ & $J_r$ & $J'$ & $J''$ & $J'''$\\
\hline
\endfirsthead
\multicolumn{8}{l}{\textsc{Table}~\ref{sm:tab:u1} (\textit{continued})}\\
\hline\hline
No. & PSG & $J_b$ & $J_g$ & $J_r$ & $J'$ & $J''$ & $J'''$\\
\hline
\endhead
\hline\hline
\endlastfoot
1 & U010-$\Phi$10110 & $-.999\ii\tau^0{-}.037\tau^3$ & $.825\ii\tau^0{-}.36\tau^3$ & $-.475\ii\tau^0{+}.644\tau^3$ &  &  & \\
2 & U110-010100 & $-\tau^3$ & $-.835\ii\tau^0{-}.335\tau^3$ & $.43\ii\tau^0{-}.675\tau^3$ &  &  & \\
3 & U110-010101 & $\tau^3$ & $.779\ii\tau^0{-}.45\tau^3$ & $.797\ii\tau^0{+}.073\tau^3$ &  &  & \\
4 & U110-110100 & $-\tau^3$ & $-.834\ii\tau^0{+}.339\tau^3$ & $.277\ii\tau^0{-}.75\tau^3$ &  &  & \\
5 & U110-110101 & $-\tau^3$ & $.111\ii\tau^0{-}.893\tau^3$ & $.431\ii\tau^0{+}.674\tau^3$ &  &  & \\
6 & U001-000000 & $-\tau^3$ & $.9\tau^3$ & $-.8\tau^3$ & $-.5\tau^3$ &  & \\
7 & U001-100001 & $\tau^3$ & $.9\tau^3$ & $.8\tau^3$ & $.5\tau^3$ &  & \\
8 & U010-$\Phi$11100 &  & $-.624\ii\tau^0{+}.648\tau^3$ & $.798\ii\tau^0{-}.055\tau^3$ & $-.354\ii\tau^0{+}.354\tau^3$ &  & \\
9 & U010-$\Phi$11110 &  & $-.234\ii\tau^0{-}.869\tau^3$ & $.525\ii\tau^0{-}.604\tau^3$ & $-.354\ii\tau^0{+}.354\tau^3$ &  & \\
10 & U011-000110 & $-\tau^3$ & $-.9\tau^3$ & $-.8\tau^3$ & $-.5\tau^3$ &  & \\
11 & U011-100111 & $\tau^3$ & $.9\tau^3$ & $.8\tau^3$ & $.5\tau^3$ &  & \\
12 & U101-000011 & $\tau^3$ & $-.9\tau^3$ & $.8\tau^3$ & $-.5\tau^3$ &  & \\
13 & U101-100010 & $\tau^3$ & $-.9\tau^3$ & $.8\tau^3$ & $-.5\tau^3$ &  & \\
14 & U110-011101 &  & $-.9\ii\tau^0{-}.012\tau^3$ & $.748\ii\tau^0{+}.285\tau^3$ & $-.309\ii\tau^0{+}.393\tau^3$ &  & \\
15 & U110-111100 &  & $.855\ii\tau^0{-}.28\tau^3$ & $-.2\ii\tau^0{+}.775\tau^3$ & $-.014\ii\tau^0{-}.5\tau^3$ &  & \\
16 & U111-000101 & $-\tau^3$ & $-.9\tau^3$ & $.8\tau^3$ & $.5\tau^3$ &  & \\
17 & U111-100100 & $\tau^3$ & $-.9\tau^3$ & $.8\tau^3$ & $-.5\tau^3$ &  & \\
18 & U001-000001 & $-\tau^3$ & $.9\tau^3$ & $-.8\tau^3$ &  & $-.3\tau^3$ & \\
19 & U001-000010 &  & $.9\tau^3$ & $.8\tau^3$ & $-.5\tau^3$ & $.3\tau^3$ & \\
20 & U001-000100 & $-\tau^3$ &  &  & $.5\tau^3$ & $-.3\tau^3$ & \\
21 & U001-100000 & $\tau^3$ & $.9\tau^3$ & $-.8\tau^3$ &  & $.3\tau^3$ & \\
22 & U001-100011 &  & $-.9\tau^3$ & $-.8\tau^3$ & $-.5\tau^3$ & $.3\tau^3$ & \\
23 & U001-100101 & $-\tau^3$ &  &  & $.5\tau^3$ & $.3\tau^3$ & \\
24 & U011-000010 & $-\tau^3$ &  &  & $.5\tau^3$ & $.3\tau^3$ & \\
25 & U011-000100 &  & $-.9\tau^3$ & $.8\tau^3$ & $.5\tau^3$ & $.3\tau^3$ & \\
26 & U011-000111 & $-\tau^3$ & $.9\tau^3$ & $.8\tau^3$ &  & $-.3\tau^3$ & \\
27 & U011-100011 & $-\tau^3$ &  &  & $.5\tau^3$ & $-.3\tau^3$ & \\
28 & U011-100101 &  & $.9\tau^3$ & $-.8\tau^3$ & $-.5\tau^3$ & $.3\tau^3$ & \\
29 & U011-100110 & $\tau^3$ & $.9\tau^3$ & $-.8\tau^3$ &  & $.3\tau^3$ & \\
30 & U101-000001 &  & $-.9\tau^3$ & $-.8\tau^3$ & $-.5\tau^3$ & $.3\tau^3$ & \\
31 & U101-000010 & $-\tau^3$ & $.9\tau^3$ & $.8\tau^3$ &  & $.3\tau^3$ & \\
32 & U101-000111 & $\tau^3$ &  &  & $-.5\tau^3$ & $.3\tau^3$ & \\
33 & U101-100000 &  & $.9\tau^3$ & $-.8\tau^3$ & $-.5\tau^3$ & $.3\tau^3$ & \\
34 & U101-100011 & $\tau^3$ & $-.9\tau^3$ & $.8\tau^3$ &  & $.3\tau^3$ & \\
35 & U101-100110 & $\tau^3$ &  &  & $-.5\tau^3$ & $-.3\tau^3$ & \\
36 & U111-000001 & $-\tau^3$ &  &  & $.5\tau^3$ & $.3\tau^3$ & \\
37 & U111-000100 & $-\tau^3$ & $-.9\tau^3$ & $.8\tau^3$ &  & $-.3\tau^3$ & \\
38 & U111-000111 &  & $.9\tau^3$ & $.8\tau^3$ & $.5\tau^3$ & $-.3\tau^3$ & \\
39 & U111-100000 & $\tau^3$ &  &  & $-.5\tau^3$ & $.3\tau^3$ & \\
40 & U111-100101 & $\tau^3$ & $.9\tau^3$ & $.8\tau^3$ &  & $.3\tau^3$ & \\
41 & U111-100110 &  & $.9\tau^3$ & $-.8\tau^3$ & $-.5\tau^3$ & $.3\tau^3$ & \\
42 & U000-001000 & $-\tau^3$ &  &  & $-.5\tau^3$ &  & $.207\ii\tau^0{+}.217\tau^3$\\
43 & U000-001001 & $\tau^3$ &  &  & $.5\ii\tau^0$ &  & $-.091\ii\tau^0{-}.286\tau^3$\\
44 & U000-011000 &  & $-.9\tau^3$ & $-.8\tau^3$ & $-.5\tau^3$ &  & $-.297\ii\tau^0{-}.04\tau^3$\\
45 & U000-011001 &  & $-.9\tau^3$ & $-.8\tau^3$ & $.5\ii\tau^0$ &  & $-.299\ii\tau^0{+}.022\tau^3$\\
46 & U000-101000 & $-\tau^3$ &  &  & $.5\ii\tau^0$ &  & $-.145\ii\tau^0{-}.263\tau^3$\\
47 & U000-101001 & $-\tau^3$ &  &  & $.5\tau^3$ &  & $-.183\ii\tau^0{-}.237\tau^3$\\
48 & U000-111000 &  & $-.9\tau^3$ & $.8\tau^3$ & $-.5\ii\tau^0$ &  & $.142\ii\tau^0{-}.264\tau^3$\\
49 & U000-111001 &  & $.9\tau^3$ & $.8\tau^3$ & $.5\tau^3$ &  & $.105\ii\tau^0{+}.281\tau^3$\\
50 & U001-000011 &  & $-.9\tau^3$ & $.8\tau^3$ &  & $-.3\tau^3$ & $-.3\tau^3$\\
51 & U001-000101 & $\tau^3$ &  &  &  & $-.3\tau^3$ & $-.3\tau^3$\\
52 & U001-100010 &  & $-.9\tau^3$ & $.8\tau^3$ &  & $.3\tau^3$ & $-.3\tau^3$\\
53 & U001-100100 & $\tau^3$ &  &  &  & $-.3\tau^3$ & $.3\tau^3$\\
54 & U010-$\Phi$01010 & $.999\ii\tau^0{+}.035\tau^3$ &  &  & $-.354\ii\tau^0{+}.354\tau^3$ &  & $.219\ii\tau^0{+}.205\tau^3$\\
55 & U010-$\Phi$01110 & $-.835\ii\tau^0{-}.55\tau^3$ &  &  & $-.354\ii\tau^0{+}.354\tau^3$ &  & $.033\ii\tau^0{+}.298\tau^3$\\
56 & U011-000011 & $\tau^3$ &  &  &  & $-.3\tau^3$ & $-.3\tau^3$\\
57 & U011-000101 &  & $.9\tau^3$ & $-.8\tau^3$ &  & $.3\tau^3$ & $.3\tau^3$\\
58 & U011-100010 & $-\tau^3$ &  &  &  & $.3\tau^3$ & $-.3\tau^3$\\
59 & U011-100100 &  & $-.9\tau^3$ & $-.8\tau^3$ &  & $.3\tau^3$ & $.3\tau^3$\\
60 & U100-001011 & $.091\ii\tau^0{-}.996\tau^3$ &  &  & $-.175\ii\tau^0{+}.468\tau^3$ &  & $.294\ii\tau^0{+}.058\tau^3$\\
61 & U100-011001 &  & $.9\tau^3$ & $-.8\tau^3$ & $.147\ii\tau^0{-}.478\tau^3$ &  & $.26\ii\tau^0{+}.149\tau^3$\\
62 & U100-011011 &  & $-.9\tau^3$ & $.8\tau^3$ & $.163\ii\tau^0{-}.473\tau^3$ &  & $-.29\ii\tau^0{+}.075\tau^3$\\
63 & U100-101010 & $.108\ii\tau^0{-}.994\tau^3$ &  &  & $.47\ii\tau^0{-}.172\tau^3$ &  & $-.12\ii\tau^0{-}.275\tau^3$\\
64 & U100-111000 &  & $-.9\tau^3$ & $.8\tau^3$ & $-.444\ii\tau^0{-}.23\tau^3$ &  & $.198\ii\tau^0{+}.225\tau^3$\\
65 & U100-111010 &  & $-.9\tau^3$ & $-.8\tau^3$ & $-.497\ii\tau^0{+}.057\tau^3$ &  & $.041\ii\tau^0{-}.297\tau^3$\\
66 & U101-000000 &  & $.9\tau^3$ & $.8\tau^3$ &  & $-.3\tau^3$ & $.3\tau^3$\\
67 & U101-000110 & $\tau^3$ &  &  &  & $-.3\tau^3$ & $.3\tau^3$\\
68 & U101-100001 &  & $-.9\tau^3$ & $.8\tau^3$ &  & $.3\tau^3$ & $.3\tau^3$\\
69 & U101-100111 & $\tau^3$ &  &  &  & $-.3\tau^3$ & $-.3\tau^3$\\
70 & U110-001001 & $\tau^3$ &  &  & $.343\ii\tau^0{+}.364\tau^3$ &  & $-.194\ii\tau^0{-}.229\tau^3$\\
71 & U110-001101 & $-\tau^3$ &  &  & $-.37\ii\tau^0{-}.336\tau^3$ &  & $.3\ii\tau^0{+}.005\tau^3$\\
72 & U110-011100 &  & $.9\ii\tau^0{-}.015\tau^3$ & $-.768\ii\tau^0{-}.223\tau^3$ &  &  & $-.247\ii\tau^0{+}.171\tau^3$\\
73 & U110-101000 & $\tau^3$ &  &  & $-.052\ii\tau^0{+}.497\tau^3$ &  & $-.129\ii\tau^0{+}.271\tau^3$\\
74 & U110-101100 & $\tau^3$ &  &  & $.292\ii\tau^0{-}.406\tau^3$ &  & $.297\ii\tau^0{+}.044\tau^3$\\
75 & U110-111101 &  & $.899\ii\tau^0{-}.045\tau^3$ & $.326\ii\tau^0{-}.73\tau^3$ &  &  & $.3\ii\tau^0{-}.014\tau^3$\\
76 & U111-000000 & $-\tau^3$ &  &  &  & $.3\tau^3$ & $-.3\tau^3$\\
77 & U111-000110 &  & $.9\tau^3$ & $-.8\tau^3$ &  & $-.3\tau^3$ & $.3\tau^3$\\
78 & U111-100001 & $\tau^3$ &  &  &  & $.3\tau^3$ & $.3\tau^3$\\
79 & U111-100111 &  & $.9\tau^3$ & $-.8\tau^3$ &  & $-.3\tau^3$ & $.3\tau^3$\\
\end{longtable}
\endgroup

\begingroup\scriptsize\setlength{\tabcolsep}{4pt}
\begin{longtable}{rlclccc}
\caption{$U(1)$ spin liquids of Table~\ref{sm:tab:u1}: on-site multiplier $a_1$ on site~1 of the reference cell (empty: no on-site field allowed) and the quantities of Table~\ref{sm:tab:z2b}.}\label{sm:tab:u1b}\\
\hline\hline
No. & PSG & $a_1$ & spectrum & $\Delta$ & $S(K')$ & $S(M')$\\
\hline
\endfirsthead
\multicolumn{7}{l}{\textsc{Table}~\ref{sm:tab:u1b} (\textit{continued})}\\
\hline\hline
No. & PSG & $a_1$ & spectrum & $\Delta$ & $S(K')$ & $S(M')$\\
\hline
\endhead
\hline\hline
\endlastfoot
1 & U010-$\Phi$10110 &  & gapped & 0.754 & 0.1059 & 0.0849\\
2 & U110-010100 &  & NL & 0 & 0.1169 & 0.0779\\
3 & U110-010101 &  & gapped & 0.780 & 0.1011 & 0.0901\\
4 & U110-110100 &  & gapped & 0.773 & 0.0926 & 0.0995\\
5 & U110-110101 &  & NL & 0 & 0.1141 & 0.0805\\
6 & U001-000000 & $-0.19288\tau^3$ & NL & 0 & 0.1160 & 0.0795\\
7 & U001-100001 & $-0.25765\tau^3$ & gapped & 0.322 & 0.1241 & 0.0718\\
8 & U010-$\Phi$11100 &  & D24 & 0 & 0.1835 & 0.0547\\
9 & U010-$\Phi$11110 &  & gapped & 0.789 & 0.1831 & 0.0340\\
10 & U011-000110 & $0.24545\tau^3$ & gapped & 0.113 & 0.1172 & 0.0964\\
11 & U011-100111 & $0.23993\tau^3$ & gapped & 0.124 & 0.1201 & 0.0916\\
12 & U101-000011 & $0.33979\tau^3$ & gapped & 0.378 & 0.1081 & 0.0844\\
13 & U101-100010 & $-0.36207\tau^3$ & gapped & 0.430 & 0.1089 & 0.0839\\
14 & U110-011101 &  & gapped & 0.216 & 0.1833 & 0.0245\\
15 & U110-111100 &  & gapped & 0.094 & 0.1834 & 0.0577\\
16 & U111-000101 & $0.31976\tau^3$ & gapped & 0.559 & 0.1094 & 0.0939\\
17 & U111-100100 & $-0.34073\tau^3$ & gapped & 0.714 & 0.1090 & 0.0925\\
18 & U001-000001 & $0.3\tau^3$ & gapped & 0.827 & 0.1026 & 0.0890\\
19 & U001-000010 & $0.3\tau^3$ & gapped & 0.896 & 0.1818 & 0.0313\\
20 & U001-000100 & $0.15\tau^3$ & gapped & 1.336 & 0.0451 & 0.1633\\
21 & U001-100000 & $-0.23681\tau^3$ & gapped & 1.081 & 0.1027 & 0.0889\\
22 & U001-100011 & $0.2683\tau^3$ & gapped & 1.030 & 0.1818 & 0.0313\\
23 & U001-100101 & $-0.08687\tau^3$ & gapped & 1.462 & 0.0451 & 0.1633\\
24 & U011-000010 & $-0.00844\tau^3$ & gapped & 0.675 & 0.0156 & 0.1615\\
25 & U011-000100 & $0.24835\tau^3$ & D6 & 0 & 0.1353 & 0.0343\\
26 & U011-000111 & $0.16144\tau^3$ & gapped & 0.868 & 0.0921 & 0.0878\\
27 & U011-100011 & $0.00813\tau^3$ & gapped & 0.675 & 0.0155 & 0.1618\\
28 & U011-100101 & $0.25942\tau^3$ & D12 & 0 & 0.1347 & 0.0338\\
29 & U011-100110 & $0.10486\tau^3$ & gapped & 1.051 & 0.1051 & 0.0798\\
30 & U101-000001 & $-0.52778\tau^3$ & D6 & 0 & 0.1728 & 0.0694\\
31 & U101-000010 & $-0.2183\tau^3$ & NL & 0 & 0.1123 & 0.0852\\
32 & U101-000111 & $0.00593\tau^3$ & gapped & 1.358 & 0.0451 & 0.1633\\
33 & U101-100000 & $0.20675\tau^3$ & gapped & 0.723 & 0.1830 & 0.0218\\
34 & U101-100011 & $0.30337\tau^3$ & NL & 0 & 0.1036 & 0.0896\\
35 & U101-100110 & $-0.00583\tau^3$ & gapped & 1.358 & 0.0451 & 0.1633\\
36 & U111-000001 & $0.15\tau^3$ & gapped & 0.655 & 0.0155 & 0.1618\\
37 & U111-000100 & $0.05257\tau^3$ & NL & 0 & 0.1060 & 0.0705\\
38 & U111-000111 & $-0.30014\tau^3$ & gapped & 0.435 & 0.1392 & 0.0424\\
39 & U111-100000 & $0.15\tau^3$ & gapped & 0.655 & 0.0156 & 0.1617\\
40 & U111-100101 & $0.0798\tau^3$ & gapped & 0.228 & 0.1135 & 0.0708\\
41 & U111-100110 & $-0.12626\tau^3$ & gapped & 1.084 & 0.1709 & 0.0271\\
42 & U000-001000 &  & gapped & 0.380 & 0.0313 & 0.1698\\
43 & U000-001001 &  & gapped & 1.327 & 0.0526 & 0.1694\\
44 & U000-011000 &  & gapped & 1.074 & 0.1841 & 0.0336\\
45 & U000-011001 &  & gapped & 0.032 & 0.1840 & 0.0620\\
46 & U000-101000 &  & gapped & 1.224 & 0.0526 & 0.1694\\
47 & U000-101001 &  & gapped & 0.242 & 0.0301 & 0.1698\\
48 & U000-111000 &  & gapped & 0.984 & 0.1830 & 0.0342\\
49 & U000-111001 &  & gapped & 0.556 & 0.1837 & 0.0434\\
50 & U001-000011 & $0.21258\tau^3$ & gapped & 1.514 & 0.1831 & 0.0270\\
51 & U001-000101 & $-0.15\tau^3$ & gapped & 1.100 & 0.0386 & 0.1626\\
52 & U001-100010 & $-0.2007\tau^3$ & gapped & 1.548 & 0.1831 & 0.0271\\
53 & U001-100100 & $-0.07363\tau^3$ & gapped & 1.341 & 0.0393 & 0.1623\\
54 & U010-$\Phi$01010 &  & gapped & 0.123 & 0.0153 & 0.1704\\
55 & U010-$\Phi$01110 &  & gapped & 2.010 & 0.0587 & 0.1692\\
56 & U011-000011 & $-0.0105\tau^3$ & gapped & 0.651 & 0.0169 & 0.1605\\
57 & U011-000101 & $0.139\tau^3$ & gapped & 0.700 & 0.1691 & 0.0093\\
58 & U011-100010 & $0.00947\tau^3$ & gapped & 0.565 & 0.0166 & 0.1609\\
59 & U011-100100 & $-0.43285\tau^3$ & gapped & 0.381 & 0.1120 & 0.0244\\
60 & U100-001011 &  & gapped & 0.941 & 0.0457 & 0.1691\\
61 & U100-011001 &  & gapped & 0.392 & 0.1838 & 0.0343\\
62 & U100-011011 &  & gapped & 0.295 & 0.1834 & 0.0398\\
63 & U100-101010 &  & gapped & 1.056 & 0.0500 & 0.1695\\
64 & U100-111000 &  & gapped & 0.640 & 0.1840 & 0.0084\\
65 & U100-111010 &  & gapped & 0.018 & 0.1833 & 0.0724\\
66 & U101-000000 & $-0.15019\tau^3$ & gapped & 0.334 & 0.1841 & 0.0082\\
67 & U101-000110 & $0.00604\tau^3$ & gapped & 1.375 & 0.0381 & 0.1634\\
68 & U101-100001 & $0.61956\tau^3$ & NL & 0 & 0.1496 & 0.0787\\
69 & U101-100111 & $0.00603\tau^3$ & gapped & 1.358 & 0.0381 & 0.1634\\
70 & U110-001001 &  & gapped & 0.065 & 0.0153 & 0.1704\\
71 & U110-001101 &  & gapped & 2.033 & 0.0588 & 0.1692\\
72 & U110-011100 &  & gapped & 0.201 & 0.1842 & 0.0150\\
73 & U110-101000 &  & gapped & 0.057 & 0.0153 & 0.1704\\
74 & U110-101100 &  & gapped & 2.070 & 0.0590 & 0.1692\\
75 & U110-111101 &  & gapped & 0.008 & 0.1841 & 0.0516\\
76 & U111-000000 & $-0.12853\tau^3$ & gapped & 0.616 & 0.0157 & 0.1619\\
77 & U111-000110 & $0.34833\tau^3$ & gapped & 0.523 & 0.1356 & 0.0236\\
78 & U111-100001 & $-0.12062\tau^3$ & gapped & 0.625 & 0.0157 & 0.1618\\
79 & U111-100111 & $0.34833\tau^3$ & gapped & 0.523 & 0.1357 & 0.0235\\
\end{longtable}
\endgroup

\begingroup\scriptsize\setlength{\tabcolsep}{4pt}
\begin{longtable}{rlccccclccc}
\caption{$SU(2)$ spin liquids of Table~\ref{tab:su2ans} of the main text: reference-bond amplitudes $u=\ii\chi\tau^0$ (the entries give $\chi$; empty: orbit not included) and the quantities of Table~\ref{sm:tab:z2b}.}\label{sm:tab:su2}\\
\hline\hline
No. & PSG & $J_b$ & $J_g$ & $J_r$ & $J'$ & $J'''$ & spectrum & $\Delta$ & $S(K')$ & $S(M')$\\
\hline
\endfirsthead
\multicolumn{11}{l}{\textsc{Table}~\ref{sm:tab:su2} (\textit{continued})}\\
\hline\hline
No. & PSG & $J_b$ & $J_g$ & $J_r$ & $J'$ & $J'''$ & spectrum & $\Delta$ & $S(K')$ & $S(M')$\\
\hline
\endhead
\hline\hline
\endlastfoot
1 & SU010110 & $1$ & $0.9$ & $-0.8$ &  &  & NL & 0 & 0.1093 & 0.0855\\
2 & SU010111 & $1$ & $0.9$ & $-0.8$ &  &  & gapped & 0.827 & 0.1058 & 0.0850\\
3 & SU110110 & $-1$ & $-0.9$ & $-0.8$ &  &  & gapped & 0.770 & 0.0905 & 0.1019\\
4 & SU110111 & $-1$ & $-0.9$ & $-0.8$ &  &  & gapped & 0.152 & 0.1195 & 0.0742\\
5 & SU011101 &  & $-0.9$ & $0.8$ & $0.5$ &  & gapped & 0.709 & 0.1838 & 0.0202\\
6 & SU011111 &  & $0.9$ & $0.8$ & $-0.5$ &  & gapped & 0.087 & 0.1825 & 0.0678\\
7 & SU111100 &  & $-0.9$ & $-0.8$ & $-0.5$ &  & D12 & 0 & 0.1835 & 0.0577\\
8 & SU111110 &  & $0.9$ & $-0.8$ & $0.5$ &  & gapped & 1.045 & 0.1832 & 0.0284\\
9 & SU001011 & $-1$ &  &  & $0.5$ & $0.3$ & NL & 0 & 0.0286 & 0.1631\\
10 & SU001111 & $1$ &  &  & $-0.5$ & $-0.3$ & gapped & 2.040 & 0.0590 & 0.1692\\
11 & SU011100 &  & $0.9$ & $0.8$ &  & $-0.3$ & gapped & 0.670 & 0.1841 & 0.0082\\
12 & SU011110 &  & $0.9$ & $-0.8$ &  & $0.3$ & gapped & 0.200 & 0.1845 & 0.0322\\
13 & SU101010 & $1$ &  &  & $0.5$ & $-0.3$ & NL & 0 & 0.0351 & 0.1655\\
14 & SU101110 & $-1$ &  &  & $-0.5$ & $-0.3$ & gapped & 2.000 & 0.0580 & 0.1692\\
15 & SU111101 &  & $-0.9$ & $-0.8$ &  & $0.3$ & gapped & 0.793 & 0.1841 & 0.0082\\
16 & SU111111 &  & $0.9$ & $0.8$ &  & $-0.3$ & gapped & 1.675 & 0.1843 & 0.0242\\
\end{longtable}
\endgroup

\begingroup\scriptsize\setlength{\tabcolsep}{4pt}
\begin{longtable}{lrllcll}
\caption{Spin liquids whose spinon spectrum differs between point A, the amplitude point of Tables~\ref{sm:tab:z2}--\ref{sm:tab:su2} (the generic point of Sec.~\ref{sm:app:dos}), and point B, with $|u|=1$ on $J_b$, $J_g$, and $J_r$, $0.5$ on $J'$, $0.3$ on $J''$, and $0.2$ on $J'''$, other random directions, the basis vector with positive first coefficient for one-dimensional allowed spaces, and $\Phi=2\pi/3$. Types and thresholds $\Delta$ as in Table~\ref{sm:tab:z2b}; numbers as in Tables~\ref{sm:tab:z2}--\ref{sm:tab:su2}.}\label{sm:tab:compare}\\
\hline\hline
IGG & No. & PSG & type (A) & $\Delta$ (A) & type (B) & $\Delta$ (B)\\
\hline
\endfirsthead
\multicolumn{7}{l}{\textsc{Table}~\ref{sm:tab:compare} (\textit{continued})}\\
\hline\hline
IGG & No. & PSG & type (A) & $\Delta$ (A) & type (B) & $\Delta$ (B)\\
\hline
\endhead
\hline\hline
\endlastfoot
$\Z_2$ & 3 & Z110000(0) & NL & 0 & gapped & 0.728\\
$\Z_2$ & 39 & Z111100(1) & gapped & 0.088 & NL & 0\\
$\Z_2$ & 51 & Z001110(1) & P & 0 & gapped & 0.265\\
$\Z_2$ & 56 & Z010111(1) & P & 0 & gapped & 0.194\\
$\Z_2$ & 64 & Z011111(1) & gapped & 0.123 & NL & 0\\
$\Z_2$ & 65 & Z100010(1) & gapped & 0.832 & D24 & 0\\
$\Z_2$ & 76 & Z101111(1) & P & 0 & NL & 0\\
$\Z_2$ & 80 & Z110110(1) & P & 0 & gapped & 0.016\\
$\Z_2$ & 83 & Z111010(1) & gapped & 0.984 & D12 & 0\\
$\Z_2$ & 86 & Z111110(1) & gapped & 0.637 & NL & 0\\
$U(1)$ & 6 & U001-000000 & NL & 0 & gapped & 0.189\\
$U(1)$ & 8 & U010-$\Phi$11100 & D24 & 0 & gapped & 0.601\\
$U(1)$ & 25 & U011-000100 & D6 & 0 & gapped & 1.032\\
$U(1)$ & 28 & U011-100101 & D12 & 0 & gapped & 1.174\\
$U(1)$ & 30 & U101-000001 & D6 & 0 & gapped & 0.093\\
$U(1)$ & 33 & U101-100000 & gapped & 0.723 & D12 & 0\\
$U(1)$ & 34 & U101-100011 & NL & 0 & gapped & 0.265\\
$U(1)$ & 37 & U111-000100 & NL & 0 & gapped & 0.045\\
$U(1)$ & 50 & U001-000011 & gapped & 1.514 & Q1 & 0\\
$U(1)$ & 64 & U100-111000 & gapped & 0.640 & D12 & 0\\
$U(1)$ & 68 & U101-100001 & NL & 0 & gapped & 0.850\\
$SU(2)$ & 1 & SU010110 & NL & 0 & Q2 & 0\\
$SU(2)$ & 9 & SU001011 & NL & 0 & gapped & 0.200\\
$SU(2)$ & 13 & SU101010 & NL & 0 & gapped & 0.490\\
\end{longtable}
\endgroup
\endgroup

\clearpage
\section{Derivation of the $\Z_2$ Projective Symmetry Groups}
\label{sm:app:psg}

This section gives the complete derivation of the $\Z_2$ PSGs of the SHD lattice that is summarized in Appendix~B of the main text, and Sec.~\ref{sm:app:psg-u1} derives the $SU(2)$, $U(1)$, and space-group PSGs.

\subsection{Conventions}
\label{sm:app:psg-conv}

\paragraph*{Sites and symmetry operations.}
A site is labeled $i=(x,y,\mu)$, where $x,y\in\Z$ label the unit cell at $x\bm{T}_1+y\bm{T}_2$ and $\mu=1,\dots,12$ labels the sublattice. Sublattice indices are understood modulo 12 and brought back to the range $1,\dots,12$ by $[k]\equiv((k-1)\bmod 12)+1$. The generators act on sites as in Eq.~(3) of the main text, and time reversal $\T$ leaves every site fixed. The sublattice map of the mirror can be written as $\mu_\sigma=[9-\mu]$, so that
\begin{align}
C_6&:\ (x,y,\mu)\mapsto(-y,\,x+y,\,[\mu-2]),\nonumber\\
\sigma&:\ (x,y,\mu)\mapsto(y,\,x,\,[9-\mu]),\nonumber\\
C_6^{-1}&:\ (x,y,\mu)\mapsto(x+y,\,-x,\,[\mu+2]),\qquad \sigma^{-1}=\sigma .
\label{sm:eq:psg-inverses}
\end{align}
The symmetry group $p6mm\times\Z_2^\T$ is defined by the thirteen relations~(4) of the main text. Each relation can be checked directly from Eq.~(3) of the main text and Eq.~\eqref{sm:eq:psg-inverses}; for instance $T_2C_6T_1^{-1}C_6^{-1}$ maps $(x,y)\mapsto(x+y,-x)\mapsto(x+y-1,-x)\mapsto(x,y-1)\mapsto(x,y)$.

\paragraph*{PSG elements and their composition.}
The mean-field ansatz $\{u_{ij}\}$ is invariant under a symmetry operation $O$ combined with a gauge transformation $G_O$, where $G_O(i)\in SU(2)$ is a site-dependent matrix. The conditions are Eq.~(9) of the main text for a spatial operation, where $R_O$ rotates the spin index of the triplet amplitudes, and Eq.~(10) of the main text for time reversal. For the algebraic classification only the multiplication law of the pairs $\hat O\equiv(G_O,O)$ matters. Applying $O_2$ and then $O_1$ to the condition above shows that the combined operation $O_1O_2$ is accompanied by the matrices of Eq.~(11) of the main text, $G_{O_1O_2}(i)=G_{O_1}(i)\,G_{O_2}(O_1^{-1}(i))$, and that the inverse operation is accompanied by
\begin{equation}
G_{O^{-1}}(i)=\big[G_O(O(i))\big]^{-1}.
\label{sm:eq:psg-inverse-G}
\end{equation}
Because $\T$ acts on the amplitudes as $u\to-G_\T uG_\T^\dagger$, a linear map, it composes with the other operations according to the same rule. For a word $O_1O_2\cdots O_n$ the rule gives
\begin{align}
G_{O_1\cdots O_n}(i)&=G_{O_1}(i_0)\,G_{O_2}(i_1)\cdots G_{O_n}(i_{n-1}),\nonumber\\
i_0&=i,\qquad i_k=O_k^{-1}(i_{k-1}),
\label{sm:eq:psg-word}
\end{align}
where a letter $O_k=O^{-1}$ contributes the factor $G_{O^{-1}}(i_{k-1})=[G_O(O(i_{k-1}))]^{-1}$.

\paragraph*{Invariant gauge group and relation values.}
We work in the canonical gauge in which the IGG consists of site-independent matrices: $\{\pm\tau^0\}$ for a $\Z_2$ spin liquid, $\{e^{\ii\theta\tau^3}\}$ for a $U(1)$ spin liquid, and all of $SU(2)$ for an $SU(2)$ spin liquid. If a word $O_1\cdots O_n$ equals the identity of the symmetry group, the combined operation leaves the ansatz invariant, so $G_{O_1\cdots O_n}(i)$ belongs to the IGG and does not depend on $i$. We call this matrix the value of the relation. In the $\Z_2$ case we denote the values of the relations~(4) of the main text by $\eta$ with a subscript. Here $\eta_{12}$ is the value of $T_2T_1T_2^{-1}T_1^{-1}$, $\eta_{T_1C_6}$ that of $T_1C_6T_1^{-1}T_2C_6^{-1}$, $\eta_{T_2C_6}$ that of $T_2C_6T_1^{-1}C_6^{-1}$, $\eta_{T_2\sigma}$ that of $T_2\sigma T_1^{-1}\sigma^{-1}$, $\eta_{T_1\sigma}$ that of $T_1\sigma T_2^{-1}\sigma^{-1}$, $\eta_{\sigma C_6}$ that of $(\sigma C_6)^2$, $\eta_{C_6}$ that of $C_6^6$, $\eta_\sigma$ that of $\sigma^2$, $\eta_\T$ that of $\T^2$, $\eta_{T_a\T}$ that of $T_a\T T_a^{-1}\T^{-1}$ ($a=1,2$), and $\eta_{\T C_6}$, $\eta_{\T\sigma}$ those of $C_6\T C_6^{-1}\T^{-1}$ and $\sigma\T\sigma^{-1}\T^{-1}$. Each $\eta$ equals $\pm\tau^0$ and is identified with the sign $\pm1$.

\paragraph*{Equivalences.}
Two sets of matrices $\{G_O\}$ describe the same PSG if they are related by (i) a gauge transformation $W_i\in SU(2)$, under which $\psi_i\to\psi_iW_i^\dagger$, $u_{ij}\to W_iu_{ij}W_j^\dagger$ and
\begin{equation}
G_O(i)\ \to\ W_i\,G_O(i)\,W^\dagger_{O^{-1}(i)},
\label{sm:eq:psg-gauge}
\end{equation}
or (ii) a re-lifting $G_O\to\epsilon_O G_O$ with $\epsilon_O$ in the IGG, which does not change the symmetry operation because $\epsilon_O$ leaves the ansatz invariant. Equation~\eqref{sm:eq:psg-gauge} follows by writing the invariance condition for the transformed amplitudes $u'_{ij}=W_iu_{ij}W_j^\dagger$: one finds $u'_{O(i)O(j)}=G'_O(O(i))\,[R_Ou']_{ij}\,G'_O(O(j))^\dagger$ with $G'_O(O(i))=W_{O(i)}G_O(O(i))W_i^\dagger$.

\paragraph*{Lemma (invariance of relation values).}
Under the gauge transformation \eqref{sm:eq:psg-gauge}, the factors of Eq.~\eqref{sm:eq:psg-word} transform as $G_{O_k}(i_{k-1})\to W_{i_{k-1}}G_{O_k}(i_{k-1})W_{i_k}^\dagger$, and the inner $W$ matrices cancel pairwise. The value of a relation therefore becomes $W_{i_0}\,\eta\,W_{i_n}^\dagger=W_i\,\eta\,W_i^\dagger$, since a relation returns to its starting site, $i_n=i$. For $\eta=\pm\tau^0$ this equals $\eta$: relation values are gauge invariant. Under a re-lifting $G_O\to-G_O$ the value of a relation is multiplied by $(-1)^{N_O}$, where $N_O$ is the number of letters $O$ or $O^{-1}$ in the word. The value of a single relation is therefore invariant under all re-liftings, and hence a label of the $\Z_2$ PSG, if and only if every generator occurs an even number of times in the relation. Products of non-invariant values can be invariant; for instance $\eta_{T_2C_6}\eta_{T_1\sigma}$ contains each translation twice, and Sec.~\ref{sm:app:psg-z2} shows that it equals $\eta_{12}$. Counting the letters in Eq.~(4) of the main text:
\begin{itemize}
\item $T_2T_1T_2^{-1}T_1^{-1}$, $(\sigma C_6)^2$, $C_6^6$, $\sigma^2$, $\T^2$, and the four $O\T O^{-1}\T^{-1}$ contain every generator an even number of times; their values are invariant.
\item $T_1C_6T_1^{-1}T_2C_6^{-1}$ contains $T_2$ once, and $T_2C_6T_1^{-1}C_6^{-1}$, $T_2\sigma T_1^{-1}\sigma^{-1}$, $T_1\sigma T_2^{-1}\sigma^{-1}$ contain $T_1$ and $T_2$ once each; their values change under re-liftings of the translations and are not labels.
\end{itemize}
This leaves nine candidate labels, $\eta_{12}$, $\eta_{\sigma C_6}$, $\eta_{C_6}$, $\eta_\sigma$, $\eta_\T$, $\eta_{T_1\T}$, $\eta_{T_2\T}$, $\eta_{\T C_6}$, and $\eta_{\T\sigma}$. We show in Sec.~\ref{sm:app:psg-z2} that $\eta_{T_1\T}=\eta_{T_2\T}=+1$ are forced, so that seven labels remain.

\subsection{Free Action and the Spanning-Tree Gauge}
\label{sm:app:psg-tree}

The twelve sites of a unit cell form a single orbit of the point group $C_{6v}=\langle C_6,\sigma\rangle$: by Eq.~(3) of the main text, $C_6^a$ maps the sublattice $\mu=1$ onto $[1-2a]$, which runs through all odd sublattices for $a=0,\dots,5$, and $C_6^a\sigma$ maps it onto $[8-2a]$, which runs through all even sublattices. The twelve elements of $C_{6v}$ thus map $\mu=1$ onto twelve different sublattices, so that no element other than the identity fixes a site. Combined with the translations, every site $i$ is the image $g_i(i_0)$ of the reference site $i_0=(0,0,1)$ under exactly one element $g_i\in p6mm$. In crystallographic language the spins occupy the general Wyckoff position $12f$, on which $p6mm$ acts freely and transitively.

This property has a direct consequence for the PSG. Draw a graph whose vertices are the sites and which contains, for every site $i$ and every spatial generator $O\in\{T_1,T_2,C_6,\sigma\}$, an edge joining $O^{-1}(i)$ to $i$; attach to this edge the matrix $G_O(i)$. By Eq.~\eqref{sm:eq:psg-gauge} these matrices transform like link variables of a lattice gauge field, and the product of the matrices along a closed path transforms by conjugation with $W$ at its starting point. Since $p6mm$ acts freely, a path that starts at $i$ and reads the letters of a word $O_1\cdots O_n$ is closed precisely when the word equals the identity of $p6mm$. Such a word is a product of conjugates of the relators in Eq.~(4) of the main text, so the ordered product of the matrices along any closed path is a product of conjugated relation values. For the $\Z_2$ IGG these values are $\pm\tau^0$, and conjugation leaves them unchanged.

Now choose a spanning tree, a set of edges that connects all sites without forming a closed path. Starting from the root $i_0$ and moving outward along the tree, one can choose $W_i$ at each newly reached site such that the transformed matrix on the tree edge equals $\tau^0$: for a tree edge from $j=O^{-1}(i)$ to $i$ with $W_j$ already fixed, take $W_i=W_jG_O(i)^{-1}$, so that $W_iG_O(i)W_j^\dagger=\tau^0$. Every edge outside the tree closes a unique path with tree edges, and its matrix equals the product of the matrices around that path, which is a product of relation values. Therefore:
\begin{itemize}
\item In the tree gauge every matrix $G_O(i)$ of a $\Z_2$ PSG equals $\pm\tau^0$, and its sign is determined by the relation values.
\item Two PSGs with the same relation values coincide in the tree gauge and are hence gauge equivalent.
\end{itemize}
Throughout this section and Sec.~\ref{sm:app:psg-u1} we use the tree consisting of all $T_1$ edges, the $T_2$ edges in the column $x=0$, and eleven edges inside the reference cell $(0,0)$. The latter are the $C_6$ edges $G_{C_6}(0,0,\mu)$, which join $[\mu+2]$ to $\mu$, for $\mu\in\{1,3,5,7,9\}$ and $\mu\in\{8,10,12,2,4\}$, together with the $\sigma$ edge $G_\sigma(0,0,1)$, which joins $8$ to $1$. These eleven edges form two paths $11\!-\!9\!-\!7\!-\!5\!-\!3\!-\!1$ and $6\!-\!4\!-\!2\!-\!12\!-\!10\!-\!8$ connected by the edge $1\!-\!8$, a spanning tree of the twelve sublattices. The Cayley graph of $C_{6v}$ with respect to $C_6$ and $\sigma$ has 12 vertices and 18 edges (twelve $C_6$ edges and six $\sigma$ edges), so seven edges lie outside the tree; their matrices are fixed by the relations $C_6^6$, $\sigma^2$, and $(\sigma C_6)^2$, as shown explicitly in Step~10 of Sec.~\ref{sm:app:psg-z2}.

The time-reversal matrices $G_\T(i)$ are not edges of the graph, because $\T$ fixes every site. They are on-site matrices, and the only constraint specific to them is the on-site relation $\T^2$; this is the origin of the distinction between $g_\T=\tau^0$ and $g_\T=\ii\tau^2$ below.

\subsection{Derivation of the $\Z_2$ PSG}
\label{sm:app:psg-z2}

The derivation follows the spanning tree of Sec.~\ref{sm:app:psg-tree}: first the translations, then the $x,y$ dependence of $G_{C_6}$ and $G_\sigma$, then the sublattice dependence, and finally time reversal.

\paragraph*{Step 1: gauge fixing of the translations.}
Choose first $W_{(0,y,\mu)}$ along the column $x=0$ by $W_{(0,0,\mu)}=\tau^0$ and $W_{(0,y,\mu)}=W_{(0,y-1,\mu)}\,G_{T_2}(0,y,\mu)^{-1}$ for $y\ge1$, $W_{(0,y-1,\mu)}=W_{(0,y,\mu)}\,G_{T_2}(0,y,\mu)$ for $y\le0$. Then choose $W$ along each row by $W_{(x,y,\mu)}=W_{(x-1,y,\mu)}G_{T_1}(x,y,\mu)^{-1}$ for $x\ge1$ and $W_{(x-1,y,\mu)}=W_{(x,y,\mu)}G_{T_1}(x,y,\mu)$ for $x\le0$. By Eq.~\eqref{sm:eq:psg-gauge}, the transformed matrices satisfy
\begin{equation}
G_{T_1}(x,y,\mu)=\tau^0,\qquad G_{T_2}(0,y,\mu)=\tau^0 .
\label{sm:eq:z2-step1}
\end{equation}

\paragraph*{Step 2: the relation $T_2T_1T_2^{-1}T_1^{-1}$.}
Starting from $i=(x,y,\mu)$, Eq.~\eqref{sm:eq:psg-word} visits the sites $i_1=T_2^{-1}(i)=(x,y-1,\mu)$, $i_2=T_1^{-1}(i_1)=(x-1,y-1,\mu)$, $i_3=T_2(i_2)=(x-1,y,\mu)$, and $i_4=T_1(i_3)=i$. The four factors are $G_{T_2}(x,y,\mu)$, $G_{T_1}(x,y-1,\mu)$, $G_{T_2}(T_2(i_2))^{-1}=G_{T_2}(x-1,y,\mu)^{-1}$, and $G_{T_1}(T_1(i_3))^{-1}=G_{T_1}(x,y,\mu)^{-1}$, so
\begin{align}
&G_{T_2}(x,y,\mu)\,G_{T_1}(x,y-1,\mu)\nonumber\\
&\quad\times G_{T_2}(x-1,y,\mu)^{-1}\,G_{T_1}(x,y,\mu)^{-1}=\eta_{12}.
\end{align}
With Eq.~\eqref{sm:eq:z2-step1} this is the recursion $G_{T_2}(x,y,\mu)=\eta_{12}\,G_{T_2}(x-1,y,\mu)$, whose solution with $G_{T_2}(0,y,\mu)=\tau^0$ is
\begin{equation}
G_{T_1}(x,y,\mu)=\tau^0,\qquad G_{T_2}(x,y,\mu)=\eta_{12}^{\,x}\,\tau^0 .
\label{sm:eq:z2-trans}
\end{equation}

\paragraph*{Step 3: $x,y$ dependence of $G_{C_6}$.}
For $T_1C_6T_1^{-1}T_2C_6^{-1}$ the visited sites are
\begin{align}
i_1&=(x-1,\,y,\,\mu),\nonumber\\
i_2&=(x+y-1,\,1-x,\,[\mu+2]),\nonumber\\
i_3&=(x+y,\,1-x,\,[\mu+2]),\nonumber\\
i_4&=(x+y,\,-x,\,[\mu+2]),\nonumber
\end{align}
and $i_5=C_6(i_4)=i$. The factors are $G_{T_1}(i)=\tau^0$, $G_{C_6}(i_1)$, $G_{T_1}(T_1(i_2))^{-1}=\tau^0$, $G_{T_2}(i_3)=\eta_{12}^{\,x+y}$, and $G_{C_6}(C_6(i_4))^{-1}=G_{C_6}(i)^{-1}$. For $T_2C_6T_1^{-1}C_6^{-1}$ the visited sites are $(x,y-1,\mu)$, $(x+y-1,-x,[\mu+2])$, $(x+y,-x,[\mu+2])$, and the factors are $\eta_{12}^{\,x}$, $G_{C_6}(x,y-1,\mu)$, $\tau^0$, and $G_{C_6}(i)^{-1}$. Since signs commute with all matrices and $\eta^{-1}=\eta$, the two relations become
\begin{align}
G_{C_6}(x,y,\mu)&=\eta_{T_1C_6}\,\eta_{12}^{\,x+y}\,G_{C_6}(x-1,y,\mu),\nonumber\\
G_{C_6}(x,y,\mu)&=\eta_{T_2C_6}\,\eta_{12}^{\,x}\,G_{C_6}(x,y-1,\mu).
\label{sm:eq:z2-c6-rec}
\end{align}
Iterating the first recursion at $y=0$ from $x=0$ produces the exponent $\sum_{k=1}^{x}k=x(x+1)/2$ of $\eta_{12}$; the function $x(x+1)/2$ satisfies $f(x)-f(x-1)=x$ for every integer $x$, so the same expression holds for negative $x$. Iterating the second recursion in $y$ produces $\eta_{12}^{\,xy}$. Hence
\begin{equation}
G_{C_6}(x,y,\mu)=\eta_{T_1C_6}^{\,x}\,\eta_{T_2C_6}^{\,y}\,\eta_{12}^{\,x(x+1)/2+xy}\;g_{C_6}(\mu),
\label{sm:eq:z2-c6-xy}
\end{equation}
with $g_{C_6}(\mu)\equiv G_{C_6}(0,0,\mu)\in SU(2)$. One verifies that Eq.~\eqref{sm:eq:z2-c6-xy} satisfies both recursions at every $(x,y)$: the change of the exponent of $\eta_{12}$ from $x-1$ to $x$ is $x+y$, and from $y-1$ to $y$ it is $x$.

\paragraph*{Step 4: $x,y$ dependence of $G_\sigma$.}
For $T_2\sigma T_1^{-1}\sigma^{-1}$ the visited sites are $(x,y-1,\mu)$, $(y-1,x,[9-\mu])$, and $(y,x,[9-\mu])$, with factors $\eta_{12}^{\,x}$, $G_\sigma(x,y-1,\mu)$, $\tau^0$, and $G_\sigma(\sigma(y,x,[9-\mu]))^{-1}=G_\sigma(x,y,\mu)^{-1}$. For $T_1\sigma T_2^{-1}\sigma^{-1}$ they are $(x-1,y,\mu)$, $(y,x-1,[9-\mu])$, and $(y,x,[9-\mu])$, with factors $\tau^0$, $G_\sigma(x-1,y,\mu)$, $G_{T_2}(y,x,[9-\mu])^{-1}=\eta_{12}^{\,y}$, and $G_\sigma(x,y,\mu)^{-1}$. Hence
\begin{align}
G_\sigma(x,y,\mu)&=\eta_{T_2\sigma}\,\eta_{12}^{\,x}\,G_\sigma(x,y-1,\mu),\nonumber\\
G_\sigma(x,y,\mu)&=\eta_{T_1\sigma}\,\eta_{12}^{\,y}\,G_\sigma(x-1,y,\mu),
\end{align}
with the solution
\begin{equation}
G_\sigma(x,y,\mu)=\eta_{T_1\sigma}^{\,x}\,\eta_{T_2\sigma}^{\,y}\,\eta_{12}^{\,xy}\;g_\sigma(\mu),
\label{sm:eq:z2-s-xy}
\end{equation}
with $g_\sigma(\mu)\equiv G_\sigma(0,0,\mu)$.

\paragraph*{Step 5: the relation $\sigma^2$.}
The word $\sigma\sigma$ visits $(y,x,[9-\mu])$, so $G_\sigma(x,y,\mu)\,G_\sigma(y,x,[9-\mu])=\eta_\sigma$. Inserting Eq.~\eqref{sm:eq:z2-s-xy} and using $\eta_{12}^{\,2xy}=1$,
\begin{equation}
(\eta_{T_1\sigma}\eta_{T_2\sigma})^{x+y}\;g_\sigma(\mu)\,g_\sigma([9-\mu])=\eta_\sigma .
\end{equation}
The right-hand side does not depend on $x+y$, so $\eta_{T_1\sigma}=\eta_{T_2\sigma}\equiv\eta_{T\sigma}$, and
\begin{equation}
g_\sigma(\mu)\,g_\sigma([9-\mu])=\eta_\sigma .
\label{sm:eq:z2-s2-sub}
\end{equation}

\paragraph*{Step 6: removing $\eta_{T\sigma}$.}
Equation~\eqref{sm:eq:z2-step1} does not exhaust the gauge freedom. Consider $W_{(x,y,\mu)}=\nu^{\,y}\tau^0$ with $\nu=\pm1$. By Eq.~\eqref{sm:eq:psg-gauge} it leaves $G_{T_1}$ invariant and multiplies $G_{T_2}$ by $\nu^{\,y}\nu^{\,y-1}=\nu$, which is undone by the re-lifting $G_{T_2}\to\nu G_{T_2}$. It multiplies $G_{C_6}(x,y,\mu)$ by $W_iW_{C_6^{-1}(i)}=\nu^{\,y}\nu^{-x}=\nu^{\,x+y}$ and $G_\sigma(x,y,\mu)$ by $\nu^{\,y}\nu^{\,x}=\nu^{\,x+y}$, and leaves $G_\T$ unchanged. Comparing with Eqs.~\eqref{sm:eq:z2-c6-xy} and \eqref{sm:eq:z2-s-xy}, the net effect is
\begin{equation}
(\eta_{T_1C_6},\,\eta_{T_2C_6},\,\eta_{T\sigma})\ \to\ (\nu\,\eta_{T_1C_6},\,\nu\,\eta_{T_2C_6},\,\nu\,\eta_{T\sigma}).
\end{equation}
We choose $\nu=\eta_{T\sigma}$, which sets $\eta_{T\sigma}=+1$. Then
\begin{equation}
G_\sigma(x,y,\mu)=\eta_{12}^{\,xy}\,g_\sigma(\mu).
\label{sm:eq:z2-s-final}
\end{equation}

\paragraph*{Step 7: the relation $(\sigma C_6)^2$.}
Starting from $i=(x,y,\mu)$, the word $\sigma C_6\sigma C_6$ visits
\begin{align}
i_1&=(y,\,x,\,[9-\mu]),\nonumber\\
i_2&=(x+y,\,-y,\,[11-\mu]),\nonumber\\
i_3&=(-y,\,x+y,\,[\mu-2]),\nonumber\\
i_4&=i,
\label{sm:eq:z2-sc6-sites}
\end{align}
and its value is $G_\sigma(i)\,G_{C_6}(i_1)\,G_\sigma(i_2)\,G_{C_6}(i_3)=\eta_{\sigma C_6}$. The $x,y$-dependent signs of the four factors follow from Eqs.~\eqref{sm:eq:z2-c6-xy} and \eqref{sm:eq:z2-s-final}. The exponents of $\eta_{12}$ are
\begin{align}
&xy \;+\; \Big[\tfrac{y(y+1)}{2}+xy\Big] \;+\; \big[-xy-y^2\big]\nonumber\\
&\quad+\;\Big[\tfrac{y(y-1)}{2}-xy-y^2\Big] \;=\; -y^2,
\end{align}
the exponents of $\eta_{T_1C_6}$ are $y$ (from $i_1$) and $-y$ (from $i_3$), and those of $\eta_{T_2C_6}$ are $x$ and $x+y$. Using $y^2\equiv y$ and $2x\equiv0$ modulo 2,
\begin{align}
&(\eta_{12}\,\eta_{T_2C_6})^{\,y}\;g_\sigma(\mu)\,g_{C_6}([9-\mu])\nonumber\\
&\quad\times g_\sigma([11-\mu])\,g_{C_6}([\mu-2])=\eta_{\sigma C_6}.
\label{sm:eq:z2-sc6}
\end{align}
Independence of $y$ requires $\eta_{T_2C_6}=\eta_{12}$. (Before Step~6 the prefactor reads $(\eta_{12}\eta_{T_2C_6}\eta_{T\sigma})^y$, since $G_\sigma(i)G_\sigma(i_2)$ contributes $\eta_{T\sigma}^{\,(x+y)+x}$.)

\paragraph*{Step 8: removing $\eta_{T_1C_6}$.}
The gauge transformation $W_{(x,y,\mu)}=\nu^{\,x+y}\tau^0$ multiplies $G_{T_1}$ and $G_{T_2}$ by $\nu$, which is undone by re-liftings. It multiplies $G_{C_6}(x,y,\mu)$ by $\nu^{\,x+y}\nu^{\,(x+y)-x}=\nu^{\,x}$, $G_\sigma(x,y,\mu)$ by $\nu^{\,x+y}\nu^{\,y+x}=+1$ and $G_\T$ by $\nu^{\,2(x+y)}=+1$. Its only effect is therefore $\eta_{T_1C_6}\to\nu\,\eta_{T_1C_6}$, and with $\nu=\eta_{T_1C_6}$ we set $\eta_{T_1C_6}=+1$. The parameter $\eta_{T_1C_6}$ is pure gauge: the PSGs with $\eta_{T_1C_6}=+1$ and $\eta_{T_1C_6}=-1$ and otherwise identical parameters are mapped onto each other by $W=(-1)^{x+y}\tau^0$. Collecting Steps 3, 7, and 8,
\begin{equation}
G_{C_6}(x,y,\mu)=\eta_{12}^{\,x(x+1)/2+xy+y}\;g_{C_6}(\mu).
\label{sm:eq:z2-c6-final}
\end{equation}

\paragraph*{Step 9: the relation $C_6^6$.}
The word $C_6^6$ visits the orbit of $(x,y)$ under $C_6^{-1}$,
\begin{align}
(X_k,Y_k)\ =\ &(x,y),\ (x+y,-x),\ (y,-x-y),\nonumber\\
&(-x,-y),\ (-x-y,x),\ (-y,x+y),
\label{sm:eq:z2-orbit}
\end{align}
for $k=0,\dots,5$, with sublattices $[\mu+2k]$. Over this orbit
\begin{align}
&\textstyle\sum_kX_k=\sum_kY_k=0,\nonumber\\
&\textstyle\sum_kX_k^2=2s,\nonumber\\
&\textstyle\sum_kX_kY_k=-s,\nonumber\\
&s\equiv x^2+y^2+(x+y)^2,
\end{align}
so the exponent of $\eta_{12}$ collected from Eq.~\eqref{sm:eq:z2-c6-final},
\begin{equation}
\sum_{k}\Big[\tfrac{X_k(X_k+1)}{2}+X_kY_k+Y_k\Big]=\tfrac12(2s)+0-s+0=0,
\label{sm:eq:z2-orbit-sum}
\end{equation}
vanishes identically. The relation $C_6^6$ is therefore a condition on the sublattice matrices alone,
\begin{equation}
g_{C_6}(\mu)\,g_{C_6}([\mu+2])\,g_{C_6}([\mu+4])\cdots g_{C_6}([\mu+10])=\eta_{C_6}.
\label{sm:eq:z2-c66-sub}
\end{equation}

\paragraph*{Step 10: the sublattice matrices.}
After Steps 1 to 9 the relations $C_6^6$, $\sigma^2$, and $(\sigma C_6)^2$ reduce to Eqs.~\eqref{sm:eq:z2-c66-sub}, \eqref{sm:eq:z2-s2-sub} and \eqref{sm:eq:z2-sc6}, which involve only the 24 matrices $g_{C_6}(\mu)$ and $g_\sigma(\mu)$. The remaining gauge freedom consists of transformations $W_{(x,y,\mu)}=W_\mu$ that do not depend on $x,y$; they leave Eqs.~\eqref{sm:eq:z2-trans}, \eqref{sm:eq:z2-s-final} and \eqref{sm:eq:z2-c6-final} intact and act as
\begin{align}
g_{C_6}(\mu)&\to W_\mu\,g_{C_6}(\mu)\,W_{[\mu+2]}^\dagger,\nonumber\\
g_\sigma(\mu)&\to W_\mu\,g_\sigma(\mu)\,W_{[9-\mu]}^\dagger.
\end{align}
Following the tree of Sec.~\ref{sm:app:psg-tree}, set $W_1=\tau^0$, $W_{[\mu+2]}=W_\mu\,g_{C_6}(\mu)$ for $\mu=1,3,5,7,9$, $W_8=W_1\,g_\sigma(1)$, and $W_{[\mu+2]}=W_\mu\,g_{C_6}(\mu)$ for $\mu=8,10,12,2,4$. This fixes all twelve $W_\mu$ and gives
\begin{equation}
g_{C_6}(\mu)=\tau^0\ \ (\mu\neq6,11),\qquad g_\sigma(1)=\tau^0 .
\end{equation}
The remaining matrices follow from the relations. Equation~\eqref{sm:eq:z2-c66-sub} at $\mu=1$ is the product of $g_{C_6}$ over the odd sublattices, $g_{C_6}(11)=\eta_{C_6}$; at $\mu=8$ it is the product over the even sublattices, $g_{C_6}(6)=\eta_{C_6}$. The matrices $g_\sigma(\mu)$ are obtained by alternating Eq.~\eqref{sm:eq:z2-s2-sub} and Eq.~\eqref{sm:eq:z2-sc6} (with $\eta_{12}\eta_{T_2C_6}=1$):
\begin{align}
\sigma^2,\ \mu=1:&\quad g_\sigma(8)=\eta_\sigma,\nonumber\\
(\sigma C_6)^2,\ \mu=1:&\quad g_\sigma(10)=\eta_{\sigma C_6}\eta_{C_6},\nonumber\\
\sigma^2,\ \mu=10:&\quad g_\sigma(11)=\eta_\sigma\eta_{\sigma C_6}\eta_{C_6},\nonumber\\
(\sigma C_6)^2,\ \mu=11:&\quad g_\sigma(12)=\eta_\sigma\eta_{C_6},\nonumber\\
\sigma^2,\ \mu=12:&\quad g_\sigma(9)=\eta_{C_6},\nonumber\\
(\sigma C_6)^2,\ \mu=9:&\quad g_\sigma(2)=\eta_{\sigma C_6}\eta_{C_6},\nonumber\\
\sigma^2,\ \mu=2:&\quad g_\sigma(7)=\eta_\sigma\eta_{\sigma C_6}\eta_{C_6},\nonumber\\
(\sigma C_6)^2,\ \mu=7:&\quad g_\sigma(4)=\eta_\sigma\eta_{C_6},\nonumber\\
\sigma^2,\ \mu=4:&\quad g_\sigma(5)=\eta_{C_6},\nonumber\\
(\sigma C_6)^2,\ \mu=5:&\quad g_\sigma(6)=\eta_{\sigma C_6}\eta_{C_6},\nonumber\\
\sigma^2,\ \mu=6:&\quad g_\sigma(3)=\eta_\sigma\eta_{\sigma C_6}\eta_{C_6}.
\label{sm:eq:z2-gs-chain}
\end{align}
For example, the second line is Eq.~\eqref{sm:eq:z2-sc6} at $\mu=1$, $g_\sigma(1)g_{C_6}(8)g_\sigma(10)g_{C_6}(11)=\eta_{\sigma C_6}$, with $g_\sigma(1)=g_{C_6}(8)=\tau^0$ and $g_{C_6}(11)=\eta_{C_6}$. The chain closes consistently: Eq.~\eqref{sm:eq:z2-sc6} at $\mu=3$ reads $g_\sigma(3)g_{C_6}(6)g_\sigma(8)g_{C_6}(1)=(\eta_\sigma\eta_{\sigma C_6}\eta_{C_6})\,\eta_{C_6}\,\eta_\sigma=\eta_{\sigma C_6}$. All other instances of the three relations are then satisfied identically. Indeed, Eq.~\eqref{sm:eq:z2-s2-sub} at $\mu$ and at $[9-\mu]$ is the same equation; Eq.~\eqref{sm:eq:z2-sc6} at $\mu$ and at $[11-\mu]$ involves the same four commuting signs, so the six even values of $\mu$ repeat the six odd ones; and Eq.~\eqref{sm:eq:z2-c66-sub} at any $\mu$ is a cyclic permutation of the products already used. The result is listed in Table~II of the main text.

After this gauge fixing the only remaining freedom is a global transformation $W_i=W$ for all $i$, with $W\in SU(2)$ arbitrary; it commutes with all matrices found so far.

\paragraph*{Step 11: time reversal.}
The relations $T_1\T T_1^{-1}\T^{-1}$ and $T_2\T T_2^{-1}\T^{-1}$ visit $T_a^{-1}(i)$ and return; with Eq.~\eqref{sm:eq:z2-trans} their values are
\begin{align}
G_\T(x-1,y,\mu)\,G_\T(x,y,\mu)^{-1}&=\eta_{T_1\T},\nonumber\\
\eta_{12}^{\,x}\,G_\T(x,y-1,\mu)\,\eta_{12}^{\,x}\,G_\T(x,y,\mu)^{-1}&=\eta_{T_2\T},
\end{align}
so that $G_\T(x,y,\mu)=\eta_{T_1\T}^{\,x}\eta_{T_2\T}^{\,y}\,g_\T(\mu)$ with $g_\T(\mu)\equiv G_\T(0,0,\mu)$. The relation $C_6\T C_6^{-1}\T^{-1}$ visits $C_6^{-1}(i)=(x+y,-x,[\mu+2])$; since $G_{C_6}(i)=\pm\tau^0$ commutes with everything,
\begin{equation}
\eta_{T_1\T}^{\,y}\,\eta_{T_2\T}^{\,x+y}\;g_\T([\mu+2])\,g_\T(\mu)^{-1}=\eta_{\T C_6},
\end{equation}
where the exponents follow from $(x+y)+x\equiv y$ and $-x+y\equiv x+y$ modulo 2. Independence of $x$ and $y$ forces $\eta_{T_2\T}=1$ and then $\eta_{T_1\T}=1$, and leaves
\begin{equation}
g_\T([\mu+2])=\eta_{\T C_6}\,g_\T(\mu).
\end{equation}
In the same way $\sigma\T\sigma^{-1}\T^{-1}$ gives $g_\T([9-\mu])=\eta_{\T\sigma}\,g_\T(\mu)$. Starting from $\mu=1$ these two rules determine $g_\T(\mu)=h(\mu)\,g_\T$ with $g_\T\equiv g_\T(1)$ and
\begin{equation}
h(\mu)=\begin{cases}\eta_{\T C_6}^{\,(\mu-1)/2}, & \mu\ \text{odd},\\[2pt] \eta_{\T\sigma}\,\eta_{\T C_6}^{\,\mu/2}, & \mu\ \text{even},\end{cases}
\label{sm:eq:z2-h}
\end{equation}
which is the last column of Table~II of the main text. The sign function $h$ is a character of $p6mm$. The map $\chi$ that assigns to each $O$ the value of $O\T O^{-1}\T^{-1}$, i.e., $\chi(T_a)=\eta_{T_a\T}$, $\chi(C_6)=\eta_{\T C_6}$, and $\chi(\sigma)=\eta_{\T\sigma}$, is a homomorphism from $p6mm$ to $\{\pm1\}$, because $\T$ commutes with every spatial operation and the relation values are central; and $h(i)=\chi(g_i)$, where $g_i$ is the unique element of $p6mm$ with $g_i(i_0)=i$ (Sec.~\ref{sm:app:psg-tree}). As shown in Sec.~III of the main text, every homomorphism $\chi:p6mm\to\{\pm1\}$ is trivial on translations, so $\chi$ is fixed by $\chi(C_6)=\eta_{\T C_6}$ and $\chi(\sigma)=\eta_{\T\sigma}$, which can be chosen freely because $C_6$ and $\sigma$ occur an even number of times in each of the relations $C_6^6$, $\sigma^2$, and $(\sigma C_6)^2$.

The relation $\T^2$ visits only $i$, so $G_\T(i)^2=h(\mu)^2g_\T^2=g_\T^2=\eta_\T$. Write $g_\T=a_0\tau^0+\ii\,\bm{a}\cdot\bm{\tau}$ with $a_0^2+\bm{a}^2=1$; then $g_\T^2=(a_0^2-\bm{a}^2)\tau^0+2\ii a_0\,\bm{a}\cdot\bm{\tau}$. For $\eta_\T=+1$ this requires $\bm{a}=0$, so $g_\T=\pm\tau^0$, and the re-lifting $G_\T\to-G_\T$ (which changes no label, since $\T$ appears an even number of times in every relation containing it) gives $g_\T=\tau^0$. For $\eta_\T=-1$ it requires $a_0=0$, so $g_\T=\ii\,\hat{\bm{n}}\cdot\bm{\tau}$ with a unit vector $\hat{\bm n}$, and the global gauge freedom $g_\T\to Wg_\T W^\dagger$ rotates $\hat{\bm n}$ onto the 2 axis: $g_\T=\ii\tau^2$.

\paragraph*{Result.}
Collecting Eqs.~\eqref{sm:eq:z2-trans}, \eqref{sm:eq:z2-s-final} and \eqref{sm:eq:z2-c6-final}, the sublattice signs of Step~10 and the time-reversal matrices of Step~11, every $\Z_2$ PSG of the SHD lattice is gauge equivalent to the form~(13) of the main text, with $g_{C_6}(\mu)$, $g_\sigma(\mu)$, and $h(\mu)$ as listed in Table~II of the main text and $g_\T\in\{\tau^0,\ii\tau^2\}$. It is labeled as in Sec.~IV\,A of the main text. Substituting Eq.~\eqref{eq:z2psg} of the main text into the thirteen relations reproduces the label values for all $2^7$ choices of the labels; the four non-invariant relation values are $\eta_{T_1C_6}=\eta_{T_2C_6}\eta_{12}=\eta_{T_1\sigma}=\eta_{T_2\sigma}=+1$, and $\eta_{T_1\T}=\eta_{T_2\T}=+1$.

\subsection{Gauge Invariance, Independence, and Cohomology}
\label{sm:app:psg-z2inv}

\paragraph*{Invariance and independence.}
By the Lemma of Sec.~\ref{sm:app:psg-conv} the seven labels are invariant; by Steps 1 to 11 they fix the PSG; and since Eq.~\eqref{eq:z2psg} of the main text satisfies all relations for each of the $2^7$ choices, they are independent, so
\begin{equation}
N_{\Z_2}=2^6\times2=128 .
\end{equation}

\paragraph*{Comparison with group cohomology.}
For a $\Z_2$ IGG the relation values define a central extension of the symmetry group by $\Z_2$, and inequivalent consistent assignments of relation values are classified by the second cohomology group with $\Z_2$ coefficients. The tree argument of Sec.~\ref{sm:app:psg-tree} shows that on the SHD lattice each cohomology class is realized by exactly one gauge class: uniqueness holds because the spatial action is free, and existence holds because the only on-site constraint, $G_\T(i)^2=\eta_\T$, has solutions in $SU(2)$ for both signs. The count $128=2^7$ must therefore agree with $|H^2(p6mm\times\Z_2^\T,\Z_2)|$, which we now compute. All coefficients are in the field $\mathbb{F}_2=\Z_2$.

The K\"unneth formula over a field gives
\begin{align}
H^2(p6mm\times\Z_2^\T)={}&H^2(p6mm)\nonumber\\
&\oplus\big[H^1(p6mm)\otimes H^1(\Z_2^\T)\big]\nonumber\\
&\oplus H^2(\Z_2^\T),
\label{sm:eq:coh-kunneth}
\end{align}
where $H^0=\Z_2$ has been used for both factors. Here $H^n(\Z_2,\Z_2)=\Z_2$ for every $n$, so $H^2(\Z_2^\T)=\Z_2$; it is detected by $\eta_\T$. The group $H^1(p6mm,\Z_2)=\mathrm{Hom}(p6mm,\Z_2)$ consists of the homomorphisms $\chi$ discussed in Step~11. Since $\chi(T_1)=\chi(T_2)=1$ and $\chi(C_6)$, $\chi(\sigma)$ are free, $H^1(p6mm,\Z_2)=\Z_2^2$, detected by $\eta_{\T C_6}$ and $\eta_{\T\sigma}$.

For $H^2(p6mm,\Z_2)$ we use the Lyndon--Hochschild--Serre spectral sequence of the extension $1\to\Z^2\to p6mm\to C_{6v}\to1$, with
\begin{equation}
E_2^{p,q}=H^p\big(C_{6v},\,H^q(\Z^2,\Z_2)\big)\ \Rightarrow\ H^{p+q}(p6mm,\Z_2).
\end{equation}
The three terms with $p+q=2$ are as follows.

(i) $E_2^{2,0}=H^2(C_{6v},\Z_2)$. The central element $C_6^3$ generates a $\Z_2$ factor and $C_{6v}\cong S_3\times\Z_2$ with $S_3=\langle C_6^2,\sigma\rangle$. Because the normal subgroup $\langle C_6^2\rangle\cong\Z_3$ of $S_3$ has odd order, its mod-2 cohomology is trivial in positive degrees, and $H^n(S_3,\Z_2)\cong H^n(S_3/\Z_3,\Z_2)=H^n(\Z_2,\Z_2)=\Z_2$ for every $n$. The K\"unneth formula gives $\dim H^2(S_3\times\Z_2,\Z_2)=1+1+1=3$. Hence $E_2^{2,0}=\Z_2^3$. These three classes are the point-group relation values $\eta_{C_6}$, $\eta_\sigma$, and $\eta_{\sigma C_6}$ of the presentation $\langle C_6,\sigma\,|\,C_6^6,\sigma^2,(\sigma C_6)^2\rangle$.

(ii) $E_2^{1,1}=H^1(C_{6v},M)$ with $M=\mathrm{Hom}(\Z^2,\Z_2)\cong\mathbb{F}_2^2$, on which $C_{6v}$ acts through its action on the lattice. In the basis $(\bm T_1,\bm T_2)$, $C_6\bm T_1=\bm T_2$, $C_6\bm T_2=\bm T_2-\bm T_1$, and $\sigma$ exchanges $\bm T_1$ and $\bm T_2$, so
\begin{equation}
M_{C_6}=\begin{pmatrix}0&-1\\1&1\end{pmatrix},\qquad M_\sigma=\begin{pmatrix}0&1\\1&0\end{pmatrix}.
\end{equation}
Modulo 2, $M_{C_6}$ has order 3 and $M_{C_6}^3=-\one\equiv\one$; the image of $C_{6v}$ is the full group $GL(2,\mathbb{F}_2)\cong S_3$ of order 6, and the central $C_6^3$ acts trivially. The module $M$ (the dual of the mod-2 lattice, isomorphic to it) is the two-dimensional natural, or Steinberg, module of $GL(2,\mathbb{F}_2)$. It is irreducible, and it is projective: restricted to the Sylow 2-subgroup $\langle\sigma\rangle$, $\sigma$ exchanges the two basis vectors, so $M$ is the regular, hence free, $\mathbb{F}_2[\Z_2]$-module, and a module that is free over a Sylow $p$-subgroup is projective. Consequently $H^n(S_3,M)=0$ for $n\ge1$, and $H^0(S_3,M)=M^{S_3}=0$ because $M$ is irreducible and nontrivial. The K\"unneth formula for $S_3\times\Z_2$, with $\Z_2=\langle C_6^3\rangle$ acting trivially, gives $H^n(C_{6v},M)=\bigoplus_{a+b=n}H^a(S_3,M)\otimes H^b(\Z_2,\Z_2)=0$ for all $n$. In particular $E_2^{1,1}=0$, and also $E_2^{2,1}=E_2^{0,1}=0$.

(iii) $E_2^{0,2}=H^2(\Z^2,\Z_2)^{C_{6v}}$. The group $H^2(\Z^2,\Z_2)=\Z_2$ is generated by the class of the commutator of the two translations; a point-group element acts on it by its determinant on the lattice, $\pm1\equiv1$ modulo 2. Hence $E_2^{0,2}=\Z_2$, detected by $\eta_{12}$.

The extension is split, $p6mm$ being symmorphic: $C_{6v}$ is a subgroup of $p6mm$ that meets $\Z^2$ only in the identity. Restricting from $p6mm$ to this subgroup undoes the inflation from $C_{6v}$, so the inflation $H^*(C_{6v})\to H^*(p6mm)$ is injective and no differential can reach the bottom row $E^{p,0}$. The differentials leaving $E^{0,2}$ end in $E_2^{2,1}=0$ or in the bottom row, so they vanish, and $E_\infty^{p,q}=E_2^{p,q}$ for $p+q=2$. Therefore
\begin{equation}
|H^2(p6mm,\Z_2)|=|E_\infty^{2,0}|\,|E_\infty^{1,1}|\,|E_\infty^{0,2}|=2^3\cdot1\cdot2=2^4,
\end{equation}
and, with Eq.~\eqref{sm:eq:coh-kunneth},
\begin{equation}
H^2(p6mm\times\Z_2^\T,\Z_2)=\Z_2^4\oplus\Z_2^2\oplus\Z_2=\Z_2^7 .
\end{equation}
The seven generators correspond one to one to the seven labels $\eta_{12}$, $\eta_{C_6}$, $\eta_\sigma$, $\eta_{\sigma C_6}$, $\eta_{\T C_6}$, $\eta_{\T\sigma}$, $\eta_\T$. The vanishing of $E_2^{1,1}$ is the cohomological counterpart of Steps 6 to 8: the mixed relation values $\eta_{T_1C_6}$, $\eta_{T_2C_6}$, and $\eta_{T\sigma}$, which tie the point group to the translations, are either gauge or fixed by $\eta_{12}$, and they carry no independent label.

\section{$SU(2)$, $U(1)$, and Space-Group PSGs}
\label{sm:app:psg-u1}

This section extends the derivation of Sec.~\ref{sm:app:psg} to the IGGs $SU(2)$ and $U(1)$, and to PSGs of the space group alone. The notation is that of Sec.~\ref{sm:app:psg-conv}. We write $c\equiv\ii\tau^1$ and use repeatedly
\begin{equation}
c\,e^{\ii\phi\tau^3}=e^{-\ii\phi\tau^3}\,c,\qquad c^2=-\tau^0,\qquad c^{-1}=-c .
\label{sm:eq:u1-c-rules}
\end{equation}

\subsection{$SU(2)$ PSGs}
\label{sm:app:psg-su2}

For an $SU(2)$ spin liquid the IGG is the group of site-independent matrices $g\in SU(2)$. A PSG element must map IGG elements to IGG elements: $\hat O\,g\,\hat O^{-1}$, which acts on site $i$ as $G_O(i)\,g\,G_O(i)^{-1}$, must be a site-independent element $g'$ for every $g$. Therefore $G_O(i)^{-1}G_O(j)$ commutes with every $g\in SU(2)$ and equals $\pm\tau^0$, i.e.,
\begin{equation}
G_O(i)=\epsilon_O(i)\,g_O,\qquad \epsilon_O(i)=\pm1,\qquad g_O\in SU(2).
\label{sm:eq:su2-form}
\end{equation}
The uniform matrix $g_O$ belongs to the IGG, and the re-lifting $G_O\to g_O^{-1}G_O=\epsilon_O(i)\,\tau^0$ removes it. The gauge transformations that preserve the form \eqref{sm:eq:su2-form} are $W_i=s_i\,W$ with $s_i=\pm1$ and a uniform $W$, and the re-liftings that preserve it are the global signs $G_O\to-G_O$. The sign cochain $\epsilon_O(i)$ is therefore a $\Z_2$ PSG with $g_\T=\tau^0$, and the $SU(2)$ PSGs are in one-to-one correspondence with its equivalence classes under sign gauge transformations and sign re-liftings. Two consequences follow.

(i) The invariants are the relation values of the sign cochain: since every $g_O$ has been removed, the relation values are products of signs, and by the Lemma of Sec.~\ref{sm:app:psg-conv} the six quantities $\eta_{12}$, $\eta_{C_6}$, $\eta_\sigma$, $\eta_{\sigma C_6}$, $\eta_{\T C_6}$, $\eta_{\T\sigma}$ are invariant. The derivation of Sec.~\ref{sm:app:psg-z2} applies verbatim and gives the form (13) of the main text with $g_\T=\tau^0$.

(ii) $\T^2$ is not an invariant. In the sign cochain $G_\T(i)^2=\epsilon_\T(i)^2=+1$ always. The $\Z_2$ PSGs with $g_\T=\ii\tau^2$ and $g_\T=\tau^0$ differ by the uniform factor $\ii\tau^2$, which belongs to the $SU(2)$ IGG and is removed by a re-lifting.

Hence there are
\begin{equation}
N_{SU(2)}=2^6=64
\end{equation}
$SU(2)$ PSGs, labeled $n_{T_2}n_{\T\sigma}n_{\T C_6}n_\sigma n_{\sigma C_6}n_{C_6}$, with the matrices of Eq.~\eqref{eq:z2psg} of the main text at $g_\T=\tau^0$.

\subsection{$U(1)$ PSGs: General Form and Sectors}
\label{sm:app:psg-u1-form}

For a $U(1)$ spin liquid in the canonical gauge the IGG is $\{e^{\ii\theta\tau^3}\}$. Conjugation by $G_O(i)$ must map it onto itself, so $G_O(i)$ lies in the normalizer of $U(1)$ in $SU(2)$, which consists of the matrices $e^{\ii\phi\tau^3}$ and $e^{\ii\phi\tau^3}c$. By Eq.~\eqref{sm:eq:u1-c-rules}, $G_O(i)\,e^{\ii\theta\tau^3}G_O(i)^{-1}=e^{\ii s\theta\tau^3}$ with $s=+1$ for the first type and $s=-1$ for the second. The image must be a site-independent IGG element, so the type cannot depend on $i$:
\begin{equation}
G_O(i)=e^{\ii\phi_O(i)\tau^3}\,c^{\,w_O},\qquad w_O\in\{0,1\},
\label{sm:eq:u1-form}
\end{equation}
and we write $s_O\equiv(-1)^{w_O}$.

\paragraph*{The integers $w_O$ are gauge invariants.}
The PSG acts on its own IGG by conjugation, $\hat O\,e^{\ii\theta\tau^3}\hat O^{-1}=e^{\ii s_O\theta\tau^3}$. This defines an automorphism of the IGG, either the identity or the inversion. A gauge transformation conjugates the PSG and the IGG simultaneously, $\hat O\to W\hat OW^\dagger$ and $e^{\ii\theta\tau^3}\to We^{\ii\theta\tau^3}W^\dagger$, and therefore does not change which automorphism $\hat O$ induces; a re-lifting multiplies $\hat O$ by an element of the Abelian IGG and does not change it either. Since $\widehat{O_1O_2}=\hat O_1\hat O_2$ up to IGG elements, the map $O\mapsto s_O$ is a homomorphism from $p6mm\times\Z_2^\T$ to $\{\pm1\}$. By the argument of Step~11 of Sec.~\ref{sm:app:psg-z2}, every such homomorphism is trivial on translations. Hence
\begin{equation}
w_{T_1}=w_{T_2}=0,
\end{equation}
and the $U(1)$ PSGs fall into eight sectors labeled by $(w_{C_6},w_\sigma,w_\T)\in\{0,1\}^3$, which cannot be connected by any gauge transformation or re-lifting.

\paragraph*{Products of $U(1)$ matrices.}
Moving all phases to the left with Eq.~\eqref{sm:eq:u1-c-rules},
\begin{align}
&e^{\ii\phi_1\tau^3}c^{w_1}\,e^{\ii\phi_2\tau^3}c^{w_2}\cdots e^{\ii\phi_n\tau^3}c^{w_n}\nonumber\\
&\quad=e^{\ii(\phi_1+s_1\phi_2+s_1s_2\phi_3+\dots)\tau^3}\,c^{\,w_1+\dots+w_n},
\label{sm:eq:u1-product}
\end{align}
with $s_k=(-1)^{w_k}$, and $[e^{\ii\phi\tau^3}c^{w}]^{-1}=e^{-\ii s\phi\tau^3}c^{-w}$. Only $C_6$, $\sigma$, and $\T$ can have $w=1$, and each of them occurs an even number of times (counting inverses) in every relation, so $w_1+\dots+w_n$ is even and $c^{w_1+\dots+w_n}=(-1)^{(w_1+\dots+w_n)/2}\tau^0$. The value of each relation is $e^{\ii\theta\tau^3}$ with a phase $\theta$ defined modulo $2\pi$. We denote by $\Phi$ the phase of $T_2T_1T_2^{-1}T_1^{-1}$, by $\theta_{T_1C_6},\theta_{T_2C_6},\theta_{T_2\sigma},\theta_{T_1\sigma}$ the phases of the four mixed translation relations, by $\theta_{T_1\T},\theta_{T_2\T}$ those of $T_a\T T_a^{-1}\T^{-1}$, and by
\begin{equation}
\theta_{C_6},\ \theta_\sigma,\ \theta_{\sigma C_6},\ \theta_{\T C_6},\ \theta_{\T\sigma},\ \theta_\T
\label{sm:eq:u1-pg-values}
\end{equation}
the phases of $C_6^6$, $\sigma^2$, $(\sigma C_6)^2$, $C_6\T C_6^{-1}\T^{-1}$, $\sigma\T\sigma^{-1}\T^{-1}$, and $\T^2$. For a $\Z_2$-like value, $e^{\ii\theta\tau^3}=\pm\tau^0$ corresponds to $\theta=0$ or $\pi$.

\subsection{Translations and the Flux $\Phi$}
\label{sm:app:psg-u1-flux}

Step~1 of Sec.~\ref{sm:app:psg-z2} uses only $G_{T_1}$ and $G_{T_2}$, which have $w=0$; starting from $W_{(0,0,\mu)}=\tau^0$, the gauge transformation it constructs consists of $U(1)$ matrices. It gives $G_{T_1}=\tau^0$ and $G_{T_2}(0,y,\mu)=\tau^0$, and Step~2 gives $G_{T_2}(x,y,\mu)=e^{\ii\Phi x\tau^3}$. The relation value $e^{\ii\Phi\tau^3}$ is the $U(1)$ flux through the unit cell seen by the spinons. It is invariant under re-liftings, because $T_1$ and $T_2$ appear once with each sign and have $w=0$.

\paragraph*{Constraint on $\Phi$.}
Write $T_2T_1T_2^{-1}T_1^{-1}=e$ for the PSG operators as $[\hat T_2,\hat T_1]=e^{\ii\Phi\tau^3}$, with $[A,B]\equiv ABA^{-1}B^{-1}$. Conjugating with $\hat C_6$ gives on the right-hand side $e^{\ii s_{C_6}\Phi\tau^3}$. On the left-hand side, $\hat C_6\hat T_1\hat C_6^{-1}$ and $\hat C_6\hat T_2\hat C_6^{-1}$ equal $\hat T_2$ and $\hat T_2\hat T_1^{-1}$ up to IGG elements, because $C_6T_1C_6^{-1}=T_2$ and $C_6T_2C_6^{-1}=T_2T_1^{-1}$. The IGG elements commute with the $w=0$ translations and drop out of the commutator, so
\begin{equation}
\hat C_6[\hat T_2,\hat T_1]\hat C_6^{-1}=[\hat T_2\hat T_1^{-1},\hat T_2]=e^{\ii\Phi\tau^3}.
\end{equation}
The last equality follows from $\hat T_1^{-1}\hat T_2=e^{\ii\Phi\tau^3}\hat T_2\hat T_1^{-1}$, a rearrangement of $[\hat T_2,\hat T_1]=e^{\ii\Phi\tau^3}$. A rotation preserves the orientation of the unit cell and maps the flux onto itself, while the PSG maps it to $s_{C_6}\Phi$. Similarly, $\sigma T_1\sigma^{-1}=T_2$ gives $\hat\sigma[\hat T_2,\hat T_1]\hat\sigma^{-1}=[\hat T_1,\hat T_2]=e^{-\ii\Phi\tau^3}$: the mirror reverses the orientation of the cell. Time reversal commutes with the translations, $\hat\T[\hat T_2,\hat T_1]\hat\T^{-1}=e^{\ii\Phi\tau^3}$. Comparing with the action on the IGG, $e^{\ii\Phi\tau^3}\to e^{\ii s_O\Phi\tau^3}$, yields
\begin{equation}
\begin{aligned}
(1-s_{C_6})\Phi&=0,\\
(1+s_\sigma)\Phi&=0,\\
(1-s_\T)\Phi&=0
\end{aligned}
\label{sm:eq:u1-flux-constraints}
\end{equation}
modulo $2\pi$. Each condition with a nonzero coefficient forces $2\Phi=0$, i.e., $\Phi\in\{0,\pi\}$. A continuous flux survives only if all three coefficients vanish, $s_{C_6}=+1$, $s_\sigma=-1$, $s_\T=+1$, that is, in the single sector
\begin{equation}
(w_{C_6},w_\sigma,w_\T)=(0,1,0).
\end{equation}

\subsection{Reduction of the $x,y$ Dependence}
\label{sm:app:psg-u1-xy}

We repeat Steps 3 to 8 of Sec.~\ref{sm:app:psg-z2} with phases. Write $\phi_O(x,y,\mu)$ for the phase in Eq.~\eqref{sm:eq:u1-form}. Evaluating the relations with Eq.~\eqref{sm:eq:u1-product} and the site sequences of Sec.~\ref{sm:app:psg-z2} gives the recursions
\begin{align}
\phi_{C_6}(x,y,\mu)-\phi_{C_6}(x-1,y,\mu)&=s_{C_6}\Phi\,(x+y)-\theta_{T_1C_6},\nonumber\\
\phi_{C_6}(x,y,\mu)-\phi_{C_6}(x,y-1,\mu)&=\Phi x-\theta_{T_2C_6},\nonumber\\
\phi_\sigma(x,y,\mu)-\phi_\sigma(x-1,y,\mu)&=-s_\sigma\Phi y-\theta_{T_1\sigma},\nonumber\\
\phi_\sigma(x,y,\mu)-\phi_\sigma(x,y-1,\mu)&=\Phi x-\theta_{T_2\sigma},\nonumber\\
\phi_\T(x,y,\mu)-\phi_\T(x-1,y,\mu)&=-\theta_{T_1\T},\nonumber\\
\phi_\T(x,y,\mu)-\phi_\T(x,y-1,\mu)&=(1-s_\T)\Phi x-\theta_{T_2\T}.
\label{sm:eq:u1-recursions}
\end{align}
For example, the first line is the phase of $G_{C_6}(x-1,y,\mu)\,G_{T_2}(x+y,1-x,[\mu+2])\,G_{C_6}(x,y,\mu)^{-1}$, which by Eq.~\eqref{sm:eq:u1-product} equals $\phi_{C_6}(x-1,y,\mu)+s_{C_6}\Phi(x+y)-\phi_{C_6}(x,y,\mu)$; the factor $s_{C_6}$ arises because the phase of $G_{T_2}$ is moved through $c^{w_{C_6}}$. Taking the difference of the first line at $y$ and $y-1$, and of the second line at $x$ and $x-1$, gives $s_{C_6}\Phi$ and $\Phi$ for the same mixed second difference of $\phi_{C_6}$, which is the first condition of Eq.~\eqref{sm:eq:u1-flux-constraints}; the other pairs give the other two. With these conditions, $s_{C_6}\Phi\equiv\Phi$ and $-s_\sigma\Phi\equiv\Phi$ modulo $2\pi$, and the recursions integrate to
\begin{align}
\phi_{C_6}&=\Phi\big[\tfrac{x(x+1)}{2}+xy\big]-\theta_{T_1C_6}x-\theta_{T_2C_6}y+\rho_{C_6}(\mu),\nonumber\\
\phi_\sigma&=\Phi\,xy-\theta_{T_1\sigma}x-\theta_{T_2\sigma}y+\rho_\sigma(\mu),\nonumber\\
\phi_\T&=-\theta_{T_1\T}x-\theta_{T_2\T}y+\rho_\T(\mu),
\label{sm:eq:u1-xy-general}
\end{align}
with $\rho_O(\mu)\equiv\phi_O(0,0,\mu)$.

\paragraph*{The relation $\sigma^2$.}
Its phase is $\phi_\sigma(x,y,\mu)+s_\sigma\phi_\sigma(y,x,[9-\mu])+\pi w_\sigma$, whose $x,y$-dependent part is $(1+s_\sigma)\Phi xy-(\theta_{T_1\sigma}+s_\sigma\theta_{T_2\sigma})x-(\theta_{T_2\sigma}+s_\sigma\theta_{T_1\sigma})y$. The first term vanishes by Eq.~\eqref{sm:eq:u1-flux-constraints}, and the others vanish if and only if $\theta_{T_1\sigma}=-s_\sigma\theta_{T_2\sigma}$.

\paragraph*{Residual gauge freedom.}
The transformation $W_{(x,y,\mu)}=e^{-\ii(\kappa_1x+\kappa_2y)\tau^3}$, followed by the re-liftings $G_{T_1}\to e^{\ii\kappa_1\tau^3}G_{T_1}$ and $G_{T_2}\to e^{\ii\kappa_2\tau^3}G_{T_2}$, preserves $G_{T_1}=\tau^0$ and $G_{T_2}=e^{\ii\Phi x\tau^3}$. By Eq.~\eqref{sm:eq:psg-gauge} it changes $\phi_O(i)\to\phi_O(i)-\alpha(i)+s_O\,\alpha(O^{-1}(i))$ with $\alpha=\kappa_1x+\kappa_2y$, which shifts the linear coefficients of Eq.~\eqref{sm:eq:u1-xy-general} as
\begin{align}
\theta_{T_1C_6}&\to\theta_{T_1C_6}+(1-s_{C_6})\kappa_1+s_{C_6}\kappa_2,\nonumber\\
\theta_{T_2C_6}&\to\theta_{T_2C_6}+\kappa_2-s_{C_6}\kappa_1,\nonumber\\
\theta_{T_1\sigma}&\to\theta_{T_1\sigma}+\kappa_1-s_\sigma\kappa_2,\nonumber\\
\theta_{T_2\sigma}&\to\theta_{T_2\sigma}+\kappa_2-s_\sigma\kappa_1,\nonumber\\
\theta_{T_a\T}&\to\theta_{T_a\T}+(1-s_\T)\kappa_a,\quad a=1,2.
\label{sm:eq:u1-kappa}
\end{align}
We use it to set $\theta_{T_1C_6}=\theta_{T_2\sigma}=0$, and then $\theta_{T_1\sigma}=0$ by the $\sigma^2$ condition. For $s_{C_6}=+1$ this requires $\kappa_2=-\theta_{T_1C_6}$ and $\kappa_1=s_\sigma(\theta_{T_2\sigma}+\kappa_2)$. For $s_{C_6}=-1$ it requires $\kappa_2=\theta_{T_1C_6}+2\kappa_1$ and $(2-s_\sigma)\kappa_1=-\theta_{T_2\sigma}-\theta_{T_1C_6}$, which can always be solved because $U(1)$ is divisible. This generalizes Steps 6 and 8.

\paragraph*{The relation $(\sigma C_6)^2$.}
With the sites \eqref{sm:eq:z2-sc6-sites} and Eq.~\eqref{sm:eq:u1-product}, its phase is
\begin{align}
&\phi_\sigma(i)+s_\sigma\,\phi_{C_6}(i_1)+s_\sigma s_{C_6}\,\phi_\sigma(i_2)+s_{C_6}\,\phi_{C_6}(i_3)\nonumber\\
&\quad+\pi(w_\sigma+w_{C_6}).
\label{sm:eq:u1-sc6-phase}
\end{align}
Insert Eq.~\eqref{sm:eq:u1-xy-general} with $\theta_{T_1C_6}=\theta_{T_{1,2}\sigma}=0$. If $2\Phi=0$, the signs in front of the $\Phi$ terms are immaterial and these terms add up to $\Phi$ times $xy+[\frac{y(y+1)}{2}+xy]+[xy+y^2]+[\frac{y(y-1)}{2}+xy+y^2]=4xy+3y^2\equiv y$ modulo 2, while the $\theta_{T_2C_6}$ terms give $-\theta_{T_2C_6}[(s_\sigma+s_{C_6})x+s_{C_6}y]$. Independence of $y$ requires $\theta_{T_2C_6}=s_{C_6}\Phi=\Phi$, and then the $x$ term $-(s_\sigma+s_{C_6})\Phi x$ vanishes as well. In the sector $(0,1,0)$, where $\Phi$ is arbitrary, $s_\sigma=-1$ and $s_{C_6}=+1$, the $\Phi$ terms are
\begin{align}
&\Phi\Big\{xy-\big[\tfrac{y(y+1)}{2}+xy\big]+\big[xy+y^2\big]\nonumber\\
&\qquad+\big[\tfrac{y(y-1)}{2}-xy-y^2\big]\Big\}=-\Phi y,
\end{align}
and the $\theta_{T_2C_6}$ terms are $\theta_{T_2C_6}x-\theta_{T_2C_6}(x+y)=-\theta_{T_2C_6}y$, so that $\theta_{T_2C_6}=-\Phi$. In every sector, therefore, $-\theta_{T_2C_6}\,y\equiv\Phi\,y$, and
\begin{equation}
\phi_{C_6}(x,y,\mu)=\Phi\Big[\tfrac{x(x+1)}{2}+xy+y\Big]+\rho_{C_6}(\mu),
\end{equation}
which is the $U(1)$ version of Eq.~\eqref{sm:eq:z2-c6-final}.

\paragraph*{The relations involving $\T$.}
Write $a\equiv\theta_{T_1\T}$ and $b\equiv\theta_{T_2\T}$. Since $1-s_\T\neq0$ implies $2\Phi=0$, the last recursion of Eq.~\eqref{sm:eq:u1-recursions} reduces to the form given in Eq.~\eqref{sm:eq:u1-xy-general}. The phase of $C_6\T C_6^{-1}\T^{-1}$ is
\begin{equation}
(1-s_\T)\,\phi_{C_6}(i)+s_{C_6}\,\phi_\T(C_6^{-1}(i))-\phi_\T(i).
\label{sm:eq:u1-c6th-phase}
\end{equation}
Its first term is independent of $x,y$: it vanishes for $s_\T=+1$, and for $s_\T=-1$ it equals $2\phi_{C_6}(i)\equiv2\rho_{C_6}(\mu)$ because $2\Phi=0$. With $C_6^{-1}(x,y)=(x+y,-x)$, the coefficients of $x$ and $y$ in the remaining terms are $a(1-s_{C_6})+s_{C_6}b$ and $b-s_{C_6}a$. In the same way, $\sigma\T\sigma^{-1}\T^{-1}$ produces the coefficients $a-s_\sigma b$ and $b-s_\sigma a$. All four must vanish. For $s_{C_6}=+1$ this gives $a=b=0$. For $s_{C_6}=-1$ it gives $b=-a$ and $3a=0$, and the mirror condition $a=s_\sigma b=-s_\sigma a$ gives $2a=0$ if $s_\sigma=+1$, hence $a=0$. If $s_\T=+1$, the relation $\T^2$, whose phase is $2\phi_\T(i)$, requires $2a=2b=0$ and hence $a=0$ as well. The only case left is the sector $(1,1,1)$, where $a\in\{0,2\pi/3,4\pi/3\}$ and $b=-a$ are allowed by the relations. There the residual transformation of Eq.~\eqref{sm:eq:u1-kappa} with $\kappa_2=2\kappa_1$ and $3\kappa_1=0$ leaves $\theta_{T_1C_6}$, $\theta_{T_2C_6}$, $\theta_{T_1\sigma}$, and $\theta_{T_2\sigma}$ unchanged (each shift is a multiple of $3\kappa_1$) and maps $a\to a+2\kappa_1$; choosing $2\kappa_1=-a$ sets $a=b=0$. Hence $\phi_\T(x,y,\mu)=\rho_\T(\mu)$ in every sector.

\paragraph*{Summary of the reduction.}
Every $U(1)$ PSG in the sector $(w_{C_6},w_\sigma,w_\T)$ is gauge equivalent to
\begin{align}
G_{T_1}(x,y,\mu)&=\tau^0,\nonumber\\
G_{T_2}(x,y,\mu)&=e^{\ii\Phi x\tau^3},\nonumber\\
G_{C_6}(x,y,\mu)&=e^{\ii\Phi[x(x+1)/2+xy+y]\tau^3}\,g_{C_6}(\mu),\nonumber\\
G_\sigma(x,y,\mu)&=e^{\ii\Phi xy\tau^3}\,g_\sigma(\mu),\nonumber\\
G_\T(x,y,\mu)&=g_\T(\mu),
\label{sm:eq:u1-reduced}
\end{align}
with $g_O(\mu)=e^{\ii\rho_O(\mu)\tau^3}c^{w_O}$. Outside the sector $(0,1,0)$, $\Phi\in\{0,\pi\}$ and $e^{\ii\Phi(\cdots)\tau^3}=n_{T_2}^{(\cdots)}$ with $n_{T_2}=e^{\ii\Phi}$, exactly as in Eq.~\eqref{eq:z2psg} of the main text. The $x,y$-dependent parts of the six relations \eqref{sm:eq:u1-pg-values} now vanish: for $C_6^6$ by Eq.~\eqref{sm:eq:z2-orbit-sum}, which is an identity in $x$ and $y$, and for the others by the computations above. The six phases \eqref{sm:eq:u1-pg-values} are therefore relations among the 36 sublattice matrices $g_O(\mu)$ alone.

\subsection{Sublattice Problem: Relation Values, Identities, and Re-liftings}
\label{sm:app:psg-u1-sub}

\paragraph*{Tree gauge.}
The transformations $W_\mu=e^{-\ii\alpha_\mu\tau^3}$ independent of $x,y$ preserve Eq.~\eqref{sm:eq:u1-reduced} and act as $\rho_O(\mu)\to\rho_O(\mu)-\alpha_\mu+s_O\,\alpha_{O^{-1}(\mu)}$. Along the eleven tree edges of Sec.~\ref{sm:app:psg-tree} one can choose $\alpha_\mu$ so that the tree matrices become $g_{C_6}(\mu)=c^{w_{C_6}}$ ($\mu\neq6,11$) and $g_\sigma(1)=c^{w_\sigma}$; the starting value $\alpha_1$ remains free. The seven non-tree matrices then follow from the relation values $\theta_{C_6}$, $\theta_\sigma$, $\theta_{\sigma C_6}$, by the chain of Eq.~\eqref{sm:eq:z2-gs-chain} with phases. The relations $C_6\T C_6^{-1}\T^{-1}$ and $\sigma\T\sigma^{-1}\T^{-1}$ determine $g_\T(\mu)$ from $g_\T(1)=e^{\ii\rho\tau^3}c^{w_\T}$ and $\theta_{\T C_6}$, $\theta_{\T\sigma}$. If $w_\T=0$, the relation $\T^2$ fixes $2\rho=\theta_\T$, which leaves only the sign of $g_\T(1)$, a re-lifting by $-\tau^0$; if $w_\T=1$, the free $\alpha_1$ changes $\rho\to\rho-2\alpha_1$ and sets $\rho=0$. Consequently:

\emph{A $U(1)$ PSG in a given sector is determined, up to gauge transformations and re-liftings, by $\Phi$ and the six phases \eqref{sm:eq:u1-pg-values}.}

The classification thus reduces to two questions: which values of the six phases are consistent, and which values are related by re-liftings.

\paragraph*{Re-liftings.}
The re-liftings $G_{C_6}\to e^{\ii\beta_{C_6}\tau^3}G_{C_6}$, $G_\sigma\to e^{\ii\beta_\sigma\tau^3}G_\sigma$, and $G_\T\to e^{\ii\beta_\T\tau^3}G_\T$ leave $\Phi$ and the mixed translation relations unchanged and shift the six phases. Writing $\hat C\equiv\hat C_6$, $\hat S\equiv\hat\sigma$, $\hat\T$ for the PSG operators and moving the phases to the left with $\hat O\,e^{\ii\beta\tau^3}=e^{\ii s_O\beta\tau^3}\hat O$, one finds, for example, $(e^{\ii\beta_{C_6}\tau^3}\hat C)^6=e^{\ii\beta_{C_6}(1+s_{C_6}+\dots+s_{C_6}^5)\tau^3}\hat C^6$. With
\begin{equation}
\Sigma\equiv\sum_{k=0}^{5}s_{C_6}^k=\begin{cases}6,&w_{C_6}=0,\\0,&w_{C_6}=1,\end{cases}
\end{equation}
the shifts are
\begin{align}
\delta\theta_{C_6}&=\Sigma\,\beta_{C_6},\nonumber\\
\delta\theta_\sigma&=(1+s_\sigma)\,\beta_\sigma,\nonumber\\
\delta\theta_{\sigma C_6}&=(1+s_\sigma s_{C_6})(\beta_\sigma+s_\sigma\beta_{C_6}),\nonumber\\
\delta\theta_{\T C_6}&=(1-s_\T)\beta_{C_6}-(1-s_{C_6})\beta_\T,\nonumber\\
\delta\theta_{\T\sigma}&=(1-s_\T)\beta_\sigma-(1-s_\sigma)\beta_\T,\nonumber\\
\delta\theta_\T&=(1+s_\T)\beta_\T.
\label{sm:eq:u1-relift}
\end{align}
For $\delta\theta_{\T C_6}$, for instance, $e^{\ii\beta\tau^3}\hat C\,\hat\T\,\hat C^{-1}e^{-\ii\beta\tau^3}\hat\T^{-1}=e^{\ii\beta\tau^3}\,e^{\ii\theta_{\T C_6}\tau^3}\hat\T\,e^{-\ii\beta\tau^3}\hat\T^{-1}=e^{\ii(\theta_{\T C_6}+\beta-s_\T\beta)\tau^3}$ for $\beta=\beta_{C_6}$.

\paragraph*{Identities among relation values.}
The six phases are not independent. Each identity below follows from evaluating one product of PSG operators in two ways, using only the defining relations and $\hat O\,e^{\ii\theta\tau^3}\hat O^{-1}=e^{\ii s_O\theta\tau^3}$.
\begin{itemize}
\item[(a)] $\hat C$ commutes with $\hat C^6=e^{\ii\theta_{C_6}\tau^3}$, but conjugation by $\hat C$ maps it to $e^{\ii s_{C_6}\theta_{C_6}\tau^3}$: $(1-s_{C_6})\theta_{C_6}=0$.
\item[(b)] In the same way, $(1-s_\sigma)\theta_\sigma=0$.
\item[(c)] $\hat S\hat C$ commutes with $(\hat S\hat C)^2$: $(1-s_\sigma s_{C_6})\theta_{\sigma C_6}=0$.
\item[(d)] $\hat\T$ commutes with $\hat\T^2$: $(1-s_\T)\theta_\T=0$.
\item[(e)] From $\hat C\hat\T\hat C^{-1}=e^{\ii\theta_{\T C_6}\tau^3}\hat\T$, $\hat C\hat\T^2\hat C^{-1}=(e^{\ii\theta_{\T C_6}\tau^3}\hat\T)^2=e^{\ii(1+s_\T)\theta_{\T C_6}\tau^3}\hat\T^2$; the left-hand side is $e^{\ii s_{C_6}\theta_\T\tau^3}$: $(1+s_\T)\theta_{\T C_6}=-(1-s_{C_6})\theta_\T$.
\item[(f)] In the same way, $(1+s_\T)\theta_{\T\sigma}=-(1-s_\sigma)\theta_\T$.
\item[(g)] From the definition of $\theta_{\T C_6}$, $\hat\T\hat C\hat\T^{-1}=e^{-\ii\theta_{\T C_6}\tau^3}\hat C$, so $\hat\T\hat C^6\hat\T^{-1}=e^{-\ii\Sigma\theta_{\T C_6}\tau^3}\hat C^6$; the left-hand side is $e^{\ii s_\T\theta_{C_6}\tau^3}$: $(1-s_\T)\theta_{C_6}=\Sigma\,\theta_{\T C_6}$.
\item[(h)] In the same way, $(1-s_\T)\theta_\sigma=(1+s_\sigma)\theta_{\T\sigma}$.
\item[(i)] $\hat\T(\hat S\hat C)^2\hat\T^{-1}=(e^{-\ii\theta_{\T\sigma}\tau^3}\hat S\,e^{-\ii\theta_{\T C_6}\tau^3}\hat C)^2$ gives $(1-s_\T)\theta_{\sigma C_6}=(1+s_\sigma s_{C_6})(\theta_{\T\sigma}+s_\sigma\theta_{\T C_6})$.
\item[(j)] From $(\hat S\hat C)^2=e^{\ii\theta_{\sigma C_6}\tau^3}$ and $\hat S^{-2}=e^{-\ii\theta_\sigma\tau^3}$, $\hat S\hat C\hat S^{-1}=e^{\ii(\theta_{\sigma C_6}-s_{C_6}\theta_\sigma)\tau^3}\hat C^{-1}$; raising to the sixth power, $\hat S\hat C^6\hat S^{-1}=e^{\ii\Sigma(\theta_{\sigma C_6}-s_{C_6}\theta_\sigma)\tau^3}\hat C^{-6}$, and the left-hand side is $e^{\ii s_\sigma\theta_{C_6}\tau^3}$: $(1+s_\sigma)\theta_{C_6}=\Sigma(\theta_{\sigma C_6}-s_{C_6}\theta_\sigma)$.
\item[(k)] On-site realizability: if $w_\T=1$, $G_\T(i)^2=e^{\ii\rho\tau^3}c\,e^{\ii\rho\tau^3}c=c^2=-\tau^0$, so $\theta_\T=\pi$.
\end{itemize}
All identities hold modulo $2\pi$. Each is invariant under the re-liftings \eqref{sm:eq:u1-relift}, as it must be. Identities (a) to (j) follow from the relations alone and hold for every class of $H^2(p6mm\times\Z_2^\T,U(1))$. Identity (k) is different: it is the on-site condition for time reversal, which fixes every site of the SHD lattice, and it removes the classes with $\theta_\T=0$ in the four sectors with $w_\T=1$ (Sec.~IV\,C of the main text).

\paragraph*{Counting and representatives.}
In each sector we use the re-liftings to set some phases to fixed values, use the identities to restrict the remaining phases, and check that no re-lifting preserving the normalization changes them. In each case the remaining phases are $0$ or $\pi$, and the normalized values coincide with those of the sign form
\begin{equation}
G_O(x,y,\mu)=G_O^{\Z_2}(x,y,\mu)\,(\ii\tau^1)^{w_O},
\end{equation}
which is Eq.~\eqref{eq:u1psg} of the main text,
where $G_O^{\Z_2}$ is Eq.~\eqref{eq:z2psg} of the main text at $g_\T=\tau^0$. The relation values of these sign forms follow from Eq.~\eqref{sm:eq:u1-product}: the signs contribute the $\Z_2$ labels, and the factors $c$ contribute $c^6=-\tau^0$ to $C_6^6$ if $w_{C_6}=1$, $c^2=-\tau^0$ to $\sigma^2$ if $w_\sigma=1$, $(-1)^{w_\sigma+w_{C_6}}$ to $(\sigma C_6)^2$, and $-\tau^0$ to $\T^2$ if $w_\T=1$, and nothing to the relations with $\T$ conjugation. Writing $\bar n\equiv0$ for $n=+1$ and $\bar n\equiv\pi$ for $n=-1$,
\begin{align}
\theta_{C_6}&=\bar n_{C_6}+\pi w_{C_6},\nonumber\\
\theta_\sigma&=\bar n_\sigma+\pi w_\sigma,\nonumber\\
\theta_{\sigma C_6}&=\bar n_{\sigma C_6}+\pi(w_\sigma+w_{C_6}),\nonumber\\
\theta_{\T C_6}&=\bar n_{\T C_6},\nonumber\\
\theta_{\T\sigma}&=\bar n_{\T\sigma},\nonumber\\
\theta_\T&=\pi w_\T,
\label{sm:eq:u1-signform-values}
\end{align}
and $\Phi=\theta_{T_2C_6}=\bar n_{T_2}$, while the other mixed translation relations and $T_a\T T_a^{-1}\T^{-1}$ have trivial values. Since the normalized phases take all values allowed by the identities, and Eq.~\eqref{sm:eq:u1-signform-values} realizes each such value, every $U(1)$ class has a representative of the form (16) of the main text.

\emph{Sector $(0,0,0)$}: $s_{C_6}=s_\sigma=s_\T=1$, $\Sigma=6$. The identities reduce to $2\theta_{\T C_6}=0$ (e), $2\theta_{\T\sigma}=0$ (f), and $2\theta_{C_6}=6(\theta_{\sigma C_6}-\theta_\sigma)$ (j). The re-liftings shift $\theta_{C_6}$ by $6\beta_{C_6}$, $\theta_\sigma$ by $2\beta_\sigma$, $\theta_{\sigma C_6}$ by $2(\beta_\sigma+\beta_{C_6})$, and $\theta_\T$ by $2\beta_\T$. Choose $\beta_\T$, $\beta_\sigma$, and then $\beta_{C_6}$ to set $\theta_\T=\theta_\sigma=\theta_{\sigma C_6}=0$. Then (j) gives $2\theta_{C_6}=0$. The re-liftings preserving this normalization have $\beta_\sigma,\beta_{C_6}\in\{0,\pi\}$ and change nothing. Classes: $\Phi,\theta_{C_6},\theta_{\T C_6},\theta_{\T\sigma}\in\{0,\pi\}$, i.e., $2^4=16$. The re-lifting-invariant combination $\theta_{C_6}-3\theta_{\sigma C_6}+3\theta_\sigma$ equals $\bar n_{C_6}+\bar n_{\sigma C_6}+\bar n_\sigma$ on sign forms.

\emph{Sector $(0,0,1)$}: $\Sigma=6$, $\theta_\T=\pi$. Identities: $2\theta_{C_6}=6\theta_{\T C_6}$ (g), $2\theta_\sigma=2\theta_{\T\sigma}$ (h), $2\theta_{\sigma C_6}=2(\theta_{\T\sigma}+\theta_{\T C_6})$ (i). Re-liftings: $\delta\theta_{C_6}=6\beta_{C_6}$, $\delta\theta_\sigma=2\beta_\sigma$, $\delta\theta_{\sigma C_6}=2(\beta_\sigma+\beta_{C_6})$, $\delta\theta_{\T C_6}=2\beta_{C_6}$, $\delta\theta_{\T\sigma}=2\beta_\sigma$. Set $\theta_{\T C_6}=\theta_{\T\sigma}=0$; then $\theta_{C_6},\theta_\sigma,\theta_{\sigma C_6}\in\{0,\pi\}$. Classes: $2^4=16$. Invariants: $\theta_{C_6}-3\theta_{\T C_6}$, $\theta_\sigma-\theta_{\T\sigma}$, $\theta_{\sigma C_6}-\theta_{\T C_6}-\theta_{\T\sigma}$, and $\Phi$.

\emph{Sector $(0,1,0)$}: $s_\sigma=-1$, $\Sigma=6$, $\Phi$ arbitrary. Identities: $2\theta_\sigma=0$ (b), $2\theta_{\sigma C_6}=0$ (c), $2\theta_{\T C_6}=0$ (e), $2\theta_{\T\sigma}=-2\theta_\T$ (f). Re-liftings: $\delta\theta_{C_6}=6\beta_{C_6}$, $\delta\theta_{\T\sigma}=-2\beta_\T$, $\delta\theta_\T=2\beta_\T$, others zero. Set $\theta_{C_6}=\theta_\T=0$; then $\theta_\sigma,\theta_{\sigma C_6},\theta_{\T C_6},\theta_{\T\sigma}\in\{0,\pi\}$. Classes: 16 branches, each with a continuous $\Phi$. Invariants: $\theta_\sigma$, $\theta_{\sigma C_6}$, $\theta_{\T C_6}$, $\theta_{\T\sigma}+\theta_\T$, and $\Phi$.

\emph{Sector $(0,1,1)$}: $s_\sigma=s_\T=-1$, $\Sigma=6$, $\theta_\T=\pi$. Identities: $2\theta_\sigma=0$ (b), $2\theta_{\sigma C_6}=0$ (c), $2\theta_{C_6}=6\theta_{\T C_6}$ (g). Re-liftings: $\delta\theta_{C_6}=6\beta_{C_6}$, $\delta\theta_{\T C_6}=2\beta_{C_6}$, $\delta\theta_{\T\sigma}=2\beta_\sigma-2\beta_\T$. Set $\theta_{\T C_6}=\theta_{\T\sigma}=0$; then $\theta_{C_6},\theta_\sigma,\theta_{\sigma C_6}\in\{0,\pi\}$. Classes: 16. Invariants: $\theta_{C_6}-3\theta_{\T C_6}$, $\theta_\sigma$, $\theta_{\sigma C_6}$, $\Phi$.

\emph{Sector $(1,0,0)$}: $s_{C_6}=-1$, $\Sigma=0$. Identities: $2\theta_{C_6}=0$ (a), $2\theta_{\sigma C_6}=0$ (c), $2\theta_{\T C_6}=-2\theta_\T$ (e), $2\theta_{\T\sigma}=0$ (f). Re-liftings: $\delta\theta_\sigma=2\beta_\sigma$, $\delta\theta_{\T C_6}=-2\beta_\T$, $\delta\theta_\T=2\beta_\T$. Set $\theta_\sigma=\theta_\T=0$; then $\theta_{C_6},\theta_{\sigma C_6},\theta_{\T C_6},\theta_{\T\sigma}\in\{0,\pi\}$. Classes: $2^5=32$. Invariants: $\theta_{C_6}$, $\theta_{\sigma C_6}$, $\theta_{\T C_6}+\theta_\T$, $\theta_{\T\sigma}$, $\Phi$.

\emph{Sector $(1,0,1)$}: $s_{C_6}=s_\T=-1$, $\Sigma=0$, $\theta_\T=\pi$. Identities: $2\theta_{C_6}=0$ (a), $2\theta_{\sigma C_6}=0$ (c), $2\theta_\sigma=2\theta_{\T\sigma}$ (h). Re-liftings: $\delta\theta_\sigma=2\beta_\sigma$, $\delta\theta_{\T C_6}=2\beta_{C_6}-2\beta_\T$, $\delta\theta_{\T\sigma}=2\beta_\sigma$. Set $\theta_{\T C_6}=\theta_{\T\sigma}=0$; then $\theta_{C_6},\theta_\sigma,\theta_{\sigma C_6}\in\{0,\pi\}$. Classes: 16. Invariants: $\theta_{C_6}$, $\theta_{\sigma C_6}$, $\theta_\sigma-\theta_{\T\sigma}$, $\Phi$.

\emph{Sector $(1,1,0)$}: $s_{C_6}=s_\sigma=-1$, $\Sigma=0$. Identities: $2\theta_{C_6}=0$ (a), $2\theta_\sigma=0$ (b), $2\theta_{\T C_6}=-2\theta_\T$ (e), $2\theta_{\T\sigma}=-2\theta_\T$ (f); (c) is empty because $s_\sigma s_{C_6}=1$. Re-liftings: $\delta\theta_{\sigma C_6}=2(\beta_\sigma-\beta_{C_6})$, $\delta\theta_{\T C_6}=\delta\theta_{\T\sigma}=-2\beta_\T$, $\delta\theta_\T=2\beta_\T$. Set $\theta_{\sigma C_6}=\theta_\T=0$; then $\theta_{C_6},\theta_\sigma,\theta_{\T C_6},\theta_{\T\sigma}\in\{0,\pi\}$. Classes: $2^5=32$. Invariants: $\theta_{C_6}$, $\theta_\sigma$, $\theta_{\T C_6}+\theta_\T$, $\theta_{\T\sigma}+\theta_\T$, $\Phi$.

\emph{Sector $(1,1,1)$}: all $s_O=-1$, $\Sigma=0$, $\theta_\T=\pi$. Identities: $2\theta_{C_6}=0$ (a), $2\theta_\sigma=0$ (b), $2\theta_{\sigma C_6}=2(\theta_{\T\sigma}-\theta_{\T C_6})$ (i). Re-liftings: $\delta\theta_{\sigma C_6}=2(\beta_\sigma-\beta_{C_6})$, $\delta\theta_{\T C_6}=2(\beta_{C_6}-\beta_\T)$, $\delta\theta_{\T\sigma}=2(\beta_\sigma-\beta_\T)$. Set $\theta_{\T C_6}=\theta_{\T\sigma}=0$; then $\theta_{C_6},\theta_\sigma,\theta_{\sigma C_6}\in\{0,\pi\}$. Classes: 16. Invariants: $\theta_{C_6}$, $\theta_\sigma$, $\theta_{\sigma C_6}+\theta_{\T C_6}-\theta_{\T\sigma}$, $\Phi$.

Gauge transformations within the canonical form are $U(1)$ matrices, which leave every phase unchanged, possibly combined with the global transformation $W_i=\ii\tau^1$, which reverses the sign of every phase and is therefore immaterial for phases equal to $0$ or $\pi$. In every sector the listed invariants are unchanged by Eq.~\eqref{sm:eq:u1-relift} and take two values each, as required by the identities, so the classes counted above are pairwise inequivalent. The results are collected in Table~\ref{sm:tab:u1-sectors}, which supplements Table~III of the main text with the identifications and invariants. The total number of $U(1)$ PSGs is $16+16+16+32+16+32+16=144$ isolated classes in the seven sectors with $\Phi\in\{0,\pi\}$, together with 16 one-parameter families in the sector $(0,1,0)$.

\begin{table*}[t]
\caption{$U(1)$ PSGs. For each sector $(w_{C_6},w_\sigma,w_\T)$ the representatives are the sign forms of Eq.~\eqref{eq:u1psg} of the main text; the labels not listed as free are set to $+1$. The third column lists the sets of $\Z_2$ labels whose simultaneous flip is a re-lifting (and hence describes the same PSG), together with the re-lifting $G_O\to\ii\tau^3G_O$ that produces it. The fourth column gives a complete set of invariants, evaluated on the sign forms. In the sector $(0,1,0)$ the flux $\Phi$ is continuous and $n_{T_2}=-1$ is the point $\Phi=\pi$ of the family.}
\label{sm:tab:u1-sectors}
\begin{ruledtabular}
\begin{tabular}{cllll}
sector & free labels & trivial flips (re-lifting) & invariants on sign forms & number\\
\hline
$(0,0,0)$ & $n_{T_2},n_{\T\sigma},n_{\T C_6},n_{C_6}$ & $\{n_\sigma,n_{\sigma C_6}\}$ ($\sigma$), $\{n_{C_6},n_{\sigma C_6}\}$ ($C_6$) & $n_{T_2}, n_{\T C_6}, n_{\T\sigma}, n_{C_6}n_\sigma n_{\sigma C_6}$ & 16\\
$(0,0,1)$ & $n_{T_2},n_\sigma,n_{\sigma C_6},n_{C_6}$ & \vtop{\hbox{$\{n_\sigma,n_{\sigma C_6},n_{\T\sigma}\}$ ($\sigma$),}\hbox{$\{n_{C_6},n_{\sigma C_6},n_{\T C_6}\}$ ($C_6$)}} & $n_{T_2}, n_{C_6}n_{\T C_6}, n_\sigma n_{\T\sigma}, n_{\sigma C_6}n_{\T C_6}n_{\T\sigma}$ & 16\\
$(0,1,0)$ & $\Phi;\ n_{\T\sigma},n_{\T C_6},n_\sigma,n_{\sigma C_6}$ & $\{n_{C_6}\}$ ($C_6$) & $\Phi, n_\sigma, n_{\sigma C_6}, n_{\T C_6}, n_{\T\sigma}$ & $16\times U(1)$\\
$(0,1,1)$ & $n_{T_2},n_\sigma,n_{\sigma C_6},n_{C_6}$ & $\{n_{\T\sigma}\}$ ($\sigma$ or $\T$), $\{n_{C_6},n_{\T C_6}\}$ ($C_6$) & $n_{T_2}, n_\sigma, n_{\sigma C_6}, n_{C_6}n_{\T C_6}$ & 16\\
$(1,0,0)$ & $n_{T_2},n_{\T\sigma},n_{\T C_6},n_{\sigma C_6},n_{C_6}$ & $\{n_\sigma\}$ ($\sigma$) & $n_{T_2}, n_{C_6}, n_{\sigma C_6}, n_{\T C_6}, n_{\T\sigma}$ & 32\\
$(1,0,1)$ & $n_{T_2},n_\sigma,n_{\sigma C_6},n_{C_6}$ & $\{n_{\T C_6}\}$ ($C_6$), $\{n_\sigma,n_{\T\sigma}\}$ ($\sigma$) & $n_{T_2}, n_{C_6}, n_{\sigma C_6}, n_\sigma n_{\T\sigma}$ & 16\\
$(1,1,0)$ & $n_{T_2},n_{\T\sigma},n_{\T C_6},n_\sigma,n_{C_6}$ & $\{n_{\sigma C_6}\}$ ($\sigma$ or $C_6$) & $n_{T_2}, n_{C_6}, n_\sigma, n_{\T C_6}, n_{\T\sigma}$ & 32\\
$(1,1,1)$ & $n_{T_2},n_\sigma,n_{\sigma C_6},n_{C_6}$ & $\{n_{\sigma C_6},n_{\T C_6}\}$ ($C_6$), $\{n_{\sigma C_6},n_{\T\sigma}\}$ ($\sigma$) & $n_{T_2}, n_{C_6}, n_\sigma, n_{\sigma C_6}n_{\T C_6}n_{\T\sigma}$ & 16\\
\end{tabular}
\end{ruledtabular}
\end{table*}

\subsection{Why Some $\Z_2$ Labels Become Redundant}
\label{sm:app:psg-u1-ident}

A $\Z_2$ label is a relation value that is invariant under re-liftings by $\pm\tau^0$. With a $U(1)$ IGG more re-liftings are available, in particular by $e^{\ii(\pi/2)\tau^3}=\ii\tau^3$, and a relation value can change by $\pi$ under such a re-lifting. By Eq.~\eqref{sm:eq:u1-relift}, the re-lifting $G_O\to\ii\tau^3G_O$ ($\beta_O=\pi/2$) shifts a phase by $\pi/2$ times an integer, which is $\pm\pi$ whenever the integer is $\pm2$ (or $\pm6$). The PSG obtained after the re-lifting describes the same symmetry operations, so it is the same PSG; it has the relation values of the sign form with the corresponding labels flipped, and by the tree-gauge statement of Sec.~\ref{sm:app:psg-u1-sub} it is gauge equivalent to that sign form. Two representative cases:
\begin{itemize}
\item Sector $(0,0,0)$, re-lifting $G_\sigma\to\ii\tau^3G_\sigma$. Since $w_\sigma=0$, $G_\sigma$ commutes with $\ii\tau^3$, so $G_\sigma(i)G_\sigma(\sigma i)$ acquires $(\ii\tau^3)^2=-\tau^0$ and $\sigma^2$ flips. The word $(\sigma C_6)^2$ contains two $G_\sigma$ factors and, with $w_{C_6}=0$, also acquires $(\ii\tau^3)^2=-\tau^0$, so $(\sigma C_6)^2$ flips as well. In $\sigma\T\sigma^{-1}\T^{-1}$ the factors $\ii\tau^3$ and $(\ii\tau^3)^{-1}$ cancel because $w_\T=0$. Hence $(n_\sigma,n_{\sigma C_6})\to(-n_\sigma,-n_{\sigma C_6})$ is a re-lifting. Equivalently, the gauge transformation $W_i=\ii\tau^3$ on the even (B) sublattices and $W_i=\tau^0$ on the odd (A) ones, followed by the re-lifting $G_\sigma\to\ii\tau^3G_\sigma$, multiplies $G_\sigma(x,y,\mu)$ by $-1$ for even $\mu$ and leaves all other matrices unchanged; this flips $g_\sigma(\mu)g_\sigma([9-\mu])$ and the product in Eq.~\eqref{sm:eq:z2-sc6}, since $\sigma$ exchanges odd and even sublattices.
\item Sector $(0,0,0)$, re-lifting $G_{C_6}\to\ii\tau^3G_{C_6}$. $C_6^6$ acquires $(\ii\tau^3)^6=-\tau^0$ and $(\sigma C_6)^2$ acquires $(\ii\tau^3)^2=-\tau^0$, so $(n_{C_6},n_{\sigma C_6})$ flip together. In the sector $(1,0,0)$ the same re-lifting changes nothing, because $\Sigma=0$ and $1+s_\sigma s_{C_6}=0$; this is why $n_{C_6}$ remains a label there.
\end{itemize}
The other entries of the third column of Table~\ref{sm:tab:u1-sectors} are read off from Eq.~\eqref{sm:eq:u1-relift} in the same way. For example, in the sector $(0,1,1)$ the re-lifting of $\T$ gives $\delta\theta_{\T\sigma}=-(1-s_\sigma)\pi/2=-\pi$ and leaves the other phases unchanged ($\delta\theta_\T=0$ because $s_\T=-1$), so $n_{\T\sigma}$ is redundant. In the sectors with $w_\T=0$ the re-lifting $G_\T\to\ii\tau^3G_\T$ changes $\theta_\T$ by $\pi$: the $U(1)$ analog of the $\Z_2$ choice $g_\T=\ii\tau^2$ is $g_\T=\ii\tau^3$, which belongs to the IGG, so $\T^2$ is not a $U(1)$ label.

\subsection{The Continuous-Flux Sector}
\label{sm:app:psg-u1-cont}

In the sector $(w_{C_6},w_\sigma,w_\T)=(0,1,0)$ the reduction~\eqref{sm:eq:u1-reduced} together with the normalization $\theta_{C_6}=\theta_\T=0$ of Sec.~\ref{sm:app:psg-u1-sub} gives the representatives of Eq.~\eqref{eq:u1flux} of the main text, completed by $G_{T_1}=\tau^0$ and $G_\T=h(\mu)\,\tau^0$, with $g_{C_6}(\mu)$, $g_\sigma(\mu)$, and $h(\mu)$ from Table~II of the main text at $n_{C_6}=+1$ and the four free signs $n_{\T\sigma},n_{\T C_6},n_\sigma,n_{\sigma C_6}$; $\Phi=0,\pi$ reproduce the sign forms with $n_{T_2}=\pm1$.

\subsection{PSGs of the Space Group Alone and Chiral Classes}
\label{sm:app:psg-sg}

\paragraph*{Chiral symmetry groups.}
The chiral classes A--D of Sec.~III of the main text have the symmetry groups $G_\chi=\{g\,\T^{(1-\chi(g))/2}:g\in p6mm\}$, which are closed if and only if $\chi$ is a homomorphism $p6mm\to\{\pm1\}$, and then $G_\chi\cong p6mm$. Because time reversal composes like a unitary operation in the gauge algebra (Sec.~\ref{sm:app:psg-conv}), the algebraic PSGs of $G_\chi$ are those of $p6mm$ for every $\chi$; $\chi$ enters only the conditions on the mean-field amplitudes.

\paragraph*{Algebraic PSGs of $p6mm$.}
Dropping $\T$ and the relations containing it, the derivations above apply unchanged. For a $\Z_2$ IGG the labels are $n_{T_2},n_\sigma,n_{\sigma C_6},n_{C_6}$, which gives $2^4=16$ classes, in agreement with $|H^2(p6mm,\Z_2)|=2^4$ (Sec.~\ref{sm:app:psg-z2inv}). For an $SU(2)$ IGG the argument of Sec.~\ref{sm:app:psg-su2} again gives 16 classes. For a $U(1)$ IGG the sectors are labeled by $(w_{C_6},w_\sigma)$, the flux constraints are the first two of Eq.~\eqref{sm:eq:u1-flux-constraints}, and the relevant identities are (a), (b), (c), and (j), with the re-liftings $\beta_{C_6},\beta_\sigma$ of Eq.~\eqref{sm:eq:u1-relift}:
\begin{itemize}
\item $(0,0)$: (j) reads $2\theta_{C_6}=6(\theta_{\sigma C_6}-\theta_\sigma)$; setting $\theta_\sigma=\theta_{\sigma C_6}=0$ leaves $\theta_{C_6},\Phi\in\{0,\pi\}$: 4 classes, labels $n_{T_2},n_{C_6}$.
\item $(0,1)$: $2\theta_\sigma=2\theta_{\sigma C_6}=0$ and $\theta_{C_6}$ is removed by $\beta_{C_6}$; $\Phi$ is continuous: 4 branches, labels $n_\sigma,n_{\sigma C_6}$, each with a flux $\Phi$.
\item $(1,0)$: $2\theta_{C_6}=2\theta_{\sigma C_6}=0$ and $\theta_\sigma$ is removed by $\beta_\sigma$: 8 classes, labels $n_{T_2},n_{\sigma C_6},n_{C_6}$.
\item $(1,1)$: $2\theta_{C_6}=2\theta_\sigma=0$ and $\theta_{\sigma C_6}$ is removed by $\beta_\sigma-\beta_{C_6}$: 8 classes, labels $n_{T_2},n_\sigma,n_{C_6}$.
\end{itemize}
Representatives are the sign forms of Eq.~\eqref{eq:u1psg} and, in the sector $(0,1)$, the continuous-flux forms of Eq.~\eqref{eq:u1flux} of the main text, both without $G_\T$.

\section{Bond Transformation Rules and Allowed Amplitudes}
\label{sm:sec:bonds}

\subsection{Bond Transformation Rules}
\label{sm:app:bondrules}

The transformation rules used to construct the ans\"atze of Sec.~\ref{sec:ansatz} of the main text are listed in Table~\ref{sm:tab:bondrules}.

\begin{table*}
\caption{Bond transformation rules for the $\Z_2$ and $SU(2)$ projective symmetry groups of Eq.~\eqref{eq:z2psg} of the main text. Every bond $i\to j$ of an orbit with $i$ in the reference unit cell carries $u^{(0)}_{ij}=s\,u^{(0)}_{\rm ref}$, with the sign $s$ given; the subscript on $j$ gives its unit cell when it is not $(0,0)$. For $U(1)$ PSGs the same signs apply, with $\tau^3\to-\tau^3$ for each factor $\ii\tau^1$ of the transforming element and, in the sector $(0,1,0)$ at generic $\Phi$, the additional phases of Eq.~\eqref{eq:u1flux} of the main text; triplet links acquire the spin rotation $R_O$ of Eq.~(9) of the main text. Bonds starting in the cell $(m,n)$ follow from $u_{i+m\bm{T}_1+n\bm{T}_2,\,j+m\bm{T}_1+n\bm{T}_2}=\eta_{12}^{\,n(x_i+x_j)}u_{ij}$.}
\label{sm:tab:bondrules}
\begin{ruledtabular}
\begin{tabular}{llc|llc}
bond & $i\to j$ & $s$ & bond & $i\to j$ & $s$\\\hline
$J_b$ & $1\to2$ & $1$ & $J'$ & $7\to1_{(1,-1)}$ & $n_{T_2}n_{\sigma}n_{\sigma C_6}n_{C_6}$\\
$J_b$ & $3\to4$ & $1$ & $J'$ & $8\to2_{(1,-1)}$ & $n_{C_6}$\\
$J_b$ & $5\to6$ & $1$ & $J'$ & $9\to3_{(0,-1)}$ & $n_{\sigma}n_{\sigma C_6}n_{C_6}$\\
$J_b$ & $7\to8$ & $n_{C_6}$ & $J'$ & $10\to4_{(0,-1)}$ & $n_{T_2}n_{C_6}$\\
$J_b$ & $9\to10$ & $n_{C_6}$ & $J'$ & $11\to5_{(-1,0)}$ & $n_{T_2}n_{\sigma}n_{\sigma C_6}n_{C_6}$\\
$J_b$ & $11\to12$ & $n_{C_6}$ & $J'$ & $12\to6_{(-1,0)}$ & $n_{C_6}$\\
$J_g$ & $2\to3$ & $1$ & $J''$ & $1\to9_{(-1,1)}$ & $n_{T_2}n_{C_6}$\\
$J_g$ & $4\to5$ & $1$ & $J''$ & $2\to6_{(-1,1)}$ & $1$\\
$J_g$ & $6\to7$ & $1$ & $J''$ & $3\to11_{(0,1)}$ & $n_{T_2}n_{C_6}$\\
$J_g$ & $8\to9$ & $n_{C_6}$ & $J''$ & $4\to8_{(0,1)}$ & $n_{C_6}$\\
$J_g$ & $10\to11$ & $n_{C_6}$ & $J''$ & $5\to1_{(1,0)}$ & $1$\\
$J_g$ & $12\to1$ & $1$ & $J''$ & $6\to10_{(1,0)}$ & $n_{T_2}n_{C_6}$\\
$J_r$ & $1\to8_{(-1,1)}$ & $n_{T_2}n_{\sigma}$ & $J''$ & $7\to3_{(1,-1)}$ & $n_{T_2}$\\
$J_r$ & $2\to7_{(-1,1)}$ & $1$ & $J''$ & $8\to12_{(1,-1)}$ & $1$\\
$J_r$ & $3\to10_{(0,1)}$ & $n_{T_2}n_{\sigma}$ & $J''$ & $9\to5_{(0,-1)}$ & $1$\\
$J_r$ & $4\to9_{(0,1)}$ & $1$ & $J''$ & $10\to2_{(0,-1)}$ & $n_{T_2}$\\
$J_r$ & $5\to12_{(1,0)}$ & $n_{\sigma}$ & $J''$ & $11\to7_{(-1,0)}$ & $n_{T_2}$\\
$J_r$ & $6\to11_{(1,0)}$ & $n_{T_2}$ & $J''$ & $12\to4_{(-1,0)}$ & $1$\\
$J_r$ & $7\to2_{(1,-1)}$ & $n_{T_2}n_{\sigma}$ & $J'''$ & $1\to11$ & $n_{\sigma}n_{\sigma C_6}n_{C_6}$\\
$J_r$ & $8\to1_{(1,-1)}$ & $1$ & $J'''$ & $2\to4$ & $1$\\
$J_r$ & $9\to4_{(0,-1)}$ & $n_{\sigma}$ & $J'''$ & $3\to1$ & $n_{\sigma}n_{\sigma C_6}$\\
$J_r$ & $10\to3_{(0,-1)}$ & $n_{T_2}$ & $J'''$ & $4\to6$ & $1$\\
$J_r$ & $11\to6_{(-1,0)}$ & $n_{T_2}n_{\sigma}$ & $J'''$ & $5\to3$ & $n_{\sigma}n_{\sigma C_6}$\\
$J_r$ & $12\to5_{(-1,0)}$ & $1$ & $J'''$ & $6\to8$ & $n_{C_6}$\\
$J'$ & $1\to7_{(-1,1)}$ & $n_{T_2}n_{\sigma}n_{\sigma C_6}$ & $J'''$ & $7\to5$ & $n_{\sigma}n_{\sigma C_6}$\\
$J'$ & $2\to8_{(-1,1)}$ & $1$ & $J'''$ & $8\to10$ & $1$\\
$J'$ & $3\to9_{(0,1)}$ & $n_{T_2}n_{\sigma}n_{\sigma C_6}$ & $J'''$ & $9\to7$ & $n_{\sigma}n_{\sigma C_6}$\\
$J'$ & $4\to10_{(0,1)}$ & $1$ & $J'''$ & $10\to12$ & $1$\\
$J'$ & $5\to11_{(1,0)}$ & $n_{\sigma}n_{\sigma C_6}$ & $J'''$ & $11\to9$ & $n_{\sigma}n_{\sigma C_6}$\\
$J'$ & $6\to12_{(1,0)}$ & $n_{T_2}$ & $J'''$ & $12\to2$ & $1$\\
\end{tabular}
\end{ruledtabular}
\end{table*}

\subsection{Allowed Amplitudes on the Reference Bonds}
\label{sm:app:refbonds}

Once the amplitudes on the reference bond $i\to j$ of an orbit are fixed, Eq.~(9) of the main text determines them on every other bond of the orbit (Table~\ref{sm:tab:bondrules}). The reference amplitudes themselves are restricted only by the symmetry operations that map the bond onto itself. Because $p6mm$ acts freely on the sites, an element that fixes both endpoints is the identity, so the only nontrivial lattice operation that maps an unoriented bond onto itself is an element $\bar O$ that exchanges its endpoints, $\bar O(i)=j$ and $\bar O(j)=i$, with $\bar O^2=e$ (Table~\ref{tab:bonds} of the main text). This subsection derives the resulting conditions, which are summarized in Table~\ref{sm:tab:refbonds}.

\begin{table*}
\caption{Symmetry-allowed singlet and triplet amplitudes on the reference bonds $i\to j$ of Table~\ref{tab:bonds} of the main text, for every PSG of Eq.~\eqref{eq:z2psg} of the main text and its $U(1)$ and $SU(2)$ counterparts, with or without time reversal. (a)~Data of each bond orbit: the element $\bar O$ that reverses the bond, the unit vector $\hat{\bm{e}}_\parallel$ along $i\to j$ in Cartesian components $(x,y)$, the spin rotation $R_{\bar O}$, the reversal sign $\epsilon=G_{\bar O}(i)G_{\bar O}(j)$ evaluated with the sign forms of Eq.~\eqref{eq:z2psg} of the main text, the time-reversal selector $\zeta=h(\mu_i)h(\mu_j)$ from Table~II of the main text, the chiral classes in which $\bar O$ is combined with $\T$, and the number $n_w$ of factors $\ii\tau^1$ in $G_{\bar O}$ for $U(1)$ PSGs. $C_2(\hat{\bm{n}})=2\hat{\bm{n}}\hat{\bm{n}}^{\rm T}-\one$ is the twofold rotation about $\hat{\bm{n}}$, $\hat{\bm{e}}_\perp=\hat{\bm{z}}\times\hat{\bm{e}}_\parallel$, and the two direction columns list the eigenvectors of $R_{\bar O}$ with eigenvalue $+1$ and $-1$. (b)~Data of each gauge channel $k$ of Eq.~\eqref{eq:links} of the main text: the parity $\rho_k=\pm1$ of the basis matrix under Hermitian conjugation, the sign $\theta_k=\pm1$ of its conjugation by $g_\T$, with $t=0$ ($1$) for $g_\T=\tau^0$ ($\ii\tau^2$) and $t\to w_\T$ for $U(1)$ PSGs, and the channels kept by the $U(1)$ and $SU(2)$ IGGs. Channel $k$ is allowed on the reference bond if and only if (i)~$\zeta=-\theta_k$, imposed only in time-reversal-symmetric classes, and (ii)~for a singlet, $\tilde\epsilon_k=\rho_k$, and for a triplet, its spin vector $\bm{t}^k=(t^{xk},t^{yk},t^{zk})$ lies in the span of the directions listed for $\tilde\epsilon_k\rho_k$. Here $\tilde\epsilon_k=\epsilon$, multiplied by $-1$ in the chiral classes listed in (a) and, for $U(1)$ PSGs, by $(-1)^{n_w}$ in the channels $\tau^3$ and $\ii\tau^3$. Condition (ii) is absent on $J''$ and $J'''$, whose triplet vectors are unconstrained. In the $U(1)$ sector $(0,1,0)$ at generic flux $\Phi$, the rule applies to $J'$ after the bond rotation $u^{(\alpha)}_{ij}\to e^{\ii\Phi\tau^3/2}u^{(\alpha)}_{ij}$. Amplitudes on the other bonds of each orbit follow from Table~\ref{sm:tab:bondrules}.}
\label{sm:tab:refbonds}
\begin{ruledtabular}
\begin{tabular}{lccccccccc}
\multicolumn{10}{l}{(a) Bond orbits}\\
 & & & & & & \multicolumn{2}{c}{directions of $\bm{t}^k$} & & \\
\cline{7-8}
bond & $\bar O$ & $\hat{\bm{e}}_\parallel$ & $R_{\bar O}$ & $\epsilon$ & $\zeta$ & $\tilde\epsilon_k\rho_k=+1$ & $\tilde\epsilon_k\rho_k=-1$ & $\bar O\T$ in class & $n_w$\\
\hline
$J_b$ & $C_6^3\sigma$ & $(1,0)$ & $C_2(\hat{\bm{e}}_\parallel)$ & $\eta_{\sigma C_6}$ & $\eta_{\T\sigma}\eta_{\T C_6}$ & $\hat{\bm{e}}_\parallel$ & $\hat{\bm{e}}_\perp,\hat{\bm{z}}$ & B, C & $w_{C_6}+w_\sigma$\\
$J_g$ & $C_6^2\sigma$ & $(\sqrt3,-1)/2$ & $C_2(\hat{\bm{e}}_\parallel)$ & $\eta_\sigma$ & $\eta_{\T\sigma}$ & $\hat{\bm{e}}_\parallel$ & $\hat{\bm{e}}_\perp,\hat{\bm{z}}$ & B, D & $w_\sigma$\\
$J_r$ & $T_1^{-1}T_2\sigma$ & $(0,1)$ & $C_2(\hat{\bm{e}}_\parallel)$ & $\eta_\sigma$ & $\eta_{\T\sigma}$ & $\hat{\bm{e}}_\parallel$ & $\hat{\bm{e}}_\perp,\hat{\bm{z}}$ & B, D & $w_\sigma$\\
$J'$ & $T_1^{-1}T_2C_6^3$ & $(-1,1)/\sqrt2$ & $C_2(\hat{\bm{z}})$ & $\eta_{12}\eta_{C_6}$ & $\eta_{\T C_6}$ & $\hat{\bm{z}}$ & $\hat{\bm{e}}_\parallel,\hat{\bm{e}}_\perp$ & C, D & $w_{C_6}$\\
$J''$ & -- & $(1,\sqrt3)/2$ & -- & -- & $+1$ & \multicolumn{2}{c}{$\hat{\bm{e}}_\parallel,\hat{\bm{e}}_\perp,\hat{\bm{z}}$} & -- & --\\
$J'''$ & -- & $(1,-1)/\sqrt2$ & -- & -- & $\eta_{\T C_6}$ & \multicolumn{2}{c}{$\hat{\bm{e}}_\parallel,\hat{\bm{e}}_\perp,\hat{\bm{z}}$} & -- & --
\end{tabular}
\end{ruledtabular}
\begin{ruledtabular}
\begin{tabular}{lcccccccc}
\multicolumn{9}{l}{(b) Gauge channels}\\
 & \multicolumn{4}{c}{singlet $u^{(0)}_{ij}$} & \multicolumn{4}{c}{triplet $u^{(a)}_{ij}$}\\
\cline{2-5}\cline{6-9}
channel $k$ & $\ii\tau^0$ ($\chi^0$) & $\tau^1$ ($\Delta^1$) & $\tau^2$ ($\Delta^2$) & $\tau^3$ ($\chi^3$) & $\tau^0$ ($t^{a0}$) & $\ii\tau^1$ ($t^{a1}$) & $\ii\tau^2$ ($t^{a2}$) & $\ii\tau^3$ ($t^{a3}$)\\
\hline
$\rho_k$ & $-1$ & $+1$ & $+1$ & $+1$ & $+1$ & $-1$ & $-1$ & $-1$\\
$\theta_k$ & $+1$ & $(-1)^t$ & $+1$ & $(-1)^t$ & $+1$ & $(-1)^t$ & $+1$ & $(-1)^t$\\
$U(1)$ & yes & no & no & yes & yes & no & no & yes\\
$SU(2)$ & yes & no & no & no & yes & no & no & no
\end{tabular}
\end{ruledtabular}
\end{table*}

\paragraph*{Bond reversal.} Taking the element $O$ of Eq.~(9) of the main text to be $\bar O$ and using $u^{(\alpha)}_{ji}=u^{(\alpha)\dagger}_{ij}$ gives
\begin{align}
u^{(0)\dagger}_{ij}&=G_{\bar O}(j)\,u^{(0)}_{ij}\,G_{\bar O}^\dagger(i),\nonumber\\
u^{(b)\dagger}_{ij}&=\sum_a\left[R_{\bar O}\right]_{ba}G_{\bar O}(j)\,u^{(a)}_{ij}\,G_{\bar O}^\dagger(i).
\label{sm:eq:rev}
\end{align}
In the gauge of Eq.~\eqref{eq:z2psg} of the main text every spatial $G_O$ is a sign times $\tau^0$, and so is $G_{\bar O}$, which is obtained from the generators by Eq.~(11) of the main text. The gauge factors in Eq.~\eqref{sm:eq:rev} then reduce to the sign $\epsilon=G_{\bar O}(i)G_{\bar O}(j)$. By Eq.~(11) of the main text, $G_{\bar O}(i)G_{\bar O}(j)=G_{\bar O}(i)G_{\bar O}(\bar O^{-1}(i))=G_{\bar O^2}(i)$ is the value of the relation $\bar O^2=e$ in the IGG, so $\epsilon$ is gauge invariant. As an example, for $J_r$, with $\bar O=T_1^{-1}T_2\sigma$, $i=(0,0,2)$, and $j=(-1,1,7)$, Eq.~(11) of the main text gives $G_{\bar O}(i)=G_{T_2}(1,0,2)\,G_\sigma(1,-1,2)=\eta_{12}\cdot\eta_{12}\,g_\sigma(2)=\eta_{C_6}\eta_{\sigma C_6}$ and $G_{\bar O}(j)=G_{T_2}(0,1,7)\,G_\sigma(0,0,7)=g_\sigma(7)=\eta_{C_6}\eta_{\sigma C_6}\eta_\sigma$, hence $\epsilon=\eta_\sigma$. The same steps give $\epsilon=\eta_{\sigma C_6}$ for $J_b$, $\eta_\sigma$ for $J_g$, and $\eta_{12}\eta_{C_6}$ for $J'$.

Each basis matrix of Eq.~\eqref{eq:links} of the main text is either Hermitian or anti-Hermitian. We write $\rho_k=+1$ for the Hermitian matrices $\tau^1$, $\tau^2$, $\tau^3$ of the singlet and $\tau^0$ of the triplet, and $\rho_k=-1$ for the anti-Hermitian matrices $\ii\tau^0$ of the singlet and $\ii\tau^{1,2,3}$ of the triplet. Since the gauge factors in Eq.~\eqref{sm:eq:rev} are signs, the channels decouple. A singlet coefficient survives if $\rho_k=\epsilon$ and vanishes otherwise: for $\epsilon=+1$ the bond carries real hopping $\chi^3$ and pairing $\Delta^{1,2}$, for $\epsilon=-1$ only imaginary hopping $\chi^0$. For the triplet, the coefficients of channel $k$ form the vector $\bm{t}^k=(t^{xk},t^{yk},t^{zk})$, and Eq.~\eqref{sm:eq:rev} becomes the eigenvalue condition
\begin{equation}
R_{\bar O}\,\bm{t}^k=\epsilon\rho_k\,\bm{t}^k.
\label{sm:eq:rev-t}
\end{equation}

\paragraph*{Spin rotations of the reversing elements.} The matrices $R_{\bar O}$ follow from $R_{C_6}$ and $R_\sigma$ (Sec.~III of the main text). With $\hat{\bm{e}}_\parallel$ the unit vector along $i\to j$ and $\hat{\bm{e}}_\perp=\hat{\bm{z}}\times\hat{\bm{e}}_\parallel$, one finds in Cartesian components
\begin{align}
J_b:&\ R_{\bar O}=R_{C_6}^3R_\sigma=\mathrm{diag}(1,-1,-1),\quad \hat{\bm{e}}_\parallel=\hat{\bm{x}},\nonumber\\
J_g:&\ R_{\bar O}=R_{C_6}^2R_\sigma=\begin{pmatrix}\frac12&-\frac{\sqrt3}{2}&0\\-\frac{\sqrt3}{2}&-\frac12&0\\0&0&-1\end{pmatrix},\nonumber\\
&\ \hat{\bm{e}}_\parallel=\big(\tfrac{\sqrt3}{2},-\tfrac12,0\big),\nonumber\\
J_r:&\ R_{\bar O}=R_\sigma=\mathrm{diag}(-1,1,-1),\quad \hat{\bm{e}}_\parallel=\hat{\bm{y}},\nonumber\\
J':&\ R_{\bar O}=R_{C_6}^3=\mathrm{diag}(-1,-1,1),\nonumber\\
&\ \hat{\bm{e}}_\parallel=\big(-\tfrac{1}{\sqrt2},\tfrac{1}{\sqrt2},0\big).
\label{sm:eq:Rh}
\end{align}
On the three nearest-neighbor orbits $\bar O$ is the mirror whose line is the perpendicular bisector of the bond. A mirror with normal $\hat{\bm{n}}$ acts on an axial vector as the twofold rotation $C_2(\hat{\bm{n}})=2\hat{\bm{n}}\hat{\bm{n}}^{\rm T}-\one$, and the three matrices above are $C_2(\hat{\bm{e}}_\parallel)$. On $J'$, $\bar O$ is the twofold rotation about the square center, which is the midpoint of the diagonal, and $R_{\bar O}=C_2(\hat{\bm{z}})$. The eigenvalue $+1$ of $C_2(\hat{\bm{n}})$ belongs to $\hat{\bm{n}}$ and the eigenvalue $-1$ to the plane perpendicular to it. Equation~\eqref{sm:eq:rev-t} therefore gives
\begin{align}
\epsilon\rho_k=+1:&\ \bm{t}^k\parallel\hat{\bm{e}}_\parallel\ (J_b,J_g,J_r),\nonumber\\
&\ \bm{t}^k\parallel\hat{\bm{z}}\ (J'),\nonumber\\
\epsilon\rho_k=-1:&\ \bm{t}^k\in\{\hat{\bm{e}}_\perp,\hat{\bm{z}}\}\ (J_b,J_g,J_r),\nonumber\\
&\ \bm{t}^k\in\{\hat{\bm{e}}_\parallel,\hat{\bm{e}}_\perp\}\ (J'),
\label{sm:eq:rev-dirs}
\end{align}
where braces denote the span. For the channel $\ii\tau^3$, whose vector $\bm{t}^3$ multiplies the imaginary spin-dependent hopping, $\epsilon=+1$ reproduces Moriya's rules: a Dzyaloshinskii--Moriya vector in the mirror plane on the nearest-neighbor bonds and perpendicular to the twofold axis on $J'$. A projective sign $\epsilon=-1$ exchanges the two options.

\paragraph*{Time reversal.} With $G_\T(i)=h(\mu_i)g_\T$, Eq.~(10) of the main text on the reference bond reads $u^{(\alpha)}_{ij}=-\zeta\,g_\T u^{(\alpha)}_{ij}g_\T^\dagger$, with $\zeta=h(\mu_i)h(\mu_j)$. Conjugation by $g_\T$ multiplies each basis matrix by a sign $\theta_k$. For $g_\T=\tau^0$ all $\theta_k=+1$, and for $g_\T=\ii\tau^2$ the matrices containing $\tau^0$ or $\tau^2$ have $\theta_k=+1$ and those containing $\tau^1$ or $\tau^3$ have $\theta_k=-1$. A channel survives if $\zeta=-\theta_k$. Time reversal does not rotate the spin index, so it selects channels without changing the directions of Eq.~\eqref{sm:eq:rev-dirs}. Reading $h(\mu)$ from Table~II of the main text gives $\zeta=\eta_{\T\sigma}\eta_{\T C_6}$ on $J_b$, $\eta_{\T\sigma}$ on $J_g$ and $J_r$, $\eta_{\T C_6}$ on $J'$ and $J'''$, and $\zeta=+1$ on $J''$. Consequently, for $g_\T=\tau^0$ every amplitude vanishes on bonds with $\zeta=+1$, in particular on all $J''$ bonds.

\paragraph*{Bonds without a reversing element.} On $J''$ and $J'''$ only time reversal constrains the reference amplitudes. A channel allowed by $\zeta=-\theta_k$ carries an arbitrary singlet coefficient or an arbitrary triplet vector $\bm{t}^k$, and its images on the twelve bonds per unit cell follow from Table~\ref{sm:tab:bondrules} together with the rotation $R_O$ of Eq.~(9) of the main text.

\paragraph*{Chiral classes.} In a chiral class time reversal is not a symmetry, and the condition $\zeta=-\theta_k$ is dropped. An element $O\T$ imposes the condition of Eq.~(9) of the main text with an extra minus sign (Sec.~III of the main text). The reversing element $\bar O$ is combined with $\T$ when its word contains an odd number of generators that are combined with $\T$: $J_b$ ($C_6^3\sigma$) in classes B and C, $J_g$ ($C_6^2\sigma$) and $J_r$ ($T_1^{-1}T_2\sigma$) in classes B and D, and $J'$ ($T_1^{-1}T_2C_6^3$) in classes C and D. In these cases $\epsilon$ is replaced by $-\epsilon$ in all of the conditions above.

\paragraph*{$U(1)$ and $SU(2)$ projective symmetry groups.} The $U(1)$ PSGs of Eq.~\eqref{eq:u1psg} of the main text multiply each sign form by $(\ii\tau^1)^{w_O}$. The bond-reversing element then carries $G_{\bar O}=\pm(\ii\tau^1)^{n_w}$, where $n_w$ counts the factors $\ii\tau^1$ in its word, $n_w=3w_{C_6}+w_\sigma\equiv w_{C_6}+w_\sigma$ (mod 2) for $J_b$, $w_\sigma$ for $J_g$ and $J_r$, and $w_{C_6}$ for $J'$. Conjugation by $\ii\tau^1$ leaves $\ii\tau^0$ and $\tau^0$ unchanged and reverses $\tau^3$ and $\ii\tau^3$, the only other channels allowed by the IGG. The rules above therefore apply with $\epsilon$ replaced by $(-1)^{n_w}\epsilon$ in the channels $\tau^3$ and $\ii\tau^3$, and with $g_\T=(\ii\tau^1)^{w_\T}$, so that $\theta_k=(-1)^{w_\T}$ for these two channels. In the sector $(0,1,0)$ at generic flux the phases of Eq.~\eqref{eq:u1flux} of the main text enter $G_{\bar O}$ on $J'$ and rotate the allowed space by the bond-uniform factor $e^{-\ii\Phi\tau^3/2}$; the link $e^{\ii\Phi\tau^3/2}u^{(\alpha)}_{ij}$ obeys the rules with $\eta_{12}=+1$. For the $SU(2)$ PSGs only the channels $\ii\tau^0$ and $\tau^0$ remain, both with $\theta_k=+1$: a singlet amplitude requires $\epsilon=-1$ and $\zeta=-1$, and a triplet amplitude requires $\zeta=-1$ and points along the directions of Eq.~\eqref{sm:eq:rev-dirs} with $\epsilon\rho_k=\epsilon$.

\paragraph*{Example.} In the class Z000000(1) all signs are $+1$ and $g_\T=\ii\tau^2$, so $\epsilon=\zeta=+1$ on every orbit. Time reversal keeps the channels with $\theta_k=-1$: $\tau^1$ and $\tau^3$ for the singlet and $\ii\tau^1$ and $\ii\tau^3$ for the triplet. Both singlet channels have $\rho_k=\epsilon$ and survive on all six orbits. The triplet channels have $\epsilon\rho_k=-1$, so $\bm{t}^1$ and $\bm{t}^3$ lie in $\{\hat{\bm{e}}_\perp,\hat{\bm{z}}\}$ on $J_b$, $J_g$, and $J_r$, in the plane on $J'$, and are unconstrained on $J''$ and $J'''$. The result is Moriya's pattern for a state in which every symmetry acts without projective signs.

A single projective sign can invert this pattern. In the $U(1)$ class U001-000001 the only negative sign is $\eta_{C_6}=-1$, so that $\epsilon=\eta_{12}\eta_{C_6}=-1$ on $J'$, while $n_w=0$ and $\zeta=\eta_{\T C_6}=+1$ there. With $w_\T=1$, time reversal keeps the channels with $\theta_k=-1$, here $\tau^3$ and $\ii\tau^3$, and removes the $\tau^0$ channel. The singlet $\chi^3\tau^3$, with $\rho_k=+1\neq\epsilon$, vanishes on $J'$, and the triplet channel $\ii\tau^3$ has $\epsilon\rho_k=+1$, so that $\bm{t}^3\parallel\hat{\bm{z}}$. The square diagonals then carry an out-of-plane Dzyaloshinskii--Moriya vector, which Moriya's rule forbids on these bonds, while the in-plane direction allowed by Moriya's rule would require the $\tau^0$ channel, which time reversal removes. This amplitude drives the spinon quantum spin Hall states of this class (Sec.~IX\,B of the main text).

\section{Closed-Form Dirac Node Positions}
\label{sm:app:nodes}

This section derives the energies and momenta of the Dirac nodes on the $\Gamma$--$M$ mirror lines. The line from $\Gamma$ to the $M$ point of Eq.~(D2) of the main text is $\bm{k}=s\bm{M}$ with $0<s<1$, on which $k_1=k_2=\pi s$, and we write
\begin{equation}
c=\cos(\pi s),\qquad z=e^{\ii\pi s}.
\label{sm:eq:cdef}
\end{equation}
The mirror $\sigma$ maps every point of this line to itself, and the other five $\Gamma$--$M$ lines are its $C_6$ images, so each node $\bm{k}_D$ on the line belongs to an orbit of six nodes; the nodes $\bm{k}_D$ and $\bm{G}-\bm{k}_D$, with $\bm{G}=2\bm{M}$, meet at $M$. Lengths are in units of the nearest-neighbor bond length, so that $|\bm{M}|=2\pi/(\sqrt3A)=0.7666$ for $A=3+\sqrt3$, and the amplitudes $\chi_o$ are those of Sec.~VIII of the main text.

\subsection{U001-000000: Mirror Blocks}

On the line, the PSG representative of $\sigma$ commutes with $h(s\bm{M})$ and splits it into a mirror-even and a mirror-odd $6\times6$ block, $h_\pm(s)$. Expanding their determinants, one finds that the phases $z^{\pm2}$ cancel and that $z$ and $z^{-1}$ enter with equal coefficients, so that the characteristic polynomials are linear in $c$,
\begin{equation}
P_\pm(E,c)\equiv\det(E-h_\pm)=A_\pm(E)+c\,B_\pm(E),
\label{sm:eq:Ppm}
\end{equation}
with $A_\pm$ monic of degree six and
\begin{align}
B_\pm={}&\mp2\chi_{23}^2(\chi_{12}\pm\chi_{28})\big[\chi_{28}E^2-2(\chi_{12}\chi_{27}\mp\chi_{23}\chi_{28})E\nonumber\\
&+\chi_{28}\left(\chi_{12}^2+\chi_{23}^2+\chi_{27}^2-\chi_{28}^2\right)\mp2\chi_{12}\chi_{23}\chi_{27}\big].
\label{sm:eq:Bpm}
\end{align}
The polynomials $A_\pm$ follow from the levels at $\Gamma$ ($c=1$), listed in Sec.~VIII of the main text: each doublet contributes one level to each sector, and the four singlets split two and two, so that
\begin{equation}
A_\pm(E)=\Sigma_\pm(E)\,G_1(E)\,G_2(E)-B_\pm(E),
\label{sm:eq:Apm}
\end{equation}
with the doublet factors $G_1=(E-\chi_{28})^2-q_1^2$ and $G_2=(E+\chi_{28})^2-q_2^2$ and the singlet factors
\begin{align}
\Sigma_+={}&(E-\chi_{12}+\chi_{23}+\chi_{27}-\chi_{28})\nonumber\\
&\times(E+\chi_{12}+\chi_{23}+\chi_{27}+\chi_{28}),\nonumber\\
\Sigma_-={}&(E-\chi_{12}-\chi_{23}-\chi_{27}+\chi_{28})\nonumber\\
&\times(E+\chi_{12}-\chi_{23}-\chi_{27}-\chi_{28}).
\label{sm:eq:sigpm}
\end{align}
At $M$ ($c=-1$), $P_\pm(E,-1)=\Sigma_\pm G_1G_2-2B_\pm$ is a product of two cubic polynomials in each sector.

\subsection{U001-000000: Node Energy}

A node on the line is a common root $(E,c)$ of $P_+$ and $P_-$, that is, a crossing of a mirror-even and a mirror-odd level. Since both polynomials are linear in $c$, a common root requires $c=-A_+/B_+=-A_-/B_-$, or $W(E)\equiv A_-B_+-A_+B_-=0$. With Eq.~\eqref{sm:eq:Apm}, $W=G_1G_2(\Sigma_-B_+-\Sigma_+B_-)$, and the bracket factorizes,
\begin{equation}
\Sigma_-B_+-\Sigma_+B_-=-4\chi_{23}^2\,Q_1(E)\,Q_2(E),
\label{sm:eq:W}
\end{equation}
with
\begin{align}
Q_1={}&\chi_{12}(E^2-\chi_{12}^2-\chi_{23}^2+\chi_{27}^2+\chi_{28}^2)\nonumber\\
&-2\chi_{27}\chi_{28}E,\nonumber\\
Q_2={}&\chi_{28}(E^2+\chi_{12}^2-\chi_{23}^2+\chi_{27}^2-\chi_{28}^2)\nonumber\\
&-2\chi_{12}\chi_{27}E.
\label{sm:eq:Q12}
\end{align}
$Q_2$ is $Q_1$ with $\chi_{12}$ and $\chi_{28}$ interchanged. At a root of $G_1G_2$, the momentum
\begin{equation}
c(E)=-\frac{A_+(E)}{B_+(E)}=1-\frac{\Sigma_+(E)\,G_1(E)\,G_2(E)}{B_+(E)}
\label{sm:eq:cE}
\end{equation}
equals one: these roots are the $\Gamma$ doublets, which are shared by both sectors at the end point of every mirror line, and they are discarded. The candidates are the at most four real roots of $Q_1Q_2$ with $-1<c(E)<1$. A candidate is the node at half filling if exactly five of the twelve levels of $h(\bm{k})$ lie below $E$, with the twofold level at $E$ the sixth and seventh; equivalently, the numbers of roots of $P_+(\cdot,c)$ and $P_-(\cdot,c)$ below $E$ add to five. This selection rule is empirical. Among $4500$ amplitude sets $(\chi_{12},\chi_{23},\chi_{27},\chi_{28})$ drawn uniformly from $[-1.5,1.5]^4$, with the chemical potential fixed by half filling, $1352$ have a band touching at the Fermi level and no pockets: $333$ have a $\Gamma$ doublet at $E_F$, $325$ are Dirac states, all with $|\chi_{28}|>|\chi_{12}|$ and with $E_F$ equal to the root of $Q_1$ closest to zero, Eq.~(22) of the main text, and $694$ have a constant-energy nodal line, all with $|\chi_{28}|<|\chi_{12}|$ and with $E_F$ equal to the root $E_0$ of $Q_2$ (Sec.~\ref{sm:app:nodes-line}). With Fermi pockets or close to phase boundaries the count must be applied directly. Whether the crossing is the Fermi level, that is, whether there are no Fermi pockets elsewhere in the zone, is a global condition that is not part of the closed form.

\subsection{U001-000000: Node Position}

With $\epsilon_D=-\operatorname{sgn}(\chi_{27}\chi_{28})$, the node energy of Eq.~(22) of the main text is $E_D=(\chi_{27}\chi_{28}+\epsilon_DD_1^{1/2})/\chi_{12}$. Inserting it into Eq.~\eqref{sm:eq:cE} and rationalizing gives
\begin{equation}
\cos(\pi s_D)=\frac{\tilde\beta+\epsilon_D\,\chi_{28}\left(\chi_{12}^2-\chi_{27}^2\right)S\,D_1^{1/2}}{2\chi_{12}^4\chi_{23}^2\left(\chi_{12}^2-\chi_{28}^2\right)T},
\label{sm:eq:cD}
\end{equation}
with the same sign $\epsilon_D$ as in $E_D$ and
\begin{align}
T&=\left(\chi_{12}^2-\chi_{28}^2\right)\left(\chi_{27}^2-\chi_{28}^2\right)-\chi_{23}^2\chi_{28}^2,\nonumber\\
S&=\chi_{12}^4\chi_{23}^3-8\chi_{27}\left(\chi_{12}^2-\chi_{28}^2\right)T,
\label{sm:eq:ST}
\end{align}
\begin{align}
\tilde\beta={}&\chi_{12}^2\chi_{23}^4\chi_{28}^2\big(\chi_{12}^4-3\chi_{12}^2\chi_{27}^2-2\chi_{12}^2\chi_{28}^2\nonumber\\
&+4\chi_{27}^2\chi_{28}^2\big)\nonumber\\
&+\chi_{12}^4\chi_{23}^3\chi_{27}\left(\chi_{12}^2-\chi_{27}^2\right)\left(\chi_{12}^2-\chi_{28}^2\right)\nonumber\\
&-\chi_{23}^2\left(\chi_{12}^2-\chi_{28}^2\right)\big(\chi_{12}^6\chi_{27}^2-\chi_{12}^6\chi_{28}^2-3\chi_{12}^4\chi_{27}^4\nonumber\\
&+5\chi_{12}^4\chi_{27}^2\chi_{28}^2+6\chi_{12}^4\chi_{28}^4-16\chi_{12}^2\chi_{27}^2\chi_{28}^4\nonumber\\
&+8\chi_{27}^4\chi_{28}^4\big)\nonumber\\
&+4\left(\chi_{12}^2-\chi_{27}^2\right)\left(\chi_{12}^2-\chi_{28}^2\right)^2\left(\chi_{27}^2-\chi_{28}^2\right)\nonumber\\
&\times\left(\chi_{12}^2\chi_{27}^2+\chi_{12}^2\chi_{28}^2-2\chi_{27}^2\chi_{28}^2\right).
\label{sm:eq:betatilde}
\end{align}
Equivalently, $\cos(\pi s_D)$ is the root, on the branch $\epsilon_D$, of the quadratic polynomial
\begin{equation}
K_1(c)=\alpha c^2+\beta c+\gamma
\label{sm:eq:K1}
\end{equation}
defined by the resultant $\mathrm{Res}_E[Q_1(E),P_+(E,c)]=-4(\chi_{12}^2-\chi_{27}^2)(\chi_{12}+\chi_{28})^2K_1(c)$, whose coefficients are
\begin{align}
\alpha&=4\chi_{12}^4\chi_{23}^4\left(\chi_{12}^2-\chi_{28}^2\right)T,\qquad\beta=-4\chi_{23}^2\tilde\beta,\nonumber\\
\beta^2-4\alpha\gamma&=16\chi_{23}^4\chi_{28}^2\left(\chi_{12}^2-\chi_{27}^2\right)^2S^2D_1.
\label{sm:eq:K1coef}
\end{align}
For $(\chi_{12},\chi_{23},\chi_{27},\chi_{28})=(1,1,1.272,2)$, Eq.~(22) of the main text and Eq.~\eqref{sm:eq:cD} give $E_D=0.854636$ and $s_D=0.320599$.

\subsection{U001-000000: End Points and Asymptotics}

The node reaches $M$ when $K_1(-1)=\alpha-\beta+\gamma=0$. This value factorizes exactly,
\begin{align}
K_1(-1)={}&R(\chi_{12},\chi_{23},\chi_{27},\chi_{28})\nonumber\\
&\times R(\chi_{12},\chi_{23},\chi_{27},-\chi_{28}),
\label{sm:eq:K1m}
\end{align}
where $R$ is the semi-Dirac polynomial of Sec.~VIII of the main text,
\begin{align}
R={}&\chi_{12}^4(\chi_{23}+2\chi_{27}-2\chi_{28})\big(\chi_{23}^2-\chi_{23}\chi_{27}-\chi_{23}\chi_{28}\nonumber\\
&+2\chi_{27}^2+4\chi_{27}\chi_{28}+2\chi_{28}^2\big)\nonumber\\
&+\chi_{12}^2\big[\chi_{23}^3\chi_{27}(4\chi_{28}-\chi_{27})\nonumber\\
&+\chi_{23}^2\chi_{27}\left(3\chi_{27}^2+7\chi_{27}\chi_{28}-4\chi_{28}^2\right)\nonumber\\
&-4(\chi_{27}+\chi_{28})\left(\chi_{27}^2+\chi_{28}^2\right)\left(\chi_{23}\chi_{28}+\chi_{27}^2-\chi_{28}^2\right)\big]\nonumber\\
&+4\chi_{27}^2\chi_{28}^2\big[\chi_{23}^3-\chi_{23}^2\chi_{28}+\chi_{23}\chi_{28}(\chi_{27}+\chi_{28})\nonumber\\
&+(\chi_{27}+\chi_{28})^2(\chi_{27}-\chi_{28})\big].
\label{sm:eq:R}
\end{align}
$R$ is the resultant of one cubic factor of $P_+(E,-1)$ and one of $P_-(E,-1)$, divided by $8\chi_{23}^2$, and the second factor in Eq.~\eqref{sm:eq:K1m} is its image under $\chi_{28}\to-\chi_{28}$, $E\to-E$. The nodes therefore reach $M$ exactly on $R=0$. At $(\chi_{12},\chi_{23},\chi_{27})=(1,1,1.272)$ this gives $\chi_{28}^c=1.808807$, where $E_D=0.750343$ is the Fermi energy of the semi-Dirac state. At $\Gamma$,
\begin{align}
K_1(1)={}&\left(\chi_{12}^2-\chi_{27}^2\right)^2(\chi_{23}+\chi_{27}-\chi_{28})\nonumber\\
&\times(\chi_{23}+\chi_{27}+\chi_{28})\,\Xi^2,\nonumber\\
\Xi={}&\chi_{12}^2(\chi_{23}-2\chi_{27})^2\nonumber\\
&-4\chi_{28}^2\left(\chi_{12}^2+\chi_{23}^2-\chi_{23}\chi_{27}+\chi_{27}^2-\chi_{28}^2\right).
\label{sm:eq:K1p}
\end{align}
The first factor vanishes at the singlet--doublet crossing $\chi_{27}=\chi_{12}$, where $E_D\to\chi_{28}-\operatorname{sgn}(\chi_{28})|\chi_{23}|$; there $K_1(1)$ and $K_1'(1)$ vanish together, so that $1-c$ is linear in $\chi_{27}-\chi_{12}$. The second vanishes at $\chi_{28}=\chi_{23}+\chi_{27}$, where the singlets $\chi_{12}-\chi_{23}-\chi_{27}+\chi_{28}$ and $\chi_{12}+\chi_{23}+\chi_{27}-\chi_{28}$ cross and $E_D\to\chi_{12}$; the third is its image under $\chi_{28}\to-\chi_{28}$. The double root at $c=1$ from $\Xi^2$ is a tangency, at which no node enters. At fixed $\chi_{27}>\chi_{12}$ the six nodes thus run from $M$ at $\chi_{28}^c$ to $\Gamma$ at $\chi_{23}+\chi_{27}$.

Close to $M$, expanding $K_1$ about $c=-1$ and using $1+\cos(\pi s)\simeq\frac{\pi^2}{2}(1-s)^2$ gives
\begin{equation}
(1-s_D)^2=\frac{2}{\pi^2}\,\frac{R\tilde R}{2\alpha-\beta}+O(R^2),
\label{sm:eq:asymM}
\end{equation}
with $\tilde R=R|_{\chi_{28}\to-\chi_{28}}$ and $2\alpha-\beta=-K_1'(-1)$.
Along $\chi_{28}$ at fixed $(\chi_{12},\chi_{23},\chi_{27})$, with $\delta=\chi_{28}-\chi_{28}^c$, this is $1-s_D\simeq(C_2\delta)^{1/2}$ with
\begin{equation}
C_2=\frac{2}{\pi^2}\,\frac{\tilde R\,\partial_{\chi_{28}}R}{2\alpha-\beta}\bigg|_{R=0}.
\label{sm:eq:C2}
\end{equation}
At $(1,1,1.272)$, $C_2=7.3077$, so the separation of the two merging nodes is $2k_0=2|\bm{M}|(1-s_D)=4.1447\,\delta^{1/2}$. Close to $\Gamma$, for $\chi_{27}\to\chi_{12}^+$,
\begin{align}
s_D\simeq{}&\frac1\pi\left[2u_\Gamma(\chi_{27}-\chi_{12})\right]^{1/2},\nonumber\\
u_\Gamma={}&(\chi_{12}+\chi_{23}-\chi_{28})\nonumber\\
&\times\left(2\chi_{12}^2-\chi_{12}\chi_{23}+2\chi_{23}\chi_{28}-2\chi_{28}^2\right)^2\nonumber\\
&\times\left[\chi_{12}\chi_{23}^2(\chi_{12}-\chi_{28})\left(\chi_{12}^2+\chi_{23}\chi_{28}-\chi_{28}^2\right)\right]^{-1},
\label{sm:eq:asymG1}
\end{align}
so that the nodes emerge from $\Gamma$ on the side $\chi_{27}>\chi_{12}$ when $u_\Gamma>0$; $u_\Gamma$ changes sign at $\chi_{28}=\chi_{12}+\chi_{23}$, where $s_D$ becomes linear in $\chi_{27}-\chi_{12}$. For $\chi_{28}\to(\chi_{23}+\chi_{27})^-$,
\begin{equation}
s_D\simeq\frac{3}{\pi}\left[\frac{\left(\chi_{27}^2-\chi_{12}^2\right)(\chi_{23}+\chi_{27}-\chi_{28})}{\chi_{23}\left(\chi_{23}\chi_{27}+\chi_{27}^2-\chi_{12}^2\right)}\right]^{1/2}.
\label{sm:eq:asymG2}
\end{equation}
Both ends are square-root laws, as expected when the two crossing levels have quadratic dispersion along the line at the merging point.

\subsection{U001-000000: Constant-Energy Nodal Line}
\label{sm:app:nodes-line}

For $|\chi_{28}|<|\chi_{12}|$ the crossing of levels six and seven on the $\Gamma$--$M$ line is a root of $Q_2$. With $\epsilon_0=-\operatorname{sgn}(\chi_{12}\chi_{27})$, its energy and position are
\begin{align}
E_0&=\frac{\chi_{12}\chi_{27}+\epsilon_0D_2^{1/2}}{\chi_{28}},\nonumber\\
D_2&=\chi_{12}^2\chi_{27}^2-\chi_{28}^2\left(\chi_{12}^2-\chi_{23}^2+\chi_{27}^2-\chi_{28}^2\right),
\label{sm:eq:E0}
\end{align}
\begin{equation}
c_0=\frac{\tilde\beta_2+\epsilon_0\,\chi_{12}U_2V_2\,D_2^{1/2}}{2\chi_{23}^3\chi_{28}^6\left(\chi_{12}^2-\chi_{28}^2\right)},
\label{sm:eq:c0}
\end{equation}
which is Eq.~\eqref{sm:eq:cE} at $E=E_0$, with
\begin{align}
U_2&=4\chi_{12}^2\chi_{27}^2-\chi_{12}^2\chi_{28}^2-3\chi_{27}^2\chi_{28}^2,\nonumber\\
U_2'&=4\chi_{12}^2\chi_{27}^2-3\chi_{12}^2\chi_{28}^2-\chi_{27}^2\chi_{28}^2,\nonumber\\
V_2&=4\left(\chi_{12}^2-\chi_{28}^2\right)\left(\chi_{27}^2-\chi_{28}^2\right)+\chi_{23}^2\chi_{28}^2,
\label{sm:eq:UV}
\end{align}
\begin{align}
\tilde\beta_2={}&4\chi_{27}\left(\chi_{12}^2-\chi_{28}^2\right)^2\left(\chi_{27}^2-\chi_{28}^2\right)U_2'\nonumber\\
&+3\chi_{23}^2\chi_{27}\chi_{28}^2\left(\chi_{12}^2-\chi_{28}^2\right)U_2'\nonumber\\
&-\chi_{23}^3\chi_{28}^6\left(\chi_{12}^2+\chi_{27}^2-2\chi_{28}^2\right).
\label{sm:eq:beta2}
\end{align}
On the mirror line $\gamma(\bm{k})=2c+1$. Since the nodal line is a contour of $\gamma(\bm{k})$ (Sec.~VIII of the main text), it is the contour $\gamma(\bm{k})=\gamma_0=2c_0+1$ at the energy $E_0$, which is the Fermi energy. For $(\chi_{12},\chi_{23},\chi_{27},\chi_{28})=(1,1,0.5,0.2)$, $E_0=0.0423589$ and $\gamma_0=-0.91518$.

\subsection{Sublattice-Symmetric States}

In SU011101, U000-011001, and U100-011001 the spinon Hamiltonian has the off-diagonal form of Sec.~VIII of the main text, with a $6\times6$ block $B$ between the sites $\mu\in\{1,2,5,6,9,10\}$ and $\mu\in\{3,4,7,8,11,12\}$, so that zero modes occur where $\det B=0$ and $E_F=0$; no branch has to be selected. On the $\Gamma$--$M$ line $\det B$ is a Laurent polynomial in $z$ of degree two, and one finds that it is palindromic with real coefficients,
\begin{align}
\det B|_{\Gamma M}&=a_0\left(z^2+z^{-2}\right)+a_1\left(z+z^{-1}\right)+a_2\nonumber\\
&=4a_0c^2+2a_1c+a_2-2a_0\equiv F(c).
\label{sm:eq:Fc}
\end{align}
Although $\det B$ is complex at generic momenta, it is real on the mirror lines, and the nodes there are the real roots of the quadratic $F$ in $-1<c<1$, each root giving an orbit of six nodes. Writing $F_\Gamma=F(1)=\det B(\Gamma)$ and $F_M=F(-1)=\det B(M)$,
\begin{equation}
F(c)=\tfrac12(1+c)F_\Gamma+\tfrac12(1-c)F_M-4a_0\left(1-c^2\right),
\label{sm:eq:Finterp}
\end{equation}
and the roots are
\begin{align}
\cos(\pi s_D)&=\frac{F_M-F_\Gamma\pm\Delta^{1/2}}{16a_0},\nonumber\\
\Delta&=(F_\Gamma-F_M)^2-32a_0(F_\Gamma+F_M)+256a_0^2.
\label{sm:eq:chiralroots}
\end{align}
The number of orbits can change only where $F_\Gamma=0$ (nodes merge at $\Gamma$), $F_M=0$ (nodes merge at $M$ into a semi-Dirac point), or $\Delta=0$ (two nodes on one line annihilate at generic $s$). The three states differ in $a_0$, $F_\Gamma$, and $F_M$.

\textit{SU011101.} With $u_{ij}=\ii\chi_o\tau^0$ on the $J_g$, $J_r$, $J'$, and $J'''$ bonds,
\begin{align}
a_0&=\chi_{24}^4\left(\chi_{28}^2-\chi_{27}^2\right),\nonumber\\
F_\Gamma&=(\chi_{28}+2\chi_{24}-\chi_{23}+\chi_{27})(\chi_{28}+2\chi_{24}+\chi_{23}-\chi_{27})\nonumber\\
&\quad\times\left[(\chi_{28}-\chi_{24})^2-\chi_{23}^2-\chi_{23}\chi_{27}-\chi_{27}^2\right]^2,
\label{sm:eq:SUaF}
\end{align}
and $F_M=F_1(\chi_{23},\chi_{27})\,F_1(-\chi_{23},-\chi_{27})$, the product $F_1F_2$ of Sec.~VIII of the main text, with the cubic
\begin{align}
F_1={}&\chi_{28}^3-(\chi_{23}+\chi_{27})\chi_{28}^2+\big(\chi_{24}^2+2\chi_{24}\chi_{23}\nonumber\\
&-2\chi_{24}\chi_{27}+\chi_{23}^2+\chi_{23}\chi_{27}-\chi_{27}^2\big)\chi_{28}\nonumber\\
&+2\chi_{24}^3+\chi_{24}^2(\chi_{23}+3\chi_{27})\nonumber\\
&-2\chi_{24}\left(\chi_{23}^2-\chi_{23}\chi_{27}-\chi_{27}^2\right)-\chi_{23}^3+\chi_{27}^3.
\label{sm:eq:SUF1}
\end{align}
The discriminant is $\Delta=16(\chi_{23}\chi_{28}+2\chi_{24}\chi_{27})^2\Delta_8$ with a polynomial $\Delta_8$ of degree eight, so that the two roots coincide on the surface $\chi_{23}\chi_{28}+2\chi_{24}\chi_{27}=0$. At $(\chi_{23},\chi_{27},\chi_{28},\chi_{24})=(-0.95,-0.92,0.94,-0.23)$ the roots are $c=0.218821$ and $c=-1108.5$, so that the six nodes lie at $s_D=0.429779$.

\textit{U000-011001.} The amplitudes are $\chi_{23}$ and $\chi_{27}$ ($\tau^3$), $\chi_{28}$ ($\ii\tau^0$), and $\chi^0_{24}$ and $\chi^3_{24}$, in the notation of Eq.~\eqref{eq:links} of the main text; we write $\delta=\chi^0_{24}+\ii\chi^3_{24}$ and $Y=|\delta|^2$. Then
\begin{align}
a_0={}&-\left(\chi_{27}^2+\chi_{28}^2\right)Y^2,\nonumber\\
F_\Gamma={}&-\left[\left(\chi_{28}+2\chi^0_{24}\right)^2+(\chi_{23}-\chi_{27})^2\right]\nonumber\\
&\times\big[\left(\chi_{28}-\chi^0_{24}\right)^2-3\left(\chi^3_{24}\right)^2\nonumber\\
&+\chi_{23}^2+\chi_{23}\chi_{27}+\chi_{27}^2\big]^2,
\label{sm:eq:U000aF}
\end{align}
\begin{align}
F_M={}&f_0+g_0Y+h_0Y^2-2Y^3+2\,\mathrm{Re}\big[(f_1+g_1Y)\delta\nonumber\\
&+(f_2+g_2Y)\delta^2+(f_3+g_3Y)\delta^3+f_4\delta^4-\delta^6\big],
\label{sm:eq:U000FM}
\end{align}
with
\begin{align}
f_0={}&-\left(\chi_{23}^3-\chi_{27}^3\right)^2+\chi_{23}^2\chi_{28}^2\left(\chi_{23}^2+\chi_{27}^2+\chi_{28}^2\right)\nonumber\\
&-\chi_{28}^2\left(\chi_{28}^4+3\chi_{27}^2\chi_{28}^2+3\chi_{27}^4\right),\nonumber\\
f_1={}&2\chi_{23}\chi_{27}\chi_{28}\left(\chi_{23}^2-3\chi_{23}\chi_{27}+\chi_{27}^2+\chi_{28}^2\right),\nonumber\\
f_2={}&-3\chi_{23}^4+\chi_{23}^2\chi_{27}^2+\chi_{23}^2\chi_{28}^2+6\chi_{23}\chi_{27}\chi_{28}^2\nonumber\\
&+\chi_{27}^2\chi_{28}^2+\chi_{27}^4,\nonumber\\
f_3={}&-2\chi_{28}\left(\chi_{28}^2-\chi_{23}\chi_{27}\right),\qquad f_4=\chi_{27}^2-3\chi_{23}^2,\nonumber\\
g_0={}&-2\big(\chi_{28}^4+\chi_{23}^2\chi_{28}^2+\chi_{27}^2\chi_{28}^2-3\chi_{23}\chi_{27}\chi_{28}^2\nonumber\\
&-\chi_{23}^3\chi_{27}-\chi_{23}^2\chi_{27}^2+3\chi_{23}\chi_{27}^3\big),\nonumber\\
g_1={}&2\chi_{28}\left(3\chi_{27}^2-\chi_{23}^2-\chi_{23}\chi_{27}\right),\nonumber\\
g_2={}&2\chi_{23}\chi_{27},\qquad g_3=-2\chi_{28},\nonumber\\
h_0={}&-\left(3\chi_{23}^2+2\chi_{23}\chi_{27}+3\chi_{27}^2+\chi_{28}^2\right).
\label{sm:eq:U000coef}
\end{align}
Here $a_0\le0$ and $F_\Gamma\le0$ for all amplitudes, and $F_M<0$ at all of $2\times10^5$ randomly sampled amplitude sets, although a general proof of $F_M\le0$ is not available. Where it holds, $F$ has an even number of roots in $(-1,1)$, and the nodes on the mirror lines form two orbits of six ($N_f=24$) or none. Nodes at generic momenta are zeros of the complex $\det B$ that $F$ does not describe; together with the two orbits on the mirror lines, one generic orbit of twelve gives the $N_f=48$ phase of Sec.~VIII\,D of the main text.

\textit{U100-011001.} Here $\chi_{28}$ also has two components, and we write $\zeta=\chi^0_{28}+\ii\chi^3_{28}$, $X=|\zeta|^2$, and $w=\zeta^*\delta$, with $\delta$ and $Y$ as above. Then
\begin{align}
a_0={}&-Y^2\left(X-\chi_{27}^2\right),\nonumber\\
F_\Gamma={}&-\left(|\zeta-\delta|^2-\chi_{23}^2-\chi_{23}\chi_{27}-\chi_{27}^2\right)^2\nonumber\\
&\times\left[|\zeta+2\delta|^2-(\chi_{23}-\chi_{27})^2\right],
\label{sm:eq:U100aF}
\end{align}
\begin{align}
F_M={}&(\chi_{23}-\chi_{27})^2\left(\chi_{23}^2+\chi_{23}\chi_{27}+\chi_{27}^2\right)^2\nonumber\\
&+2\big(\chi_{27}^4-3\chi_{23}\chi_{27}^3+2\chi_{23}^2\chi_{27}^2+\chi_{23}^3\chi_{27}\nonumber\\
&-3\chi_{23}^4\big)Y+\left(9\chi_{23}^2-2\chi_{23}\chi_{27}+\chi_{27}^2\right)Y^2-4Y^3\nonumber\\
&+\left(\chi_{23}^4+\chi_{23}^2\chi_{27}^2-3\chi_{27}^4\right)X\nonumber\\
&-2\left(\chi_{23}^2+3\chi_{23}\chi_{27}+\chi_{27}^2\right)XY-XY^2\nonumber\\
&+\left(3\chi_{27}^2-\chi_{23}^2\right)X^2-X^3\nonumber\\
&+4\big[\chi_{23}\chi_{27}\left(\chi_{23}^2-3\chi_{23}\chi_{27}+\chi_{27}^2\right)\nonumber\\
&+\left(\chi_{23}^2-3\chi_{27}^2\right)Y-Y^2-\chi_{23}\chi_{27}X\big]\,\mathrm{Re}\,w\nonumber\\
&+2\left(\chi_{23}^2-6\chi_{23}\chi_{27}+\chi_{27}^2-X\right)\mathrm{Re}\,w^2-4\,\mathrm{Re}\,w^3.
\label{sm:eq:U100FM}
\end{align}
Depending on the amplitudes, $F$ has zero, one, or two roots in $(-1,1)$. The twelvefold orbit of nodes at generic momenta found in the survey of Sec.~VIII of the main text lies off the mirror lines and is not described by $F$.

\subsection{Doubled-Cell States}

For $n_{T_2}=-1$ the Bloch Hamiltonian is a $24\times24$ matrix on the doubled cell, and its coefficients on the mirror line are not available in closed form; only the form and degree of the node conditions are fixed. In U011-100101 the mirror $\sigma$ is realized with $\ii\tau^1$, and on the line a momentum-independent involution $\Lambda$ anticommutes with $h(s\bm{M})$ and acts as a sublattice symmetry. Solving $\Lambda h_c+h_c\Lambda=0$ for the hopping matrices $h_c$ of all amplitude components, including the staggered multiplier, at four momenta on the line gives a one-dimensional solution space, so that $\Lambda$ is unique up to sign and independent of the amplitudes. In its eigenbasis $h$ is off-diagonal with a $12\times12$ block $B$. The form of $\det B$ on the line is established numerically: for $60$ random sets of amplitudes up to $J'''$ and of $a^3$, $\det B$ divided by a constant phase is real to a relative accuracy of $10^{-14}$, and its expansion in Chebyshev polynomials of $\cos(2\pi s)$ has no component beyond second order, to a relative accuracy of $3\times10^{-15}$, so that
\begin{equation}
\det B|_{\Gamma M}=e^{\ii\varphi}(q_0+q_1u+q_2u^2),\quad u=\cos(2\pi s),
\label{sm:eq:dblchiral}
\end{equation}
with a constant phase $\varphi$ and real $q_j$, homogeneous of degree $12$ in the amplitudes and the staggered multiplier. We have no analytic proof of this form. Each root $u_D\in(-1,1)$ gives the pair of nodes $s_D$ and $1-s_D$ at $E=0$. On the other two spectral mirror lines, the line from $M/2$ to the midpoint $K''$ of the zone edge between $M$ and $K$ and the zone edge from $M$ to $K''$, a unique involution anticommuting with $h$ exists as well, but $\det B$ is complex: after removal of the phase at its maximum, $\max|\mathrm{Im}\det B|/\max|\det B|$ ranges from $0.01$ to $0.99$ over the same $60$ amplitude sets, and zeros there are not protected.

In U001-100001 a momentum-independent unitary mirror splits $h(s\bm{M})$ into two $12\times12$ sectors whose characteristic polynomials are of degree four in $c$ and satisfy $P_-(E,c)=P_+(E,-c)$. Writing $P_+=\sum_{j=0}^4p_j(E)\,c^j$, with $p_j$ of degree $12$, $9$, $6$, $4$, and $2$ in $E$, a crossing between the sectors requires $P_+(E,c)=P_+(E,-c)=0$, that is, $p_1+p_3c^2=0$ and $p_0+p_2c^2+p_4c^4=0$. Eliminating $c^2$,
\begin{equation}
p_0p_3^2-p_1p_2p_3+p_4p_1^2=0,\qquad c_D^2=-\frac{p_1(E_D)}{p_3(E_D)},
\label{sm:eq:dblW}
\end{equation}
a polynomial condition of degree at most $20$ in $E$. The node is the root with $0<c_D^2<1$ at which eleven of the twenty-four levels lie below $E_D$. With amplitudes up to $J'$ only, $p_3=p_4=0$, $E_D$ is a root of $p_1$, of degree eight, and $c_D^2=-p_0/p_2$. In both doubled-cell states the nodes at generic momenta of the reduced zone are not described by these conditions.

\section{Spinon Densities of States}
\label{sm:app:dos}

We computed the spinon density of states of every symmetric spin liquid realized with singlet amplitudes up to $J'''$, the states of Tables~\ref{tab:z2ans}--\ref{tab:su2ans} of the main text, at one generic point of its amplitude space. The type of the low-energy spectrum found in this way belongs to a region of amplitude space that contains this point, not to the PSG: the same PSG can be gapped in one region and gapless in another (Sec.~VIII of the main text). Section~\ref{sm:sec:ansatz} lists the amplitudes and multipliers of every state at this point, Sec.~\ref{sm:sec:sf} gives its spin structure factors computed at the same point, and Table~\ref{sm:tab:compare} compares the types with those at a second amplitude point. Each ansatz contains singlet amplitudes on all orbits allowed by its PSG up to $J'$, and up to its range in these tables when this is $J''$ or $J'''$. Since the range is the shortest one at which the bond amplitudes alone have the IGG of the PSG, every ansatz has this IGG (Sec.~\ref{sec:ansatz} of the main text). On the reference bond of each orbit the link is $u=|u|\,\hat{\bm{n}}\cdot(\ii\tau^0,\tau^1,\tau^2,\tau^3)$, where $\hat{\bm{n}}$ is a unit vector in the allowed space drawn at random and $(|u_b|,|u_g|,|u_r|,|u_{J'}|,|u_{J''}|,|u_{J'''}|)=(1,0.9,0.8,0.5,0.3,0.3)$. For a one-dimensional allowed space $\hat{\bm{n}}$ is fixed up to a random sign, which selects a flux sector; SU010111, for instance, is taken at $(\chi_{12},\chi_{23},\chi_{27})=(1,0.9,-0.8)$, in the Lieb-flux sector, whereas Fig.~\ref{fig:spec-sq} of the main text shows its $\pi$-flux state. Unequal nearest-neighbor magnitudes avoid fine-tuned points such as the phase boundaries of Sec.~VIII of the main text, and energies are in units of $|u_b|$. In the $U(1)$ sector $(0,1,0)$ we take $\Phi=\pi/2$, which makes the ansatz periodic on four unit cells. The Lagrange multipliers are then solved at half filling. A uniform $U(1)$ multiplier $a^3$ is the chemical potential and places the Fermi level $E_F$ at half filling; a staggered $a^3$ is fixed by $\sum_ip_i(\langle n_i\rangle-1)=0$, with $p_i=\pm1$ its sign pattern; the $\Z_2$ multipliers are fixed by $\langle\psi_i^\dagger\tau^k\psi_i\rangle=0$, $k=1,2,3$, on a $60\times60$ mesh. For six gapless $\Z_2$ states these equations have no solution at zero temperature on the mesh, because the occupations jump with the multipliers; for them we solved the constraints with a Fermi smearing, lowered to $T=10^{-3}$, which leaves residuals below $4\times10^{-3}$ at $T=0$.
The density of states $\rho(E)$ is given per site, includes both spins, and is normalized to $\int\rho\,dE=2$; $E$ is measured from $E_F$. For $\Z_2$ states, $\rho(E)$ is the density of states of the Bogoliubov--de Gennes matrix $M(\bm{k})$ of Appendix~E of the main text, which is even in $E$ and counts every quasiparticle of energy $|E|$ once, so that the quasiparticle density at energy $E>0$ is $2\rho(E)$ and the thermodynamic formulas below hold for all IGGs. We used the linear triangle method on a $300\times300$ mesh of the Brillouin zone of the (super)cell, refined to $900\times900$ near $E_F$ for gapless states, together with refined band extrema and minima of the direct gap. A state is gapped if the band edges adjacent to $E_F$ are separated, and we quote the quasiparticle gap $\Delta$, half the distance between them, so that $\rho(E)=0$ for $|E|<\Delta$. For a vanishing gap we searched $128$ lines across the zone for zeros of the direct gap, which distinguishes nodal lines from point nodes, and at each point node we determined the power of the gap along the principal axes of its quadratic form. Linear dispersion in both directions defines a Dirac node, and the prefactor $c$ of $\rho=c|E|$ follows from the cone velocities including their tilt; it agrees to within $1\%$ with a separate triangle integration on node-centered meshes, except for two states discussed below. The low-energy laws and the resulting specific heat $C_V$ are as follows. A gap gives $C_V\propto e^{-\Delta/T}$. Dirac nodes give $\rho=c|E|$ and $C_V=9\zeta(3)\,c\,T^2$. A nodal line or a Fermi surface gives a finite $\rho(0)$ and $C_V=(\pi^2/3)\rho(0)T$, as does a quadratic band touching, with constant $\rho$ near the touching. A semi-Dirac node gives $\rho\propto|E|^{1/2}$ and $C_V\propto T^{3/2}$, and a flat band at $E_F$ gives a $\delta$ peak and a residual entropy.

\begin{figure*}[p]
\includegraphics[width=\textwidth]{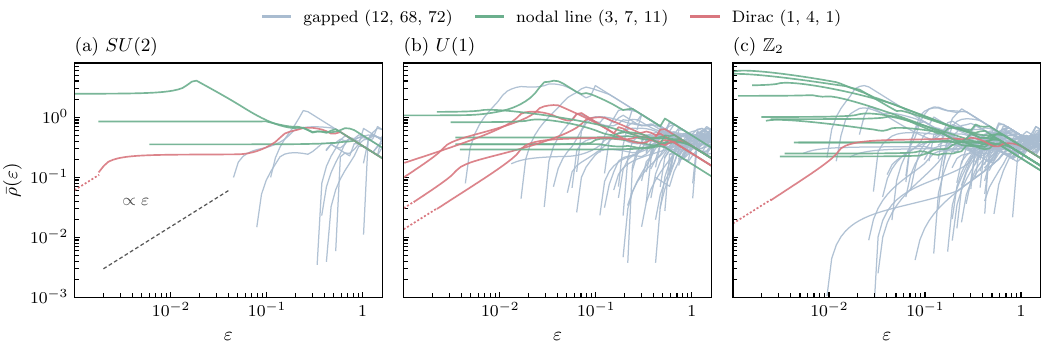}
\caption{Low-energy spinon densities of states of the symmetric spin liquids realized up to $J'''$ at the generic amplitude point of Sec.~\ref{sm:app:dos}: (a) the 16 $SU(2)$, (b) the 79 $U(1)$, and (c) 84 of the 88 $\Z_2$ states (the four states of type P are not shown). Each curve shows the mean density of states $\bar\rho(\varepsilon)=(2\varepsilon)^{-1}\int_{-\varepsilon}^{\varepsilon}\rho(E)\,dE$ in a window of half-width $\varepsilon$ around the Fermi level, per site and including both spins, colored by the low-energy type; the numbers in the legend count the states of each type in panels (a), (b), and (c). Gapped states (blue) drop to zero at $\varepsilon=\Delta$, nodal lines (green) approach the constant $\rho(0)$, and Dirac states (rose) approach $\bar\rho=c\varepsilon/2$, shown dotted below the resolved range with the prefactor $c$ of Table~\ref{sm:tab:dos1} or~\ref{sm:tab:dos2}; the dashed line has slope one. Curves end at three times the energy resolution of the momentum mesh. The values of $\Delta$, $\rho(0)$, and $c$ are listed in Tables~\ref{sm:tab:dos1} and~\ref{sm:tab:dos2}.}
\label{sm:fig:dos}
\end{figure*}

Tables~\ref{sm:tab:dos1} and~\ref{sm:tab:dos2} list the results, and Fig.~\ref{sm:fig:dos} shows the low-energy densities of states. Of the $16$ $SU(2)$ states, $12$ are gapped, SU010110, SU001011, and SU101010 have nodal lines, and SU111100 is a Dirac spin liquid with $N_f=24$ two-component cones. Of the $79$ $U(1)$ states, $68$ are gapped, $7$ have nodal lines, and $4$ are Dirac spin liquids, with $N_f=12$ (U011-000100 and U101-000001), $N_f=24$ (U011-100101), and $N_f=48$ (U010-$\Phi$11100). Of the $88$ $\Z_2$ states, $72$ are gapped, $11$ have nodal lines of the Bogoliubov spectrum, and Z011011(1) has $N_f=12$ Dirac cones. In the remaining four $\Z_2$ states, Z001110(1), Z010111(1), Z101111(1), and Z110110(1) (type P in Table~\ref{sm:tab:dos2}), the constraints pin zero-energy quasiparticles at $E_F$: with the smeared multipliers the quasiparticle gap is $0.03$ at $T=10^{-2}$ and $0.0008$--$0.003$ at $T=10^{-3}$, and it closes faster than $T$ as $T\to0$, so that the self-consistent state is gapless, with zero modes whose occupation is fixed by the constraints. These four states are not shown in Fig.~\ref{sm:fig:dos}. Type P is a property of this amplitude point, not of the PSGs: at the second amplitude point of Table~\ref{sm:tab:compare}, with equal nearest-neighbor magnitudes, Z001110(1), Z010111(1), and Z110110(1) are gapped with constraints satisfied to $10^{-14}$, and Z101111(1) has small nodal loops with residuals below $10^{-4}$. At the present point their mean-field states have no gapped or nodal self-consistent solution, and we leave their low-energy type open. Fermi pockets, quadratic band touchings, semi-Dirac points, and flat bands at $E_F$ do not occur at this point; they occur in other regions or on special lines of parameter space (Sec.~VIII of the main text). A nodal line at $E_F$ is thermodynamically a Fermi surface of zero enclosed area: it gives a finite $\rho(0)$ and $C_V\propto T$, although the number of filled bands is the same at every momentum. The linear law holds only below a crossover energy $E_c$, which we define as the energy above which the local exponent $d\ln\rho/d\ln|E|$, obtained from state counts on node-centered meshes, deviates from one by more than $10\%$. For the cones of U011-000100 and U101-000001, $E_c\approx6\times10^{-3}$ and $2\times10^{-3}$. The cones of SU111100 and U011-100101 are strongly anisotropic, with $E_c\approx7\times10^{-4}$ and $1\times10^{-3}$, respectively. Those of Z011011(1) are anisotropic as well, and the cone velocities do not resolve them accurately; for this state we quote the node-centered integration, which gives a value $4\%$ below the cone formula. The $24$ nodes per spin of U010-$\Phi$11100 come in $12$ close pairs, separated by $3.6\times10^{-3}|\bm{b}_1|$, so that the linear law holds only below $E_c\approx1\times10^{-5}$; for $7\times10^{-5}\lesssim|E|\lesssim3\times10^{-4}$ the density of states follows $\rho\approx|E|^{1/2}$, as for semi-Dirac points. For $T\gtrsim E_c$ these three states therefore do not show the $C_V\propto T^2$ of a Dirac spin liquid, and their labels D$_{24}$ and D$_{48}$ refer to the limit $|E|\ll E_c$; these crossovers are visible in Fig.~\ref{sm:fig:dos}.

For $16$ $U(1)$ and $20$ $\Z_2$ states the bond amplitudes up to $J'$ have the larger IGG $SU(2)$ or $U(1)$, respectively, and for $24$ $\Z_2$ states those up to $J''$ have IGG $U(1)$; their ans\"atze therefore contain the $J''$ and $J'''$ bonds, respectively.

For $76$ states the PSG allows further orbits up to $J'''$ that are not included at this point. For the $15$ of them that are gapless here, we repeated the calculation with all allowed orbits up to $J'''$, with the same magnitudes and the same directions on the orbits already present. Nine keep their type: U011-000100 and U101-000001 (Dirac), and U101-000010, U111-000100, Z001001(1), Z001011(1), Z011101(1), Z101000(1), and Z101010(1) (nodal lines). Six change it (marked in Tables~\ref{sm:tab:dos1} and~\ref{sm:tab:dos2}). In U001-000000 the $J''$ and $J'''$ bonds move the nodal line away from $E_F$ and leave small Fermi pockets, so that its nodal line is a property of the truncated ansatz; U101-100011 and the Dirac states SU111100, U010-$\Phi$11100, and U011-100101 become gapped, and the nodal line of Z010110(1) gives way to point nodes. The remaining $16$ gapless states contain all orbits allowed up to $J'''$. Of the $61$ gapped states with further allowed orbits, $57$ stay gapped when these orbits are added, whereas U111-000001 and U111-100000 acquire nodal lines and Z000000(1) and Z111011(0) point nodes (also marked).

\begin{table*}
\caption{Low-energy spinon spectra of the $SU(2)$ and $U(1)$ spin liquids of Tables~\ref{tab:su2ans}, IX, and~X of the main text at the generic amplitude point of Sec.~\ref{sm:app:dos}, in the order of those tables ($SU(2)$ first). Type G: gapped, value $\Delta$; NL: nodal line, value $\rho(0)$; D$_{N_f}$: Dirac with $N_f$ two-component cones including spin, value $c$ in $\rho=c\,|E|$. Energies in units of $|u_b|$, $\rho$ per site including both spins. $^a$Type changes when all orbits allowed up to $J'''$ are included (see text). $^b$Linear law only below the crossover energy given in the text.}
\label{sm:tab:dos1}
\begin{ruledtabular}
\begin{tabular}{lcr@{\hspace{1.6em}}lcr@{\hspace{1.6em}}lcr}
PSG & type & value & PSG & type & value & PSG & type & value\\
\hline
SU010110 & NL & 0.354 & U111-100100 & G & 0.357 & U000-111001 & G & 0.278\\
SU010111 & G & 0.414 & U001-000001 & G & 0.414 & U001-000011 & G & 0.757\\
SU110110 & G & 0.385 & U001-000010 & G & 0.448 & U001-000101 & G & 0.550\\
SU110111 & G & 0.076 & U001-000100 & G & 0.668 & U001-100010 & G & 0.774\\
SU011101 & G & 0.354 & U001-100000 & G & 0.540 & U001-100100 & G & 0.670\\
SU011111 & G & 0.043 & U001-100011 & G & 0.515 & U010-$\Phi$01010 & G & 0.062\\
SU111100 & D$_{24}^{a,b}$ & 122 & U001-100101 & G & 0.731 & U010-$\Phi$01110 & G & 1.005\\
SU111110 & G & 0.522 & U011-000010 & G & 0.338 & U011-000011 & G & 0.325\\
SU001011 & NL & 0.849 & U011-000100 & D$_{12}$ & 27.3 & U011-000101 & G & 0.350\\
SU001111 & G & 1.020 & U011-000111 & G & 0.434 & U011-100010 & G & 0.283\\
SU011100 & G & 0.335 & U011-100011 & G & 0.338 & U011-100100 & G & 0.191\\
SU011110 & G & 0.100 & U011-100101 & D$_{24}^{a,b}$ & 200 & U100-001011 & G & 0.470\\
SU101010 & NL & 2.459 & U011-100110 & G & 0.526 & U100-011001 & G & 0.196\\
SU101110 & G & 1.000 & U101-000001 & D$_{12}$ & 61.3 & U100-011011 & G & 0.147\\
SU111101 & G & 0.397 & U101-000010 & NL & 1.198 & U100-101010 & G & 0.528\\
SU111111 & G & 0.838 & U101-000111 & G & 0.679 & U100-111000 & G & 0.320\\
U010-$\Phi$10110 & G & 0.377 & U101-100000 & G & 0.361 & U100-111010 & G & 0.0088\\
U110-010100 & NL & 0.353 & U101-100011 & NL$^a$ & 0.819 & U101-000000 & G & 0.167\\
U110-010101 & G & 0.390 & U101-100110 & G & 0.679 & U101-000110 & G & 0.688\\
U110-110100 & G & 0.386 & U111-000001 & G$^a$ & 0.327 & U101-100001 & NL & 1.072\\
U110-110101 & NL & 0.355 & U111-000100 & NL & 0.291 & U101-100111 & G & 0.679\\
U001-000000 & NL$^a$ & 0.460 & U111-000111 & G & 0.217 & U110-001001 & G & 0.033\\
U001-100001 & G & 0.161 & U111-100000 & G$^a$ & 0.327 & U110-001101 & G & 1.016\\
U010-$\Phi$11100 & D$_{48}^{a,b}$ & 2032 & U111-100101 & G & 0.114 & U110-011100 & G & 0.101\\
U010-$\Phi$11110 & G & 0.395 & U111-100110 & G & 0.542 & U110-101000 & G & 0.028\\
U011-000110 & G & 0.056 & U000-001000 & G & 0.190 & U110-101100 & G & 1.035\\
U011-100111 & G & 0.062 & U000-001001 & G & 0.663 & U110-111101 & G & 0.0042\\
U101-000011 & G & 0.189 & U000-011000 & G & 0.537 & U111-000000 & G & 0.308\\
U101-100010 & G & 0.215 & U000-011001 & G & 0.016 & U111-000110 & G & 0.261\\
U110-011101 & G & 0.108 & U000-101000 & G & 0.612 & U111-100001 & G & 0.312\\
U110-111100 & G & 0.047 & U000-101001 & G & 0.121 & U111-100111 & G & 0.261\\
U111-000101 & G & 0.279 & U000-111000 & G & 0.492 &  &  & \\
\end{tabular}
\end{ruledtabular}
\end{table*}

\begin{table*}
\caption{Low-energy spectra of the $\Z_2$ spin liquids of Tables~\ref{tab:z2ans} and~VIII of the main text, in the order of those tables, from the Bogoliubov--de Gennes spectrum; columns as in Table~\ref{sm:tab:dos1}, and type P: zero-energy quasiparticles pinned at $E_F$ by the constraints (see text); superscript $a$ as in Table~\ref{sm:tab:dos1}.}
\label{sm:tab:dos2}
\begin{ruledtabular}
\begin{tabular}{lcr@{\hspace{1.6em}}lcr@{\hspace{1.6em}}lcr}
PSG & type & value & PSG & type & value & PSG & type & value\\
\hline
Z010000(0) & G & 0.437 & Z100011(1) & G & 0.058 & Z011100(0) & G & 0.135\\
Z010001(0) & G & 0.395 & Z100101(1) & G & 0.350 & Z011110(0) & G & 0.072\\
Z110000(0) & NL & 0.249 & Z110001(1) & G & 0.126 & Z011110(1) & G & 0.383\\
Z110001(0) & G & 0.223 & Z110011(1) & G & 0.225 & Z011111(1) & G & 0.061\\
Z000000(1) & G$^a$ & 0.024 & Z110101(1) & G & 0.133 & Z100010(1) & G & 0.416\\
Z001000(1) & G & 0.185 & Z110111(1) & G & 0.019 & Z100100(1) & G & 0.326\\
Z001001(1) & NL & 0.380 & Z111000(1) & G & 0.011 & Z101000(0) & G & 0.685\\
Z001010(1) & G & 0.0011 & Z111001(1) & G & 0.370 & Z101001(0) & G & 0.621\\
Z001011(1) & NL & 0.222 & Z111100(1) & G & 0.044 & Z101011(0) & G & 0.723\\
Z011000(0) & G & 0.410 & Z111101(1) & G & 0.064 & Z101100(0) & G & 0.127\\
Z011010(0) & G & 0.144 & Z000011(1) & G & 0.061 & Z101100(1) & G & 0.446\\
Z100001(1) & G & 0.037 & Z000101(1) & G & 0.326 & Z101101(0) & G & 0.249\\
Z101000(1) & NL & 0.381 & Z001000(0) & G & 0.766 & Z101101(1) & G & 0.449\\
Z101001(1) & G & 0.435 & Z001001(0) & G & 0.995 & Z101110(1) & G & 0.300\\
Z101010(1) & NL & 0.949 & Z001010(0) & G & 0.918 & Z101111(0) & G & 0.100\\
Z101011(1) & G & 0.031 & Z001100(0) & G & 0.210 & Z101111(1) & P & --\\
Z111001(0) & G & 0.379 & Z001100(1) & G & 0.679 & Z110000(1) & G & 0.285\\
Z111011(0) & G$^a$ & 0.011 & Z001101(0) & G & 0.184 & Z110010(1) & NL & 2.259\\
Z000001(1) & G & 0.175 & Z001101(1) & G & 0.488 & Z110100(1) & NL & 1.006\\
Z000010(1) & G & 0.382 & Z001110(0) & G & 0.033 & Z110110(1) & P & --\\
Z000100(1) & G & 0.349 & Z001110(1) & P & -- & Z111000(0) & G & 0.757\\
Z010000(1) & G & 0.089 & Z001111(1) & G & 0.306 & Z111010(0) & G & 0.0090\\
Z010010(1) & G & 0.122 & Z010001(1) & G & 0.497 & Z111010(1) & G & 0.492\\
Z010100(1) & G & 0.169 & Z010011(1) & NL & 0.930 & Z111011(1) & G & 0.818\\
Z010110(1) & NL$^a$ & 4.925 & Z010101(1) & NL & 3.520 & Z111101(0) & G & 0.210\\
Z011000(1) & G & 0.390 & Z010111(1) & P & -- & Z111110(1) & G & 0.318\\
Z011001(1) & G & 0.029 & Z011001(0) & G & 0.227 & Z111111(0) & G & 0.352\\
Z011100(1) & G & 0.029 & Z011010(1) & G & 0.422 & Z111111(1) & G & 0.393\\
Z011101(1) & NL & 5.196 & Z011011(0) & G & 0.0092 &  &  & \\
Z100000(1) & G & 0.180 & Z011011(1) & D$_{12}$ & 34.1 &  &  & \\
\end{tabular}
\end{ruledtabular}
\end{table*}

\clearpage

\section{Regular Magnetic Orders}
\label{sm:app:regular}

In a regular order every element $g$ of $p6mm$ is undone by a global $O(3)$ operation $G_g$, and $g\mapsto G_g$ is a homomorphism~\SMcite{SM:MessioLhuillierMisguich2011}. Since $p6mm$ acts freely on the SHD sites, each site is $g(i_0)$ for a unique $g$, and the order is fixed by the reference spin $\bm{S}_0$ on $i_0=(0,0,1)$ through $\bm{S}_{g(i_0)}=G_g\bm{S}_0$. The site $(x,y,\mu)$ is $T_1^xT_2^yC_6^{k_\mu}\sigma^{e_\mu}(i_0)$, with
\begin{center}
\begin{tabular}{c|cccccccccccc}
$\mu$ & 1 & 2 & 3 & 4 & 5 & 6 & 7 & 8 & 9 & 10 & 11 & 12\\\hline
$k_\mu$ & 0 & 3 & 5 & 2 & 4 & 1 & 3 & 0 & 2 & 5 & 1 & 4\\
$e_\mu$ & 0 & 1 & 0 & 1 & 0 & 1 & 0 & 1 & 0 & 1 & 0 & 1
\end{tabular}
\end{center}
so that $\bm{S}_{(x,y,\mu)}=G_{T_1}^xG_{T_2}^yG_{C_6}^{k_\mu}G_\sigma^{e_\mu}\bm{S}_0$. All bonds of one orbit carry the same energy, and the energy per site is
\begin{equation}
\frac{E}{N}=\bm{S}_0^{T}M\bm{S}_0,\qquad M=\sum_{\beta}J_\beta A_\beta,
\label{sm:eq:regE}
\end{equation}
with $A_\beta=\frac14\sum_{j}(G_{h_j}+G_{h_j}^{T})$, where $j$ runs over the neighbors of $i_0$ coupled by $J_\beta\in\{J_b,J_g,J_r,J',J'',J'''\}$ and $h_j(i_0)=j$. The optimal $\bm{S}_0$ is the lowest eigenvector of $M$, and the bond cosines are $\langle\bm{S}_i\cdot\bm{S}_j\rangle_\beta=\bm{S}_0^TA_\beta\bm{S}_0/z_\beta$, with $z_\beta=\frac12$ bonds per site for $J_b$, $J_g$, $J_r$, $J'$ and $z_\beta=1$ for $J''$, $J'''$. The eigenvalues of $2M$ over all irreducible orders coincide with the twelve Luttinger--Tisza eigenvalues at $\Gamma$ and at $M$ and with four of the twelve at $K$.

\textit{One-dimensional orders.} With $G_{C_6}=\chi(C_6)$, $G_\sigma=\chi(\sigma)$ and $G_{T_{1,2}}=1$, the spins are collinear, $\bm{S}_{(x,y,\mu)}=\chi(C_6)^{k_\mu}\chi(\sigma)^{e_\mu}\bm{S}_0$, and
\begin{align}
E_{\Gamma_1}&=\tfrac12(J_b+J_g+J_r+J')+J''+J''',\nonumber\\
E_{\Gamma_2}&=-\tfrac12(J_b+J_g+J_r)+\tfrac12J'+J''+J''',\nonumber\\
E_{\Gamma_3}&=\tfrac12(-J_b+J_g+J_r-J')+J''-J''',\nonumber\\
E_{\Gamma_4}&=\tfrac12(J_b-J_g-J_r-J')+J''-J''',
\label{sm:eq:reg1d}
\end{align}
for $(\chi(C_6),\chi(\sigma))=(+,+)$, $(+,-)$, $(-,+)$, and $(-,-)$.

\textit{Two-dimensional orders.} The spins lie in a plane, $\bm{S}_0=(\cos\theta,\sin\theta)$, and $M=a+c\,\tau^z+s\,\tau^x$ with
\begin{align}
E_1:&\ a=\tfrac12(J'''-J'-J''),\ c=\tfrac12J_r-\tfrac12J_b-\tfrac14J_g,\nonumber\\
&\ s=-\tfrac{\sqrt3}{4}J_g,\nonumber\\
E_2:&\ a=\tfrac12(J'-J''-J'''),\ c=\tfrac12J_b+\tfrac12J_r-\tfrac14J_g,\nonumber\\
&\ s=\tfrac{\sqrt3}{4}J_g,\nonumber\\
K_1:&\ a=\tfrac12(J_b-J''),\ c=\tfrac12J_g-\tfrac14(J_r+J')+J''',\nonumber\\
&\ s=\tfrac{\sqrt3}{4}(J_r+J'),\nonumber\\
K_2:&\ a=-\tfrac12(J_b+J''),\ c=-\tfrac12J_g+\tfrac14(J_r-J')+J''',\nonumber\\
&\ s=\tfrac{\sqrt3}{4}(J'-J_r).
\label{sm:eq:reg2d}
\end{align}
The energy is $E/N=a-\rho$ with $\rho=(c^2+s^2)^{1/2}$, at the angle $(\cos2\theta,\sin2\theta)=-(c,s)/\rho$. The generators are $G_{C_6}=R(60^\circ)$ ($E_1$) or $R(120^\circ)$ ($E_2$) and $G_\sigma=\mathrm{diag}(1,-1)$, so that the spin angle is $\phi_\mu=60^\circ k_\mu+(-1)^{e_\mu}\theta$ in $E_1$ and $\phi_\mu=120^\circ k_\mu+(-1)^{e_\mu}\theta$ in $E_2$. For $K_1$ and $K_2$, $G_{T_1}=R(120^\circ)$, $G_{T_2}=R(-120^\circ)$, $G_{C_6}=\mathrm{diag}(1,-1)$, and $G_\sigma=\pm\mathrm{diag}(1,-1)$, so that
\begin{align}
K_1:&\ \phi_{(x,y,\mu)}=120^\circ(x-y)+(-1)^{k_\mu+e_\mu}\theta,\nonumber\\
K_2:&\ \phi_{(x,y,\mu)}=120^\circ(x-y)+180^\circ e_\mu+(-1)^{k_\mu+e_\mu}\theta.
\label{sm:eq:regK}
\end{align}
In $E_1$ the $J'$, $J''$, and $J'''$ bonds are at $180^\circ$, $120^\circ$, and $60^\circ$ for all couplings, and $\langle\bm{S}_i\cdot\bm{S}_j\rangle=-\cos2\theta$, $-\cos(2\theta-60^\circ)$, and $\cos2\theta$ on the $J_b$, $J_g$, and $J_r$ bonds. In $K_2$ the $J_b$ pairs are antiparallel and the $J''$ bonds at $120^\circ$, while $\langle\bm{S}_i\cdot\bm{S}_j\rangle=-\cos2\theta$, $\cos(2\theta+60^\circ)$, $-\cos(2\theta+60^\circ)$, and $\cos2\theta$ on the $J_g$, $J_r$, $J'$, and $J'''$ bonds.

\textit{Three-dimensional orders.} The four $M$ orders have $G_{T_1}=\mathrm{diag}(-1,-1,1)$ and $G_{T_2}=\mathrm{diag}(-1,1,-1)$, so that the spin component along $\bm{e}_a$ is modulated with the wave vector $\bm{q}_a$ of one of the three $M$ points, $\bm{q}_1=(\bm{b}_1+\bm{b}_2)/2$, $\bm{q}_2=\bm{b}_1/2$, $\bm{q}_3=\bm{b}_2/2$. Their point-group generators are
\begin{align}
M_{1,2}:&\ G_{C_6}=\left(\begin{smallmatrix}0&1&0\\0&0&1\\1&0&0\end{smallmatrix}\right),\ G_\sigma=\pm\left(\begin{smallmatrix}1&0&0\\0&0&1\\0&1&0\end{smallmatrix}\right),\nonumber\\
M_{3,4}:&\ G_{C_6}=\left(\begin{smallmatrix}0&1&0\\0&0&1\\-1&0&0\end{smallmatrix}\right),\ G_\sigma=\pm\left(\begin{smallmatrix}1&0&0\\0&0&-1\\0&-1&0\end{smallmatrix}\right).
\label{sm:eq:regMgen}
\end{align}
The bond matrices of $M_1$ are
\begin{align}
A_b&=\tfrac12\left(\begin{smallmatrix}1&0&0\\0&0&1\\0&1&0\end{smallmatrix}\right),\quad
A_g=\tfrac12\left(\begin{smallmatrix}0&0&1\\0&1&0\\1&0&0\end{smallmatrix}\right),\nonumber\\
A_r&=\tfrac12\left(\begin{smallmatrix}1&0&0\\0&0&-1\\0&-1&0\end{smallmatrix}\right),\quad
A'=\tfrac12\left(\begin{smallmatrix}1&0&0\\0&-1&0\\0&0&-1\end{smallmatrix}\right),\nonumber\\
A''&=\tfrac12\left(\begin{smallmatrix}0&-1&1\\-1&0&-1\\1&-1&0\end{smallmatrix}\right),\quad
A'''=\tfrac12\left(\begin{smallmatrix}0&1&1\\1&0&1\\1&1&0\end{smallmatrix}\right),
\label{sm:eq:regAM1}
\end{align}
and those of $M_3$ are
\begin{align}
A_b&=\tfrac12\left(\begin{smallmatrix}-1&0&0\\0&0&1\\0&1&0\end{smallmatrix}\right),\quad
A_g=\tfrac12\left(\begin{smallmatrix}0&0&1\\0&1&0\\1&0&0\end{smallmatrix}\right),\nonumber\\
A_r&=\tfrac12\left(\begin{smallmatrix}1&0&0\\0&0&1\\0&1&0\end{smallmatrix}\right),\quad
A'=\tfrac12\left(\begin{smallmatrix}-1&0&0\\0&1&0\\0&0&1\end{smallmatrix}\right),\nonumber\\
A''&=\tfrac12\left(\begin{smallmatrix}0&1&1\\1&0&1\\1&1&0\end{smallmatrix}\right),\quad
A'''=\tfrac12\left(\begin{smallmatrix}0&1&-1\\1&0&1\\-1&1&0\end{smallmatrix}\right).
\label{sm:eq:regAM3}
\end{align}
The matrices of $M_2$ ($M_4$) follow from those of $M_1$ ($M_3$) by reversing the sign of $A_b$, $A_g$, and $A_r$. With $D_\pm=J'''\pm J''$,
\begin{align}
2M_{M_1}&=\begin{pmatrix}J_b+J_r+J' & D_- & J_g+D_+\\ D_- & J_g-J' & J_b-J_r+D_-\\ J_g+D_+ & J_b-J_r+D_- & -J'\end{pmatrix},\nonumber\\
2M_{M_3}&=\begin{pmatrix}J_r-J_b-J' & D_+ & J_g-D_-\\ D_+ & J_g+J' & J_b+J_r+D_+\\ J_g-D_- & J_b+J_r+D_+ & J'\end{pmatrix},
\label{sm:eq:regM}
\end{align}
and $M_{M_2}$ ($M_{M_4}$) is $M_{M_1}$ ($M_{M_3}$) with $J_{b,g,r}\to-J_{b,g,r}$. The characteristic polynomials do not factorize for generic couplings. With $m=2M$, $q=\mathrm{tr}\,m/3$, $p=[\mathrm{tr}(m-q)^2/6]^{1/2}$, and $r=\det(m-q)/2p^3$, the energy is
\begin{equation}
\frac{E}{N}=\frac12\left[q+2p\cos\left(\tfrac13\arccos r+\tfrac{2\pi}{3}\right)\right],
\label{sm:eq:regViete}
\end{equation}
and $\bm{S}_0$ is parallel to the cross product of two rows of $m-2E/N$. The traces are $\mathrm{tr}\,m=\pm(J_b+J_g+J_r)-J'$ for $M_{1,2}$ and $\pm(J_b-J_g-J_r)+J'$ for $M_{4,3}$. On the flat-band line $J'=J_b=J_g=J_r=1$ the polynomial of $M_{M_2}$ factorizes, and the energies of $M_2$ are $-\frac12(J''+J''')$ and $\frac14[J''+J'''-4\pm(9J''^2-14J''J'''+9J'''^2)^{1/2}]$; at $J'''=0$ the lowest of them is the flat-band energy $E/N=-1-\frac12J''$, so that $M_2$ belongs to the degenerate ground-state manifold there. The spins are signed permutations of $\bm{S}_0=(u,v,w)$. The images of $p6mm$ are the groups $T_d$ ($M_1$) and $O$ ($M_2$) of order $24$ and $O_h$ ($M_3$, $M_4$) of order $48$, so that for generic $\bm{S}_0$ the magnetic unit cell of $48$ sites carries $24$ or $48$ directions, at the vertices of a polyhedron with $8$ hexagonal and $6$ square faces ($M_1$), of a snub cube ($M_2$), or of a polyhedron with $6$ octagonal, $8$ hexagonal, and $12$ rectangular faces ($M_3$, $M_4$).

\textit{Sublattice-sign duality.} Reversing the spins of one sublattice of the bipartite nearest-neighbor graph maps $J_{b,g,r}\to-J_{b,g,r}$ and leaves $J'$, $J''$, $J'''$ unchanged. It exchanges $\Gamma_1\leftrightarrow\Gamma_2$, $\Gamma_3\leftrightarrow\Gamma_4$, $K_1\leftrightarrow K_2$, $M_1\leftrightarrow M_2$, and $M_3\leftrightarrow M_4$ with the same $\bm{S}_0$, and maps $E_1$ and $E_2$ onto themselves.

\textit{Exact ground states.} With nearest-neighbor couplings alone, the squares and dodecagons are never frustrated, and the hexagons, whose edges alternate between $J_g$ and $J_r$, are frustrated only for $J_gJ_r<0$. For $J_gJ_r>0$ one of $\Gamma_1$--$\Gamma_4$ satisfies every bond. For $J_gJ_r<0$, every $J_g$ and $J_r$ bond belongs to exactly one hexagon, the lowest energy of a frustrated six-site ring is $-3(J_g^2+|J_gJ_r|+J_r^2)^{1/2}$, and every site belongs to one $J_b$ bond, so that
\begin{equation}
\frac{E}{N}\geq-\frac12|J_b|-\frac12\left(J_g^2+|J_gJ_r|+J_r^2\right)^{1/2}.
\label{sm:eq:reghexbound}
\end{equation}
Equation~\eqref{sm:eq:reg2d} shows that $K_2$ ($J_b>0$) and $K_1$ ($J_b<0$) reach this bound, so they are exact ground states for all magnitudes of the nearest-neighbor couplings in this sector. With further neighbors and $J_b=J_g=J_r=1$, several level crossings between regular orders are independent of one coupling: N\'eel ($\Gamma_2$) and $E_2$ have equal energies on $J''+J'''=1-1/\sqrt3$, $E_2$ and $K_2$ on $J'=-1+J'''(2-\sqrt3J''')/(2-\sqrt3)$, $K_2$ and $E_1$ on $J'=1-2J'''+3J'''^2$, and N\'eel and $\Gamma_4$ on $J'+2J'''=1$. A level crossing is a phase boundary only where both orders saturate the Luttinger--Tisza bound. The $K_2$/$E_1$ line is such a boundary for $J''=0$, $J'''<0$, and the N\'eel/$\Gamma_4$ line for $J'''=0$, $J''<0$, where it is the flat-band line $J'=1$ discussed below. The $E_2$/$K_2$ line is a boundary for $J'''\geq1$, but on its small-$J'''$ branch only for large $J''$, and the N\'eel/$E_2$ line only at $J'''=0$, since for $J'''>0$ the orders $M_4$ or $K_2$ lie lower nearby. Accordingly, the lower edge of the $E_2$ region at $J'=0$ depends on $J''$: it decreases from $J'''=0.292$ at $J''=0.17$ to $0.156$ at $J''=0.3$ and $0.1548$ at $J''=0.6$, and approaches the crossing value $2/\sqrt3-1$ only near $J''\approx1$; the upper edge is $J'''=1$. On the line $J'=1$ (at $J'''=0$) the lowest Luttinger--Tisza band is flat for both signs of $J''$, and on the line $J'=-1$ for $J''>1-1/\sqrt3$. For $J'=1$ and $J''>0$ the orders $E_1$, $K_2$, and $M_2$, and for $J'=-1$ the orders $E_2$, $K_2$, and $M_3$, are degenerate members of the ground-state manifold. For $J'=1$ and $J''<0$ the two lowest bands are flat at $E/N=-1+J''$, the energy of N\'eel, $\Gamma_4$, $M_2$, and $M_4$, which are all degenerate there. The flat bands of this case have compact eigenvectors on single hexagons. Since $J'=J_b=J_r$, every square is a complete graph, whose energy depends only on its total spin. Each square contains one $J_r$ edge of each of two hexagons, so a N\'eel pattern of independent amplitude on every hexagon, antiparallel on the $J_g$ and $J_r$ edges and parallel on the $J''$ bonds, leaves the total spin of every square zero; its energy does not depend on the relative amplitudes, and the classical states built from independent N\'eel vectors on the hexagons, which include N\'eel and $\Gamma_4$, are degenerate; for $J'<-1$ and $J''>1-1/\sqrt3$ the order $E_2$ is the exact ground state. For $J''=0$ and $-2\leq J'\leq-1$, the order $M_4$ is exact between its crossings with the N\'eel state, at $J'=-(2J'''^3-11J'''^2+10J'''-2)/(3J'''^2-6J'''+2)$, and with $E_2$, at $J'=-[2J'''^3-2J'''^2-5J'''+1+\sqrt3]/[(3+2\sqrt3)J'''-1-\sqrt3]$; at $J'=-1$ this gives $0.354<J'''<0.457$. For $J'>-1$ the $M_4$ strip separates from both regions and narrows, for instance to $0.301<J'''<0.332$ at $J'=0$. The $M_2$ strip lies at $0<J'<1$, for instance $0.376<J'''<0.394$ at $J'=0.2$, and narrows toward the flat-band point $(J',J''')=(1,0)$; at $J'=0$ it requires $J''<0$, with $0.496<J'''<0.570$ at $J''=-0.5$.

\textit{Scalar chirality.} Under a global operation $G$, the chirality $\chi_\triangle=\bm{S}_i\cdot(\bm{S}_j\times\bm{S}_k)$ of a triangle changes by the factor $\det G$. For a triangle with counterclockwise vertices, a rotation preserves and a mirror reverses the orientation. Hence $\chi_{g(\triangle)}=\det(G_g)\,\chi_\triangle$ for rotations and $\chi_{g(\triangle)}=-\det(G_g)\,\chi_\triangle$ for mirrors, and a regular order realizes the chiral class with $(\chi(C_6),\chi(\sigma))=(\det G_{C_6},\det G_\sigma)$ of Sec.~VII of the main text. Since $\det G_{T_{1,2}}=1$, $\chi_\triangle$ is translation invariant even where the spins are not. Equation~\eqref{sm:eq:regMgen} gives class A for $M_2$, B for $M_1$, C for $M_4$, and D for $M_3$; collinear and coplanar orders have $\chi_\triangle=0$. In class B the chirality is uniform and $M_1$ carries a net chirality, whereas in classes A, C, and D a mirror or the sixfold rotation maps every triangle onto one of opposite chirality, and the net chirality of $M_2$, $M_4$, and $M_3$ vanishes.

\section{Descendants Not Shown in Fig.~4 of the Main Text}
\label{sm:app:hierarchy}

This section completes the hierarchy of Sec.~VI\,C of the main text up to fourth neighbors. Labels, the symbolic digit $x$, and the bond abbreviations are those of Fig.~4 of the main text: $b,g,r$ stand for $J_b,J_g,J_r$. In the $U(1)$ sector $(0,1,0)$, and in the chiral $U(1)$ sector $(w_{C_6},w_\sigma)=(0,1)$, the first digit gives the flux, $\Phi=0$ or $\pi$, and the digit $n_{C_6}$ is redundant and set to $0$. Minimal range and genuineness are defined in Sec.~VI\,C of the main text; for a chiral $X$, genuine also requires that the generic bond amplitudes are not invariant under time reversal combined with any gauge transformation. Each listed descent was verified at its minimal range by checking that a generic ansatz of $P$ is invariant under the PSG of $X$ written in the gauge of $P$, that the generic ansatz of $X$ has the IGG of $X$, and that at no shorter range at which $P$ is realized is the IGG lowered. Comma-separated bonds are alternatives, each of which lowers the IGG when its amplitudes are added by themselves to a generic ansatz of $P$; for chiral descendants the listed bonds are those that carry time-reversal-odd amplitudes at the minimal range, and in these lists $J_1$ stands for all three nearest-neighbor bonds $b,g,r$.

Tables~\ref{sm:tab:hier-late} and~\ref{sm:tab:hier-late2} list, for the four pairs of parents of Fig.~4 of the main text, the descents that first become genuine with $J''$ or $J'''$ amplitudes. Each $\Z_2$ state reached from a $U(1)$ state is also a direct descendant of the $SU(2)$ parent, with minimal range equal to the shortest range at which it is reached, in Fig.~4 of the main text or in the table. In every parent the $\Z_2$ descendant with the trivial pattern, whose gauge matrices are those of the parent itself, is not genuine up to $J'''$ (with singlet amplitudes; the Kitaev ansatz of Sec.~X of the main text realizes the trivial-pattern descendant of SU010111 with bond-dependent triplet links that transform with $R_O=\one$). For the parents realized with $J_1$ amplitudes the same holds for the $U(1)$ descendants in the sectors $(0,0,0)$ and $(1,0,0)$: their amplitudes are confined to the nearest-neighbor bonds, and their generic ans\"atze keep the IGG $SU(2)$.

The eight $SU(2)$ parents that are realized only with $J'''$ amplitudes, on the bonds $J_b,J',J'''$ or $J_g,J_r,J'''$, form the four pairs (e)--(h) of Tables~\ref{sm:tab:hier-new} and~\ref{sm:tab:hier-new2}; the two parents of a pair again have identical hierarchies up to the flip of $n_{T_2}$ and $n_{C_6}$. Every descent from these parents has minimal range $J'''$. Every non-octahedral $\Z_2$ descendant of these parents appears under one of their $U(1)$ descendants, and the octahedral one, reached through no $U(1)$ state, is listed separately, so that the tables contain all of their descendants.

A chiral descendant of a time-reversal-symmetric state $X$ in class $\chi=(c_{C_6},c_\sigma)$, with A $=(0,0)$, B $=(0,1)$, C $=(1,0)$, and D $=(1,1)$, has the PSG generated by $G_{T_1}$, $G_{T_2}$, $G_{C_6\T^{c_{C_6}}}$, and $G_{\sigma\T^{c_\sigma}}$, with $G_{O\T}(i)=G_O(i)G_\T(O^{-1}i)$, and the IGG of $X$. Its label follows from that of $X$. For $X=\mathrm{Z}\,n_{T_2}n_{\T\sigma}n_{\T C_6}n_\sigma n_{\sigma C_6}n_{C_6}(t)$ the chiral label $n'_{T_2}n'_\sigma n'_{\sigma C_6}n'_{C_6}$ is, modulo 2,
\begin{align}
n'_{T_2}&=n_{T_2},\nonumber\\
n'_\sigma&=n_\sigma+c_\sigma(n_{\T\sigma}+t),\nonumber\\
n'_{\sigma C_6}&=n_{\sigma C_6}+(c_\sigma+c_{C_6})(n_{\T\sigma}+n_{\T C_6}+t),\nonumber\\
n'_{C_6}&=n_{C_6}+c_{C_6}(n_{\T C_6}+t),
\label{sm:eq:chiral-label}
\end{align}
which follows from evaluating $(\sigma\T^{c_\sigma})^2$, $(\sigma\T^{c_\sigma}C_6\T^{c_{C_6}})^2$, and $(C_6\T^{c_{C_6}})^6$, and the same rule with $t=0$ gives the chiral descendants of the $SU(2)$ parents. A $U(1)$ state in the sector $(w_{C_6},w_\sigma,w_\T)$ has its chiral descendants in the space-group sector $(w_{C_6}+c_{C_6}w_\T,\,w_\sigma+c_\sigma w_\T)$; their sign labels, in which the redundant digits are set to $0$, are given in Table~\ref{sm:tab:hier-chiral-u1}. Tables~\ref{sm:tab:hier-chiral-u1} and~\ref{sm:tab:hier-chiral-z2} give, for each of the 76 $U(1)$ and 86 $\Z_2$ spin liquids of the hierarchy, the chiral descendant in each class, its minimal range, and the bonds that carry time-reversal-odd amplitudes at that range. The hierarchy contains 66 of the 74 discrete $U(1)$ spin liquids of Table~VI of the main text together with ten points $\Phi\in\{0,\pi\}$ of the $(0,1,0)$ families, and 86 of the 88 $\Z_2$ spin liquids; U001-000101, U001-100100, U011-000011, U011-100010, U101-000110, U101-100111, U111-000000, U111-100001, Z000101(1), and Z100100(1) do not descend from an $SU(2)$ parent connected up to $J'''$. Each row covers the two states related by the flip of $n_{T_2}$ and $n_{C_6}$ (of $n_{T_2}$ alone where $n_{C_6}$ is redundant), written with the symbolic digit $x$, where $x=0$ for the state with $n_{T_2}=+1$; the entries of the two states coincide. All $4\times162$ chiral descendants are genuinely chiral by $J'''$, none of them with $J_1$ amplitudes alone; $SU(2)$ to chiral steps are included in Tables~\ref{sm:tab:hier-late}--\ref{sm:tab:hier-new2}.

\begin{table*}
\caption{Descents from the parents of Fig.~4 of the main text that first become genuine with $J''$ or $J'''$ amplitudes, pairs (a) and (b) side by side; the amplitude column lists the bonds carrying $\chi^3$ hopping in $SU(2)\to U(1)$ steps and pairing in $U(1)\to\Z_2$ steps. Last lines of each block: algebraic descendants not genuine up to $J'''$.}
\label{sm:tab:hier-late}
\begin{ruledtabular}
\begin{tabular}{lcllcl}
arrow & range & amplitude & arrow & range & amplitude\\
\hline
\multicolumn{3}{l}{(a) SU$x1011x$, realized from $J_1$} & \multicolumn{3}{l}{(b) SU$x1011\bar{x}$, realized from $J_1$}\\
SU$x1011x$$\,\to\,$U101-$x0001x$ & $J''$ & $\chi^3$: $J''$ & SU$x1011\bar{x}$$\,\to\,$U001-$x0000\bar{x}$ & $J''$ & $\chi^3$: $J''$\\
SU$x1011x$$\,\to\,$U111-$x0010x$ & $J''$ & $\chi^3$: $J''$ & SU$x1011\bar{x}$$\,\to\,$U011-$x0011\bar{x}$ & $J''$ & $\chi^3$: $J''$\\
U001-$x0000x$$\,\to\,$Z$x1011x$(1) & $J''$ & $J'{,}J''$ & U001-$x0000\bar{x}$$\,\to\,$Z$x0000\bar{x}$(1) & $J''$ & $b{,}g{,}r{,}J''$\\
U001-$x0000x$$\,\to\,$Z$x1110\bar{x}$(1) & $J''$ & $b{,}J''$ & U001-$x0000\bar{x}$$\,\to\,$Z$x0101x$(1) & $J''$ & $g{,}r{,}J'{,}J''$\\
U010-$x10110$$\,\to\,$Z$x0000\bar{x}$(1) & $J''$ & $J''$ & U001-$x0000\bar{x}$$\,\to\,$Z$x1110x$(1) & $J''$ & $b{,}J'{,}J''$\\
U011-$x0011x$$\,\to\,$Z$x1000x$(1) & $J''$ & $J'{,}J''$ & U010-$x10110$$\,\to\,$Z$x0000\bar{x}$(1) & $J''$ & $J''$\\
U011-$x0011x$$\,\to\,$Z$x1100\bar{x}$(1) & $J''$ & $b{,}J''$ & U011-$x0011\bar{x}$$\,\to\,$Z$x0000\bar{x}$(1) & $J''$ & $b{,}g{,}r{,}J''$\\
U101-$x0001x$$\,\to\,$Z$x0000\bar{x}$(1) & $J''$ & $b{,}g{,}r{,}J''$ & U011-$x0011\bar{x}$$\,\to\,$Z$x0100x$(1) & $J''$ & $g{,}r{,}J'{,}J''$\\
U101-$x0001x$$\,\to\,$Z$x0100\bar{x}$(1) & $J''$ & $g{,}r{,}J'{,}J''$ & U011-$x0011\bar{x}$$\,\to\,$Z$x1100x$(1) & $J''$ & $b{,}J'{,}J''$\\
U101-$x0001x$$\,\to\,$Z$x1110\bar{x}$(1) & $J''$ & $b{,}J'{,}J''$ & U101-$x0001\bar{x}$$\,\to\,$Z$x1010x$(1) & $J''$ & $J'{,}J''$\\
U111-$x0010x$$\,\to\,$Z$x0000\bar{x}$(1) & $J''$ & $b{,}g{,}r{,}J''$ & U101-$x0001\bar{x}$$\,\to\,$Z$x1110x$(1) & $J''$ & $b{,}J''$\\
U111-$x0010x$$\,\to\,$Z$x0101\bar{x}$(1) & $J''$ & $g{,}r{,}J'{,}J''$ & U111-$x0010\bar{x}$$\,\to\,$Z$x1001x$(1) & $J''$ & $J'{,}J''$\\
U111-$x0010x$$\,\to\,$Z$x1100\bar{x}$(1) & $J''$ & $b{,}J'{,}J''$ & U111-$x0010\bar{x}$$\,\to\,$Z$x1100x$(1) & $J''$ & $b{,}J''$\\
SU$x1011x$$\,\to\,$A:SU$x11x$ & $J''$ & $\Theta$-odd: $J''$ & SU$x1011\bar{x}$$\,\to\,$C:SU$x10\bar{x}$ & $J''$ & $\Theta$-odd: $J''$\\
SU$x1011x$$\,\to\,$B:SU$x00x$ & $J''$ & $\Theta$-odd: $J''$ & SU$x1011\bar{x}$$\,\to\,$D:SU$x01\bar{x}$ & $J''$ & $\Theta$-odd: $J''$\\
U101-$x0001x$$\,\to\,$Z$x1010\bar{x}$(1) & $J'''$ & $J''{,}J'''$ & U001-$x0000\bar{x}$$\,\to\,$Z$x1011\bar{x}$(1) & $J'''$ & $J''{,}J'''$\\
U111-$x0010x$$\,\to\,$Z$x1001\bar{x}$(1) & $J'''$ & $J''{,}J'''$ & U011-$x0011\bar{x}$$\,\to\,$Z$x1000\bar{x}$(1) & $J'''$ & $J''{,}J'''$\\
\multicolumn{3}{l}{\quad not genuine up to $J'''$: U000-$x1000x$,} & \multicolumn{3}{l}{\quad not genuine up to $J'''$: U000-$x1000\bar{x}$,}\\
\multicolumn{3}{l}{\quad\quad U100-$x1001x$, Z$x1001\bar{x}$(0), Z$x1010\bar{x}$(0),} & \multicolumn{3}{l}{\quad\quad U100-$x1001\bar{x}$, Z$x1001x$(0), Z$x1010x$(0),}\\
\multicolumn{3}{l}{\quad\quad Z$x1011x$(0)} & \multicolumn{3}{l}{\quad\quad Z$x1011\bar{x}$(0)}\\
\end{tabular}
\end{ruledtabular}
\end{table*}

\begin{table*}
\caption{As Table~\ref{sm:tab:hier-late}, for the pairs (c) and (d).}
\label{sm:tab:hier-late2}
\begin{ruledtabular}
\begin{tabular}{lcllcl}
arrow & range & amplitude & arrow & range & amplitude\\
\hline
\multicolumn{3}{l}{(c) SU$x1110\bar{x}$, realized from $J'$} & \multicolumn{3}{l}{(d) SU$x1111\bar{x}$, realized from $J'$}\\
SU$x1110\bar{x}$$\,\to\,$U011-$x0010x$ & $J''$ & $\chi^3$: $J''$ & SU$x1111\bar{x}$$\,\to\,$U001-$x0001x$ & $J''$ & $\chi^3$: $J''$\\
SU$x1110\bar{x}$$\,\to\,$U101-$x0000\bar{x}$ & $J''$ & $\chi^3$: $J''$ & SU$x1111\bar{x}$$\,\to\,$U111-$x0011\bar{x}$ & $J''$ & $\chi^3$: $J''$\\
U001-$x0000x$$\,\to\,$Z$x1011x$(1) & $J''$ & $J'{,}J''$ & U001-$x0001x$$\,\to\,$Z$x0001x$(1) & $J''$ & $g{,}r{,}J'{,}J''$\\
U001-$x0000x$$\,\to\,$Z$x1110\bar{x}$(1) & $J''$ & $b{,}J''$ & U001-$x0001x$$\,\to\,$Z$x0100\bar{x}$(1) & $J''$ & $b{,}g{,}r{,}J''$\\
U011-$x0010x$$\,\to\,$Z$x0001x$(1) & $J''$ & $g{,}r{,}J'{,}J''$ & U001-$x0001x$$\,\to\,$Z$x1010x$(1) & $J''$ & $b{,}J'{,}J''$\\
U011-$x0010x$$\,\to\,$Z$x0101\bar{x}$(1) & $J''$ & $b{,}g{,}r{,}J''$ & U011-$x0011x$$\,\to\,$Z$x1000x$(1) & $J''$ & $J'{,}J''$\\
U011-$x0010x$$\,\to\,$Z$x1001x$(1) & $J''$ & $b{,}J'{,}J''$ & U011-$x0011x$$\,\to\,$Z$x1100\bar{x}$(1) & $J''$ & $b{,}J''$\\
U101-$x0000\bar{x}$$\,\to\,$Z$x0001x$(1) & $J''$ & $g{,}r{,}J'{,}J''$ & U101-$x0001\bar{x}$$\,\to\,$Z$x1010x$(1) & $J''$ & $J'{,}J''$\\
U101-$x0000\bar{x}$$\,\to\,$Z$x0101x$(1) & $J''$ & $b{,}g{,}r{,}J''$ & U101-$x0001\bar{x}$$\,\to\,$Z$x1110x$(1) & $J''$ & $b{,}J''$\\
U101-$x0000\bar{x}$$\,\to\,$Z$x1011x$(1) & $J''$ & $b{,}J'{,}J''$ & U110-$x1110\bar{x}$$\,\to\,$Z$x0001x$(1) & $J''$ & $J''$\\
U110-$x1110\bar{x}$$\,\to\,$Z$x0001x$(1) & $J''$ & $J''$ & U111-$x0011\bar{x}$$\,\to\,$Z$x0001x$(1) & $J''$ & $g{,}r{,}J'{,}J''$\\
U111-$x0010\bar{x}$$\,\to\,$Z$x1001x$(1) & $J''$ & $J'{,}J''$ & U111-$x0011\bar{x}$$\,\to\,$Z$x0100x$(1) & $J''$ & $b{,}g{,}r{,}J''$\\
U111-$x0010\bar{x}$$\,\to\,$Z$x1100x$(1) & $J''$ & $b{,}J''$ & U111-$x0011\bar{x}$$\,\to\,$Z$x1000x$(1) & $J''$ & $b{,}J'{,}J''$\\
SU$x1110\bar{x}$$\,\to\,$A:SU$x10\bar{x}$ & $J''$ & $\Theta$-odd: $J''$ & SU$x1111\bar{x}$$\,\to\,$B:SU$x01\bar{x}$ & $J''$ & $\Theta$-odd: $J''$\\
SU$x1110\bar{x}$$\,\to\,$D:SU$x00x$ & $J''$ & $\Theta$-odd: $J''$ & SU$x1111\bar{x}$$\,\to\,$C:SU$x11x$ & $J''$ & $\Theta$-odd: $J''$\\
SU$x1110\bar{x}$$\,\to\,$U000-$x1100x$ & $J'''$ & $\chi^3$: $J'''$ & SU$x1111\bar{x}$$\,\to\,$U000-$x1100\bar{x}$ & $J'''$ & $\chi^3$: $J'''$\\
SU$x1110\bar{x}$$\,\to\,$U100-$x1100\bar{x}$ & $J'''$ & $\chi^3$: $J'{,}J'''$ & SU$x1111\bar{x}$$\,\to\,$U100-$x1101\bar{x}$ & $J'''$ & $\chi^3$: $J'{,}J'''$\\
U000-$x1100x$$\,\to\,$Z$x1100x$(0) & $J'''$ & $g{,}r{,}J'{,}J'''$ & U000-$x1100\bar{x}$$\,\to\,$Z$x1100\bar{x}$(0) & $J'''$ & $g{,}r{,}J'''$\\
U000-$x1100x$$\,\to\,$Z$x1100x$(1) & $J'''$ & $b{,}J''$ & U000-$x1100\bar{x}$$\,\to\,$Z$x1100\bar{x}$(1) & $J'''$ & $b{,}J''$\\
U000-$x1100x$$\,\to\,$Z$x1101\bar{x}$(0) & $J'''$ & $g{,}r{,}J'''$ & U000-$x1100\bar{x}$$\,\to\,$Z$x1101x$(0) & $J'''$ & $g{,}r{,}J'{,}J'''$\\
U000-$x1100x$$\,\to\,$Z$x1101\bar{x}$(1) & $J'''$ & $J''$ & U000-$x1100\bar{x}$$\,\to\,$Z$x1101x$(1) & $J'''$ & $J''$\\
U000-$x1100x$$\,\to\,$Z$x1110\bar{x}$(1) & $J'''$ & $b{,}J''$ & U000-$x1100\bar{x}$$\,\to\,$Z$x1110x$(0) & $J'''$ & $J'{,}J'''$\\
U000-$x1100x$$\,\to\,$Z$x1111x$(0) & $J'''$ & $J'{,}J'''$ & U000-$x1100\bar{x}$$\,\to\,$Z$x1110x$(1) & $J'''$ & $b{,}J''$\\
U000-$x1100x$$\,\to\,$Z$x1111x$(1) & $J'''$ & $J''$ & U000-$x1100\bar{x}$$\,\to\,$Z$x1111\bar{x}$(1) & $J'''$ & $J''$\\
U010-$x11100$$\,\to\,$Z$x1101\bar{x}$(0) & $J'''$ & $g{,}r{,}J'''$ & U001-$x0001x$$\,\to\,$Z$x1111\bar{x}$(1) & $J'''$ & $J'''$\\
U011-$x0010x$$\,\to\,$Z$x1101\bar{x}$(1) & $J'''$ & $J'''$ & U010-$x11110$$\,\to\,$Z$x1100\bar{x}$(0) & $J'''$ & $g{,}r{,}J'''$\\
U100-$x1100\bar{x}$$\,\to\,$Z$x1001x$(1) & $J'''$ & $b{,}J''$ & U100-$x1101\bar{x}$$\,\to\,$Z$x1000x$(1) & $J'''$ & $b{,}J''$\\
U100-$x1100\bar{x}$$\,\to\,$Z$x1011x$(1) & $J'''$ & $b{,}J''$ & U100-$x1101\bar{x}$$\,\to\,$Z$x1010x$(1) & $J'''$ & $b{,}J''$\\
U100-$x1100\bar{x}$$\,\to\,$Z$x1101x$(0) & $J'''$ & $g{,}r{,}J'{,}J'''$ & U100-$x1101\bar{x}$$\,\to\,$Z$x1100x$(0) & $J'''$ & $g{,}r{,}J'{,}J'''$\\
U100-$x1100\bar{x}$$\,\to\,$Z$x1111x$(0) & $J'''$ & $J'''$ & U100-$x1101\bar{x}$$\,\to\,$Z$x1110x$(0) & $J'''$ & $J'''$\\
U101-$x0000\bar{x}$$\,\to\,$Z$x1111x$(1) & $J'''$ & $J'''$ & U111-$x0011\bar{x}$$\,\to\,$Z$x1101x$(1) & $J'''$ & $J'''$\\
\multicolumn{3}{l}{\quad not genuine up to $J'''$: Z$x1110\bar{x}$(0)} & \multicolumn{3}{l}{\quad not genuine up to $J'''$: Z$x1111\bar{x}$(0)}\\
\end{tabular}
\end{ruledtabular}
\end{table*}

\begin{table*}
\caption{The $SU(2)$ parents realized only with $J'''$ amplitudes and their descendants, pairs (e) and (f); every descent from these parents has minimal range $J'''$. Second column: bonds whose $\chi^3$ hopping lowers $SU(2)$ to $U(1)$, or whose $SU(2)$-breaking hopping and pairing produce the octahedral $\Z_2$ descendant. Third column: genuine $\Z_2$ descendants of the $U(1)$ state, each with the range at which the $U(1)\to\Z_2$ step is genuine for the $U(1)$ state itself and the bonds carrying the pairing that effects it; for the chiral descendants, the bonds carrying $\T$-odd amplitudes.}
\label{sm:tab:hier-new}
\begin{ruledtabular}
\begin{tabular}{lll}
descendant & amplitude & $\Z_2$ descendants of the $U(1)$ state, or chiral descendants\\
\hline
\multicolumn{3}{l}{(e) SU$x0101\bar{x}$\quad ($x=0$: SU001011, $x=1$: SU101010)}\\
U000-$x0100x$ & $\chi^3$: $J'''$ & Z$x0100x$(0) [$J'''$: $b{,}J'{,}J'''$]; Z$x0100x$(1) [$J'''$: $g{,}r{,}J''$]; Z$x0101\bar{x}$(1) [$J'''$: $g{,}r{,}J''$];\\
 &  & Z$x0110\bar{x}$(0) [$J'''$: $b{,}J'''$]; Z$x0110\bar{x}$(1) [$J'''$: $J''$]; Z$x0111x$(0) [$J'''$: $J'{,}J'''$];\\
 &  & Z$x0111x$(1) [$J'''$: $J''$]\\
U001-$x0000x$ & $\chi^3$: $g{,}r{,}J''$ & Z$x0000x$(1) [$J'$: $b{,}g{,}r{,}J'$]; Z$x0101\bar{x}$(1) [$J'$: $g{,}r$]; Z$x1011x$(1) [$J''$: $J'{,}J''$]; Z$x1110\bar{x}$(1) [$J''$: $b{,}J''$]\\
U010-$x01010$ & $\chi^3$: $J'''$ & Z$x0110x$(0) [$J'''$: $b{,}J'{,}J'''$]; Z$x0110\bar{x}$(0) [$J'''$: $b{,}J'''$]; Z$x1110x$(1) [$J'''$: $g{,}r{,}J''$];\\
 &  & Z$x1110\bar{x}$(1) [$J'''$: $g{,}r{,}J''$]\\
U011-$x0001x$ & $\chi^3$: $J''$ & Z$x0010x$(1) [$J''$: $b{,}J'{,}J''$]; Z$x1010x$(1) [$J''$: $g{,}r{,}J'{,}J''$]; Z$x1110\bar{x}$(1) [$J''$: $b{,}g{,}r{,}J''$];\\
 &  & Z$x0110\bar{x}$(1) [$J'''$: $J'''$]\\
U100-$x0101\bar{x}$ & $\chi^3$: $b{,}J'{,}J'''$ & Z$x0000x$(1) [$J'''$: $g{,}r{,}J''$]; Z$x0010x$(1) [$J'''$: $J''$]; Z$x0100x$(0) [$J'''$: $b{,}J'{,}J'''$];\\
 &  & Z$x0110x$(0) [$J'''$: $b{,}J'{,}J'''$]\\
U101-$x0001\bar{x}$ & $\chi^3$: $g{,}r{,}J''$ & Z$x0000x$(1) [$J'$: $b{,}g{,}r{,}J'$]; Z$x0100x$(1) [$J'$: $g{,}r$]; Z$x1010x$(1) [$J''$: $J'{,}J''$]; Z$x1110x$(1) [$J''$: $b{,}J''$]\\
U110-$x0100\bar{x}$ & $\chi^3$: $J'{,}J'''$ & Z$x0110x$(0) [$J'''$: $b{,}J'{,}J'''$]; Z$x0111x$(0) [$J'''$: $J'''$]; Z$x1010x$(1) [$J'''$: $g{,}r{,}J''$];\\
 &  & Z$x1011x$(1) [$J'''$: $g{,}r{,}J''$]\\
U111-$x0000\bar{x}$ & $\chi^3$: $J''$ & Z$x0010x$(1) [$J''$: $b{,}J'{,}J''$]; Z$x1011x$(1) [$J''$: $g{,}r{,}J'{,}J''$]; Z$x1110x$(1) [$J''$: $b{,}g{,}r{,}J''$];\\
 &  & Z$x0111x$(1) [$J'''$: $J'''$]\\
Z$\bar{x}0110\bar{x}$(0) & $\chi^3{+}\Delta$: $b{,}J'{,}J'''$ & octahedral, no $U(1)$ intermediate\\
chiral & $\Theta$-odd & A:SU$x01\bar{x}$ [$J''$]; B:SU$x00\bar{x}$ [$g{,}r{,}J''$]; C:SU$x00x$ [$J''$]; D:SU$x01x$ [$g{,}r{,}J''$]\\
\multicolumn{3}{l}{not genuine up to $J'''$: Z$x0101\bar{x}$(0)}\\
\hline
\multicolumn{3}{l}{(f) SU$x0111\bar{x}$\quad ($x=0$: SU001111, $x=1$: SU101110)}\\
U000-$x0100\bar{x}$ & $\chi^3$: $J'''$ & Z$x0100\bar{x}$(0) [$J'''$: $b{,}J'''$]; Z$x0100\bar{x}$(1) [$J'''$: $g{,}r{,}J''$]; Z$x0101x$(0) [$J'''$: $J'{,}J'''$];\\
 &  & Z$x0101x$(1) [$J'''$: $g{,}r{,}J''$]; Z$x0110x$(0) [$J'''$: $b{,}J'{,}J'''$]; Z$x0110x$(1) [$J'''$: $J''$];\\
 &  & Z$x0111\bar{x}$(1) [$J'''$: $J''$]\\
U001-$x0010x$ & $\chi^3$: $J''$ & Z$x0010x$(1) [$J''$: $b{,}J'{,}J''$]; Z$x1001x$(1) [$J''$: $g{,}r{,}J'{,}J''$]; Z$x1100\bar{x}$(1) [$J''$: $b{,}g{,}r{,}J''$];\\
 &  & Z$x0111\bar{x}$(1) [$J'''$: $J'''$]\\
U010-$x01110$ & $\chi^3$: $J'''$ & Z$x0100x$(0) [$J'''$: $b{,}J'{,}J'''$]; Z$x0100\bar{x}$(0) [$J'''$: $b{,}J'''$]; Z$x1100x$(1) [$J'''$: $g{,}r{,}J''$];\\
 &  & Z$x1100\bar{x}$(1) [$J'''$: $g{,}r{,}J''$]\\
U011-$x0011x$ & $\chi^3$: $g{,}r{,}J''$ & Z$x0000x$(1) [$J'$: $b{,}g{,}r{,}J'$]; Z$x0100\bar{x}$(1) [$J'$: $g{,}r$]; Z$x1000x$(1) [$J''$: $J'{,}J''$]; Z$x1100\bar{x}$(1) [$J''$: $b{,}J''$]\\
U100-$x0101\bar{x}$ & $\chi^3$: $b{,}J'{,}J'''$ & Z$x0000x$(1) [$J'''$: $g{,}r{,}J''$]; Z$x0010x$(1) [$J'''$: $J''$]; Z$x0100x$(0) [$J'''$: $b{,}J'{,}J'''$];\\
 &  & Z$x0110x$(0) [$J'''$: $b{,}J'{,}J'''$]\\
U101-$x0011\bar{x}$ & $\chi^3$: $J''$ & Z$x0010x$(1) [$J''$: $b{,}J'{,}J''$]; Z$x1000x$(1) [$J''$: $g{,}r{,}J'{,}J''$]; Z$x1100x$(1) [$J''$: $b{,}g{,}r{,}J''$];\\
 &  & Z$x0110x$(1) [$J'''$: $J'''$]\\
U110-$x0110\bar{x}$ & $\chi^3$: $J'{,}J'''$ & Z$x0100x$(0) [$J'''$: $b{,}J'{,}J'''$]; Z$x0101x$(0) [$J'''$: $J'''$]; Z$x1000x$(1) [$J'''$: $g{,}r{,}J''$];\\
 &  & Z$x1001x$(1) [$J'''$: $g{,}r{,}J''$]\\
U111-$x0010\bar{x}$ & $\chi^3$: $g{,}r{,}J''$ & Z$x0000x$(1) [$J'$: $b{,}g{,}r{,}J'$]; Z$x0101x$(1) [$J'$: $g{,}r$]; Z$x1001x$(1) [$J''$: $J'{,}J''$]; Z$x1100x$(1) [$J''$: $b{,}J''$]\\
Z$\bar{x}0100\bar{x}$(0) & $\chi^3{+}\Delta$: $b{,}J'{,}J'''$ & octahedral, no $U(1)$ intermediate\\
chiral & $\Theta$-odd & A:SU$x11\bar{x}$ [$g{,}r{,}J''$]; B:SU$x10\bar{x}$ [$J''$]; C:SU$x10x$ [$g{,}r{,}J''$]; D:SU$x11x$ [$J''$]\\
\multicolumn{3}{l}{not genuine up to $J'''$: Z$x0111\bar{x}$(0)}\\
\end{tabular}
\end{ruledtabular}
\end{table*}

\begin{table*}
\caption{As Table~\ref{sm:tab:hier-new}, for the pairs (g) and (h).}
\label{sm:tab:hier-new2}
\begin{ruledtabular}
\begin{tabular}{lll}
descendant & amplitude & $\Z_2$ descendants of the $U(1)$ state, or chiral descendants\\
\hline
\multicolumn{3}{l}{(g) SU$x1110x$\quad ($x=0$: SU011100, $x=1$: SU111101)}\\
U000-$x1100\bar{x}$ & $\chi^3$: $J'{,}J'''$ & Z$x1100\bar{x}$(0) [$J'''$: $g{,}r{,}J'''$]; Z$x1100\bar{x}$(1) [$J'''$: $b{,}J''$]; Z$x1101x$(0) [$J'''$: $g{,}r{,}J'{,}J'''$];\\
 &  & Z$x1101x$(1) [$J'''$: $J''$]; Z$x1110x$(0) [$J'''$: $J'{,}J'''$]; Z$x1110x$(1) [$J'''$: $b{,}J''$];\\
 &  & Z$x1111\bar{x}$(1) [$J'''$: $J''$]\\
U001-$x0000\bar{x}$ & $\chi^3$: $b{,}J''$ & Z$x0000\bar{x}$(1) [$J''$: $b{,}g{,}r{,}J''$]; Z$x0101x$(1) [$J''$: $g{,}r{,}J'{,}J''$]; Z$x1110x$(1) [$J''$: $b{,}J'{,}J''$];\\
 &  & Z$x1011\bar{x}$(1) [$J'''$: $J''{,}J'''$]\\
U010-$x11100$ & $\chi^3$: $g{,}r{,}J'{,}J'''$ & Z$x0101x$(1) [$J'$: $b$]; Z$x0101\bar{x}$(1) [$J'$: $b$]; Z$x1101x$(0) [$J'$: $g{,}r{,}J'$]; Z$x1101\bar{x}$(0) [$J'''$: $g{,}r{,}J'''$]\\
U011-$x0010\bar{x}$ & $\chi^3$: $J''$ & Z$x0001\bar{x}$(1) [$J'''$: $g{,}r{,}J''{,}J'''$]; Z$x0101x$(1) [$J'''$: $b{,}g{,}r{,}J'{,}J''{,}J'''$];\\
 &  & Z$x1001\bar{x}$(1) [$J'''$: $b{,}J''{,}J'''$]; Z$x1101x$(1) [$J'''$: $J'{,}J'''$]\\
U101-$x0000x$ & $\chi^3$: $J''$ & Z$x0001\bar{x}$(1) [$J'''$: $g{,}r{,}J''{,}J'''$]; Z$x0101\bar{x}$(1) [$J'''$: $b{,}g{,}r{,}J'{,}J''{,}J'''$];\\
 &  & Z$x1011\bar{x}$(1) [$J'''$: $b{,}J''{,}J'''$]; Z$x1111\bar{x}$(1) [$J'''$: $J'{,}J'''$]\\
U110-$x1110x$ & $\chi^3$: $g{,}r{,}J'''$ & Z$x0000\bar{x}$(1) [$J'''$: $b{,}J''$]; Z$x0001\bar{x}$(1) [$J'''$: $J''$]; Z$x1100\bar{x}$(0) [$J'''$: $g{,}r{,}J'{,}J'''$];\\
 &  & Z$x1101\bar{x}$(0) [$J'''$: $g{,}r{,}J'{,}J'''$]\\
U111-$x0010x$ & $\chi^3$: $b{,}J''$ & Z$x0000\bar{x}$(1) [$J''$: $b{,}g{,}r{,}J''$]; Z$x0101\bar{x}$(1) [$J''$: $g{,}r{,}J'{,}J''$]; Z$x1100\bar{x}$(1) [$J''$: $b{,}J'{,}J''$];\\
 &  & Z$x1001\bar{x}$(1) [$J'''$: $J''{,}J'''$]\\
Z$\bar{x}1101x$(0) & $\chi^3{+}\Delta$: $g{,}r{,}J'{,}J'''$ & octahedral, no $U(1)$ intermediate\\
chiral & $\Theta$-odd & A:SU$x10x$ [$J''$]; B:SU$x00x$ [$b{,}J''$]; C:SU$x10\bar{x}$ [$b{,}J''$]; D:SU$x00\bar{x}$ [$J''$]\\
\multicolumn{3}{l}{not genuine up to $J'''$: U100-$x1100x$, Z$x1111\bar{x}$(0)}\\
\hline
\multicolumn{3}{l}{(h) SU$x1111x$\quad ($x=0$: SU011110, $x=1$: SU111111)}\\
U000-$x1100x$ & $\chi^3$: $J'{,}J'''$ & Z$x1100x$(0) [$J'''$: $g{,}r{,}J'{,}J'''$]; Z$x1100x$(1) [$J'''$: $b{,}J''$]; Z$x1101\bar{x}$(0) [$J'''$: $g{,}r{,}J'''$];\\
 &  & Z$x1101\bar{x}$(1) [$J'''$: $J''$]; Z$x1110\bar{x}$(1) [$J'''$: $b{,}J''$]; Z$x1111x$(0) [$J'''$: $J'{,}J'''$];\\
 &  & Z$x1111x$(1) [$J'''$: $J''$]\\
U001-$x0001\bar{x}$ & $\chi^3$: $J''$ & Z$x0001\bar{x}$(1) [$J'''$: $g{,}r{,}J''{,}J'''$]; Z$x0100x$(1) [$J'''$: $b{,}g{,}r{,}J'{,}J''{,}J'''$];\\
 &  & Z$x1010\bar{x}$(1) [$J'''$: $b{,}J''{,}J'''$]; Z$x1111x$(1) [$J'''$: $J'{,}J'''$]\\
U010-$x11110$ & $\chi^3$: $g{,}r{,}J'{,}J'''$ & Z$x0100x$(1) [$J'$: $b$]; Z$x0100\bar{x}$(1) [$J'$: $b$]; Z$x1100x$(0) [$J'$: $g{,}r{,}J'$]; Z$x1100\bar{x}$(0) [$J'''$: $g{,}r{,}J'''$]\\
U011-$x0011\bar{x}$ & $\chi^3$: $b{,}J''$ & Z$x0000\bar{x}$(1) [$J''$: $b{,}g{,}r{,}J''$]; Z$x0100x$(1) [$J''$: $g{,}r{,}J'{,}J''$]; Z$x1100x$(1) [$J''$: $b{,}J'{,}J''$];\\
 &  & Z$x1000\bar{x}$(1) [$J'''$: $J''{,}J'''$]\\
U101-$x0001x$ & $\chi^3$: $b{,}J''$ & Z$x0000\bar{x}$(1) [$J''$: $b{,}g{,}r{,}J''$]; Z$x0100\bar{x}$(1) [$J''$: $g{,}r{,}J'{,}J''$]; Z$x1110\bar{x}$(1) [$J''$: $b{,}J'{,}J''$];\\
 &  & Z$x1010\bar{x}$(1) [$J'''$: $J''{,}J'''$]\\
U110-$x1110x$ & $\chi^3$: $g{,}r{,}J'''$ & Z$x0000\bar{x}$(1) [$J'''$: $b{,}J''$]; Z$x0001\bar{x}$(1) [$J'''$: $J''$]; Z$x1100\bar{x}$(0) [$J'''$: $g{,}r{,}J'{,}J'''$];\\
 &  & Z$x1101\bar{x}$(0) [$J'''$: $g{,}r{,}J'{,}J'''$]\\
U111-$x0011x$ & $\chi^3$: $J''$ & Z$x0001\bar{x}$(1) [$J'''$: $g{,}r{,}J''{,}J'''$]; Z$x0100\bar{x}$(1) [$J'''$: $b{,}g{,}r{,}J'{,}J''{,}J'''$];\\
 &  & Z$x1000\bar{x}$(1) [$J'''$: $b{,}J''{,}J'''$]; Z$x1101\bar{x}$(1) [$J'''$: $J'{,}J'''$]\\
Z$\bar{x}1100x$(0) & $\chi^3{+}\Delta$: $g{,}r{,}J'{,}J'''$ & octahedral, no $U(1)$ intermediate\\
chiral & $\Theta$-odd & A:SU$x11x$ [$b{,}J''$]; B:SU$x01x$ [$J''$]; C:SU$x11\bar{x}$ [$J''$]; D:SU$x01\bar{x}$ [$b{,}J''$]\\
\multicolumn{3}{l}{not genuine up to $J'''$: U100-$x1101x$, Z$x1110\bar{x}$(0)}\\
\end{tabular}
\end{ruledtabular}
\end{table*}

\begin{table*}
\caption{Chiral descendants of the $U(1)$ spin liquids of the hierarchy. Columns: the $U(1)$ state $X$ (with the symbolic digit $x$), the range from which $X$ is genuine, and for each class A--D the labels $n_{T_2}n_\sigma n_{\sigma C_6}n_{C_6}$ of the chiral PSG, whose space-group sector is given in the line heading each block, followed by the minimal range and the bonds that carry $\T$-odd amplitudes at that range; $J_1$ stands for $b,g,r$ together.}
\label{sm:tab:hier-chiral-u1}
\begin{ruledtabular}
\begin{tabular}{lcllllllll}
$X$ & range & \multicolumn{2}{c}{A} & \multicolumn{2}{c}{B} & \multicolumn{2}{c}{C} & \multicolumn{2}{c}{D}\\
\hline
\multicolumn{10}{l}{sector $w_{C_6}w_\sigma w_\T=000$; space-group sectors $w_{C_6}w_\sigma$ of the chiral PSGs: A 00, B 00, C 00, D 00}\\
U000-$x0100x$ & $J'''$ & $x00x$ & $J'''$: $g{,}r{,}J''$ & $x00\bar{x}$ & $J'''$: $g{,}r{,}J''$ & $x00x$ & $J'''$: $g{,}r{,}J''$ & $x00\bar{x}$ & $J'''$: $g{,}r{,}J''$\\
U000-$x0100\bar{x}$ & $J'''$ & $x00\bar{x}$ & $J'''$: $g{,}r{,}J''$ & $x00x$ & $J'''$: $g{,}r{,}J''$ & $x00\bar{x}$ & $J'''$: $g{,}r{,}J''$ & $x00x$ & $J'''$: $g{,}r{,}J''$\\
U000-$x1100x$ & $J'''$ & $x00x$ & $J'''$: $b{,}J''$ & $x00\bar{x}$ & $J'''$: $b{,}J''$ & $x00\bar{x}$ & $J'''$: $b{,}J''$ & $x00x$ & $J'''$: $b{,}J''$\\
U000-$x1100\bar{x}$ & $J'''$ & $x00\bar{x}$ & $J'''$: $b{,}J''$ & $x00x$ & $J'''$: $b{,}J''$ & $x00x$ & $J'''$: $b{,}J''$ & $x00\bar{x}$ & $J'''$: $b{,}J''$\\
\colrule
\multicolumn{10}{l}{sector $w_{C_6}w_\sigma w_\T=001$; space-group sectors $w_{C_6}w_\sigma$ of the chiral PSGs: A 00, B 01, C 10, D 11}\\
U001-$x0000x$ & $J'$ & $x00x$ & $J''$: $J''$ & $x000$ & $J'$: $J_1$ & $x00x$ & $J'$: $b{,}J'$ & $x00x$ & $J'$: $g{,}r{,}J'$\\
U001-$x0000\bar{x}$ & $J''$ & $x00\bar{x}$ & $J''$: $J'{,}J''$ & $x000$ & $J''$: $J_1{,}J'{,}J''$ & $x00\bar{x}$ & $J''$: $b{,}J''$ & $x00\bar{x}$ & $J''$: $g{,}r{,}J''$\\
U001-$x0001x$ & $J''$ & $x00\bar{x}$ & $J''$: $b{,}J''$ & $x010$ & $J''$: $g{,}r{,}J''$ & $x01x$ & $J''$: $J'{,}J''$ & $x00x$ & $J''$: $J_1{,}J'{,}J''$\\
U001-$x0001\bar{x}$ & $J'''$ & $x00x$ & $J'''$: $b{,}J'{,}J''{,}J'''$ & $x010$ & $J'''$: $g{,}r{,}J'{,}J''{,}J'''$ & $x01\bar{x}$ & $J'''$: $J''{,}J'''$ & $x00\bar{x}$ & $J'''$: $J_1{,}J''{,}J'''$\\
U001-$x0010x$ & $J''$ & $x00\bar{x}$ & $J''$: $g{,}r{,}J''$ & $x100$ & $J''$: $b{,}J''$ & $x00x$ & $J''$: $J_1{,}J'{,}J''$ & $x10x$ & $J''$: $J'{,}J''$\\
\colrule
\multicolumn{10}{l}{sector $w_{C_6}w_\sigma w_\T=010$; space-group sectors $w_{C_6}w_\sigma$ of the chiral PSGs: A 01, B 01, C 01, D 01}\\
U010-$x01010$ & $J'''$ & $x010$ & $J'''$: $J''$ & $x000$ & $J'''$: $g{,}r{,}J''$ & $x000$ & $J'''$: $J''$ & $x010$ & $J'''$: $g{,}r{,}J''$\\
U010-$x01110$ & $J'''$ & $x110$ & $J'''$: $g{,}r{,}J''$ & $x100$ & $J'''$: $J''$ & $x100$ & $J'''$: $g{,}r{,}J''$ & $x110$ & $J'''$: $J''$\\
U010-$x10110$ & $J_1$ & $x110$ & $J'$: $J'$ & $x000$ & $J'$: $J'$ & $x100$ & $J'$: $J'$ & $x010$ & $J'$: $J'$\\
U010-$x11100$ & $J'$ & $x100$ & $J''$: $J''$ & $x000$ & $J'$: $b$ & $x100$ & $J'$: $b$ & $x000$ & $J''$: $J''$\\
U010-$x11110$ & $J'$ & $x110$ & $J'$: $b$ & $x010$ & $J''$: $J''$ & $x110$ & $J''$: $J''$ & $x010$ & $J'$: $b$\\
\colrule
\multicolumn{10}{l}{sector $w_{C_6}w_\sigma w_\T=011$; space-group sectors $w_{C_6}w_\sigma$ of the chiral PSGs: A 01, B 00, C 11, D 10}\\
U011-$x0001x$ & $J''$ & $x010$ & $J''$: $b{,}J''$ & $x00\bar{x}$ & $J''$: $g{,}r{,}J''$ & $x00x$ & $J''$: $J'{,}J''$ & $x01x$ & $J''$: $J_1{,}J'{,}J''$\\
U011-$x0010x$ & $J''$ & $x100$ & $J''$: $g{,}r{,}J''$ & $x00\bar{x}$ & $J''$: $b{,}J''$ & $x10x$ & $J''$: $J_1{,}J'{,}J''$ & $x00x$ & $J''$: $J'{,}J''$\\
U011-$x0010\bar{x}$ & $J'''$ & $x100$ & $J'''$: $g{,}r{,}J'{,}J''{,}J'''$ & $x00x$ & $J'''$: $b{,}J'{,}J''{,}J'''$ & $x10\bar{x}$ & $J'''$: $J_1{,}J''{,}J'''$ & $x00\bar{x}$ & $J'''$: $J''{,}J'''$\\
U011-$x0011x$ & $J'$ & $x110$ & $J'$: $J_1$ & $x00x$ & $J''$: $J''$ & $x10x$ & $J'$: $g{,}r{,}J'$ & $x01x$ & $J'$: $b{,}J'$\\
U011-$x0011\bar{x}$ & $J''$ & $x110$ & $J''$: $J_1{,}J'{,}J''$ & $x00\bar{x}$ & $J''$: $J'{,}J''$ & $x10\bar{x}$ & $J''$: $g{,}r{,}J''$ & $x01\bar{x}$ & $J''$: $b{,}J''$\\
\colrule
\multicolumn{10}{l}{sector $w_{C_6}w_\sigma w_\T=100$; space-group sectors $w_{C_6}w_\sigma$ of the chiral PSGs: A 10, B 10, C 10, D 10}\\
U100-$x0101\bar{x}$ & $J'''$ & $x01\bar{x}$ & $J'''$: $g{,}r{,}J''$ & $x00\bar{x}$ & $J'''$: $g{,}r{,}J''$ & $x00x$ & $J'''$: $g{,}r{,}J''$ & $x01x$ & $J'''$: $g{,}r{,}J''$\\
U100-$x1100\bar{x}$ & $J'''$ & $x00\bar{x}$ & $J'''$: $J''$ & $x00\bar{x}$ & $J'''$: $b{,}J''$ & $x00x$ & $J'''$: $b{,}J''$ & $x00x$ & $J'''$: $J''$\\
U100-$x1101\bar{x}$ & $J'''$ & $x01\bar{x}$ & $J'''$: $b{,}J''$ & $x01\bar{x}$ & $J'''$: $J''$ & $x01x$ & $J'''$: $J''$ & $x01x$ & $J'''$: $b{,}J''$\\
\colrule
\multicolumn{10}{l}{sector $w_{C_6}w_\sigma w_\T=101$; space-group sectors $w_{C_6}w_\sigma$ of the chiral PSGs: A 10, B 11, C 00, D 01}\\
U101-$x0000x$ & $J'''$ & $x00x$ & $J'''$: $J''{,}J'''$ & $x00x$ & $J'''$: $J_1{,}J''{,}J'''$ & $x00x$ & $J'''$: $b{,}J'{,}J''{,}J'''$ & $x000$ & $J'''$: $g{,}r{,}J'{,}J''{,}J'''$\\
U101-$x0000\bar{x}$ & $J''$ & $x00\bar{x}$ & $J''$: $J'{,}J''$ & $x00\bar{x}$ & $J''$: $J_1{,}J'{,}J''$ & $x00\bar{x}$ & $J''$: $b{,}J''$ & $x000$ & $J''$: $g{,}r{,}J''$\\
U101-$x0001x$ & $J''$ & $x01x$ & $J''$: $b{,}J''$ & $x00x$ & $J''$: $g{,}r{,}J''$ & $x00\bar{x}$ & $J''$: $J'{,}J''$ & $x010$ & $J''$: $J_1{,}J'{,}J''$\\
U101-$x0001\bar{x}$ & $J'$ & $x01\bar{x}$ & $J'$: $b{,}J'$ & $x00\bar{x}$ & $J'$: $g{,}r{,}J'$ & $x00x$ & $J''$: $J''$ & $x010$ & $J'$: $J_1$\\
U101-$x0011\bar{x}$ & $J''$ & $x01\bar{x}$ & $J''$: $J_1{,}J'{,}J''$ & $x10\bar{x}$ & $J''$: $J'{,}J''$ & $x00\bar{x}$ & $J''$: $g{,}r{,}J''$ & $x110$ & $J''$: $b{,}J''$\\
\colrule
\multicolumn{10}{l}{sector $w_{C_6}w_\sigma w_\T=110$; space-group sectors $w_{C_6}w_\sigma$ of the chiral PSGs: A 11, B 11, C 11, D 11}\\
U110-$x0100\bar{x}$ & $J'''$ & $x00\bar{x}$ & $J'''$: $J''$ & $x00\bar{x}$ & $J'''$: $g{,}r{,}J''$ & $x00x$ & $J'''$: $J''$ & $x00x$ & $J'''$: $g{,}r{,}J''$\\
U110-$x0110\bar{x}$ & $J'''$ & $x10\bar{x}$ & $J'''$: $g{,}r{,}J''$ & $x10\bar{x}$ & $J'''$: $J''$ & $x10x$ & $J'''$: $g{,}r{,}J''$ & $x10x$ & $J'''$: $J''$\\
U110-$x1010x$ & $J_1$ & $x10x$ & $J''$: $J''$ & $x00x$ & $J''$: $J''$ & $x10x$ & $J'$: $J'$ & $x00x$ & $J'$: $J'$\\
U110-$x1010\bar{x}$ & $J_1$ & $x10\bar{x}$ & $J'$: $J'$ & $x00\bar{x}$ & $J'$: $J'$ & $x10\bar{x}$ & $J''$: $J''$ & $x00\bar{x}$ & $J''$: $J''$\\
U110-$x1110x$ & $J'''$ & $x10x$ & $J'''$: $b{,}J''$ & $x00x$ & $J'''$: $b{,}J''$ & $x10\bar{x}$ & $J'''$: $b{,}J''$ & $x00\bar{x}$ & $J'''$: $b{,}J''$\\
U110-$x1110\bar{x}$ & $J'$ & $x10\bar{x}$ & $J'$: $b$ & $x00\bar{x}$ & $J'$: $b$ & $x10x$ & $J'$: $b$ & $x00x$ & $J'$: $b$\\
\colrule
\multicolumn{10}{l}{sector $w_{C_6}w_\sigma w_\T=111$; space-group sectors $w_{C_6}w_\sigma$ of the chiral PSGs: A 11, B 10, C 01, D 00}\\
U111-$x0000\bar{x}$ & $J''$ & $x00\bar{x}$ & $J''$: $J'{,}J''$ & $x00\bar{x}$ & $J''$: $J_1{,}J'{,}J''$ & $x000$ & $J''$: $b{,}J''$ & $x00\bar{x}$ & $J''$: $g{,}r{,}J''$\\
U111-$x0010x$ & $J''$ & $x10x$ & $J''$: $g{,}r{,}J''$ & $x00x$ & $J''$: $b{,}J''$ & $x100$ & $J''$: $J_1{,}J'{,}J''$ & $x00\bar{x}$ & $J''$: $J'{,}J''$\\
U111-$x0010\bar{x}$ & $J'$ & $x10\bar{x}$ & $J'$: $g{,}r{,}J'$ & $x00\bar{x}$ & $J'$: $b{,}J'$ & $x100$ & $J'$: $J_1$ & $x00x$ & $J''$: $J''$\\
U111-$x0011x$ & $J'''$ & $x10x$ & $J'''$: $J_1{,}J''{,}J'''$ & $x01x$ & $J'''$: $J''{,}J'''$ & $x110$ & $J'''$: $g{,}r{,}J'{,}J''{,}J'''$ & $x00x$ & $J'''$: $b{,}J'{,}J''{,}J'''$\\
U111-$x0011\bar{x}$ & $J''$ & $x10\bar{x}$ & $J''$: $J_1{,}J'{,}J''$ & $x01\bar{x}$ & $J''$: $J'{,}J''$ & $x110$ & $J''$: $g{,}r{,}J''$ & $x00\bar{x}$ & $J''$: $b{,}J''$\\
\end{tabular}
\end{ruledtabular}
\end{table*}

\begin{table*}
\caption{Chiral descendants of the $\Z_2$ spin liquids of the hierarchy, as in Table~\ref{sm:tab:hier-chiral-u1}; the labels $n_{T_2}n_\sigma n_{\sigma C_6}n_{C_6}$ of the chiral PSG follow Eq.~\eqref{sm:eq:chiral-label}.}
\label{sm:tab:hier-chiral-z2}
\begin{ruledtabular}
\begin{tabular}{lcllllllll}
$X$ & range & \multicolumn{2}{c}{A} & \multicolumn{2}{c}{B} & \multicolumn{2}{c}{C} & \multicolumn{2}{c}{D}\\
\hline
Z$x0000x$(1) & $J'$ & $x00x$ & $J'$: $J_1{,}J'$ & $x11x$ & $J'$: $J_1{,}J'$ & $x01\bar{x}$ & $J'$: $J_1{,}J'$ & $x10\bar{x}$ & $J'$: $J_1{,}J'$\\
Z$x0000\bar{x}$(1) & $J''$ & $x00\bar{x}$ & $J''$: $J_1{,}J'{,}J''$ & $x11\bar{x}$ & $J''$: $J_1{,}J'{,}J''$ & $x01x$ & $J''$: $J_1{,}J'{,}J''$ & $x10x$ & $J''$: $J_1{,}J'{,}J''$\\
Z$x0001x$(1) & $J''$ & $x01x$ & $J''$: $J_1{,}J'{,}J''$ & $x10x$ & $J''$: $J_1{,}J'{,}J''$ & $x00\bar{x}$ & $J''$: $J_1{,}J'{,}J''$ & $x11\bar{x}$ & $J''$: $J_1{,}J'{,}J''$\\
Z$x0001\bar{x}$(1) & $J'''$ & $x01\bar{x}$ & $J'''$: $J_1{,}J'{,}J''{,}J'''$ & $x10\bar{x}$ & $J'''$: $J_1{,}J'{,}J''{,}J'''$ & $x00x$ & $J'''$: $J_1{,}J'{,}J''{,}J'''$ & $x11x$ & $J'''$: $J_1{,}J'{,}J''{,}J'''$\\
Z$x0010x$(1) & $J''$ & $x10x$ & $J''$: $J_1{,}J'{,}J''$ & $x01x$ & $J''$: $J_1{,}J'{,}J''$ & $x11\bar{x}$ & $J''$: $J_1{,}J'{,}J''$ & $x00\bar{x}$ & $J''$: $J_1{,}J'{,}J''$\\
Z$x0100x$(0) & $J'''$ & $x00x$ & $J'''$: $g{,}r{,}J''$ & $x01x$ & $J'''$: $g{,}r{,}J''$ & $x01\bar{x}$ & $J'''$: $g{,}r{,}J''$ & $x00\bar{x}$ & $J'''$: $g{,}r{,}J''$\\
Z$x0100x$(1) & $J'$ & $x00x$ & $J'$: $J_1{,}J'$ & $x10x$ & $J'$: $g{,}r{,}J'$ & $x00x$ & $J''$: $g{,}r{,}J''$ & $x10x$ & $J'$: $J_1$\\
Z$x0100\bar{x}$(0) & $J'''$ & $x00\bar{x}$ & $J'''$: $g{,}r{,}J''$ & $x01\bar{x}$ & $J'''$: $g{,}r{,}J''$ & $x01x$ & $J'''$: $g{,}r{,}J''$ & $x00x$ & $J'''$: $g{,}r{,}J''$\\
Z$x0100\bar{x}$(1) & $J'$ & $x00\bar{x}$ & $J'$: $J_1$ & $x10\bar{x}$ & $J''$: $g{,}r{,}J''$ & $x00\bar{x}$ & $J'$: $g{,}r{,}J'$ & $x10\bar{x}$ & $J'$: $J_1{,}J'$\\
Z$x0101x$(0) & $J'''$ & $x01x$ & $J'''$: $g{,}r{,}J''$ & $x00x$ & $J'''$: $g{,}r{,}J''$ & $x00\bar{x}$ & $J'''$: $g{,}r{,}J''$ & $x01\bar{x}$ & $J'''$: $g{,}r{,}J''$\\
Z$x0101x$(1) & $J'$ & $x01x$ & $J'$: $g{,}r{,}J'$ & $x11x$ & $J'$: $J_1{,}J'$ & $x01x$ & $J'$: $J_1$ & $x11x$ & $J''$: $g{,}r{,}J''$\\
Z$x0101\bar{x}$(1) & $J'$ & $x01\bar{x}$ & $J''$: $g{,}r{,}J''$ & $x11\bar{x}$ & $J'$: $J_1$ & $x01\bar{x}$ & $J'$: $J_1{,}J'$ & $x11\bar{x}$ & $J'$: $g{,}r{,}J'$\\
Z$x0110x$(0) & $J'''$ & $x10x$ & $J'''$: $g{,}r{,}J''$ & $x11x$ & $J'''$: $g{,}r{,}J''$ & $x11\bar{x}$ & $J'''$: $g{,}r{,}J''$ & $x10\bar{x}$ & $J'''$: $g{,}r{,}J''$\\
Z$x0110x$(1) & $J'''$ & $x10x$ & $J'''$: $J_1{,}J'{,}J''{,}J'''$ & $x00x$ & $J'''$: $g{,}r{,}J'{,}J''{,}J'''$ & $x10x$ & $J'''$: $g{,}r{,}J''{,}J'''$ & $x00x$ & $J'''$: $J_1{,}J''{,}J'''$\\
Z$x0110\bar{x}$(0) & $J'''$ & $x10\bar{x}$ & $J'''$: $g{,}r{,}J''$ & $x11\bar{x}$ & $J'''$: $g{,}r{,}J''$ & $x11x$ & $J'''$: $g{,}r{,}J''$ & $x10x$ & $J'''$: $g{,}r{,}J''$\\
Z$x0110\bar{x}$(1) & $J'''$ & $x10\bar{x}$ & $J'''$: $J_1{,}J''{,}J'''$ & $x00\bar{x}$ & $J'''$: $g{,}r{,}J''{,}J'''$ & $x10\bar{x}$ & $J'''$: $g{,}r{,}J'{,}J''{,}J'''$ & $x00\bar{x}$ & $J'''$: $J_1{,}J'{,}J''{,}J'''$\\
Z$x0111x$(0) & $J'''$ & $x11x$ & $J'''$: $g{,}r{,}J''$ & $x10x$ & $J'''$: $g{,}r{,}J''$ & $x10\bar{x}$ & $J'''$: $g{,}r{,}J''$ & $x11\bar{x}$ & $J'''$: $g{,}r{,}J''$\\
Z$x0111x$(1) & $J'''$ & $x11x$ & $J'''$: $g{,}r{,}J'{,}J''{,}J'''$ & $x01x$ & $J'''$: $J_1{,}J'{,}J''{,}J'''$ & $x11x$ & $J'''$: $J_1{,}J''{,}J'''$ & $x01x$ & $J'''$: $g{,}r{,}J''{,}J'''$\\
Z$x0111\bar{x}$(1) & $J'''$ & $x11\bar{x}$ & $J'''$: $g{,}r{,}J''{,}J'''$ & $x01\bar{x}$ & $J'''$: $J_1{,}J''{,}J'''$ & $x11\bar{x}$ & $J'''$: $J_1{,}J'{,}J''{,}J'''$ & $x01\bar{x}$ & $J'''$: $g{,}r{,}J'{,}J''{,}J'''$\\
Z$x1000x$(0) & $J_1$ & $x00x$ & $J'$: $J'$ & $x11x$ & $J'$: $J'$ & $x01x$ & $J'$: $J'$ & $x10x$ & $J'$: $J'$\\
Z$x1000x$(1) & $J''$ & $x00x$ & $J''$: $J_1{,}J'{,}J''$ & $x00x$ & $J''$: $J'{,}J''$ & $x00\bar{x}$ & $J''$: $g{,}r{,}J'{,}J''$ & $x00\bar{x}$ & $J''$: $b{,}J'{,}J''$\\
Z$x1000\bar{x}$(0) & $J_1$ & $x00\bar{x}$ & $J'$: $J'$ & $x11\bar{x}$ & $J'$: $J'$ & $x01\bar{x}$ & $J'$: $J'$ & $x10\bar{x}$ & $J'$: $J'$\\
Z$x1000\bar{x}$(1) & $J'''$ & $x00\bar{x}$ & $J'''$: $J_1{,}J'{,}J''{,}J'''$ & $x00\bar{x}$ & $J'''$: $J'{,}J''{,}J'''$ & $x00x$ & $J'''$: $g{,}r{,}J'{,}J''{,}J'''$ & $x00x$ & $J'''$: $b{,}J'{,}J''{,}J'''$\\
Z$x1001x$(1) & $J''$ & $x01x$ & $J''$: $g{,}r{,}J'{,}J''$ & $x01x$ & $J''$: $b{,}J'{,}J''$ & $x01\bar{x}$ & $J''$: $J_1{,}J'{,}J''$ & $x01\bar{x}$ & $J''$: $J'{,}J''$\\
Z$x1001\bar{x}$(1) & $J'''$ & $x01\bar{x}$ & $J'''$: $g{,}r{,}J'{,}J''{,}J'''$ & $x01\bar{x}$ & $J'''$: $b{,}J'{,}J''{,}J'''$ & $x01x$ & $J'''$: $J_1{,}J'{,}J''{,}J'''$ & $x01x$ & $J'''$: $J'{,}J''{,}J'''$\\
Z$x1010x$(1) & $J''$ & $x10x$ & $J''$: $b{,}J'{,}J''$ & $x10x$ & $J''$: $g{,}r{,}J'{,}J''$ & $x10\bar{x}$ & $J''$: $J'{,}J''$ & $x10\bar{x}$ & $J''$: $J_1{,}J'{,}J''$\\
Z$x1010\bar{x}$(1) & $J'''$ & $x10\bar{x}$ & $J'''$: $b{,}J'{,}J''{,}J'''$ & $x10\bar{x}$ & $J'''$: $g{,}r{,}J'{,}J''{,}J'''$ & $x10x$ & $J'''$: $J'{,}J''{,}J'''$ & $x10x$ & $J'''$: $J_1{,}J'{,}J''{,}J'''$\\
Z$x1011x$(1) & $J''$ & $x11x$ & $J''$: $J'{,}J''$ & $x11x$ & $J''$: $J_1{,}J'{,}J''$ & $x11\bar{x}$ & $J''$: $b{,}J'{,}J''$ & $x11\bar{x}$ & $J''$: $g{,}r{,}J'{,}J''$\\
Z$x1011\bar{x}$(1) & $J'''$ & $x11\bar{x}$ & $J'''$: $J'{,}J''{,}J'''$ & $x11\bar{x}$ & $J'''$: $J_1{,}J'{,}J''{,}J'''$ & $x11x$ & $J'''$: $b{,}J'{,}J''{,}J'''$ & $x11x$ & $J'''$: $g{,}r{,}J'{,}J''{,}J'''$\\
Z$x1100x$(0) & $J'$ & $x00x$ & $J'$: $b$ & $x10x$ & $J'$: $b$ & $x00\bar{x}$ & $J'$: $b$ & $x10\bar{x}$ & $J'$: $b$\\
Z$x1100x$(1) & $J''$ & $x00x$ & $J''$: $J_1{,}J'{,}J''$ & $x01x$ & $J''$: $b{,}J'{,}J''$ & $x01x$ & $J''$: $J_1{,}J''$ & $x00x$ & $J''$: $b{,}J''$\\
Z$x1100\bar{x}$(0) & $J'''$ & $x00\bar{x}$ & $J'''$: $b{,}J''$ & $x10\bar{x}$ & $J'''$: $b{,}J''$ & $x00x$ & $J'''$: $b{,}J''$ & $x10x$ & $J'''$: $b{,}J''$\\
Z$x1100\bar{x}$(1) & $J''$ & $x00\bar{x}$ & $J''$: $J_1{,}J''$ & $x01\bar{x}$ & $J''$: $b{,}J''$ & $x01\bar{x}$ & $J''$: $J_1{,}J'{,}J''$ & $x00\bar{x}$ & $J''$: $b{,}J'{,}J''$\\
Z$x1101x$(0) & $J'$ & $x01x$ & $J'$: $b$ & $x11x$ & $J'$: $b$ & $x01\bar{x}$ & $J'$: $b$ & $x11\bar{x}$ & $J'$: $b$\\
Z$x1101x$(1) & $J'''$ & $x01x$ & $J'''$: $J_1{,}J'{,}J''{,}J'''$ & $x00x$ & $J'''$: $b{,}J'{,}J''{,}J'''$ & $x00x$ & $J'''$: $J_1{,}J''{,}J'''$ & $x01x$ & $J'''$: $b{,}J''{,}J'''$\\
Z$x1101\bar{x}$(0) & $J'''$ & $x01\bar{x}$ & $J'''$: $b{,}J''$ & $x11\bar{x}$ & $J'''$: $b{,}J''$ & $x01x$ & $J'''$: $b{,}J''$ & $x11x$ & $J'''$: $b{,}J''$\\
Z$x1101\bar{x}$(1) & $J'''$ & $x01\bar{x}$ & $J'''$: $J_1{,}J''{,}J'''$ & $x00\bar{x}$ & $J'''$: $b{,}J''{,}J'''$ & $x00\bar{x}$ & $J'''$: $J_1{,}J'{,}J''{,}J'''$ & $x01\bar{x}$ & $J'''$: $b{,}J'{,}J''{,}J'''$\\
Z$x1110x$(0) & $J'''$ & $x10x$ & $J'''$: $b{,}J''$ & $x00x$ & $J'''$: $b{,}J''$ & $x10\bar{x}$ & $J'''$: $b{,}J''$ & $x00\bar{x}$ & $J'''$: $b{,}J''$\\
Z$x1110x$(1) & $J''$ & $x10x$ & $J''$: $b{,}J'{,}J''$ & $x11x$ & $J''$: $J_1{,}J'{,}J''$ & $x11x$ & $J''$: $b{,}J''$ & $x10x$ & $J''$: $J_1{,}J''$\\
Z$x1110\bar{x}$(1) & $J''$ & $x10\bar{x}$ & $J''$: $b{,}J''$ & $x11\bar{x}$ & $J''$: $J_1{,}J''$ & $x11\bar{x}$ & $J''$: $b{,}J'{,}J''$ & $x10\bar{x}$ & $J''$: $J_1{,}J'{,}J''$\\
Z$x1111x$(0) & $J'''$ & $x11x$ & $J'''$: $b{,}J''$ & $x01x$ & $J'''$: $b{,}J''$ & $x11\bar{x}$ & $J'''$: $b{,}J''$ & $x01\bar{x}$ & $J'''$: $b{,}J''$\\
Z$x1111x$(1) & $J'''$ & $x11x$ & $J'''$: $b{,}J'{,}J''{,}J'''$ & $x10x$ & $J'''$: $J_1{,}J'{,}J''{,}J'''$ & $x10x$ & $J'''$: $b{,}J''{,}J'''$ & $x11x$ & $J'''$: $J_1{,}J''{,}J'''$\\
Z$x1111\bar{x}$(1) & $J'''$ & $x11\bar{x}$ & $J'''$: $b{,}J''{,}J'''$ & $x10\bar{x}$ & $J'''$: $J_1{,}J''{,}J'''$ & $x10\bar{x}$ & $J'''$: $b{,}J'{,}J''{,}J'''$ & $x11\bar{x}$ & $J'''$: $J_1{,}J'{,}J''{,}J'''$\\
\end{tabular}
\end{ruledtabular}
\end{table*}

\clearpage
\section{Parameters of the Ans\"atze in the Figures of the Main Text}
\label{sm:app:params}
\label{sm:app:chiral}

The class-B ans\"atze of Fig.~5 of the main text are specified by their singlet links on the reference bonds of Table~\ref{tab:bonds} of the main text, $u_{ij}=\ii\chi^0\tau^0+\Delta^1\tau^1+\Delta^2\tau^2+\chi^3\tau^3$, written as $(\chi^0,\Delta^1,\Delta^2,\chi^3)$ for $\Z_2$ states, $(\chi^0,\chi^3)$ for $U(1)$ states, and $\ii\chi\tau^0$ for $SU(2)$ states; the links on all other bonds follow from the PSG. With $x$, the signed ratio of the $J_r$ and $J_b$ amplitudes, and $\lambda$ the coordinates of the maps:
\begin{itemize}\sloppy
\item $\Z_2$ PSG 0100 [Fig.~5(a) of the main text]: $u_b=(-1,0,0,0)$, $u_g=(0,0.088,-0.405,-0.910)$, $u_r=x\,(0,-0.864,-0.501,0.055)$, $u_{J'}=\lambda\,(0,0.520,-0.755,-0.399)$; the state of panel (d) has $(x,\lambda)=(0.25,1)$.
\item $U(1)$ PSG 0001, sector $(1,0)$ [Fig.~5(b) of the main text]: $u_b=(0.004,1.000)$, $u_g=(1,0)$, $u_r=x\,(1,0)$, $u_{J'}=\lambda\,(-0.417,-0.909)$; panel (e) has $(x,\lambda)=(-0.5,1)$.
\item $SU(2)$ PSG 0001 [Fig.~5(c) of the main text]: $\chi_b=1$, $\chi_g=2.4$, $\chi_r=x$, $\chi_{J'}=-\lambda$; panel (f) has $(x,\lambda)=(1,2.25)$.
\end{itemize}
For the $\Z_2$ states the on-site multipliers $a_i=a^k\tau^k$ are fixed by maximizing the mean-field ground-state energy over the allowed on-site space, which imposes $\langle\psi_i^\dagger\tau^k\psi_i\rangle=0$. The $SU(2)$ states have no multipliers, and their half-filled spectrum has one fermion per site. For the $U(1)$ PSG of panel (b), with $w_{C_6}=1$, only a staggered on-site $a^3$ is allowed (Sec.~VIII of the main text); the map is computed without it. At the state of panel (e) the local occupation then deviates from one by $\pm6\%$; imposing $\langle\psi_i^\dagger\tau^3\psi_i\rangle=0$ with the staggered $a^3$ reduces the gap from $0.115W$ to $0.107W$ and leaves $C$ unchanged, as it must as long as the gap stays open. Chern numbers are computed with the Fukui--Hatsugai--Suzuki method on a $24\times24$ momentum grid and agree with the Wilson-loop winding on a $30\times30$ grid at every gapped point of the maps, which are computed on $41\times41$ [(a), (b)] and $41\times46$ (c) grids. The ribbons of panels (d)--(f) contain 168 sites across, bounded by straight cuts parallel to $\bm{T}_1$, with sites of coordination less than two removed; the edge-mode counts follow from the spectral flow of the occupied weight in the lower half of the ribbon and are the same for edges parallel to $\bm{T}_1+\bm{T}_2$. The thermal Hall integrals use a $96\times96$ grid.

The four $\Z_2$ states of Fig.~6 of the main text have singlet amplitudes up to $J'$, written as $(\chi^0,\Delta^1,\Delta^2,\chi^3)$ on the reference bonds: $u_b=(0,0.428,0.600,-0.676)$, $u_g=(0,-0.703,0.102,-0.704)$, $u_r=1.5\times(0,0.576,0.741,-0.344)$, and $u_{J'}=0.5\times(0,-0.258,0.874,0.411)$ in (a) and $u_{J'}=0.5\times(1,0,0,0)$ in (c) and (d); (b) is the $\nu=-4$ state of Fig.~5(d) of the main text. In each state the on-site multipliers are fixed as above, which imposes one fermion per site and no on-site pairing on average. All four states are gapped, with spinon gap $\Delta$ (the indirect gap of the spinon spectrum at zero energy) equal to $0.117W$ (A), $0.016W$ (B), $0.118W$ (C), and $0.061W$ (D), where $W$ is the spinon bandwidth; the Chern number is $\nu=-4$ in (b) and $\nu=0$ in (a), (c), and (d). The chirality is evaluated with Wick's theorem on a $24\times24$ momentum grid. In Fig.~6(e) of the main text, $S(\bm{q},\omega)=3S^{zz}(\bm{q},\omega)$ is the two-spinon continuum including all normal and anomalous coherence factors, computed on a $48\times48$ momentum grid with Lorentzian broadening $0.25\Delta$; it starts at $\omega=\Delta$.

The Kitaev ribbons of Figs.~18(a) and 18(b) of the main text have the geometry of the class-B ribbons; the thermal Hall integral of Fig.~18(c) of the main text is evaluated on a $60\times60$ grid.

The spinon quantum spin Hall state of Fig.~15 of the main text belongs to U001-000000. Its singlet links on the reference bonds of $(J_b,J_g,J_r,J',J'',J''')$ are $u^{(0)}=\chi^3\tau^3$ with $\chi^3=(-0.4427,-0.0217,-0.2904,0.0993,0.4508,-0.1945)$, and its triplet links are $u^{(z)}=\ii t^{z3}\tau^3$ with $t^{z3}=(0.1120,0.0455,-0.0153,0,0.0652,-0.1287)$; all other components vanish, and the uniform chemical potential is fixed by half filling. The $C_s=2$ state of Fig.~15(d) of the main text belongs to U001-000100 and has $\chi^3=(0.0795,0.1004,-0.2096,-0.0814)$ on $(J_b,J',J'',J''')$ and $t^{z3}=(-0.1530,0.1111,-0.0117)$ on $(J_b,J'',J''')$; its singlet parent has a quadratic band touching at $\Gamma$, and its gap is $0.407$. The trivial reference state is a nearest-neighbor-only ansatz with $\chi^3=1$ and $t^{z3}=0.05$ and a gap of $0.394$; its IGG is larger than $U(1)$. Wilson loops use $120\times121$ momenta, Chern numbers a $60\times60$ grid, and the spin thermal Hall conductivity the Kubo Berry curvature on a $72\times72$ grid. The ribbon of Fig.~15(c) of the main text has 288 sites across; the Zeeman term is $B\sum_if^\dagger_i\sigma^xf_i$.

The spinon quantum spin Hall state of U001-000001 in Sec.~IX\,B of the main text has singlet links $\chi^3\tau^3$ with $\chi^3=(-0.10,1.60,0.46,-0.53,-0.38)$ on the reference bonds of $(J_b,J_g,J_r,J'',J''')$, no singlet link on $J'$, where this class forbids it, and a single triplet link, $u^{(z)}=\ii t^{z3}\tau^3$ with $t^{z3}=0.6$ on $J'$. The uniform chemical potential is fixed by half filling, and $\langle n_i\rangle=1$ holds on every site. The singlet parent, $t^{z3}=0$, is gapped, with gap $1.149$ at bandwidth $11.24$ and $\xi=+1$. As $t^{z3}$ grows, the spin-up gap at $\Gamma$ closes at $t^{z3}=0.348$ and reopens with $\xi=-1$. At $t^{z3}=0.6$ the gap is $0.979$ at bandwidth $13.02$, $C_\uparrow=-C_\downarrow=1$ on $36\times36$, $60\times60$, and $96\times96$ grids, and the $\Z_2$ invariant from the Wannier-center flow on $81\times200$ momenta and from the Fukui--Hatsugai method on a $120\times120$ grid is $1$. Ribbons 24 unit cells wide, with edges parallel to $\bm{T}_1$ or to $\bm{T}_1+\bm{T}_2$, have one Fermi-level crossing per edge for $0<k_\parallel<\pi$ at each of five chemical potentials spread over the central $60\%$ of the gap, and their in-gap edge levels at $k_\parallel=0$ are Kramers degenerate.

The $N_f=48$ phase of Fig.~11 of the main text belongs to U011-100101, whose allowed singlet links are $\chi\tau^3$ on the reference bonds of $J_g$, $J_r$, $J'$, $J''$, and $J'''$ ($J_b$ is not allowed), with the staggered multiplier fixed by $\langle n_i\rangle=1$ and half filling. The multiplier is $a^3\tau^3$ on sublattice $A$ (odd $\mu$) and $-a^3\tau^3$ on sublattice $B$, so that, with the hopping $-\chi$ for $u=\chi\tau^3$, it shifts the spinon energies on $A$ by $-a^3$ and on $B$ by $+a^3$. The mirrors, which exchange $A$ and $B$ and carry a particle--hole gauge transformation, map $\langle n_i\rangle-1$ on $A$ to $-(\langle n_i\rangle-1)$ on $B$, so that the averaged condition $\sum_i\epsilon_i(\langle n_i\rangle-1)=0$ is equivalent to the local constraint; at both points below all 24 site occupations equal one to within $5\times10^{-5}$ on a $72\times72$ mesh. Panel (a) uses $\chi_{23}/\chi_{28}=-0.247$ and $\chi_{27}/\chi_{28}=0.692$; panel (b) uses $\chi_{23}/\chi_{24}=-0.023$ and $\chi_{26}/\chi_{24}=0.530$. The survey point (star) is $(\chi_{23},\chi_{27},\chi_{28},\chi_{26},\chi_{24})=(-0.209,0.587,0.848,0.757,1.476)$ with $a^3=0.430$, and the centered point (diamond) is $(-0.038,0.615,0.284,0.863,1.629)$ with $a^3=0.101$; at the latter the two node orbits lie at $|\bm{k}-\bm{M}/2|=0.137|\bm{M}|$ and $0.288|\bm{M}|$, and at the survey point at $0.140|\bm{M}|$ and $0.188|\bm{M}|$.

The second $N_f=48$ phase belongs to U000-011001, whose allowed singlet links on the reference bonds are $\chi_{23}\tau^3$ ($J_g$), $\chi_{27}\tau^3$ ($J_r$), $\ii\chi_{28}\tau^0$ ($J'$), and $\ii\chi^0_{24}\tau^0+\chi^3_{24}\tau^3$ ($J'''$); $J_b$, $J''$, and on-site multipliers are not allowed, and the sublattice symmetry gives $\langle n_i\rangle=1$ on every site at $E_F=0$. The point quoted in Sec.~VIII\,D of the main text is $(\chi_{23},\chi_{27},\chi_{28},\chi^0_{24},\chi^3_{24})=(-0.138,1.197,1.459,0.559,1.158)$; its 24 nodes per spin lie at $|\bm{k}|=0.399|\bm{M}|$ and $0.553|\bm{M}|$ on the $\Gamma$--$M$ lines and at $|\bm{k}|=0.590|\bm{M}|$ on a generic orbit. The $N_f=36$ point of SU011101 is $(\chi_{23},\chi_{27},\chi_{28},\chi_{24})=(-0.985,-1.066,-0.209,0.824)$, with $u=\ii\chi_o\tau^0$ on the reference bonds.

The $N_f=24$ phase diagrams of Fig.~10 of the main text use singlet amplitudes $\chi_o$ on the reference bonds ($u=\chi_o\tau^3$ for the $U(1)$ classes and $u=\ii\chi_o\tau^0$ for SU111100) and half filling. In (a), U011-100101 with $(\chi_{23},\chi_{27},\chi_{28})=(-0.2396,-0.2513,0.569)$ and the staggered $a^3$ fixed by $\langle n_i\rangle=1$; the star is $(\chi_{26},\chi_{24})=(-0.333,0.333)$, with $a^3=-0.267$. In (b), SU111100 with $(\chi_{23},\chi_{27})=(-0.95,-0.92)$ and star $(\chi_{28},\chi_{24})=(0.94,-0.23)$. In (c), U001-100001 with $(\chi_{12},\chi_{23},\chi_{27},\chi_{28})=(0.6912,0.9889,1.4105,1.1267)$, a uniform chemical potential fixed by half filling, and star $(\chi_{26},\chi_{24})=(1.0878,1.0242)$, at which $E_F=0.0170$. Nodes are counted per spin in the reduced zone of the 24-site cell.

\begingroup
\makeatletter
\let\SM@label\label\let\SM@ref\ref\def\SM@last{LastBibItem}%
\renewcommand\label[1]{\def\SM@arg{#1}\ifx\SM@arg\SM@last\SM@label{SM:LastBibItem}\else\SM@label{#1}\fi}%
\renewcommand\ref[1]{\def\SM@arg{#1}\ifx\SM@arg\SM@last\SM@ref{SM:LastBibItem}\else\SM@ref{#1}\fi}%
\makeatother
\endgroup

\end{document}